\documentstyle[12pt,epsf]{article}

0

\begin{document}
\baselineskip 6mm

%define commands
\newcommand{\MqV}{$\cal M$$_{q}$($\cal V$)}
\def\II{\relax{\rm 1\kern-.35em1}}
\def\IP{\relax{\rm I\kern-.18em P}}
\renewcommand{\theequation}{\thesection.\arabic{equation}}
\csname @addtoreset\endcsname{equation}{section}

\begin{flushright}
hep-th/9711102 \\
IFT-UAM/CSIC 97-4
\end{flushright}
\vglue 1cm

\begin{center}

{}~\vfill

{\large \bf  Fields, Strings and Branes}$^{\dag}$

\vspace{20 mm}

{\bf C\'{e}sar G\'{o}mez and Rafael Hern\'{a}ndez}

\vspace{1cm}

{\em Instituto de Matem\'{a}ticas y F\'{\i}sica Fundamental, CSIC \protect \\
Serrano 123, 28006 Madrid, Spain}
\vspace{1cm}

and
  
\vspace{1cm}

{\em Instituto de F\'{\i}sica Te\'{o}rica, C-XVI, Universidad Aut\'{o}noma de
Madrid \protect \\ Cantoblanco, 28049 Madrid, Spain}

\end{center}

\vspace{6 cm}

\dag  \hspace{2 mm}CIME (Summer International Center of 
Mathematics) lectures given by C.G. at Cetraro, Italy, 1997.

\pagebreak

%%%%%%%%%%%%%%%%%%%%%%%%%%%%%%%%%%%%%%%%%%%%%%%%%%%%%%%%%
%%%%%%%%%%%%%%%%%%%%%%%%%%%%%%%%%%%%%%%%%%%%%%%%%%%%%%%%%

\tableofcontents

\pagebreak

%%%%%%%%%%%%%%%%%%%%%%%%%%%%%%%%%%%%%%%%%%%%%%%%%%%%%%%%%
%%%%%%%%%%%%%%%%%%%%%%%%%%%%%%%%%%%%%%%%%%%%%%%%%%%%%%%%%

\begin{flushright}
What is your aim in phylosophy? \\
To show the fly the way out of the fly-bottle.
  
\vspace{4 mm}
  
{\bf Wittgenstein}. {\em Philosophycal Investigations, $309$}.

\end{flushright}

\vspace{30 mm}

{\large {\bf Introduction}}
  
\vspace{10 mm}

The great challenge of high energy theoretical physics is finding
a consistent theory of quantum gravity. For the time being,
string theory is the best candidate at hand. Many phisicists
think that the solution to quantum gravity will have little, if
any, practical implications in our daily way of doing physics;
others, more optimistic, or simply with a less practical approach
to science, hope that the forthcoming theory of quantum gravity
will provide a new way of thinking of quantum physics.
At present, string theory is an easy target for the criticisms of
pragmatics, as no experimental evidence is yet available; however, but it is
also a rich and deep conceptual construction where new ways of
solving longstanding theoretical problems in quantum field theory
are starting to emerge. Until now, progress in string theory is
mostly ``internal'', a way to evolve very similar to the one
underlying evolution in pure mathematics. This is not
necessarily a symptom of decadence in what is traditionally
considered as an experimeantal science, but maybe the only
possible way to improve physical intuition in the quantum realm.  
  
Until very recently, most of the work in string theory was
restricted to perturbation theory. Different string theories are,
from this perturbative point of view, defined by two dimensional
field theories, satisfying a certain set of constraints such as
conformal and modular invariance. Different orders in the string
perturbative expansion are obtained by working out these two
dimensional conformal field theories on Riemann surfaces of
different genus, and string amplitudes become good measures on
the moduli space of these surfaces. This set of rules constitutes
what we now call the ``world-sheet'' approach to string theory.
From this perturbative point of view, we can think of many
different string theories, as many as two dimensional conformal
field theories, with an appropiate value of the central
extension, which is determined by the generic constraint that
amplitudes should define good measures on the moduli of Riemann
surfaces. Among these conformal field theories, of special
interested are the ones possessing a spacetime interpretation,
which means that we can interpret them as describing the dynamics
of strings moving in a definite target spacetime. Different
string theories will then be defined as different types of
strings moving in the same spacetime. Using this definition,
we find, for instance, four different types of closed superstring
theories (type IIA, type IIB, $E_8 \times E_8$ heterotic and
$SO(32)$ heterotic) and one open superstring. However, this image
of string theory has been enormously modified in the last few
years, due to the clear emergence of duality symmetries. These
symmetries, of two different species, perturbative and non
perturbative, relate through equivalence relations a string
theory on a particular spacetime to a string theory on some
different spacetime. When this equivalence is perturbative, it
can be proved in the genus expansion, which in practice means a
general type of Montonen-Olive duality for the two dimensional
conformal field theory. These duality symmetries are usually
refered to as $T$-duality. A more ambitious type of duality
relation between string theories is known as $S$-duality, where
the equivalence is pretended to be non perturbative, and where a
transformation from strongly to weakly coupled string theory is
involved. Obviously, the first thing needed in order to address
non perturbative duality symmetries is searching for a definition
of string theory beyond perturbation theory, i. e., beyond the
worldsheet approach; it is in this direction where the most
ambitious program of research in string theory is focussing.
  
An important step in this direction comes of course from the
discovery of D-branes. These new objects, which appear as
necessary ingredients for extending $T$-duality to open strings,
are sources for the Ramond fields in string theory, a part of the
string spectrum not coupling, at the worldsheet level, to the
string, and that are therefore not entering the allowed set of
backgrounds used in the definition of the two dimensional
conformal field theory. Thus, adding this backgrounds is already
going beyond the worldsheet point of view and, therefore,
constitutes an open window for the desired non perturbative
definition of string theory.
  
Maybe the simplest way to address the problem of how a non
perturbative definition of string theory will look like is
wondering about the strong coupled behaviour of strings. This
question becomes specially neat if the string theory chosen is 
the closed string of type IIA, where the string
coupling constant can be related to the metric of eleven dimensional
supergravity, so that the strongly coupled string
theory can be understood as a new eleven dimensional theory,
M-theory. When thinking about the relation between D-branes and
M-theory or, more precisely, trying to understand the way
D-branes dynamics should be used in order to understand the
eleven dimensional dynamics describing the strong coupling regime
of string theory, a good answer comes again from the misterious,
for a while, relation between type IIA strings and
eleven dimensional supergravity: the Kaluza-Klein modes in ten
dimensions are the D-$0$brane sources for the Ramond $U(1)$
field. What makes this, superficially ordinary Kaluza-Klein
modes, very special objects is its nature of D-branes. In fact,
D-branes are sources for strings, powerful enough to provide the
whole string spectrum.
  
A very appealing way to think of these D-$0$branes comes recently
under the name of M(atrix) theory. The phylosophical ground for
M(atrix) theory goes back to the holographic principle, based on
black hole bounds on quantum information packing in space. From
this point of view, the hologram of eleven dimensional M-theory
is a ten dimensional theory for the peculiar set of ten
dimensional degrees of freedom in terms of which we can codify
all eleven dimensional physics. M(atrix) theory is the conjecture
that D-$0$brane dynamics, which is a very special type of matrix quantum
mechanics, is the correct hologram of the unknown eleven
dimensional M-theory. We do not know the non perturbative region
of string theory, but it seems we have already its healthy
radiography.
  
\vspace{2 mm}
  
These lectures were originally adressed to mathematics audience.
The content covered along them is of course only a very small
part of the huge amount of material growing around string theory
on these days, and needless to say that it reflects the personal
point of view of the authors. References are certainly not
exhaustive, so that we apologize for this in advance. 
  
Last, but not least, C. G. would like to thank the organizers and
participants of the CIME school for suggestions and interesting
questions, most of them yet unanswered in the text.

\pagebreak

\section{Chapter I}

\subsection{Dirac Monopole.}

Maxwell's equations in the absence of matter,
\begin{eqnarray}
	\nabla \vec{E} = 0, & \: \: \: \: \: \: & \nabla \vec{B} = 0,
				\nonumber \\
	\nabla \times \vec{B} - \frac {\partial \vec{E}}{\partial
t} = 0, & \: \: \: \: \: \: & \nabla \times \vec{E} + \frac
{\partial \vec{B}}{\partial t} = 0,
\label{eq:I1} 
\end{eqnarray}
are invariant under the duality transformation
\begin{eqnarray}
\vec{E} & \longrightarrow & - \vec{B}, \nonumber \\
\vec{B} & \longrightarrow & \vec{E},
\label{eq:I2}
\end{eqnarray}
or, equivalently,
\begin{eqnarray}
F^{\mu \nu} & \longrightarrow &  ^{\ast}F^{\mu
	\nu}, \nonumber \\
^{\ast}F^{\mu \nu} & \longrightarrow & - F^{\mu
	\nu},
\label{eq:I3}
\end{eqnarray}
with $^{\ast}F^{\mu \nu} \equiv \tilde{F}^{\mu \nu}= \frac {1}{2} \epsilon ^{\mu \nu \rho \sigma}
F_{\rho \sigma}$ the Hodge dual of $F^{\mu \nu} =
\partial^{\mu} A^{\nu}- \partial^{\nu} A^{\mu}$. In the pressence
of both electric and magnetic matter, Maxwell's equations become
\begin{eqnarray}
	   \partial_{\nu} F^{\mu \nu} & = & - j^{\mu} \nonumber \\
\partial_{\nu} \, ^{\ast} F^{\mu \nu} & = & - k^{\mu},
\label{eq:I4}
\end{eqnarray}
and (\ref{eq:I2}) must be generalized with a transformation law
for the currents,
\begin{eqnarray}
F^{\mu \nu}  \rightarrow  ^{\ast} F^{\mu \nu} & \> \> &
j^{\mu}  \rightarrow   k^{\mu}  \nonumber \\
^{\ast} F^{\mu \nu}  \rightarrow - F^{\mu \nu} &
\> \> &
k^{\mu}  \rightarrow  - j^{\mu}.
\label{eq:I5}
\end{eqnarray}
  
As is clear from the definition of $F^{\mu \nu}$, the
existence of magnetic sources (monopoles) \cite{D1931} requires dealing with
singular vector potentials. The appropiate mathematical language
for describing these vector potentials is that of fiber bundles \cite{WY}. 
  
To start with, we will consider $U(1)$ bundles on the two sphere
$S^2$. Denoting $H^{\pm}$ the two hemispheres, with $H^+ \cap
H^- = S^1$, the $U(1)$ bundle is defined by
\begin{equation}
g_{\pm} = e^{i\psi_{\pm}}
\label{eq:I6}
\end{equation}
$U(1)$ valued functions on the two hemispheres and such that on
the $S^1$ equator
\begin{equation}
e^{i \psi_{+}}=e^{i n \varphi}e^{i \psi_{-}},
\label{eq:I7}
\end{equation}
with $\varphi$ the equatorial angle, and $n$ some integer number
characterizing the $U(1)$ bundle. Notice that $n$ defines the
winding number of the map
\begin{equation}
e^{i n \varphi} : S^1 \longrightarrow U(1),
\label{eq:I8}
\end{equation}
classified under the first homotopy group
\begin{equation}
\Pi_1(U(1)) \simeq \Pi_1(S^1) \simeq {\bf Z}.
\label{eq:I9}
\end{equation}
  
Using the $U(1)$ valued functions $g_{\pm}$, we can define pure
gauge connections, $A_{\mu}^{\pm}$, on $H^{\pm}$ as follows:
\begin{equation}
A_{\mu}^{\pm} = g_{\pm}^{-1} \partial _{\mu} g_{\pm}.
\label{eq:I10}
\end{equation}
From (\ref{eq:I7}) we easily get, on the equator,
\begin{equation}		
A^{+}=A^{-}+n  \varphi,
\label{eq:I11}
\end{equation}
and, through Stokes theorem, we get
\begin{equation}
\int_{S^{2}} F = 
 \frac {1}{2 \pi}[\int_{H^{+}} dA^{+} + \int_{H^{-}} dA^{-}] =
 \frac {1}{2 \pi} \int_{S^{1}} A^{+} - A^{-} = n
\label{eq:I12} 
\end{equation}
identifying the winding number $n$ with the magnetic charge of
the monopole. 
  
In quantum mechanics, the presence of a magnetic charge implies a
quantization rule for the electric charge. In fact, as we require
that the Schr\"odinger wave function, for an electric field in a
monopole background, be single valued, we get 
\begin{equation}
\hbox {exp} \frac {ie}{2 \pi {\hbar}} \oint_{\Gamma} A = 1,
\label{eq:I13}
\end{equation}
with $\Gamma$ a non contractible loop. In the presence of a
magnetic charge $m \equiv \frac {1}{2\pi} \oint_{\Gamma} A $ we
get Dirac's quantization rule \cite{D1931},
\begin{equation}
em = n h.
\label{eq:I14}
\end{equation}
Notice that the quantization rule (\ref{eq:I14}) is equivalent to
the definition (\ref{eq:I12}) of the magnetic charge as a winding
number or, more precisely, as minus the first Chern class 
of a $U(1)$ principal bundle on $S^2$. In fact, the single valuedness of the Schr\"odinger wave
function is equivalent to condition (\ref{eq:I7}), where we have
required $n$ to be integer for the transition function, in order
to get a manifold. The gauge connection used in (\ref{eq:I12})
was implicitely defined as $eA$, with $A$ standing for the
physical gauge configuration appearing in the Schr\"odinger
equation. From now on, we will use units with $\hbar=1$.
  
The main problem with Dirac monopoles is that they are not part
of the spectrum of standard QED. In order to use the idea of
duality as a dynamical symmetry, we need to search for more general
gauge theories, containing in the spectrum magnetically charged
particles \cite{GO,GNO,MO}.

\subsection{The `t Hooft-Polyakov Monopole.}

Let us consider the Georgi-Glashow model \cite{GG} for $SU(2)$,
\begin{equation}
{\cal L}= - \frac {1}{4}F_{a}^{\mu \nu} F_{a \mu \nu} +
\frac {1}{2} {\cal D}^{\mu} \phi \cdot {\cal D}_{\mu} \phi - V(\phi),
\: \: \: \: a=1,2,3,
\label{eq:I15}
\end{equation}
with the Higgs field in the adjoint representation, 
${\cal D}^{\mu}\phi_{a} \equiv \partial ^{\mu} \phi_a - g \epsilon_{abc}
A_b ^{\mu} \phi_c $ the covariant derivative, and $V(\phi)$ the
Higgs potential,
\begin{equation}
V(\phi)= \frac {1}{4} \lambda (\phi^{2}-a^{2})^{2}.
\label{eq:I16}
\end{equation}
with $\lambda >0$ and $a$ arbitrary constants. 
  
A classical vacuum configuration is given by
\begin{equation}
\phi_{a}=a \delta_{a3}, \: \: \: \: \: \: A_{\mu}^{a}=0.
\label{eq:I17}
\end{equation}
We can now define the vacuum manifold ${\cal V}$ as 
\begin{equation}
{\cal V} = \{ \phi, V(\phi)=0 \},
\label{eq:I18}
\end{equation}
which in this case is a $2$-sphere of radius equal $a$. A necessary
condition for a finite energy configuration is that at infinity,
the Higgs field $\phi$ takes values in the vacuum manifold ${\cal
V}$,
\begin{equation}
\phi: S_{\infty}^2 \longrightarrow {\cal V}
\label{eq:I19}
\end{equation}
and that ${\cal D}_{\mu} \phi |_{S_{\infty}^2}=0$. Maps of
the type (\ref{eq:I19}) are classified by the second homotopy
group, $\Pi_2({\cal V})$, which for the Georgi-Glashow model
(with ${\cal V}=S^2$) is non trivial, and equal to the set of
integer numbers. These maps are characterized by their winding
number,
\begin{equation}
N= \frac {1}{4 \pi a^{3}} \int_{S_{\infty}^2} dS^{i} \frac {1}{2} \epsilon_{ijk}
\phi \cdot ( \partial^{j} \phi \wedge \partial^{k} \phi ).
\label{eq:I20}
\end{equation}
  
Once we impose the finite energy condition ${\cal D}_{\mu} \phi
|_{S_{\infty}^2}=0$, the gauge field at infinity is given by 
\begin{equation}
A^{\mu}= \frac {1}{a^{2}g} \phi \wedge \partial^{\mu} \phi
+ \frac {1}{a} \phi f^{\mu},
\label{eq:I21}
\end{equation}
where $f_{\mu}$ is an arbitrary function. The corresponding stress
tensor is given by
\begin{equation}
F_{a}^{\mu \nu}= \frac {1}{a} \phi_{a} F^{\mu \nu}=
\frac {1}{a} \phi_{a} \, \left( \frac {1}{a^{2}g} \phi \cdot (\partial^{\mu} \phi
\wedge \partial^{\nu} \phi) + \partial^{\mu}f^{\nu}-
\partial^{\nu}f^{\mu} \right),
\label{eq:I22}
\end{equation}
which implies that the magnetic charge
\begin{equation}
m = - \frac {1}{2ga^{3}} \int_{S_{\infty}^2}
\epsilon_{ijk} \phi \cdot (\partial^{j} \phi \wedge \partial^{k}
\phi) d S^{i},
\label{eq:I23}
\end{equation}
for a finite energy configuration is given in terms of the
winding number (\ref{eq:I20}) as \cite{tp}
\begin{equation}
m= - \frac {4 \pi N}{g}.
\label{eq:I24}
\end{equation}
In order to combine (\ref{eq:I24}) with Dirac's quantization rule
we should define the $U(1)$ electric charge. The $U(1)$ photon
field is defined by 
\begin{equation}
A^{\mu} = ( A^{\mu} \cdot \phi ) \frac {1}{a}.
\label{eq:I25}
\end{equation}
Thus, the electric charge of a field of isotopic spin $j$ is
given by
\begin{equation}
e= g \: j.
\label{eq:I26}
\end{equation}
From (\ref{eq:I24}) and (\ref{eq:I26}) we recover, for $j= \frac
{1}{2}$, Dirac's quantization rule.
  
For a generic Higgs model, with gauge group $G$ spontaneously
broken to $H$, the vacuum manifold ${\cal V}$ is given by 
\begin{equation}
{\cal V} = G/H,
\label{eq:I27}
\end{equation}
with
\begin{equation}
\Pi_2(G/H) \simeq \Pi_1(H)_G,
\label{eq:I28}
\end{equation}
where $\Pi_1(H)_G$ is the set of paths in $H$ that can be
contracted to a point in $G$, which again contains Dirac's
condition in the form (\ref{eq:I9}).
  
The mass of the monopole is given by 
\begin{equation}
M = \int d^3 x \frac {1}{2}  [ (E_a^{i})^2+ (B_a^{i})^2+ (D^0
\phi_a)^2+(D^{i} \phi_a)^2] + V(\phi).
\end{equation}
For a static monopole, the mass becomes
\begin{equation}
M = \int d^3 x \frac {1}{2}  [ (B_a^{i})^2+(D^{i} \phi_a)^2] +
V(\phi);
\end{equation}
then, in the Prasad-Sommerfeld \cite{PS} limit $\lambda=0$ (see equation
(\ref{eq:I16})), we get
\begin{equation}
M = \int d^3 x \frac {1}{2}  [ ((B_a^{i} + D^{i} \phi_a)^2 -2
B_a^{i} D^{i} \phi_a],
\end{equation}
which implies the Bogomolny \cite{B} bound $M \geq a m$. The Bogomolny bound
is saturated if $B_a^k = D^k \phi_a$, which are known as the
Bogomolny equations.

\subsection{Instantons.}
  
Let us now consider pure $SU(N)$ Yang-Mills theory,
\begin{equation}
{\cal L} = - \frac {1}{4} F^{a \mu \nu} F^{a}_{\mu \nu}.
\label{eq:I29}
\end{equation}
In euclidean spacetime ${\bf R}^4$, the region at infinity can be
identified with the $3$-sphere $S^3$. A necessary condition for
finite euclidean action of configurations is 
\begin{equation}
F^{a}_{\mu \nu} |_{S^3_{\infty}} = 0,
\label{eq:I30}
\end{equation}
or, equivalently, that the gauge configuration $A^{\mu}$ at
infinity is a pure gauge,
\begin{equation}
A^{\mu} |_{S_{\infty}^3} = g(x)^{-1} \partial ^{\mu} g(x).
\label{eq:I31}
\end{equation}
Hence, finite euclidean action configurations are associated with
maps 
\begin{equation}
g: S^3 \rightarrow SU(N),
\label{eq:I32}
\end{equation}
which are topologically classified in terms of the third
homotopy group,
\begin{equation}
\Pi_3(SU(N)) \simeq {\bf Z}.
\label{eq:I33}
\end{equation}
  
The winding number of the map $g$ defined by (\ref{eq:I32}) is
given by
\begin{equation}
n = \frac {1}{24 \pi^2} \int_{S^3} d^3 x \epsilon_{ijk} 
tr [ g^{-1} \nabla_i g(x) g^{-1} \nabla_j g(x) g^{-1} \nabla_k g(x) ].
\label{eq:I34}
\end{equation}
  
As for the Dirac monopole construction, we can use the map $g$ in order 
to define $SU(N)$ bundles on $S^4$. In this case, $g$ defines the transition function on 
the equator. So, for the simplest group, $SU(2)$, we will 
have different bundles, depending on 
the value of $n$; in particular, for $n=1$, we obtain the 
Hopft bundle
\begin{equation}
S^7 \longrightarrow S^4.
\label{eq:I35}
\end{equation}
Interpreting $S^4$ as the compactification of euclidean space 
${\bf R}^4$, we can define a gauge configuration on $S^4$ such 
that on the equator, which now has the topology of 
$S^3$, we have
\begin{equation}
A_{\mu}^+ = g A_{\mu}^- g^{-1} + g^{-1} \partial_{\mu} g,
\label{eq:I36}
\end{equation}
with $A^+$ and $A^-$ the gauge configurations on the two hemispheres. Using 
now the relation
\begin{equation}
\hbox {tr} (F_{\mu \nu} \tilde{F}^{\mu \nu}) = 
d \hbox {tr} ( F \wedge A - \frac {1}{3} A \wedge A \wedge A),
\label{eq:I37}  
\end{equation}
we get
\begin{equation}
- \frac {1}{8 \pi^2} \int_{S^4} \hbox {tr} (F_{\mu \nu}
\tilde{F}^{\mu \nu}) = \frac {1}{24 \pi^2} \int_{S^3}
\epsilon_{ijk} \hbox {tr} [g^{-1} \partial_i g g^{-1} \partial_j
g g^{-1} \partial_k g] = n,
\label{eq:I38}
\end{equation}
which is the generalization to $S^4$ of the relation we have
derived above between the magnetic charge of the monopole and the
winding number of the transition function defining the $U(1)$
bundle on $S^2$. The topological charge defined by
(\ref{eq:I38}) is a bound for the total euclidean action. In
fact,
\begin{equation}
\frac {1}{4} \int F^{a \mu \nu} F^{a}_{\mu \nu} \equiv \frac
{1}{2} \int \hbox {tr} (F^{\mu \nu} F_{\mu \nu}) \geq \left|
\frac {1}{2} \int \hbox {tr} (F^{\mu \nu} \tilde{F}_{\mu \nu}) \right|.
\label{eq:I39}
\end{equation}
The instanton configuration will be defined by the gauge field
saturating the bound (\ref{eq:I39}),
\begin{equation}
F_{\mu \nu} = \tilde{F}_{\mu \nu},
\label{eq:I40}
\end{equation}
and with topological charge equal one. Bianchi identity, $DF=0$,
together with  the field equations, implies $D \tilde{F}=0$; in fact, the self duality condition (\ref{eq:I40}) 
can be related to the Bogomolny equation. If we start with euclidean Yang-Mills, and reduce 
dimensionally to three dimensions through the definition $A_4 \equiv \phi$, we 
get the three dimensional Yang-Mills-Higgs lagrangian. Then, the self duality relation 
(\ref{eq:I40}) becomes the Bogomolny equation. 
  
A solution to (\ref{eq:I40}) for $SU(2)$ was discovered by Belavin
et al \cite{BPST}. Including the explicit dependence on the bare coupling
constant $g$,
\begin{equation}
F_{\mu \nu} = \partial_{\mu} A_{\nu} - \partial_{\nu} A_{\mu} +
g[A_{\mu}, A_{\nu}],
\label{eq:I41}
\end{equation}
the BPST solution to (\ref{eq:I40}) is given by
\begin{eqnarray}
A_{\mu}^{a} & = &  - \frac {2i}{g} \frac {\eta_{a \mu \nu}
x^{\nu}}{x^2+ \rho^2}, \nonumber \\
F_{\mu \nu}^{a} & = & \frac {4i}{g} \frac {\eta_{a \mu \nu}
\rho^2}{(x^2+\rho^2)^2},
\label{eq:I43}
\end{eqnarray}
with $\eta_{a \mu \nu}$ satisfying $\eta_{a \mu \nu}=\eta_{a
ij}=\epsilon_{aij}$, $\eta_{ai0}=\delta_{ai}$, $\eta_{a\mu \nu}=-
\eta_{a \mu \nu}$, and $\bar{\eta}_{a \mu \nu} =
(-1)^{\delta_{\mu 0}+\delta_{\nu 0}} \eta_{a \mu \nu}$, where
$a,i,j$ take values $1,2,3$.
  
The value of the action for this configuration is 
\begin{equation}
S= \frac {8 \pi^2}{g^2},
\label{eq:I44}
\end{equation}
with Pontryagin number
\begin{equation}
\frac {g^2}{32 \pi^2} \int F_{\mu \nu}^{a} \tilde{F}^{a \mu \nu}
d^4 x =1.
\label{eq:I45}
\end{equation}
  
Notice that the instanton solution (\ref{eq:I43}) depends on a
free parameter $\rho$, that can be interpreted as the classical
size of the configuration. In particular, we can consider the
gauge zero modes of the instanton solution, i. e., small self
dual fluctuations around the instanton solution. From
(\ref{eq:I43}), it is clear that the action is invariant under changes of
the size $\rho$, and under translations $x^{\mu} \rightarrow
x^{\mu} + a^{\mu}$. This means that we will have five independent
gauge zero modes. The number of gauge zero modes is called, in
the mathematical literature, the dimension of the moduli space of
self dual solutions. This number can be computed \cite{tHpseudo,JR,At} using index
theorems \cite{AHS}; the result for $SU(N)$ instantons on $S^4$ is 
\begin{equation}
\hbox {dim Instanton Moduli} = 4 n k - n^2+1,
\label{eq:I46}
\end{equation}
with $k$ the Pontryagin number of the instanton\footnote{$k$ must
satisfy the irreducibility condition $k \geq \frac {n}{2}$. This condition 
must hold if we require the gauge configuration to be irreducible, i. e., that 
the connection can not be obtained by embedding the connection of a smaller group.}. For
$k=1$ and $n=2$ we recover the five zero modes corresponding to
translations and dilatations of the solution (\ref{eq:I43})\footnote{Observe that the total
number of gauge zero modes is $4$, and that $n^2-1$ are simply
gauge rotations of the instanton configuration.}. The
generalization of equation (\ref{eq:I46}) to instantons on a
manifold ${\cal M}$ is
\begin{equation}
\hbox {dim} = 4 n k - \frac {1}{2} (N^2-1) [\chi- \tau],
\label{eq:I47}
\end{equation}
with $\chi$ and $\tau$ the Euler number and the signature of the
manifold ${\cal M}$.
  
In order to get a clear physical interpretation of instantons, it
is convenient to work in the $A^0=0$ temporal gauge \cite{cl,jackiw,CDG}. If we
compactify ${\bf R}^3$ to $S^3$ by impossing the boundary
condition
\begin{equation}
A_i(\vec{r})|_{|\vec{r}| \rightarrow \infty} \rightarrow 0,
\label{eq:I48}
\end{equation}
the vacuum configurations in this gauge are pure gauge
configurations, $A_{\mu} = g^{-1} \partial_{\mu} g$, with $g$ a
map from $S^3$ into the gauge group $SU(N)$. We can now define
different vacuum states $|n>$, characterized by the winding
number of the corresponding map $g$. In the temporal gauge, an
instanton configuration of Pontryagin number equal one satisfies
the following boundary conditions:
\begin{eqnarray}
A_i (t = - \infty) & = & 0, \nonumber \\
A_i (t = + \infty) & = & g_1^{-1} \partial_i g_1,
\label{eq:I49}
\end{eqnarray}
with $g_1$ a map from $S^3$ into $SU(N)$, of winding number equal
one. We can now interpret the instanton configuration
(\ref{eq:I49}) as defining a tunnelling process between the $|0>$
and $|1>$ vacua. 
  
Moreover, the vacuum states $|n>$ are not invariant under gauge
transformations with non vanishing winding number. A vacuum state
invariant under all gauge transformations would be defined by the
coherent state
\begin{equation}
|\theta> = \sum_{n}e^{in \theta}|n>,
\label{eq:I50}
\end{equation}
with $\theta$ a free parameter taking values in the interval
$[0,2\pi]$. Under gauge transformations of
winding number $m$, the vacuum states $|n>$ transform as
\begin{equation}
{\cal U}(g_m) |n> = | n+m>, 
\end{equation}
and therefore the $\theta$-vacua will transform as 
\begin{equation}
{\cal U}(g_m) |\theta> = e^{im \theta }|\theta>, 
\end{equation}
which means invariance in the projective sense, i. e., on the
Hilbert space of rays. 
  
The generating functional now becomes
\begin{equation}
<\theta|\theta> = \sum_n <0 |n>e^{in \theta} = \int dA \exp -
\left( i \int {\cal L}(A) \right),
\label{eq:I51}
\end{equation}
with the Yang-Mills lagrangian
\begin{equation}
{\cal L} = - \frac {1}{4} F^{a \mu \nu} F^{a}_{\mu \nu} + \frac
{\theta g^2}{32 \pi^2} F^{a \mu \nu} \tilde{F}^{a}_{ \mu \nu}.
\label{eq:I52}
\end{equation}
The $\theta$-topological term in (\ref{eq:I52}) breaks
explicitely the CP invariance of the lagrangian. Notice
that if we consider the euclidean functional integral
\begin{equation}
\int dA \exp - \left( \int - \frac {1}{4}F^{a \mu \nu}F^{a}_{\mu \nu} + \frac {i \theta
g^2}{32 \pi^2} F \tilde{F} \right) d^4x, 
\end{equation}
the instanton euclidean action becomes
\begin{equation}
S= \frac {8 \pi^2}{g^2} + i \theta. 
\end{equation}

\subsection{Dyon Effect.}

Let us now add the topological $\theta$-term of (\ref{eq:I52}) to
the Georgi-Glashow model (\ref{eq:I15}). At this level, we are
simply considering the $\theta$-angle as an extra coupling
constant, multiplying the topological density $F \tilde{F}$. In
order to define the $U(1)$ electric charge, we can simply apply
Noether's theorem for a gauge transformation in the unbroken
$U(1)$ direction \cite{Wdyon}. An infinitesimal gauge transformation in the
$\phi$ direction would be defined by
\begin{eqnarray}
\delta A_{\mu} & = & \frac {1}{ag} {\cal D}_{\mu} \phi, \nonumber \\
\delta \phi    & = & 0.
\label{eq:I53}
\end{eqnarray}
The corresponding Noether charge,
\begin{equation}
N = \frac {\delta {\cal L}}{\delta \partial_0 A} \cdot \delta A +
\frac {\delta {\cal L}}{\delta \partial _0 \phi} \cdot \phi,
\label{eq:I54}
\end{equation}
will be given, after the $\theta$-term is included, by 
\begin{equation}
N = \frac {1}{ag} \int d^3x \partial_i (\phi \cdot F_{0i}) +
\frac {\theta g}{8 \pi^2 a} \int d^3x \partial_i ( \phi \frac
{1}{2} \epsilon_{ijk} F_{jk})
\label{eq:I55}
\end{equation}
or, in terms of the electric charge, as
\begin{equation}
N = \frac {e}{g} + \frac {\theta g}{8 \pi^2}m.
\label{eq:I56}
\end{equation}
  
Notice from (\ref{eq:I55}) that the $\theta$-term only
contributes to $N$ in the background of the monopole field. If we
now require invariance under a $U(1)$ rotation, we get
\begin{equation}
e^{2 \pi i N} = e^{2 \pi i \left( \frac {e}{g} + \frac {\theta
g}{8 \pi^2}m \right)} = 1,
\label{eq:I57}
\end{equation}
and the electric charge becomes equal to \cite{Wdyon}
\begin{equation}
e = ng - \frac {\theta g^2}{8 \pi^2}m,
\label{eq:I58}
\end{equation}
which implies that a magnetic monopole of charge $m$ becomes a
dyon with electric charge $- \frac {\theta g^2}{8 \pi^2}m$. 
  
We can reach the same result, (\ref{eq:I57}), without incuding a
$\theta$ term in the lagrangian, if we require, for the monopole
state,
\begin{equation}
e^{2 \pi i N} |m> = e^{i\theta} |m>,
\label{eq:I59}
\end{equation}
for $N=\frac {e}{g}$. Equation (\ref{eq:I59}) implies that the
monopole state transforms under $e^{2 \pi i N}$ as the $\theta$
vacua with respect to gauge transformations of non vanishing
winding number. However, $e^{2 \pi i N}$ can be continously
connected with the identity which, in physical terms, means that
the induced electric charge of the monopole is independent of instantons, and is
not suppresed by a tunnelling factor of the order of $\exp -
\frac {8 \pi^2}{g^2}$ \cite{Wdyon,Rubakov}.

\subsection{Yang-Mills Theory on $T^4$.}

We will consider now $SU(N)$ pure Yang-Mills on a $4$-box \cite{tHtw}, with
sides of length $a_0,a_1,a_2,a_3$. Let us impose periodic boundary
conditions for gauge invariant quantities,
\begin{eqnarray}
A^{\mu} (x_0+a_0,x_1,x_2,x_3) & = & \Omega_0
A^{\mu}(x_0,x_1,x_2,x_3), \nonumber \\
A^{\mu} (x_0,x_1+a_1,x_2,x_3) & = & \Omega_1
A^{\mu}(x_0,x_1,x_2,x_3), \nonumber \\
A^{\mu} (x_0,x_1,x_2+a_2,x_3) & = & \Omega_2
A^{\mu}(x_0,x_1,x_2,x_3), \nonumber \\
A^{\mu} (x_0,x_1,x_2,x_3+a_3) & = & \Omega_3
A^{\mu}(x_0,x_1,x_2,x_3),
\label{eq:I60}
\end{eqnarray}
where 
\begin{equation}
\Omega_{\rho} A^{\mu} \equiv \Omega_{\rho} A^{\mu}
\Omega^{-1}_{\rho} + \Omega^{-1}_{\rho}
\partial ^{\mu} \Omega_{\rho}.
\label{eq:I61}
\end{equation}
  
As the gauge field transforms in the adjoint
representation, we can allow the existence of ${\bf Z}(N)$
twists,
\begin{equation}
\Omega_{\mu} \Omega_{\nu} = \Omega_{\nu} \Omega_{\mu} e^{2 \pi i
n_{\mu \nu}/N},
\label{eq:I62}
\end{equation}
and therefore we can characterize different configurations in
$T^4$ by the topological numbers $n_{\mu \nu}$. Three of
these numbers, $n_{12}, n_{13}$ and $n_{23}$, can be interpreted
as magnetic fluxes in the $3$, $2$ and $1$ directions,
respectively. In order to characterize these magnetic fluxes, we
introduce the numbers 
\begin{equation}
m_i = \epsilon_{ijk} n_{jk},
\label{eq:I63}
\end{equation}
These magnetic fluxes carry ${\bf Z}(N)$ charge, and their
topological stability is due to the fact that
\begin{equation}
\Pi_1(SU(N)/{\bf Z}(N)) \simeq {\bf Z}(N).
\label{eq:I64}
\end{equation}
  
In order to characterize the physical Hilbert space of the
theory, let us again work in the temporal gauge $A^0=0$. For the
three dimensional box $T^3$, we impose twisted boundary
conditions, corresponding to magnetic flow
$\vec{m}=(m_1,m_2,m_3)$. The residual gauge symmetry 
is defined by the set of gauge transformations preserving these
boundary conditions. We may distinguish the following different
types of gauge transformations:

\begin{itemize}
	\item[{i)}] Periodic gauge transformations, which as
usual are characterized by their winding number in $\Pi_3(SU(N))
\simeq {\bf Z}$.
	\item[{ii)}] Gauge transformations periodic, up to
elements in the center:
\begin{eqnarray}
\Omega(x_1+a_1,x_2,x_3) & = & \Omega(x_1,x_2,x_3) e^{2 \pi i
k_1/N}, \nonumber \\
\Omega(x_1,x_2+a_2,x_3) & = & \Omega(x_1,x_2,x_3) e^{2 \pi i
k_2/N}, \nonumber \\
\Omega(x_1,x_2,x_3+a_3) & = & \Omega(x_1,x_2,x_3) e^{2 \pi i
k_3/N}.
\label{eq:I65}
\end{eqnarray}
These transformations are characterized by the vector
\begin{equation}
\vec{k} = (k_1,k_2,k_3),
\label{eq:I66}
\end{equation}
and will be denoted by $\Omega_{\vec{k}}(\vec{x})$. Among this
type of transformations we can extract an extra classification:
		\begin{itemize}
		\item[{ii-1)}] Those such that
$(\Omega_{\vec{k}}(\vec{x}))^N$ is periodic, with vanishing
Pontryagin number.
		\item[{ii-2)}] Those such that
$(\Omega_{\vec{k}}(\vec{x}))^N$ is periodic, with non vanishing
Pontryagin number.
		\end{itemize}
\end{itemize}
  
In the temporal gauge, we can represent the
transformations in ii-2) in terms of unitary operators. Let $|\Psi>$ be a
state in the Hilbert space ${\cal H}(\vec{m})$; then, we get
\begin{equation}
\Omega_{\vec{k}}(\vec{x}) |\Psi> = e^{2 \pi i \frac {\vec{e}
\cdot \vec{k}}{N}} e^{ i \theta \frac {\vec{k}
\cdot \vec{m}}{N}} |\Psi>,
\label{eq:I67}
\end{equation}
where $\vec{e}$ and $\theta$ are free parameters. Notice that the second term in (\ref{eq:I67}) 
is equivalent, for ${\bf Z}(N)$ magnetic vortices, to the Witten dyon effect described in the 
previous section. In fact, we can write (\ref{eq:I67}) in terms of 
an effective $\vec{e}_{eff}$,
\begin{equation}
\vec{e}_{eff} = \vec{e} + \frac {\theta \vec{m}}{2 \pi}.
\label{nue}
\end{equation}
Moreover, as $\theta \rightarrow \theta + 2 \pi$, we change
$\vec{e}_{eff} \rightarrow \vec{e}_{eff} + \vec{m}$. 
On the other hand, the Pontryagin number of a
gauge field configuration with twisted boundary conditions,
determined by a set $n_{\mu \nu}$, is given by \cite{tHcmp}
\begin{equation}
\frac {g^2}{16 \pi^2} \int \hbox{tr} (F^{\mu \nu} \tilde{F}_{\mu
\nu}) d^4 x= k- \frac {n}{N}, 
\end{equation}
where \( n \equiv \frac {1}{4} n_{\mu \nu} n_{\mu \nu} \).
A simple way to understand the origin of the fractional piece in
the above expression is noticing that, for instance, a twist
$n_{12}$ corresponds to magnetic flux in the $3$-direction, with
value $\frac {2 \pi n_{12}}{N}$, which can be formally described
by $F_{12} \sim \frac {2 \pi n_{12}}{Na_1 a_2}$, and a twist
$n_{03}$, which corresponds to an electric field in the
$3$-direction, is described by $F_{03} \sim \frac {2 \pi
n_{03}}{Na_0a_3}$. Using now the integral representation of the
Pontryagin number we easily get the fractional piece, with the
right dependence on the twist coefficients (see section
\ref{sec:52}). Moreover, $(\Omega_{\vec{k}}(\vec{x}))^N$ acting
on the state $|\Psi>$ produces
\begin{equation}
(\Omega_{\vec{k}}(\vec{x}))^N |\Psi> = e^{i \theta \vec{k} \cdot
\vec{m}} |\Psi>,
\label{eq:I68}
\end{equation}
which means that $\vec{k} \cdot \vec{m}$ is the Pontryagin number
of the periodic gauge configuration
$(\Omega_{\vec{k}}(\vec{x}))^N$. For a generic gauge
configuration with Pontryagin number $n$ we will get, as usual,
\begin{equation}
\Omega(\vec{x};n) |\Psi> = e^{in \theta } |\Psi>.
\label{eq:I69}
\end{equation}
  
Using (\ref{eq:I65}), it is easy to see that the $\vec{k}$'s
characterizing the residual gauge transformations are nothing
else but the $n_{0i}$ twists. The physical interpretation of the
parameter $\vec{e}$ introduced in (\ref{eq:I67}), in the very
same way as the $\theta$-term, is that of an electric flux. In fact,
we can define the Wilson loop
\begin{equation}
A(C) = \frac {1}{N} \hbox {tr} \exp \int_C i g A(\xi) d \xi,
\label{eq:I70}
\end{equation}
with $C$ a path in the $3$-direction. Under
$\Omega_{\vec{k}}(\vec{x})$, $A(C)$ transforms as 
\begin{equation}
A(C) \rightarrow e^{2 \pi i k_3/N} A(C);
\label{eq:I71}
\end{equation}
therefore, we get
\begin{equation}
\Omega_{\vec{k}}(\vec{x}) A(C) |\Psi> = e^{2 \pi i \frac {\vec{e}
\cdot {k}}{N}} A(C) |\Psi>,
\label{eq:I72}
\end{equation}
which means that $A(C)$ creates a unit of electric flux in the $3$-direction.

\subsubsection{The Toron Vortex.}
\label{sec:51}

We will now consider a vacuum configuration with non vanishing
magnetic flux. It may a priori come as a surprise that we can
have magnetic flux for a classical vauum configuration. What we
need, in order to achieve this goal, is to find two constant
matrices in the gauge group, such that \cite{tHcmp}
\begin{equation}
PQ = QP {\cal Z}
\label{eq:I73}
\end{equation}
with ${\cal Z}$ a non trivial element in the center of the group.
If such matrices exist, we can use them to define twisted
boundary conditions in two directions in the box. The trivial
configuration $A=0$ automatically satisfies these boundary
conditions, and we will get a classical vacuum with a non
vanishing magnetic flux, characterized by the center element
${\cal Z}$ in (\ref{eq:I73}). For the gauge group $SU(N)$ those
matrices exist; they are 
\begin{eqnarray}
P & = & \left( \begin{array}{cccccc} 0 & 1 &   &   &   &    \\
			       & 0 & 1 &   &   &    \\
			       &   &   &   &   &    \\
			       &   &   &   &   &    \\
			       &   &   &   &   & 1  \\
			     1 &   &   &   &   & 0
	\end{array} \right), \nonumber \\
Q & = & \left( \begin{array}{cccccc} 1 &  &   &   &   &    \\
			       &   & e^{2\pi i/N} &   &   &    \\
			       &   &   &   &   &    \\
			       &   &   &   &   &    \\
			       &   &   &   &   &   \\
			       &   &   &   &   & e^{2 \pi
i(N-1)/N}
	\end{array} \right) \cdot e^{\pi i(1-N)/N},
\label{eq:I74}
\end{eqnarray}
satisfying $PQ=QPe^{2 \pi i/N}$. If we impose twisted boundary
conditions,
\begin{eqnarray}
A(x_1+a_1,x_2,x_3) & = & P A(x_1,x_2,x_3) P^{-1}, \nonumber \\
A(x_1,x_2+a_2,x_3) & = & Q A(x_1,x_2,x_3) Q^{-1}, \nonumber \\
A(x_1,x_2,x_3+a_3) & = & A(x_1,x_2,x_3),
\label{eq:I75}
\end{eqnarray}
in the temporal gauge $A^0=0$, then the classical vacuum $A=0$
is in the sector with non vanishing magnetic flux, $m_3=1$. 
  
Classical vacuum configurations, $A_i(\vec{x}) = g^{-1}(\vec{x})
\partial_i g(\vec{x})$, satisfying (\ref{eq:I75}), would be
defined by gauge transformations $g(\vec{x})$ satisfying
\begin{eqnarray}
g(x_1+a_1,x_2,x_3) & = & P g(x_1,x_2,x_3) P^{-1} e^{2 \pi i k_1/N}, \nonumber \\
g(x_1,x_2+a_2,x_3) & = & Q g(x_1,x_2,x_3) Q^{-1} e^{2 \pi i k_2/N}, \nonumber \\
g(x_1,x_2,x_3+a_3) & = & g(x_1,x_2,x_3) e^{2 \pi i k_3/N},
\label{eq:I76}
\end{eqnarray}
for generic $(k_1,k_2,k_3)$. Now, any gauge transformation
satisfying (\ref{eq:I76}) can be written as
\begin{equation}
g=T_1^{k_1} T_2^{k_2} T_3^{k_3} \tilde{g},
\label{eq:I77}
\end{equation}
with
\begin{equation}
T_1 = Q, \: \: \: \: \: \: T_2 = P^{-1},
\label{eq:I78}
\end{equation}
and $\tilde{g}$ satisfying (\ref{eq:I76}), with $k_1=k_2=k_3=0$.
Acting on the vacuum $|A_i=0>$, we get, from (\ref{eq:I78}),
\begin{eqnarray}
T_1 |A_i =0> & = & |A_i=0>, \nonumber \\
T_2 |A_i =0> & = & |A_i=0>, 
\label{eq:I79}
\end{eqnarray}
which implies, using (\ref{eq:I67}), that the different vacua
have $e_1=e_2=0$. On the other hand, we get, acting with $T_3$, 
\begin{equation}
T_3^{k_3} |A_i=0> \equiv |A_i=0; k_3>,
\label{eq:I80}
\end{equation}
and, therefore, we get $N$ different vacua defined by 
\begin{equation}
|e_3> \equiv \frac {1}{N} \sum_{k_3} e^{2 \pi i \frac {k_3
e_3}{N}} |A_i=0 ; k_3>,
\label{eq:I81}
\end{equation}
with $e_3=0, \ldots, N-1$. Acting now with $T_3^{k_3}$ on
$|e_3>$, we get
\begin{equation}
T_3^{k_3} |e_3> = e^{2 \pi i \frac {k_3 e_3}{N}} e^{i \theta
\frac {k_3 m_3}{N}} |e_3>,
\label{eq:I82}
\end{equation}
from which we observe that 
\begin{equation}
T_3^{N} |e_3> = e^{i \theta} |e_3>,
\label{eq:I83}
\end{equation}
i. e., $T_3^{N}$ is periodic, with winding number equal one.
Notice that in the definition of $|e_3>$ we have included the
$\theta$-parameter and the magnetic flux $m_3=1$, associated with
the boundary conditions (\ref{eq:I75}).
  
From the previous discussion we learn two basic things: first,
that we can get zero energy states, with both electric and
magnetic flux, provided both fluxes are parallel; secondly, that
the number of vacuum states with twisted boundary conditions
(\ref{eq:I75}) is equal to $N$. In fact, what has been computed
above is the well known Witten index, $\hbox {tr }(-1)^F$ \cite{Wind}.

\subsubsection{`t Hooft's Toron Configurations.}
\label{sec:52}

We will now try to find configurations on $T^4$ with fractional
Pontryagin number, satisfying the equations of motion.
Configurations of this type were initially discovered by `t Hooft
for $SU(N)$ \cite{tHcmp}. In order to describe this configurations, we first
choose a subgroup $SU(k) \times SU(l) \times U(1)$ of $SU(N)$, with
$k+l=N$. Let $\omega$ be the matrix corresponding to the $U(1)$
generators of $SU(k) \times SU(l) \times U(1)$,
\begin{equation}
\omega = 2 \pi \left( \begin{array}{cccccc} l &   &   &    &   &    \\
					      &   &   &    &   &    \\
					      &   & l &    &   &    \\
					      &   &   & -k &   &    \\
					      &   &   &    &   &   \\
					      &   &   &    &   & -k
	\end{array} \right), 
\label{eq:I84}
\end{equation}
with $\hbox {tr} \: \omega=0$. The toron
configuration is defined by
\begin{equation}
A_{\mu}(x) = - \omega \sum_{\lambda} \frac {\alpha_{\mu \lambda}
x_{\lambda}}{a_{\lambda}a_{\mu}},
\label{eq:I85}
\end{equation}
with 
\begin{equation}
\alpha_{\mu \nu}-\alpha_{\nu \mu} = \frac {n_{\mu \nu}^{(2)}}{Nl}
- \frac {n_{\mu \nu}^{(1)}}{Nk},
\label{eq:I86}
\end{equation}
and
\begin{equation}
n_{\mu \nu} = n_{\mu \nu}^{(1)} + n_{\mu \nu}^{(2)}.
\label{eq:I87}
\end{equation}
The stress tensor for configuration (\ref{eq:I85}) is given by 
\begin{equation}
F_{\mu \nu}= \ \omega \frac {\alpha_{\mu \nu}-\alpha_{\nu
\mu}}{a_{\mu} a_{\nu}}.
\label{eq:I88}
\end{equation}
If we consider the simplest case, $n_{12}=n_{12}^{(1)}=1$, and
$n_{30}=n_{30}^{(2)}=1$, we will be led to
\begin{eqnarray}
F_{12} & = & + \omega \frac {1}{Nk a_1 a_2}, \nonumber \\
F_{30} & = & - \omega \frac {1}{Nl a_3 a_4}, 
\label{eq:I89}
\end{eqnarray}
and therefore
\begin{equation}
\frac {g^2}{16 \pi^2} \int \hbox {tr} (F_{\mu \nu} \tilde{F}^{\mu
\nu}) = - \frac {1}{N}.
\label{eq:I90}
\end{equation}
If we now impose the self duality condition, we get
\begin{equation}
\frac {a_1 a_2}{a_3 a_4} = \frac {l}{k} = \frac {N-k}{k},
\label{eq:I91}
\end{equation}
which constrains the relative sizes of the box. 
  
The gauge zero modes for the toron configuration (\ref{eq:I85})
can be derived from the general relation (\ref{eq:I47}), with
$k=\tau=0$ for $T^4$. Thus, for Pontryagin number equal $\frac
{1}{N}$, we only get four translational zero modes for gauge
group $SU(N)$. In this sense, we can think of the toron as having
a size equal to the size of the box.
  
The toron of Pontryagin number equal $\frac {1}{N}$ can be
interpreted, as we did for the instanton, as a tunnelling process
between states $|m_3=1,\vec{k}>$ and  $|m_3=1,\vec{k}+(0,0,1)>$.

Let us fix a concrete distribution of electric and magnetic
fluxes, characterized by $\vec{e}$ and $\vec{m}$. The functional
integral for this background is given by \cite{tHcmp}
\begin{equation}
<\vec{e},\vec{m} | \vec{e},\vec{m}> = \sum_{\vec{k}} e^{2 \pi i
\frac {\vec{k} \cdot \vec{e}}{N}} W(\vec{k},\vec{m}),
\label{eq:I92}
\end{equation}
where
\begin{equation}
W(\vec{k},\vec{m}) = \int [dA]_{\vec{k},\vec{m}} \exp - \int 
{\cal L}(A),
\label{eq:I93}
\end{equation}
with the integral in (\ref{eq:I93}) over gauge field
configurations satisfying the twisted boundary conditions
defined by the twists $(\vec{k},\vec{m})$. We can consider the
particular case $\vec{m}=(0,0,1)$ to define the effective action
for the toron configuration,
\begin{equation}
S = \frac {8 \pi^2}{g^2 N} + \frac {2 \pi i e_3}{N}.
\label{eq:I94}
\end{equation}
  
A possible generalization is obtained when using configurations
with Pontryagin number equal $\frac {1}{N}$, but with
$\vec{k}=(k_1,k_2,1)$. In this case, the action (\ref{eq:I94}) becomes 
\begin{equation}
S = \frac {8 \pi^2}{g^2 N} + \frac {2 \pi i (\vec{k} \cdot \vec{e})}{N}.
\label{eq:I95}
\end{equation}
  
It must be noticed that we have not included in (\ref{eq:I94})
the effect of $\theta$, which contributes to the action with a
factor $\frac {i \theta}{N}$.

\subsection{Instanton Effective Vertex.}
\label{sec:ins}

Next, we will consider the effect of instantons on fermions \cite{cl,tHpseudo}. For
the time being, we will work on compactified euclidean spacetime, 
$S^4$. The Dirac matrices satisfy
\begin{equation}
\{\gamma^{\mu},\gamma^{\nu}\}=-2 \delta^{\mu \nu},
\label{eq:I96}
\end{equation}
and the chiral operator $\gamma_5$,
\begin{equation}
\gamma_5 \equiv \gamma^0 \gamma^1 \gamma^2 \gamma^3 = \left(
	\begin{array}{cc} {\bf 1} & {\bf 0} \\
			  {\bf 0} & {\bf 1}
	\end{array} \right).
\label{eq:I97}
\end{equation}
The space of Dirac fermions splits into two spaces of opposite
chirality,
\begin{equation}
\gamma_5 \psi_{\pm} = \pm \psi_{\pm}.
\label{eq:I98}
\end{equation}
Let us work with massless Dirac fermions coupled to an
instanton gauge configuration. We consider normalized
solutions to Dirac's equation,
\begin{equation}
\gamma^{\mu} {\cal D}_{\mu}(A) \psi = 0.
\label{eq:I99}
\end{equation}
As a consequence of the index theorem, the number $\nu_+$ of solutions 
to (\ref{eq:I99}) with positive chirality, minus the number of solutions 
with negative chirality, $\nu_-$, is given by
\begin{equation}
\nu_+ -  \nu_- = \frac {g^2 N_f}{32 \pi^2} \int F^{a}_{\mu \nu} \tilde{F}
^{\mu \nu a} d^4 x,
\label{eq:I100}
\end{equation}
i. e., by the topological charge of the instanton gauge configuration. 
Thus, the change of chirality induced by an instanton configuration 
is given by
\begin{equation}
\Delta Q_5 = 2 N_f k,
\label{eq:I101}
\end{equation}
with $k$ the Pontryagin number, and $N_f$ the number of different 
massless Dirac fermions, transforming in the fundamental representation 
of the gauge group. We can generalize equation (\ref{eq:I100}) to work with 
instanton configurations on a generic four dimensional euclidean 
manifold ${\cal M}$. The index theorem then becomes
\begin{equation}
\nu_+ - \nu_- = \frac {N}{24 \cdot 8 \pi^2} \int_{\cal M} 
\hbox {tr} (R \wedge R) - \frac {g^2 N_f}{32 \pi^2} \int_{\cal M}
 F^{a}_{\mu \nu} \tilde{F}^{\mu \nu a} d^4 x,
\label{eq:I102}
\end{equation}
where again we consider fermions in the fundamental
representation of $SU(N)$. Equation (\ref{eq:I101}) implies that
instanton configurations induce effective vertices, with change
of chirality given by (\ref{eq:I101}). In order to compute these
effective vertices, we will use a semiclassical approximation to
the generating functional,
\begin{equation}
Z(J,\bar{J}) = \int [dA] [d\bar{\psi}][d\psi] \exp - \int {\cal L}(A,\bar{\psi},\psi)+ 
J\bar{\psi}+\psi\bar{J},
\label{eq:I103}
\end{equation}
around the instanton configuration. Let us first perform the
gaussian integration of fermions in (\ref{eq:I103}):
\[
Z(J,\bar{J}) = \int [dA] \hbox {det}' D \! \! \! \! / (A) \exp \int
\bar{J}(x) G(x,y;A) J(y) dx dy \cdot \exp - \int {\cal L}(A) \cdot
\]
\begin{equation}
\prod_{n(A)} \int \bar{\psi}_0^{(n)}(x) J(x) d^4x \int
\bar{J}(y)\psi_0^{(n)}(y) d^4y,
\label{eq:I104}
\end{equation}
where $\psi_0^{(n)}$ are the fermionic zero modes for the
configuration $A$, $\hbox {det}' D \! \! \! /(A)$ is the
regularized determinant, and $G(x,y;A)$ is the regularized Green's 
function,
\begin{equation}
D \! \! \! \! /(A)G(x,y;A) = - \delta(x-y) + \sum_n \psi_0^{n}(x) \psi_0^{n}(y).
\label{eq:I105}
\end{equation}
In semiclassical approximation around the instanton, we get
\[
Z(J,\bar{J}) = \int [dQ] \hbox {det}' D \! \! \! /(A_{inst}) \exp - \frac 
{8 \pi^2}{g^2} \cdot \exp \int \bar{J}(x) G(x,y;A) J(y) d^4x d^4y \cdot 
\]
\begin{equation}
\exp \int {\cal L}_0''(A_{inst}) Q^2 \cdot \prod_{i=1}^{m} \int \bar{\psi}_0^{i}
(x) J(x) d^4 x \int \bar{J}(y)\psi_0^{i}(y) d^4y,
\label{eq:I106}
\end{equation}
where 
\begin{equation}
{\cal L}_0''(A_{inst}) = \left( \frac {\delta^2 {\cal L}_0}{\delta A \delta A} \right)_{A=
A_{inst}},
\label{eq:I107}
\end{equation}
for ${\cal L}_0 = - \frac {1}{4} F^{a \mu \nu} F_{\mu \nu}^{a}$,
and $Q$ the small fluctuation. It is clear from (\ref{eq:I107})
that the only non vanishing amplitudes are those with
\begin{equation}
\left( \frac {\delta^{2m} Z(J,\bar{J})}{\delta J(x_1) \delta \bar{J}(x_1) \ldots 
\delta J(x_m) \delta \bar{J}(x_m)} \right)_{J=\bar{J}=0},
\label{eq:I108}
\end{equation}
for $m=\nu^++\nu^-$. In order to perform the integration over
$Q$, we need to consider the gauge zero modes. Each gauge zero
mode contributes with a factor $\frac {1}{g}$. So, as we have
$4N$ zero modes, we get
\[
<\psi(x_1) \bar{\psi}(x_1) \ldots \psi(x_m) \bar{\psi}(x_m)> = \]
\begin{equation}
C \int \left( \frac {1}{g} \right)^{4N} \frac {1}{\rho^5} \rho^{3
N_f} \exp - \frac {8 \pi^2}{g^2(\mu)} [\mu]^{\beta_1} \prod_{i=1}
^{N_f} (\bar{\psi}_0^{i} {\psi}_0^{i}) d^4 z d \rho,
\label{eq:I109}
\end{equation}
where $\beta_1$ is the coefficient of the $\beta$-function, in
such a way that the result (\ref{eq:I109}) is independent of the
renormalization point $\mu$. It must be stressed that $d^4z d\rho
\rho^{-5}$ is a translation and dilatation invariant measure. The
factor $\rho^{3N_f}$ comes from the fermionic zero modes\footnote{In 
fact, $\rho^{3N_f}$ is the factor that appears in the fermionic Berezin 
measure for the fermionic zero modes.},
\begin{equation}
\psi_0^{i} = \frac {\rho^{3/2}}{(x^2+\rho^2)^{3/2}} \left( \frac
{2}{\pi^2} \right)^{1/2} \omega.
\label{eq:I110}
\end{equation}
(A chiral symmetry breaking condensate is obtained in the $N_f=1$
case). The proportionality factor $C$ in (\ref{eq:I109}) comes
from the determinants for fermions, gauge bosons and
Faddeev-Popov ghosts.
  
The previous computation was carried out for $\theta=0$. The
effect of including $\theta$ is simply
\begin{equation}
<\psi \bar{\psi}(x_1) \ldots \psi \bar{\psi}(x_m)>_{\theta} = <
\ldots >_{\theta=0} \cdot e^{i \theta}
\label{eq:I111}
\end{equation}
  
It is important to stress that the integration over the instanton
size in (\ref{eq:I109}) is infrared divergent; thus, in order to
get finite instanton contributions, we should cut off the
integration size, something that can be implemented if we work
with a Higgs model. The so defined instantons are known as
constrained instantons \cite{tHpseudo}.

\subsection{Three Dimensional Instantons.}
\label{sec:7}

An instanton in three dimensions is a finite
euclidean action configuration. This necesarily implies, in
order to have topological stability, that the second homotopy
group of the vacuum manifold is different from zero. This can not
be realized for pure gauge theories, as $\Pi_2(SU(N)) \simeq 0$,
so we will consider a Higgs model with spontaneous symmetry
breaking from the $G$ gauge group to a subgroup $H$, such that
$\Pi_2(G/H)) \neq 0$. Think of $G=SU(N)$ and
$H=U(1)^{N-1}$, then $\Pi_2={\bf Z}^{N-1}$. Thus, we see that
three dimensional instantons are nothing but `t Hooft-Polyakov
monopoles (see table).

\begin{center}

\begin{tabular}{|c|c|c|c|}     \hline\hline
	 {\bf Dimension}        & {\bf Energy Density}  &  {\bf Energy}  & {\bf Action}  \\ \hline
	 $1+1$                  &                       &  $\Pi_0$       & $\Pi_1$  \\ 
	 $2+1$                  & $\Pi_0$               &  $\Pi_1$       & $\Pi_2$  \\  
	 $3+1$                  & $\Pi_1$               &  $\Pi_2$       & $\Pi_3$  \\  \hline
	 {\bf Name}             & {\bf Vortex}          & {\bf Monopole} & {\bf Instanton} \\ \hline\hline

\end{tabular}
\label{tab:1}
\end{center}

The first thing to be noticed in three dimensions is that the
dual to the photon is a scalar field,
\begin{eqnarray}
H_{\mu} & = & * F_{\rho \sigma} \equiv \frac {1}{2}\epsilon_{\mu \rho
\sigma} F^{\rho \sigma}, \nonumber \\
H_{\mu} & = & \partial_{\mu}\chi.
\label{eq:I112}
\end{eqnarray}
In the weak coupling regime, we can describe the dilute gas of
instantons and anti-instantons as a Coulomb gas. The partition
function is given by \cite{P3d}
\[
Z= \sum_{n} \int \prod_{i=1}^{n_{\pm}} \frac {dx_i^+ dx_i^-}{n^+ ! \: n^- !}
[\exp - S_0]^{n_++n_-} \cdot \]
\begin{equation}
\exp - \frac {1}{2} \left( \frac {4\pi}{g} \right)^2 \int \rho(x)
\left( - \frac {1}{\partial^2} \right) \rho(y) d^3x d^3y,
\label{eq:I113}
\end{equation}
with $n^++n^-=n$, $S_0$ the instanton action, and $\rho$ the
instanton density,
\begin{equation}
\rho(x) = \sum_i \delta(x-x_i^+) - \sum_i \delta (x-x_i^-).
\label{eq:I114}
\end{equation}
The Coulomb interaction term admits the following gaussian
representation, in terms of the dual photon \cite{P3d}:
\begin{equation}
\exp - \frac {1}{2} \left( \frac {4\pi}{g} \right)^2 \int d^3x d^3y
\rho(x) - \frac {1}{\partial^2}\rho(y) = \int [d\chi] \exp - \int
\frac {1}{2} (\partial \chi)^2 + \frac {4\pi i\chi \rho}{g}.
\label{eq:I115}
\end{equation}
When we sum up the instanton and anti-instanton contributions, we
get the effective lagrangian for $\chi$,
\begin{equation}
{\cal L}_{eff}(\chi) = \frac {1}{2} (\partial \chi)^2+e^{-S_0}
\cos \frac {4\pi \chi}{e},
\label{eq:I116}
\end{equation}
which implies a mass for the dual photon $\chi$ equal to
$e^{-S_0}$. That $\chi$ is the dual photon becomes clear from the
$\chi- \rho$ coupling in (\ref{eq:I115}), between $\chi$ and the magnetic
density $\rho$. The generation of a mass for the dual photon in a
dilute gas of instantons is a nice example of confinement in the
sense of dual Higgs phenomena.  
  
The inclusion of massless fermions will drastically change the
physical picture. In particular, as will be shown, the photon
will become a massless Goldstone boson \cite{AHW}. This will be due to the
existence of effective fermionic vertices induced by the three
dimensional instanton, of similar type to the ones studied in
previous section. In order to analyze instanton induced effective
interactions in three dimensions, we should first consider the
problem of fermionic zero modes in the background of a monopole.

\subsubsection{Callias Index Theorem.}

Consider Dirac matrices in euclidean three
dimensional spacetime,
\begin{equation}
\gamma^{i} \gamma^{j}+ \gamma^{j} \gamma^{i} = 2 \delta^{ij}.
\label{eq:I117}
\end{equation}
We can get a representation of (\ref{eq:I117}) using constant
$2\times 2$ matrices. In general, for euclidean space of dimension $n$, the
corresponding $\gamma^{i}$ are constant $2^{(n-1)/2}$ matrices.
  
Now, we define the Dirac operator,
\begin{equation}
L= i \gamma^{i}\partial_i + \gamma^{i} A_i + i \Phi(x),
\label{eq:I119}
\end{equation}
with $A_i = g T^{a} A_i^{a}$, and $\Phi(x) = \phi^{a}(x) T^{a}$, 
for $T^{a}$ the generators of the gauge group in some particular representation. We can 
now consider a Dirac fermion in Minkowski $3+1$ spacetime. This is a four component 
spinor,
\begin{equation}
\psi =  \left( \begin{array}{c} \psi_+ \\ \psi_- \end{array} \right).
\label{eq:I120}
\end{equation}
Then, Dirac's equation in $3+1$ dimensions becomes
\begin{equation}
\left( \begin{array}{cc} {\bf 0} & L \\
			 L^+     & {\bf 0}  \end{array} \right) 
		\left( \begin{array}{c} \psi_+ \\ \psi_- \end{array} \right) 
		= E \left( \begin{array}{c} \psi_+ \\ \psi_- \end{array} 
		\right),
\label{eq:I121}
\end{equation}
for fermion fields $\psi(x,t)=\psi(x) e^{iEt}$, and where $L^+$ is the adjoint of $L$. If we consider
solutions to (\ref{eq:I121}) with $E=0$, we get
\begin{eqnarray}
  L \psi_- & = & 0, \nonumber \\
L^+ \psi_+ & = & 0,
\label{eq:I122}
\end{eqnarray}
i. e., $\psi_-$ and $\psi_+$ are zero modes of the euclidean
Dirac equation in three dimensions, defined by (\ref{eq:I119}).
  
Now,
we can define the index
\begin{equation}
I(L) = k_--k_+,
\label{eq:I123}
\end{equation}
where $k_-$ and $k_+$ are, respectively, the dimensions fo $\hbox
{Ker} (L)$ and $\hbox {Ker} (L^+)$. By generalizing the
Atiyah-Singer index theorem, Callias \cite{Callias} got the following formula
for $I(L)$:
\begin{equation}
I(L) = \frac {1}{2 \left( \frac {n-1}{2} \right) \!} \left( \frac {i}{8 \pi} 
\right)^{\frac {n-1}{2}} \int _{S^{n-1}_{\infty}} \hbox {tr} [ U(x) (dU(x))^{n-1} ],
\label{eq:I124}
\end{equation}
with $n$ the dimension of euclidean spacetime, and
\begin{equation}
U(x) \equiv | \Phi(x) |^{-1} \Phi(x).
\label{eq:I125}
\end{equation}
In our case, $n=3$. In terms of the magnetic charge of the monopole, (\ref{eq:I20}),
\begin{equation}
N = \frac {1}{8\pi} \int \epsilon_{ijk} \phi^{i} \partial
\phi^{j} \partial \phi^{k},
\label{eq:I126}
\end{equation}
where we have normalized $a=1$ in equation (\ref{eq:I20}), and using
(\ref{eq:I120}) for $\Phi$ we get, for $SU(2)$,
\begin{equation}
I(L) = 2 N,
\label{eq:I127}
\end{equation}
for fermions in the adjoint representation. Notice that in odd 
dimensions, the index is zero for compact spaces. The contribution in (\ref{eq:I124}) 
appears because we are working in a non compact space, with special 
boundary conditions at infinity, which are the ones defining the monopole configuration. 
We can also consider
the more general case of massive fermions replacing
(\ref{eq:I120}) by
\begin{equation}
\Phi = \phi^{a} T^{a} +m.
\label{eq:I128}
\end{equation}
In this case, we get, from (\ref{eq:I124}),
\begin{equation}
I(L) = (j(j+1)-\{m\}(\{m\}+1))N,
\label{eq:I129}
\end{equation}
with $\{m\}$ the largest eigenvalue of $\phi^{a} T^{a}$ smaller
than $m$ or, if there is no such eigenvalue, the smallest minus one. Thus, for 
massless fermions in the fundamental representation we have $\{m\}= - \frac {1}{2}$, 
and $I(L)=N$. It is important to observe that by changing the bare mass, the index 
also changes (we are using the normalization $a=1$). Thus, for $m> \frac {1}{2}$, and fermions 
in the fundamental representation, we get $I(L)=0$.

\subsubsection{The Dual Photon as Goldstone Boson.}

We will consider the $SU(2)$ lagrangian 
\begin{equation}
{\cal L} = - \frac {1}{4} F_{\mu \nu} ^2 + \frac {1}{2} ({\cal
D}_{\mu} \phi)^2+ V(\phi) + \psi_+ (i {\cal D} \! \! \! \! / + g \phi 
)\psi_-,
\label{eq:I130}
\end{equation}
where we have used notation (\ref{eq:I120}), and the Dirac
operator (\ref{eq:I119}). Lagrangian (\ref{eq:I130}) is invariant
under the $U(1)$ trasnformation
\begin{eqnarray}
\psi_- & \rightarrow & e^{i \theta} \psi_-, \nonumber \\
\psi_+ & \rightarrow & e^{i \theta} \psi_+.
\label{eq:I131}
\end{eqnarray}
We will assume that the $\psi_{\pm}$ transform in the adjoint
representation of $SU(2)$. Using (\ref{eq:I127}), the induced
instanton couple $\psi_-$ fermions to $\psi_-^{T}\gamma_0$,
through an instanton, while $\psi_+$ is coupled to $\psi_+^T \gamma$ 
in the anti-instanton case (the number $\psi_+(\psi_-) $ of zero modes 
for spherically symmetric monopoles in the instanton (anti-instanton) 
configuration is zero, and the two zero modes are $\psi_{\pm}$
and $\psi_{\pm}^T\gamma_0$. These vertices induce effective mass
terms for fermions with mass ${\cal O}(e^{-S_0})$\footnote{These
mass terms clearly break the $U(1)$ symmetry (\ref{eq:I131}).}.
  
Now, we should include the Coulomb interaction between
instantons; then, the effective lagrangian becomes
\begin{equation}
{\cal L} = \frac {1}{2} (\partial \chi)^2+m \psi_-^T \gamma_0
e^{\frac {4\pi i \chi}{g^2}} \psi_- + m \psi_+ \gamma_0 e^{- \frac
{4 \pi i \chi}{g^2}} \psi_+ + \cdots,
\label{eq:I133}
\end{equation}
so that now the old vertices coupling $\psi_{\pm}$ to
$\psi_{\pm}^T \gamma_0$ become vertices where the instanton or
anti-instanton couple $\psi_{\pm}$ and $\psi_{\pm}^T \gamma_0$ to 
the dual photon $\chi$\footnote{The effective lagrangian
(\ref{eq:I133}) will not be interpreted in the wilsonian sense,
but simply as the generating functional of the effective vertices
induced by instantons.}. From (\ref{eq:I133}) it is now clear
that $\chi$ becomes a Goldstone boson for the $U(1)$ symmetry
\cite{AHW} (\ref{eq:I131}). In fact, ${\cal L}$ is invariant under
(\ref{eq:I131}) if 
\begin{equation}
\chi \rightarrow \chi + \frac {g^2 \theta}{2 \pi}.
\label{eq:I135}
\end{equation}
Notice that now $\chi$ is massless, and that no potential for
$\chi$ is generated by instanton effects. It is also important to
stress that the symmetry (\ref{eq:I131}) is not anomalous in
$2+1$ dimensions, which explains, from a different point of view,
the Goldstone boson nature of $\chi$.

\subsection{$N\!=\!1$ Supersymmetric Gauge
Theories.}

As a first example, we will consider the $N\!=\!1$\footnote{For a complete reference on supersymmetry, see
\cite{WB}.} extension
of pure Yang-Mills theory. This model is defined in terms of a
vector superfield, containing the gluon and the gluino. The
gluino will be represented by a real Majorana spinor,
transforming in the adjoint representation. The lagrangian is
given by
\begin{equation}
{\cal L} = - \frac {1}{4} F^{a \mu \nu} F^{a}_{\mu \nu} + 
\frac {1}{2} i \bar{\lambda}^{a} \gamma_{\mu} {\cal D}^{\mu}(A)
\lambda^{a} + \frac {\theta g^2}{32 \pi^2} F^{a \mu
\nu}F^{a}_{\mu \nu}.
\label{eq:I136}
\end{equation}
As it can be easily checked, (\ref{eq:I136}) is invariant under
the supersymmetry transformations
\begin{eqnarray}
\delta A_{\mu}^{a} & = & i \bar{\alpha} \gamma_{\mu} \lambda^{a},
\nonumber \\
\delta \lambda^{a} & = & \frac
{1}{4}[\gamma_{\mu},\gamma_{\nu}]\alpha F^{a \mu \nu}, \nonumber
\\
\delta \bar{\lambda}^{a} & = & - \bar{\alpha} \frac {1}{4}
[\gamma_{\mu},\gamma_{\nu}] F^{a \mu \nu},
\label{eq:I137}
\end{eqnarray}
with $\alpha$ a Majorana spinor. Notice that, for $\lambda^{a}$ in
(\ref{eq:I136}), we can use either real Majorana or complex Weyl
spinors.
  
We will now study instanton effects for (\ref{eq:I136}) \cite{NSVZ,SVZ1,SVZ2,A,ADS1,ADS2}. For
$SU(N)$ gauge group, the total number of fermionic zero modes is 
\begin{equation}
\# \hbox {zero modes} = 2 Nk, 
\label{eq:I138}
\end{equation}
with $k$ the Pontryagin number of the instanton. For $SU(2)$
and Dirac fermions in the isospin representation, of dimension
$2j+1$, the generalization of (\ref{eq:I100}) is
\begin{equation}
\nu_+ -\nu_- = \frac {2}{3} (j+1)(2j+1)k,
\label{eq:I139}
\end{equation}
from which we certainly get (\ref{eq:I138}) for $j=1$, using Majorana
fermions.
  
The $2N$ zero modes for $k=1$ decompose, relative to the $SU(2)$
subgroup where the instanton lies, into
\begin{eqnarray}
4 & \hbox {triplets}, & \nonumber \\
2(N-2) & \hbox {doublets}. &
\label{eqnarray}
\end{eqnarray}
The meaning of the $4$ triplet zero modes is quite clear from
supersymmetry. Namely, two of them are just the result of acting
with the supersymmetric charges on the instanton configuration.
For $N\!=\!1$ we have four supersymmetric charges,
two of which anhilate the instanton configuration. The two other
triplets result from superconformal transformations on the
instanton. In fact, lagrangian (\ref{eq:I136}) is not only
invariant under supersymmetry, but also under the superconformal
group. Now, we can repeat the computation of section
\ref{sec:ins}. The only non vanishing amplitudes will be of the
type
\begin{equation}
<\lambda \lambda(x_1) \cdots \lambda \lambda(x_N)>.
\label{eq:I141}
\end{equation}
Impossing the instanton measure on collective coordinates to
be translation and dilatation invariant, we get
\begin{equation}
\int \frac {d^4 z d\rho \rho^{2N}}{\rho^{5}},
\label{eq:I142}
\end{equation}
where the factor $\rho^{2N}$ comes from the $2N$ fermionic zero
modes, that scale as $\frac {1}{\rho^2}$ (see table). We
must include the instanton action, and the renormalization
point, $\mu$,
\begin{equation}
\mu^{4N-\frac {2N}{2}} \exp - \frac {8 \pi^2}{g(\mu)^2},
\label{eq:I143}
\end{equation}
where the power of $\mu$ is given by $+1$ for each gauge zero
mode, and $- \frac {1}{2}$ for each Majorana fermionic zero mode.
Defining the scale,
\begin{equation}
\Lambda = \mu \exp - \int \frac {dg'}{\beta(g')},
\label{eq:I144}
\end{equation}
and using the $\beta$-function for $SU(N)$ supersymmetric
Yang-Mills,
\begin{equation}
\beta(g') = - \frac {g^3}{16\pi^2} 3N,
\label{eq:I145}
\end{equation}
(\ref{eq:I143}) becomes
\begin{equation}
\Lambda^{3N},
\label{eq:I146}
\end{equation}
with 
\begin{equation}
\Lambda = \mu \exp - \frac {8 \pi^2}{3Ng(\mu)^2}.
\label{eq:I147}
\end{equation}
Combining all these pieces, we get
\[
<\lambda \lambda(x_1) \cdots \lambda \lambda(x_N)> = \int \frac
{d^4 z d\rho \rho^{2N}}{\rho^{5}} \Lambda^{3N} \cdot \]
\begin{equation}
\sum_{\hbox {permutations}} (-1)^{P} \hbox {tr} (\lambda_{i_1} \lambda_{i_2}(x_1)) \cdots \hbox 
{tr} (\lambda_{i_{2N-1}} \lambda_{i_{2N}}(x_N)).
\label{eq:I148}
\end{equation}
In order to now perform the integration over the collective
coordinates, we need the expression for the zero modes given in
the table\footnote{The function $f(x)$ is the instanton factor $f(x)=
\frac {1}{(x-z)^+\rho^2}$.}.

\begin{center}

\begin{tabular}{|c|c|}     \hline\hline
	 {\bf Supersymmetric triplet}    & $\sim \rho^2 (f(x))^2$ \\ \hline
	 {\bf Superconformal triplet}    & $\sim \rho x (f(x))^2$ \\ \hline 
	 {\bf Doblets}                   & $\rho (f(x))^{1/2}$ \\ \hline\hline

\end{tabular}
\label{tab:2}
\end{center}

The fermionic zero modes in the above table are given for the
instanton in the singular gauge,
\begin{equation}
A_{\mu}^{inst} = \frac {2}{g} \frac {\rho^2}{(x-z)^2+\rho^2}
\eta_{\mu \nu}^{a} (x-z)_{\nu},
\label{eq:I149}
\end{equation}
with $z$ the instanton position. Using the expressions given above, we can perform the integration over $z$ and $\rho$,
to obtain the result
\begin{equation}
<\lambda \lambda(x_1) \cdots \lambda \lambda(x_N)> \sim \hbox
{constant} \Lambda^{3N},
\label{eq:I150}
\end{equation}
which is a very amusing and, a priori, surprising result. The
reason leading to (\ref{eq:I150}) is that the integral
(\ref{eq:I148}) is saturated by instantons with size of the same
order as the $|x_1-x_N|$ distance. If we now use cluster
decomposition in (\ref{eq:I150}), we get
\begin{equation}
<\lambda \lambda> \sim \hbox {constant} \Lambda^3 e^{2\pi i n/N},
\label{eq:I151}
\end{equation}
with $n=0,\ldots,N-1$. Notice that result (\ref{eq:I151}) is not
generated by instanton configurations, and that we get it
assuming clustering or, equivalently, the existence of mass gap
in the theory. This map gap should be
interpreted as confinement.
  
A different approach for computing the $<\lambda \lambda>$
condensate starts with massive supersymmetric QCD, and requires a
decoupling limit, $m \rightarrow \infty$. So, for $SU(2)$ with
one flavor of mass $m$ we get, from the instanton computation
\begin{equation}
< \lambda \lambda(x_1) \lambda \lambda(x_2)> \sim \hbox {
constant } \Lambda^5 m,
\end{equation}
with $\Lambda$ the scale of the $N=1$ QCD theory. Relying now
upon clustering, we get
\begin{equation}
< \lambda \lambda> \sim \hbox {
constant } \Lambda^{5/2} m^{1/2} e^{2 \pi i n/2}.
\end{equation}
We can now take the $m \rightarrow \infty$ limit, and define the
scale $\Lambda$ of pure $N=1$ supersymmetric Yang-Mills as 
\begin{equation}
\Lambda^3 \sim \Lambda^{5/2} m^{1/2}.
\end{equation}
The only difference with the previous computation is that now we
perform cluster decomposition before definig the decoupling
limit.
  
Until now we have consider $<\lambda \lambda>$ condensates for
vacuum angle $\theta$ equal zero. We will now show the dependence
of the condensate on $\theta$, through an argument given by
Shifman and Vainshtein. For $SU(N)$ gauge group, the axial
anomaly is given by
\begin{equation}
\partial j_{\mu}^5 = \frac {N}{16 \pi^2} F \tilde{F}. 
\end{equation}
This means that under the chiral transformation
\begin{equation}
\lambda \rightarrow e^{i \alpha} \lambda,
\end{equation}
the lagrangian changes as
\begin{equation}
{\cal L} \rightarrow {\cal L} + \frac {\alpha  N}{16 \pi^2} F
\tilde{F}.
\end{equation}
Thus, $<\lambda \lambda>$ at a non zero value of $\theta$ is the
same as $<\lambda' \lambda'>_{\theta=0}$, with
\begin{equation}
\lambda'=e^{i \alpha} \lambda,
\end{equation}
where now 
\begin{equation}
2 \pi \alpha = \theta.
\end{equation}
Hence \cite{SV},
\begin{equation}
<\lambda \lambda>_{\theta=0} = <\lambda' \lambda'>_{\theta=0} =
<\lambda \lambda>_{\theta=0} e^{i \frac {\theta}{n}}.
\label{smi}
\end{equation}

\subsection{Instanton Generated Superpotentials in Three
Dimensional $N\!=\!2$.}
\label{sec:91}

To start with, we will consider dimensional reduction of
lagrangian (\ref{eq:I136}) to three dimensions. In this case, we
arrive to the Higgs lagrangian in $2+1$ discussed in section
\ref{sec:7}. We can then define a complex Higgs field, with the
real part given by the fourth component of $A_{\mu}$ in $3+1$,
and the imaginary part by the photon field $\chi$. If, as was
the case in section \ref{sec:7}, we consider $<\phi>=0$ for the
real Higgs field, then we automatically break superconformal
invariance, and for the $SU(2)$ case we will find only two
fermionic zero modes in the instanton background (`t
Hooft-Polyakov monopole). The action of the three dimensional
instanton is
\begin{equation}
S_{\hbox {inst}} = \frac {4\pi \phi}{g^2},
\label{eq:I154}
\end{equation}
with $\phi$ standing for the vacuum expectation value of the
Higgs field\footnote{Notice that the gauge coupling constant, in
three dimensions, has $\hbox {length}^{-1/2}$ dimensions.}. The
effective lagrangian (\ref{eq:I133}) becomes
\[
{\cal L} = \frac {1}{2} ((\partial \chi)^2+(\partial \phi)^2) +
\bar{\psi}i {\partial}\! \! \!  / \psi + m e^{-4\pi \phi/g^2} \psi^T
\gamma_0 \psi e^{i 4 \pi \chi/g^2} + \]
\begin{equation}
m e^{4 \pi \phi/g^2} \bar{\psi}\gamma_0\bar{\psi}^T e^{-i4 \pi
\chi/g^2},
\label{eq:I155}
\end{equation}
where we have included the kinetic term for both the real Higgs
field $\phi$, and the dual photon $\chi$. In (\ref{eq:I155}) we
can define a complex Higgs field,
\begin{equation}
\Phi = \phi + i \chi,
\label{eq:I156}
\end{equation}
in order to notice that the instanton is certainly generating a
Yukawa coupling, which is nothing but the vertex coupling $\psi$
fields to the dual photon $\chi$. In order to write (\ref{eq:I155}) 
as a supersymmetric lagrangian,
we need to add a superpotential term of the type \cite{AHW}
\begin{equation}
W(\Phi)_{\theta \theta} = \exp - \Phi + \hbox {hc},
\label{eq:I157}
\end{equation}
which induces an effective potential for $\phi$ of the type
\begin{equation}
V(\phi) = \frac {\partial W}{\partial \Phi} \frac {\partial
\bar{W}}{\partial \Phi^*} = \exp - \phi,
\label{eq:I158}
\end{equation}
i. e., no potential, as expected for the dual photon field,
$\chi$. The minima for the potential (\ref{eq:I158}) is at
$\phi=\infty$\footnote{The reader might be slightly surprised
concerning potential (\ref{eq:I158}) for the Higgs field. The
crucial issue for the correct understanding of this potential
requires noticing that the $N\!=\!2$ three dimensional
theory has been obtained through dimensional reduction of
$N\!=\!1$ four dimensional Yang-Mills, which contains a flat
direction (in next chapter we will define these flat directions
more precisely, as Coulomb branches of moduli of vacua).}.
  
It is important to stress some aspects of the previous
computation: first of all, the superpotential (\ref{eq:I157}) is
simply given by the instanton action, with the extra term $\frac
{i 4 \pi \chi}{g^2}$ the analog of a topological $\theta$ term
in four dimensions. Secondly, the fermions appearing in
(\ref{eq:I155}), the effective lagrangian, are the ones in the
hypermultiplet of the $N\!=\!2$ theory. Finally, the
superpotential for $\Phi$ is defined on a flat direction.
  
The generalization of the previous picture to the four
dimensional case is certainly not straightforward, as in that
case we have not flat directions, and the effective lagrangian
can not be written in terms of chiral superfields containing the
gluino, but the gluino-gluino pair.

\subsubsection{A Toron Computation.}

A direct way to obtain $<\lambda \lambda>$ condensates in four
dimensional $N\!=\!1$ Yang-Millsis using self dual gauge
configurations, with Pontryagin number $\frac {1}{N}$\footnote{It 
should already be noticed that topological configurations directly 
contributing to $<\lambda \lambda>$ are most probably the relevant 
configurations for confinement, as $<\lambda \lambda>$ was 
derived through a cluster argument assuming the existence of a mass 
gap.} \cite{CG}. In subsection \ref{sec:52} we
have described these configurations. The main point in using
these torons is that the number of fermionic zero modes
automatically reduces to two, which we can identify with the two
triplets defined by supersymmetry transformations of one
instanton configurations. We will perform the computation in a
box, sending at the end its size to infinity. The size of the box
is the size of the toron, but we will avoid the dilatation zero
mode and the two triplet zero modes defined by superconformal
transformations. The toron measure now becomes, simply,
\begin{equation}
\int d^4 z
\label{eq:I159}
\end{equation}
for the translation collective coordinate. Now, we have a power
of $\mu$, given by the four translation zero modes, and two
fermionic zero modes,
\begin{equation}
\mu^3 \exp - \frac {8 \pi^2}{g(\mu)^2 N},
\label{eq:I160}
\end{equation}
where we have included the toron action $\frac {8\pi^2}{g^2N}$.
Notice that (\ref{eq:I160}) is simply $\Lambda^3$. Now, we
integrate $z$ over the box of size $L$. The two fermionic zero
modes are obtained by the supersymmetry transformation
(\ref{eq:I137}) over the toron configuration (\ref{eq:I85}),
which means that each fermionic zero mode behaves as $\frac
{1}{L^2}$, and therefore no powers of $L$ should be included in
the measure. The final result is
\begin{equation}
<\lambda \lambda> \sim \hbox {constant} \Lambda^3 e^{2 \pi i
e/N},\: \: \: \: e= 0,1,\ldots,N-1
\label{eq:I161}
\end{equation}
in agreement with the cluster derivation. How should this result
be interpreted? First of all, the expectation value
(\ref{eq:I161}) corresponds to the amplitude
\[
<\vec{e},\vec{m}=(0,0,1) | \lambda \lambda |
\vec{e},\vec{m}=(0,0,1)>=\]
\begin{equation}
<\vec{k}+(0,0,1),\vec{m}=(0,0,1) | \lambda
\lambda | \vec{k},\vec{m}=(0,0,1)> e^{2\pi i \frac {\vec{e} \cdot
(0,0,1)}{N}}.
\label{eq:I162}
\end{equation}
Then, the $e$ in (\ref{eq:I161}) is $e_3$, and the different
values in (\ref{eq:I161}) correspond to the set of $N$ different
vacua described in subsection \ref{sec:51}. 
  
Notice that a change $\theta \rightarrow \theta + 2 \pi$ in
equation (\ref{smi}) produces a change 
\begin{equation}
<\lambda \lambda>_{\theta} \rightarrow <\lambda \lambda>_{\theta}
e^{2\pi i/N},
\end{equation}
i. e., a ${\bf Z}(N)$ rotation. In other words, $\theta
\rightarrow \theta + 2 \pi$ exchanges the different vacua. Let us
now try the same argument for (\ref{eq:I162}). Using (\ref{nue}),
we observe that
\begin{equation}
<\lambda \lambda> \sim \Lambda^3 e^{2 \pi i e_{eff}/N} =
\Lambda^3 e^{2 \pi i e/N} e^{i \theta/N},
\end{equation}
in agreement with (\ref{smi}). So, under $\theta \rightarrow
\theta + 2 \pi$, we go, using (\ref{nue}), from $\vec{e}_{eff}$
to $\vec{e}_{eff}+ \vec{m}$. Notice that for the toron compuation
we are using $\vec{m}={\bf 1}$.

%%%%%%%%%%%%%%%%%%%%%%%%%%%%%%%%%%%%%%%%%%%%%%%%%%%%%%%%%%%%%%%%%%%%%%%%%%%%%%%%%%
%%%%%%%%%%%%%%%%%%%%%%%%%%%%%%%%%%%%%%%%%%%%%%%%%%%%%%%%%%%%%%%%%%%%%%%%%%%%%%%%%%

\section{Chapter II}

\subsection{Moduli of Vacua.}

In this part of the lectures, we will consider gauge theories
possessing potentials with flat directions. The existence of flat
potentials will motivate the definition of moduli of vacua, which
we will understand as the quotient manifold
\begin{equation}
{\cal M} = {\cal V}/{\cal G},
\label{eq:II1}
\end{equation}
obtained from the modding of the vacuum manifold ${\cal V}$ by
gauge symmetries. In the first chapter, an example has already
been discussed, namely three dimensional $N\!=\!2$ Yang-Mills,
defined as dimensional reduction of $N\!=\!1$ Yang-Mills in four
dimensions. Denoting by $\phi^{a}=A_4^{a}$ the fourth component of
the gauge field, the dimensionally reduced lagrangian is 
\begin{equation}
{\cal L} = - \frac {1}{4} F_{ij}^{a} F^{aij} + \frac {1}{2} {\cal
D}_i \phi^{a} {\cal D}^{i} \phi^{a} + i \bar{\chi}^{a} \gamma_i
{\cal D}^{i} \chi + i f_{abc} \bar{\chi}^{b} \chi^{c} \phi^{a}.
\label{eq:II2}
\end{equation}
This is the Yang-Mills-Higgs lagrangian in the Prasad-Sommerfeld
limit $V(\phi)=0$. At tree level, the vacuum expectation value
for the field $\phi$ is undetermined; therefore, at the classical
level we can define a moduli of (real) dimension one, 
parametrizing the different values of $<\phi>$. As we already
know, in addition to the scalar $\phi$ we have yet another scalar
field, $\chi$, the dual photon field. No potential can be defined
for $\chi$, neither classically nor quantum mechanically. If we
took into account the action of the Weyl group, $\phi \rightarrow
- \phi$, $\chi \rightarrow - \chi$, the classical moduli manifold
should be
\begin{equation}
{\bf R} \times S^{1} / {\bf Z}_2.
\label{eq:II3}
\end{equation}
The fields $\phi$ and $\chi$ can be combined into a complex
scalar, $\Phi = \phi + i \chi$. As discussed in chapter I,
instantons generate a superpotential of type $e^{-\Phi}$, which
induces a potential for the $\phi$ fields with its minimum at
$\infty$. This potential lifts the classical degeneracy of vacua.
The vacuum expectation value of $\chi$ still remains
undetermined, but can be changed by just shifting the coefficient
of the topological term. The physics of this first example is
what we expect from physical grounds: quantum effects breaking
the classical vacuum degeneracy. However, there are cases where
the amount of supersymmetry prevents, a priori, the generation of
superpotential terms; it is in these cases, where we should be able
to define the most general concept of quantum moduli \cite{SW,SW2}, where
quantum effects will modify the topology and geometry of the
classical moduli manifold.

\subsection{$N\!=\!4$ Three Dimensional Yang-Mills.}
\label{sec:22}

$N\!=\!4$ three dimensional Yang-Mills will be defined through
dimensional reduction of $N\!=\!1$ six dimensional Yang-Mills \cite{SW3d}.
The three real scalars $\phi_i^{a}$, with $i=1,2,3$,
corresponding to the $\mu=3,4,5$ vector components of
$A_{\mu}^{a}$, are in the adjoint representation of the gauge group, and
will transform as a vector with respect to the $SO(3)_R$ group of
rotations in the $3,4,5$-directions. The fermions in the model
will transform, with respect to the $SU(2)_R$ double cover, as
doublets, i. e., as spin one half particles. If we now consider
the $SU(2)_E$ rotation group of euclidean space, ${\bf R}^3$,
then fermions transform again as doublets, while scalars,
$\phi_i^{a}$, are singlets. By dimensional reduction, we get the
following potential for the $\phi_i$:
\begin{equation}
V(\phi) = \frac {1}{4g^2} \sum_{i<j} \hbox {tr}
[\phi_i,\phi_j]^2, 
\label{eq:II4}
\end{equation}
where we have used a six dimensional lagrangian 
$- \frac {1}{4g^2} F_{\mu \nu}^{a} F^{a \mu \nu}$. Obviously,
the potential (\ref{eq:II4}) possesses flat directions, obtained
as those whose $\phi_i$ fields are in the Cartan subalgebra of
the gauge group. For $SU(N)$, we will get
\begin{equation}
\phi_i = \left( \begin{array}{ccccc} a_i^1   &   &    &   &    \\
					     &   &    &   &    \\
					     &   &  \ddots  &   &    \\
					     &   &    &   &    \\
					     &   &    &   & a_i^N 
	\end{array} \right), 
\label{eq:II5}
\end{equation}
with $\sum_{j=1}^{N} a_i^{j} = 0$, so that $3(N-1)$
parameters characterize a point in the flat directions of
(\ref{eq:II4}). In the general case of a gauge group of rank $r$,
$3r$ coordinates will be required. A vacuum expectation value
like (\ref{eq:II5}) breaks $SU(N)$ to $U(1)^{N-1}$. As each
$U(1)$ has associated a dual photon field, $\chi_j$, with
$j=1,\ldots, N-1$, the classical moduli has a total
dimension equal to $3r+r=4r$. The simplest case of $SU(2)$,
corresponds to a four dimensional moduli, which classically is
\begin{equation}
{\cal M} = ({\bf R}^3 \times S^1)/{\bf Z}_2.
\label{eq:II6}
\end{equation}
We have quotiented by the ${\bf Z}_2$ Weyl action, and
$S^1$ parametrizes the expectation value of the dual photon field.
The group $SU(2)_R$ acts on the ${\bf R}^3$ piece of
(\ref{eq:II6}). In the same sense as for the $N\!=\!2$ theory
considered in chapter I, the $N\!=\!4$ model posseses instanton
solutions to the Bogomolny equations which are simply the
dimensional reduction to three dimensional euclidean space of
four dimensional self dual Yang-Mills equations. These
Bogomolny-Prasad-Sommerfeld instantons involve only one scalar
field, $\phi_i$, out of the three available (we will therefore 
choose one of them, say $\phi_3$), and satisfy the
equation 
\begin{equation}
F = * {\cal D} \phi_3.
\label{eq:II7}
\end{equation}
Once we choose a particular vacuum expectation value for the
$\phi_i$ fields, we break $SU(2)_R$ to an $U(1)_R$ subgroup. In
particular, we can choose $\phi_1=\phi_2=0$, and $\phi_3$
different from zero, with $\phi_3$ the field used in the
construction of the BPS instanton. The remaining $U(1)_R$ stands
for rotations around the $\phi_3$ direction. 
   
Now, as discussed above, the BPS instanton can induce effective
fermionic vertices. We first will consider the case of no extra
hypermultiplets. In this case, we have four fermionic zero modes,
corresponding to the four supersymmetry transformations that do
not annihilate the instanton solution. In addition, we know, from the results in chapter I, 
that the effective vertex should behave like
\begin{equation}
\psi \psi \psi \psi \exp - (I+i\chi),
\label{eq:II7b}
\end{equation}
with $\chi$ the dual photon field, and $I$ the classical instanton action, behaving 
like $\frac {\phi_3}{g^2}$. The term (\ref{eq:II7}) breaks the $U(1)_R$ 
symmetry, as the $\psi$ fermions transform under the $U(1)_R$ subgroup of 
$SU(2)_R$ like
\begin{equation}
\psi \rightarrow e^{i \theta/2} \psi.
\label{eq:II8}
\end{equation}
However, the dual photon field plays the role of a Goldstone
boson, to compensate this anomalous behaviour; $\chi$
transforms under $U(1)_R$ as
\begin{equation}
\chi \rightarrow \chi - 4 \left( \frac {\theta}{2} \right).
\label{eq:II9}
\end{equation}
This is a quantum effect, that will have important consequences
on the topology of the classical moduli. In fact, the $U(1)_R$ is
not acting only on the ${\bf R}^3$ part of ${\cal M}$, but also
(as expressed by (\ref{eq:II9})) on the $S^1$ piece.
  
The topological meaning of (\ref{eq:II9}) can be better
understood if we work \cite{SW3d} on the boundary of ${\bf R}^3 \times S^1$,
namely $S^2 \times S^1$. In this region at infinity,
(\ref{eq:II9}) will define a non trivial $S^1$ bundle on $S^2$.
In order to see the way this is working, it is useful to use
the spinorial notation. Henceforth, let $u_1$ and $u_2$ be two
complex variables, satisfying
\begin{equation}
|u_1|^2+|u_2|^2=1.
\label{eq:II10}
\end{equation}
This defines the sphere $S^3$. Parametrizing points in
$S^2$ by a vector $\vec{n}$, defined as follows:
\begin{equation}
\vec{n} = \bar{u} \vec{\sigma} u,
\label{eq:II11}
\end{equation}
with $\vec{\sigma}$ the Pauli matrices, and using (\ref{eq:II11}), we
can define a projection,
\begin{equation}
\Phi: S^3 \rightarrow S^2,
\label{eq:II12}
\end{equation}
associating to $u$, in $S^3$, a point $\vec{n}$ in $S^2$. The
fiber of the projection (\ref{eq:II12}) is $S^1$. In fact,
$u_{\alpha}$ and $e^{i \theta}u_{\alpha}$ yield the same value
of $\vec{n}$; therefore, we conclude
\begin{equation}
S^2 = S^3 /U(1),
\label{eq:II13}
\end{equation}
with the $U(1)$ action
\begin{equation}
u_{\alpha} \rightarrow e^{i \theta} u_{\alpha}.
\label{eq:II14}
\end{equation}
A $U(1)$ rotation around the point $\vec{n}$ preserves
$\vec{n}$, but changes $u$ as
\begin{equation}
u_{\alpha} \rightarrow e^{i \theta /2} u_{\alpha},
\label{eq:II15}
\end{equation}
where now $\theta$ in (\ref{eq:II15}) is the $U(1)_R$ rotation
angle. Moreover, it also changes $\chi$ to $\chi - 4 \left(
\frac {\theta}{2} \right)$. So, the infinity of our quantum
moduli looks like
\begin{equation}
(S^3 \times S^1)/U(1),
\label{eq:II16}
\end{equation}
with the $U(1)$ action
\begin{equation}
u_{\alpha} \rightarrow e^{i \theta} u_{\alpha}, \: \: \: \: \chi
\rightarrow \chi - 4 \theta.
\label{eq:II17}
\end{equation}
The space (\ref{eq:II16}) is the well known Lens space $L_{-4}$.
Generically, $L_s$ spaces can be defined through (\ref{eq:II16}),
with the $U(1)$ action defined by
\begin{equation}
\chi \rightarrow \chi + s \theta.
\label{eq:II18}
\end{equation}
The spaces $L_s$ can also be defined as
\begin{equation}
L_s = S^3/{\bf Z}_s,
\label{eq:II19}
\end{equation}
with the ${\bf Z}_s$ action
\begin{equation}
\beta : u_{\alpha} \rightarrow e^{2 \pi i/s} u_{\alpha}.
\label{eq:II20}
\end{equation}
To finish the construction of the infinity of the quantum moduli,
we need to include the Weyl action, ${\bf Z}_2$, so that we get
\begin{equation}
L_s/{\bf Z}_2
\label{eq:II21}
\end{equation}
or, equivalently, $S^3/\Gamma_s$, where $\Gamma_s$ is the group
generated by $\beta$ given in (\ref{eq:II20}), and the Weyl
action 
\begin{equation}
\alpha : (u_1,u_2) \rightarrow (\bar{u}_2, - u_1),
\label{eq:II22}
\end{equation}
which reproduces $\vec{n} \rightarrow - \vec{n}$. The relations
defining the group $\Gamma_s$ are
\begin{equation}
\alpha^2 = \beta^s = 1, \: \: \: \: \alpha \beta = \beta^{-1}
\alpha.
\label{eq:II23}
\end{equation}
Moreover, for $s=2k$, $\Gamma_s$ is the dihedral group $D_{2k}$.
  
Before entering into a more careful analysis of the moduli space
for these $N\!=\!4$ theories, let us come back to the discussion
on the $SU(2)_R$ symmetry, and the physical origin of the
parameter $s$. In order to do so, we will first compactify six
dimensional super Yang-Mills down to four dimensions. The
resulting $N\!=\!2$ supersymmetric Yang-Mills theory has only two scalar fields,
$\phi_1$ and $\phi_2$. The rotation symmetry
in the compactified $(4,5)$-plane becomes now a $U(1)$ symmetry
for the $\phi_i$ fields, which is the well known $U(1)_R$
symmetry of $N\!=\!2$ supersymmetric four dimensional Yang-Mills.
As in the case for the $N\!=\!4$ theory in three dimensions,
instantons in four dimensions generate effective fermionic
vertices which break this $U(1)_R$ symmetry. Following the steps
of the instanton computations presented in chapter I, we easily
discover a breakdown of $U(1)_R$ to ${\bf Z}_8$. In the case of
$N\!=\!2$ in four dimensions, the potential obtained from
dimensional reduction is
\begin{equation}
V(\phi) = \frac {1}{4g^2} \hbox {tr}
[\phi_1,\phi_2]^2, 
\label{eq:II24}
\end{equation}
with the flat direction, for $SU(2)$,
\begin{equation}
\phi_i = \left( \begin{array}{cc} a_i   &   \\
					& -a_i 
	\end{array} \right).
\label{eq:II25}
\end{equation}
The moduli coordinate is therefore $u = \hbox {tr} \phi^2$, where
we now define $\phi$ as a complex field. The $U(1)_R$
transformation of $\phi$ is given by
\begin{equation}
\phi \rightarrow e^{2 i \alpha} \phi,
\label{eq:II26}
\end{equation}
so that $u$ is ${\bf Z}_4$ invariant, and the ${\bf Z}_8$
symmetry acts as ${\bf Z}_2$ on the moduli space.
  
The difference with the three dimensional case is that now the
instanton contribution does not contain any dual photon field,
that can play the role of a Goldstone boson. From the three
dimensional point of view, the $U(1)_R$ symmetry of the four
dimensional theory is the rotation group acting on fields
$\phi_1$ and $\phi_2$, and can therefore be identified with the
$U(1)_R$ part of $SU(2)_R$ fixing the $\phi_3$ direction. We
should wonder about the relationship between the effect observed in
three dimensions, and the breakdown of $U(1)_R$ in four
dimensions. A qualitative answer is simple to obtain. The
breakdown of $U(1)_R$ in four dimensions can be studied in the
weak coupling limit (corresponding to $u \rightarrow \infty$),
since the theory is assymptotically free, using instantons. Then, we
get an effective vertex of the type
\begin{equation}
<\psi \psi \psi \psi>_0 e^{i \theta} = < \psi \psi \psi \psi
>_{\theta}.
\label{eq:II27}
\end{equation}
We have only four fermionic zero modes, since for $u \neq 0$ we
break the superconformal invariance of the instanton. It is
clear, from (\ref{eq:II27}), that a $U(1)_R$ transformation is
equivalent to the change
\begin{equation}
\theta \rightarrow \theta - 4 \alpha,
\label{eq:II28}
\end{equation}
with $\alpha$ the $U(1)_R$ parameter. Now, this change in
$\theta$ is, in fact, the perfect analog of transformation rule
(\ref{eq:II9}), for the dual photon field.
This should not come as a surprise; in fact, the four dimensional
topological term
\begin{equation}
\frac {i \theta}{32 \pi^2} F *F
\label{eq:II29}
\end{equation}
produces, by dimensional reduction, the three dimensional
topological term
\begin{equation}
\frac {i \theta}{32 \pi^2} \epsilon_{ijk}F_{jk} \partial_i
\phi_3.
\label{eq:II30}
\end{equation}
This is precisely the type of coupling of the dual photon, in
three dimensions, with the topological charge, and thus we again
recover the result of section \ref{sec:7}. 
  
From the previous discussion, we can discover something else,
specially interesting from a physical point of view. The 
transformation law of $\chi$ was derived counting instanton
fermionic zero modes; however, the effect we are
describing is a pure perturbative one loop effect, as is the
$U(1)_R$ anomaly in four dimensions. Consider 
the wilsonian \cite{Shifwils,Sbeta} effective coupling
constant for the $N\!=\!2$ theory, without hypermultiplets.
Recall that in the wilsonian approach \cite{Wilson}, the effective coupling
constant is defined in terms of the scale we use to
integrate out fluctuations with wave length smaller than that
scale (this is the equivalent to the Kadanoff approach for
lattice models). In a Higgs model, the natural scale is the vacuum
expectation value of the Higgs field. Using the above notation,
the wilsonian coupling constant in the four dimensional model is
$\frac {1}{g(u)^2}$, with $u$ the moduli parameter defined by
$\hbox {tr} \phi^2$. Now, let us write the lagrangian as follows:
\begin{equation}
{\cal L} = \frac {1}{64 \pi} \hbox {Im} \int \tau (F+i*F)^2,
\label{eq:II31}
\end{equation}
with $\tau$ defined by
\begin{equation}
\tau \equiv \frac {i 8 \pi}{g^2} + \frac {\theta}{\pi}.
\label{eq:II32}
\end{equation}
Using $F^2= -*F^2$ we get, from (\ref{eq:II31}), the standard
lagrangian in Minkowski space,
\begin{equation}
{\cal L} = \int \frac {1}{4g^2} FF + \frac {\theta}{32 \pi^2} F*F.
\label{eq:II33}
\end{equation}
Now, we use the one loop effective beta function for the
theory,
\begin{equation}
\frac {8\pi}{g(u)^2} = \frac {2}{\pi} \ln u + {\cal O}(1).
\label{eq:II34}
\end{equation}
In general, if we add $n$ hypermultiplets, we get, for the four
dimensional theory,
\begin{equation}
\frac {8\pi}{g(u)^2} = \frac {4-n}{2\pi} \ln u + {\cal O}(1),
\label{eq:II35}
\end{equation}
recovering the well known result for $N\!=\!2$ supersymmetric
$SU(2)$ gauge theories in four dimensions,
of finiteness of the theory when $n=4$, and infrared freedom for $n>4$. For
$n<4$ the theory is assymptotically free, so that
the perturbative computation (\ref{eq:II35}) is only valid at 
small distances, for $u$ in the assymptotic infinity. 
  
Now, let us perform a rotation on $u$,
\begin{equation}
u \rightarrow e^{2 \pi i}u
\label{eq:II36}
\end{equation}
From (\ref{eq:II34}) we get, for $n=0$,
\begin{equation}
\frac {8\pi}{g(u)^2} \rightarrow \frac {8\pi}{g(u)^2} + 4i,
\label{eq:II37}
\end{equation}
so, using (\ref{eq:II32}), we get
\begin{equation}
\theta \rightarrow \theta - 4\pi,
\label{eq:II38}
\end{equation}
in perfect agreement with equation (\ref{eq:II28}). Thus, we
observe that the $s$ term (at least for the case without
hypermultiplets) that we have discovered above using three
dimensional instanton effects, is exactly given by the one loop
effect of the four dimensional $N\!=\!2$ theory. But what about
higher order loop effects? As the argument we have presented is
nothing but the non renormalization theorem \cite{Sbeta} in supersymmetric
theories, the $U(1)_R$ action on the wilsonian scale $u$ forces
the renormalization of the coupling constant to be consistent
with the $U(1)$ anomalous behaviour of the lagrangian, which is
determined by the Adler-Bardeen theorem \cite{AB} to be exact at one loop.
What happens as we include hypermultiplets? First of all, and
from the point of view of the three dimensional theory, the
instanton effect will now be a vertex of type,
\begin{equation}
\psi \psi \psi \psi \prod^{2N_f} \chi e^{-(I+i\chi)},
\label{eq:II39}
\end{equation}
with the $2N_f$ fermionic zero modes appearing as a consequence
of Callias index theorem \cite{Callias}, (\ref{eq:I129}), for $j=1/2$ and
$\{m\}=-1/2$. From (\ref{eq:II39}), we get $s=-4+2N_f$, which
means a dihedral group $\Gamma_s$ of type $D_{2N_f-4}$ or,
equivalently, a Dynkin diagram of type $D_{N_f}$. Notice that in
deriving this diagram we have already taken into account the Weyl
action, ${\bf Z}_2$. 
  
The connection between the
dihedral group, characterizing the moduli of the three
dimensional $N\!=\!4$ theory, and the beta function for the
four dimensional $N\!=\!2$ theory can be put on more solid
geometrical grounds. The idea is simple. Let us work with the
$N\!=\!2$ four dimensional theory on ${\bf R}^3 \times S^1$,
instead of on euclidean space, ${\bf R}^4$. The massless fields,
from the three dimensional point of view, contain the fourth
component of the photon in four dimensions, and the standard dual
photon $\chi$ in three dimensions. Requiring, as in a
Kaluza-Klein compactification, all fields to be independent of
$x_4$, we still have residual gauge transformations of the type
\begin{equation}
A_4 \rightarrow A_4 + \partial_4 \Lambda(x_4),
\label{eq:II41}
\end{equation}
with $\Lambda(x_4)= \frac {x_4 b}{\pi R}$, with $b$ an angular
variable, $b \in [0,2\pi]$. This is equivalent to saying that we
have non trivial Wilson lines in the $S^1$ direction, if the
gauge field is in $U(1)$, as happens to be the case for a
generic value of $u$\footnote{We are not impossing, to the 
magnetic flux through the $S^1$, to be topologically stable in the sense of
$\Pi_1(U(1)) \simeq {\bf Z}$. The crucial point is that the value
of $b$ at this point of the game is completely undetermined, and
in that sense is a moduli parameter.}. Now, at each point $u$ in
the four dimensional $N\!=\!2$ moduli, we have a two torus $E_u$, 
parametrized by the dual photon field $\chi$, and the field $b$.
This $E_u$ is obtained from the $S^1$ associated with $\chi$, and
the $S^1$ associated with $b$. Its volume, in units defined by
the three dimensional coupling constant, is of order $\frac
{1}{R}$,
\begin{equation}
\hbox {Vol} \: E \sim \frac {1}{R}.
\label{eq:II42}
\end{equation}
In fact, the volume is $\frac {1}{R} \cdot g_3^2$, where $g_3^2$,
the three dimensional coupling constant, is the size of the $S^1$
associated to the dual photon (notice that the coupling
constant $g_3^2$, in three dimensions, has units of inverse
length). Equation (\ref{eq:II42}) shows how, in the four
dimensional limit, $E_u$ goes to zero volume. Now, we have a
picture of the theory in ${\bf R}^3 \times S_R^1$, if we keep $R$
finite, namely that of an elliptic fibration over the $u$-plane,
parametrizing the vacuum expectation values of the
$N\!=\!2$ four dimensional theory. 
  
If we keep ourselves at one particular point $u$, the torus $E_u$
should be the target space for the effective lagrangian for the
fields $b$ and $\chi$. There is a simple way to derive this
lagrangian by means of a general procedure, called dualization,
that we will now describe. To show the steps to follow, we will
consider the four dimensional lagrangian (\ref{eq:II31}). In
order to add a dual photon field, let us say $A_D^{\mu}$, we must
couple $A_D^{\mu}$ to the monopole charge,
\begin{equation}
\epsilon^{0ijk} \partial_i F_{jk} = 4\pi \delta^{(3)}(x).
\label{eq:II43}
\end{equation}
Thus, we add a term
\begin{equation}
\frac {1}{4} \int A_D^{\mu} \epsilon_{\mu \nu \rho \sigma}
\partial^{\nu} F^{\rho \sigma} \equiv \frac {1}{4\pi} \int *F_D
F.
\label{eq:II44}
\end{equation}
Using the same notation as in (\ref{eq:II31}), we get
\begin{equation}
\frac {1}{4\pi} \int *F_D F = \frac {1}{8 \pi} \hbox {Re} \int
(*F_D-iF_D)(F+i*F),
\label{eq:II45}
\end{equation}
so that our lagrangian is
\begin{equation}
{\cal L} = \frac {1}{64 \pi} \hbox {Im} \int \tau (F+i*F)^2 +
\frac {1}{8\pi} \hbox {Re} \int (*F_D-iF_D)(F+i*F).
\label{eq:II46}
\end{equation}
After gaussian integration, we finally get
\begin{equation}
{\cal L} = \frac {1}{64\pi} \hbox {Im} \int \left( \frac
{-1}{\tau} \right) (*F_D-iF_D)^2,
\label{eq:II47}
\end{equation}
i. e., lagrangian (\ref{eq:II31}), with $\tau$ replaced by $\frac
{-1}{\tau}$. The reader should take into account that these
gaussian integrations are rather formal manipulations. Now, we
use the same trick to get an effective lagrangian for the fields
$b$ and $\chi$. Start with the four dimensional lagrangian,
\begin{equation}
{\cal L} = \int \frac {1}{4g^2}FF+ \frac {i\theta}{32 \pi^2} F*F,
\label{eq:II48}
\end{equation}
where we now work in euclidean space. In three dimensions,
using $d^4x = 2 \pi R d^3x$, we get
\begin{equation}
{\cal L} = \int d^3x \frac {1}{\pi Rg^2} |db|^2+ \frac {\pi
R}{2g^2} F_{ij}^2 + \frac {i \theta}{8\pi^2} \epsilon^{ijk}F_{jk}
\partial_k b.
\label{eq:II49}
\end{equation}
Now, as we did before, we couple the dual photon field to the
monopole charge,
\begin{equation}
\partial_i H^{i} = 4\pi \delta^{(3)} (x),
\label{eq:II50}
\end{equation}
with $H^{i} \equiv \epsilon^{ijk}F_{jk}$, to get a term
\begin{equation}
\frac {i}{8\pi} \epsilon^{ijk}F_{jk}\partial_i \chi,
\label{eq:II51}
\end{equation}
so that we can perform a gaussian integration,
\begin{equation}
{\cal L} = \int d^3 x \frac {1}{\pi Rg^2} |db|^2 + \frac
{g^2}{\pi R(8\pi)^2} | d \chi - \frac {\theta}{\pi} db|^2.
\label{eq:II52}
\end{equation}
What we get is precisely a target space for $\chi$ and $b$
fields, which is the torus of moduli $\tau$ given by
(\ref{eq:II32}). Observe that the
complex structure of the torus $E_u$ is given in terms of the
four dimensional coupling constant $g$ \cite{SW3d}, and the four dimensional
$\theta$-parameter, while its volume, (\ref{eq:II42}), depends on the three
dimensional coupling constant $g_3$, that acts as unit. When we
go to the four dimensional $R \rightarrow \infty$ limit, this
volume becomes zero, but the complex structure remains the same.
The fact that the complex structure of $E_u$ is given by the four
dimensional effective coupling will make more transparent the
meaning of equation (\ref{eq:II35}). In fact, the monodromy
around $u=\infty$ for (\ref{eq:II35}) is given by a matrix
\begin{equation}
\left( \begin{array}{cc} a & b  \\
			 c & d 
	\end{array} \right) = 
\left( \begin{array}{cc} -1 & -n+4  \\
			 0 & -1 
	\end{array} \right),
\label{eq:II55}
\end{equation}
with $\tau$ transforming as
\begin{equation}
\tau \rightarrow \frac {a\tau +b}{c\tau +d},
\label{eq:II56a}
\end{equation}
so that for $n=0$ we get transformation (\ref{eq:II38}). Next, we
will see that transformation (\ref{eq:II55}) is precisely what we
need, in order to match the dihedral group characterization of
the $N\!=\!4$ three dimensional moduli space; however, in order to do
that we need a few words on Atiyah-Hitchin spaces \cite{AH}.

\subsection{Atiyah-Hitchin Spaces.}

Atiyah-Hitchin spaces appear in the study of moduli spaces for
static multi-monopole configurations. Static solutions are
defined by the BPS equations, (\ref{eq:II7}), which 
are simply the dimensional
reduction to ${\bf R}^3$ of euclidean self-dual
equations for instantons. Next, we simply summarize some of the
relevant results on Atiyah-Hitchin spaces for our problem (we
refer the interested reader to the book by M. Atiyah and N. Hitchin, 
\cite{AH}). First of all, the Atiyah-Hitchin spaces are
hyperk\"{a}hler manifolds of dimension $4r$, on which a rotation
$SO(3)$ is acting in a specific way. This is part of what we need
to define the moduli space of $N\!=\!4$ three dimensional
Yang-Mills theory for gauge group of rank $r$. In fact, in order to
define $N\!=\!4$ supersymmetry on this space, interpreted as a
$\sigma$-model target space of the low energy effective
lagrangian, we have to require hyperk\"{a}hler structure. Recall
here that hyperk\"{a}hler simply means that we have three
different complex structures, $I$, $J$ anf $K$, and therefore
three different K\"{a}hler forms, $\omega_i$, $\omega_j$ and
$\omega_k$, which are closed. Following the notation used by
Atiyah and Hitchin, we define $N_k$ as the moduli space of a $k$
monopole configuration. The dimension of $N_k$ is $4k-1$, so for
$k=1$ we get dimension $3$, corresponding to the position of the
monopole center. If we mode out by the translation of the center
of mass, we get the space
\begin{equation}
M_k^0 = N_k/{\bf R}^3,
\label{eq:II57}
\end{equation}
of dimension $4(k-1)$. For two monopoles, we get $\hbox {dim}
M_2^0=4$. Now, the spaces $M_k^0$ are generically non simply
connected,
\begin{equation}
\Pi_1(M_k^0) \simeq {\bf Z}_k,
\label{eq:II56}
\end{equation}
so we can define its $k$-fold covering $\tilde{M}_k^0$. The known
results, for $k=2$, are the following: the spaces $M_2^0$ and
$\tilde{M}_2^0$ are, at infinity, respectively of type
$L_{-4}/{\bf Z}_2$, and $L_{-2}/{\bf Z}_2$, which strongly
indicates that $M_2^0$ is a good candidate for the moduli of the
$N_f=0$ case, and $\tilde{M}_2^0$ is the adequate for the
$N_f=1$ case. Moreover, the spaces $\tilde{M}_2^0$ can be
represented by a surface in ${\bf C}^3$, defined by
\begin{equation}
y^2 = x^2 v +1.
\label{eq:II57b}
\end{equation}
The space $M_2^0=\tilde{M}_2^0/{\bf Z}_2$, can be obtained using
variables $\chi=x^2$ and $y=x$, so that we get
\begin{equation}
y^2 = \chi^2 v + \chi.
\label{eq:II58}
\end{equation}
The spaces $\tilde{M}_2^0$, defined by (\ref{eq:II57}), can be
interpreted as a limit of the family of spaces
\begin{equation}
y^2 = x^2 v +v^l,
\label{eq:II59}
\end{equation}
where $l$ should, in our case, be identified with $N_f-1$.
Surfaces (\ref{eq:II59}) are well known in singularity theory;
they give rise to the type of singularities obtained from ${\bf
C}^3/\Gamma$, with $\Gamma$ a discrete subgroup of $SO(3)$, and
are classified according to the following table \cite{Arnold},

\begin{center}

\begin{tabular}{|c|c|c|}     \hline\hline
	 ${\bf \Gamma}$         & {\bf Name}  &  {\bf Singularity}     \\ \hline
	 ${\bf Z}_n$            & $A_{n-1}$   &  $v^n+xy=0$            \\ 
	 ${\bf D}_{2n}$         & $D_{n+2}$   &  $vx^2-v^{n+1}+y^2=0$  \\  
	 ${\bf T}_{12}$         & $E_6$       &  $v^4+x^3+y^2=0$       \\  
	 ${\bf O}_{24}$         & $E_7$       &  $v^3+vx^3+y^2=0$      \\ 
	 ${\bf I}_{60}$         & $E_8$       &  $v^5+x^3+y^2=0$       \\ \hline\hline

\end{tabular}
\label{tab:3}
\end{center}

As can be seen from this table, the manifold (\ref{eq:II59})
corresponds to a $D_{n+2}$ singularity, with $n=N_f-2$, and
dihedral group $D_{2N_f-4}$, i. e., the group $\Gamma$ we have
discussed in the previous section.
  
It is important to stress that the type of singularities we are
describing in the above table are the so called rational
singularities \cite{Artin}. The geometrical meaning of the associated Dynkin
diagram is given by the resolution of the corresponding
singularity as the intersection matrix of the irreducible
components obtained by blowing up the singularity. In this
interpretation, each mode of the diagram corresponds to an
irreducible component, which is a rational curve $X_i$, with self
dual intersection $X_i . X_i=-2$, and each line to the
intersection $X_i . X_j$ between different irreducible
components.  
  
In the previous section we have modelled the $N=4$ moduli space
as an elliptic fibration, with fiber $E_u$ of volume $\frac
{1}{R}$, and moduli $\tau$ given by the coupling constant of the
four dimensional $N=2$ gauge theory. Next, we will try to connect
the dihedral group, characterizing the Atiyah-Hitchin space
describing the $N=4$ moduli, with the monodromy at infinity of
the elliptic modulus of $E_u$. But before doing this, we will
briefly review Kodaira's theory on elliptic singularities.

\subsection{Kodaira's Classification of Elliptic Fibrations.}

According to Kodaira's notation \cite{Kodaira}, we define an elliptic fibration
$V$ onto $\Delta$, where $\Delta$ will be chosen as a compact
Riemann surface. In general, we take $\Delta$ to be of genus
equal zero. The elliptic fibration,
\begin{equation}
\Phi: V \rightarrow \Delta,
\label{eq:II60}
\end{equation}
will be singular at some discrete set of points, $a_{\rho}$. The
singular fibers, $C_{a\rho}$, are given by
\begin{equation}
C_{a\rho} = \sum n_{s} \Theta_{\rho s},
\label{eq:II61}
\end{equation}
with $\Theta_{\rho s}$ irreducible curves. According to Kodaira's
theorem (see section \ref{sec:elliptic} for more details), all posible types of singular curves are of the
following types:
\begin{itemize}

	\item \begin{equation}
I_{n+1}: \: \: C_{a\rho} = \Theta_0+\Theta_1+ \cdots + \Theta_n,
\: \: \: \: n+1 \geq 3,
\label{eq:K1}
\end{equation}
where $\Theta_i$ are non singular rational curves with intersections
$(\Theta_0,\Theta_1)=(\Theta_1,\Theta_2)=\cdots=(\Theta_n,\Theta_0)=1$. 
  
The $A_{n}$ affine Dynkin diagram can be associated to $I_{n+1}$.
Different cases are 
\begin{itemize}
	\item[{i)}] $I_0$, with $C_{\rho}=\Theta_0$ and $\Theta_0$
elliptic and non singular.
	\item[{ii)}] $I_1$, with $C_{\rho}=\Theta_0$ and $\Theta_0$ a
rational curve, with one ordinary double point.
	\item[{iii)}] $I_2$, with $C_{\rho}=\Theta_0+\Theta_1$ and
$\Theta_0$ and $\Theta_1$ non singular rational points, with
intersection $(\Theta_0,\Theta_1)=p_1+p_2$, i. e., two points.
\end{itemize}
Notice that $I_1$ and $I_2$ correspond to diagrams $A_0$ and
$A_1$, respectively.
  
	\item Singularities of type $I_{n-4}^*$ are characterized by
\begin{equation}
I_{n-4}^*: \: \: C_{\rho} = \Theta_0 + \Theta_1 + \Theta_2 +
\Theta_3 + 2 \Theta_4 + 2 \Theta_5 + \cdots + 2 \Theta_n,
\label{eq:K2}
\end{equation}
with intersections $(\Theta_0,\Theta_4)=(\Theta_1,\Theta_4)= 
(\Theta_2,\Theta_4)=(\Theta_3,\Theta_4)=(\Theta_4,\Theta_5)=
(\Theta_5,\Theta_6)= \cdots =1$, these singularities correspond to the $D_n$
Dynkin diagram.
  
	\item Singularities of type $II^*$, $III^*$ and $IV^*$ correspond to
types $E_6$, $E_7$ and $E_8$. 
  
In addition to these singularities,
we have also the types
	\item $II: \: \: C_{\rho} = \Theta_0$, with $\Theta_0$ a
rational curve with a cusp.
	\item $III: \: \: C_{\rho} = \Theta_0 + \Theta_1$, with
$\Theta_0$ and $\Theta_1$ non singular rational curves, with
intersection $(\Theta_0,\Theta_1)=2p$.
	\item $IV: \: \: C_{\rho} = \Theta_0 + \Theta_1 +
\Theta_2$, with $\Theta_0$, $\Theta_1$ and $\Theta_2$ non
singular rational curves, with intersections $(\Theta_0,\Theta_1)=
(\Theta_1,\Theta_2)=(\Theta_2,\Theta_0)=p$.
\end{itemize}

In contrast to the singularities described in last section (the
rational ones), these singularities are associated to affine
Dynkin diagrams. Observe that for all these singularities we have 
\begin{equation}
C.C =0,
\end{equation}
while in the rational case the corresponding maximal cycle
satisfies 
\begin{equation}
C.C=-2.
\end{equation}
  
The origin for the affinization of the Dynkin diagram is the
elliptic fibration structure. In fact, we can think of a rational
singularity of ADE type in a surface, and get the affinization of
the Dynkin diagram whenever there is a singular curve passing
through the singularity. In the case of an elliptic fibration,
this curve is the elliptic fiber itself. So, the extra node in
the Dynkin diagram can be interpreted as the elliptic fiber. This
can be seen more clearly as we compute the Picard of the surface.
In fact, for the elliptic fibration the contribution to the
Picard comes from the fiber, the basis, and the contribution from
each singularity. Now, in the contribution to Picard from each
singularity, we should not count the extra node, since this has
already been taken into account when we count the fiber as an
element in the Picard.
  
The previous discussion is already telling us what happens when
we go to the $R=0$ limit, i. e., to the three dimensional $N=4$
gauge theory. In this limit, the elliptic fiber $E_u$ becomes of
infinite volume, and therefore we can not consider it anymore as
a compact torus, i. e., as an elliptic curve. Thus, in this limit
the corresponding singularity should become rational, and the
Dynkin diagram is not affine.
  
However, before entering that discussion, let us work out the
monodromies for the elliptic fibrations of Kodaira's
classification.

We will then define
\begin{equation}
\tau(u) = \frac {\int_{u \times \gamma_1} \varphi(u)}{\int_{u
\times \gamma_2} \varphi(u)},
\label{eq:II62}
\end{equation}
with $\varphi(u)$ the holomorphic one form on $C_u$. From
(\ref{eq:II62}), it follows that $\tau(u)$ is a holomorphic
function of $u$. Next, we define the elliptic modular function,
$j(\tau(u))$, on the upper half plane,
\begin{equation}
j(\tau(u)) = \frac {1728 g_2^3}{\Delta},
\label{eq:II63}
\end{equation}
where $\Delta$ is the discriminant $\Delta=g_2^3-27g_3^2$, with
the modular functions
\begin{eqnarray}
g_2 & = & 60 \sum_{n_1,n_2 \: \in {\bf Z}} \frac {1}{(n_1+n_2
\tau)^4}, \nonumber \\
g_3 & = & 140 \sum_{n_1,n_2 \: \in {\bf Z}} \frac {1}{(n_1+n_2
\tau)^6}.
\label{eq:II64}
\end{eqnarray}
  
Defining ${\cal F}(u) \equiv j(w(u))$ as a function of $\Delta$,
it turns out to be a meromorphic function. To each pole
$a_{\rho}$, and each non contractible path $\gamma_{\rho}$ 
in $\Delta$ (that is, an element in $\Pi_1(\Delta)$), we want to
associate a monodromy matrix, $A_{\rho}$,
\begin{equation}
A_{\rho} \tau \rightarrow \frac {a\tau+b}{c\tau+d}.
\label{eq:II65}
\end{equation}
If $a_{\rho}$ is a pole of ${\cal F}(u)$, of order $b_{\rho}$,
then it can be proved that $A_{\rho}$ is of type
\begin{equation}
A_{\rho} \tau \rightarrow \tau + b_{\rho},
\label{eq:II66}
\end{equation}
for some $b_{\rho}$. The matrix $A_{\rho}$, of finite order, 
$A_{\rho}^m=1$, for some $m$, corresponds to singularities which can be removed. 
Moreover, if $A_{\rho}$ is of infinite order, then it is always possible 
to find numbers $p$, $q$, $r$ and $s$ such that
\begin{equation}
\left( \begin{array}{cc} s & -r  \\
			 -q & p 
	\end{array} \right)  
\left( \begin{array}{cc} a & b  \\
			 c & d 
	\end{array} \right)
\left( \begin{array}{cc} p & r  \\
			 q & s 
	\end{array} \right) = 
\left( \begin{array}{cc} 1 & b_{\rho}  \\
			 0 & 1 
	\end{array} \right),
\label{eq:II67}
\end{equation}
with $ps-qr=1$. Next, we relate matrices $A_{\rho}$ with the
different types of singularities. The classification, according
to Kodaira's work, is as shown in the table below.

\begin{center}

\begin{tabular}{|c|c|}     \hline\hline
	 {\bf Matrix}         & {\bf Type of singularity}     \\ \hline
	 $\left( \begin{array}{cc} 1 & 1  \\
			 0 & 1 
	\end{array} \right)$     & $I_1$   \\
	 $\left( \begin{array}{cc} 1 & b  \\
			 0 & 1 
	\end{array} \right)$     & $I_b$   \\
	 $\left( \begin{array}{cc} -1 & -b  \\
			 0 & -1 
	\end{array} \right)$     & $I_b^*$   \\
	 $\left( \begin{array}{cc} 1 & 1  \\
			 -1 & 0 
	\end{array} \right)$     & $II$   \\
	 $\left( \begin{array}{cc} 0 & 1  \\
			 -1 & 0 
	\end{array} \right)$     & $III$   \\
	 $\left( \begin{array}{cc} 0 & 1  \\
			 -1 & -1 
	\end{array} \right)$     & $IV$   \\ \hline\hline
	 
\end{tabular}
\label{tab:4}
\end{center}

Now, we can compare the monodromy (\ref{eq:II55}) with the ones
in the table. It corresponds to the one associated with a
singularity of type $I_b^*$, with $b=n-4$, i. e., a Dynkin
diagram of type $D_n$. In the rational case, this corresponds to
a dihedral group $D_{2n-4}$. In (\ref{eq:II55}), $n$ represents
the number of flavors, so that we get the dihedral group of the
corresponding Atiyah-Hitchin space.
  
Summarizing, we get that the dihedral group of $N=4$ in three
dimensions is the one associated with the type of elliptic
singularity at infinity of the elliptic fibration defined by the
$N=2$ four dimensional theory. In other words, the picture we get
is the following: in the $R \rightarrow 0$ three dimensional
limit we have, at infinity, a rational singularity of type ${\bf
C}^3/D_{2N_f-4}$. When we go to the $R \rightarrow \infty$ limit
we get, at infinity, an elliptic singularity with Dynkin diagram
$D_{N_f}$. Both types of singularities describe, respectively,
one loop effects in three dimensional $N=4$ and four dimensional
$N=2$.

\subsection{The Moduli Space of the Four Dimensional $N\!=\!2$
Supersymmetric Yang-Mills Theory. The Seiberg-Witten Solution.}

From our previous discussion, we have observed that the complex
structure of the moduli space of three dimensional $N\!=\!4$
supersymmetric Yang-Mills theory is given by the elliptic
fibration on the moduli space of the four dimensional $N\!=\!2$
theory, where the elliptic modulus is identified with the
effective complexified coupling constant $\tau$, as defined in
(\ref{eq:II32}). This result will in practice mean that the complete 
solution to the four dimensional $N\!=\!2$ theory can be directly 
read out form the complex structure of the Atiyah-Hitchin spaces 
(\ref{eq:II59}), with $l=N_f-1$. In previous sections, we have 
already done part of this job, comparing the monodromy of $\tau$ 
around $u=\infty$, i. e., in the assymptotic freedom regime, with the 
dihedral group characterizing the infinity of the three dimensional 
$N\!=\!4$ moduli space. In this section, we will briefly review the 
Seiberg-Witten solution \cite{SW,SW2,rep1,rep2,rep3,rep4,rep5,rep6,IS,Reviews} 
for four dimensional $N\!=\!2$ Yang-Mills theory, 
and compare the result with the complex structure of 
Atiyah-Hitchin spaces. Recall that the 
Atiyah-Hitchin spaces are hyperk\"{a}hler, and therefore possess three 
different complex structures. The complex structure determined by the 
four dimensional $N\!=\!2$ solution is one of these complex structures, 
namely the one where the Atiyah-Hitchin space becomes elliptically 
fibered. The analysis of Seiberg and Witten was originally based on the 
following argument: the moduli space parametrized by $u$ should be compactified 
to a sphere (we will first of all consider the $N_f=0$ case, for $SU(2)$ gauge 
group). According to Kodaira's notation, $\Delta$ is taken to be of 
genus equal zero. Next, the behaviour of $\tau$ at 
$u=\infty$ is directly obtained from the one loop beta function (see 
equation (\ref{eq:II35})); this leads to a monodromy around infinity 
of the type (\ref{eq:II55}). Next, if $\tau(u)$ is a holomorphic function of $u$, 
which is clear from the elliptic fibration mathematical point of view (see 
equation (\ref{eq:II62})), and is a direct consequence of $N\!=\!2$ 
supersymmetry, then the real and imaginary parts are 
harmonic functions. As the coupling constant is the imaginary part of the 
complex structure $\tau(u)$, which is on physical grounds always positive, 
we are dealing with an elliptic fibration, so we already know all 
posible types of singularities. That some extra singularities 
should exist, in addition to the one at infinity, is clear form the 
harmonic properties of $\hbox {Im} \tau(u)$, and the fact that it is 
positive, but in principle we do not how many of them we should 
expect, and of what type. The answer to this question can not, in principle, 
be derived from Kodaira's theory. In fact, all what we can obtain 
from Kodaira's approach, using the adjunction formula, is a relation 
between the canonical bundle $K$ of the elliptic fibration, the 
$K$ of the base space, which we can take as $\IP^1$, and the type 
of singularities,
\begin{equation}
K_V = \Pi^* (K_{\Delta} + \sum a_i P_i),
\label{eq:II68}
\end{equation}
where the $a_i$, for each type of singularity, are given below.

\begin{center}

\begin{tabular}{|c|c|}     \hline\hline
	 {\bf Singularity}      & ${\bf a_i}$     \\ \hline
	 $I_1$                  & $1/12$   \\
	 $I_b^*$                & $1/2+b/12$   \\
	 $I_b$                  & $b/12$   \\
	 $II$                   & $1/6$   \\
	 $III$                  & $1/4$   \\
	 $IV$                   & $1/3$   \\ 
	 $II^*$                 & $5/6$   \\         
	 $III^*$                & $3/4$   \\
	 $IV^*$                 & $2/3$   \\ \hline\hline

\end{tabular}
\label{tab:5}
\end{center}

However, (\ref{eq:II68}) is not useful, at this point, since we do not 
know the $V$ manifold, which is what we are looking for. 
We will therefore proceed according to physical arguments.
  
The singularities we are looking for are singularities in the strong coupling 
regime of the moduli space of the theory, so it is hopeless to try to use 
a naive perturbative analysis; instead, we can rely on a duality approach. 
In dual variables (see equation (\ref{eq:II47})), the effective coupling constant 
behaves like $\frac {-1}{\tau}$, i. e., we have performed an $S$ transformation, 
with
\begin{equation}
	S= \left( \begin{array}{cc} 0 & -1  \\
			 1 & 0 
	\end{array} \right).
\label{eq:II69}
\end{equation}
Thinking of $\frac {-1}{\tau}$ as the effective magnetic coupling, 
$\tau^{mag}$, we can reduce our analysis to looking for perturbative 
monodromies of type
\begin{equation}
\tau^{mag} \rightarrow \tau^{mag} + b.
\label{eq:II70}
\end{equation}
Indeed, we know that any singularity of Kodaira's 
type is related to a monodromy of type (\ref{eq:II70}), up to a unitary 
transformation, (see equation (\ref{eq:II67})).
  
Now, and on physical grounds, we can expect a transformation of the type (\ref{eq:II70}) 
as the monodromy singularity for the effective coupling constant of an effective 
$U(1)$ theory, with $b$ equal to the number of massless hypermultiplets. In fact, 
the beta function for the $U(1)$ $N\!=\!2$ theory, with $n$ hypermultiplets, 
is given by
\begin{equation}
\tau^{mag} (u) = - \frac {ik}{2\pi} \ln (u),
\label{eq:II71}
\end{equation}
with $k$ the number of massless hypermultiplets. This yields the 
monodromy 
\begin{equation}
	\left( \begin{array}{cc} 1 & k  \\
			 0 & 1 
	\end{array} \right),
\label{eq:II72}
\end{equation}
or, in Kodaira's notation, a monodromy of type $A_{k-1}$. Notice
that the difference in sign between the type $D$, and the type
$A$ monodromies, reflects that we are obtaining type $A$ for
infrared free theories, and type $D$ (that is $D_0$, $D_1$,
$D_2$, and $D_3$) for assymtotically free theories (notice the
sign in (\ref{eq:II71})) \cite{GMS}. Now, we should wonder about the meaning
of (\ref{eq:II71}). Recall that our analysis relies upon the
wilsonian coupling constant, so the meaning of $u$ in
(\ref{eq:II71}) must be related to the scale in the $U(1)$
theory, i. e. the vacuum expectation value for the scalar field
in the photon multiplet or, more properly, in the dual photon
multiplet. This vacuum expectation value gives a mass to the
hypermultiplets through the standard Yukawa coupling, so the
singularity of (\ref{eq:II71}) should be expected at $u=0$, with $u$
proportional to the mass of the hypermultiplet. Fortunately, we
do know which hypermultiplet we should consider: the one
defined by the monopole of the theory. In fact, we should
rewrite (\ref{eq:II71}) as
\begin{equation}
\tau^{mag}(u) = - \frac {ik}{2\pi} \ln (M(u)),
\label{eq:II73}
\end{equation}
with $M(u)$ the mass of the monopole, and consider (\ref{eq:II73})
perturbatively around the point $u_0$, where 
\begin{equation}
M(u_0)=0.
\label{eq:II74}
\end{equation}
Therefore, we conclude that a singularity of $A_0$ type will appear whenever
the mass of the monopole equals zero. The nature of the point $u_0$ 
is quite clear from a
physical point of view: the magnetic effective coupling constant
is zero, as can be seen from (\ref{eq:II73}), so that the dual
electric coupling should become infinity. But the point where the
coupling constant is infinity is by definition the
scale $\Lambda$ of the theory; then, $u_0=\Lambda$. 
  
Now, it
remains to discover how many singularities of $A_0$ type are there. In
principle, a single point where the monopole becomes massless
should be expected (the $u_0=\Lambda$ point); however, as
mentioned in section \ref{sec:22}, the $U(1)_R$ symmetry is
acting on the moduli space as a ${\bf Z}_2$ transformation.
Therefore, in order to implement this symmetry, an extra singularity of $A_0$
type must exist. The simplest solution for the $N_f=0$ theory,
with $SU(2)$ gauge group, corresponds to an elliptic fibration
over $\IP^1$, the compactified $u$-plane, with three singular
points, of type 
\begin{equation}
D_0; \: \: A_0, \: A_0,
\label{eq:II75}
\end{equation}
with $D_0$ the singularity at infinity, and the two $A_0$
singularities at the points $\pm \Lambda$, with $\Lambda$ the
scale of the theory. 
  
What about the inclusion of flavors? In
this case, we know that $D_0$ in (\ref{eq:II75}) is replaced by
$D_{N_f}$. The case $N_f=2$ should be clear, as $D_2$ is
equivalent to two $A_1$ singularities and therefore, we should expect
\begin{equation}
D_2; \: \: A_1, \: A_1.
\label{eq:II76}
\end{equation}
The singularities of $A_1$ type indicate that two hypermultiplets
become massless. Another simple case is that with $N_f=4$,
where there is a trivial monodromy $D_4$, which is now the
monodromy around the origin. The two other cases of
assymptotically free theories can be obtained through decoupling
arguments, and taking into account the residual $U(1)_R$ symmetry. 
The results are \cite{GMS}
\begin{eqnarray}
D_1 & ; & \: \: A_0, \: A_0, \: A_0, \nonumber \\
D_3 & ; & \: \: A_0, \: A_3.
\label{eq:II77}
\end{eqnarray}

Now, with these elliptic fibrations, we shoud consider the
complex structure. As we know from Kodaira's argument for the
$N_f=0$ case, the $A_0$ singularities correspond to a rational
curve with a double singular point; as we know that this double
singularity appears at $u=\pm \Lambda$, the simplest guess for
the corresponding complex structure is, with $\Lambda=1$,
\begin{equation}
y^2= x^3-x^2u+x.
\label{eq:282}
\end{equation}
  
The curve (\ref{eq:283}), for generic $u$, does not have singular
points. Recall that for a curve defined by $f(x,y;u)=0$, the
singular points are those such that
\begin{eqnarray}
F & = & 0, \nonumber \\
F_x = F_y & = & 0,
\label{eq:283}
\end{eqnarray}
with $F_x$ and $F_y$ the derivatives with respect to $x$ and $y$,
respectively. The genus of the curve can be obtained using
Riemann's theorem,
\begin{equation}
g = \frac {(n-1)(n-2)}{2} - \sum_p \frac {r_p(r_p-1)}{2}, 
\label{eq:284}
\end{equation}
where the sum is over singular points, $r_p$ is the order of the
singularity, and $n$ in (\ref{eq:284}) is the degree of the
polynomial $F$, defining the curve. So, for generic $u$, we get,
for (\ref{eq:282}), $g=1$.
  
Now, for $u=\pm 2$, we have a singular point satisfying
(\ref{eq:283}), namely
\begin{eqnarray}
y & = & 0, \nonumber \\
x & = & \frac {u}{2}.
\label{eq:285}
\end{eqnarray}
This is a double point and therefore, using (\ref{eq:284}), we
get $g=0$. From Kodaira's classification, we know that at this
points we get two singularities of $A_0$ type. Notice also that
at the origin, $u=0$, we have the curve $y^2=x^3+x$, which is of
genus one, since there are no singular points. Moreover, if we
take $\Lambda=0$, we get the curve
\begin{equation}
y^2 = x^3 - x^2 u.
\label{eq:286}
\end{equation}
This curve has a double point at $x=y=0$ for generic $u$. Using
(\ref{eq:284}), we now get genus equal zero. Thus, the curve
(\ref{eq:282}) satisfies all the properties derived above.
  
The curve (\ref{eq:282}) has a point at $x=y=\infty$. In order to
compactify the curve, we must add the point at infinity. This can
be done going to the projective curve
\begin{equation}
zy^2 = x^3 - zx^2 u + z^2 x.
\label{eq:287}
\end{equation}
The region at infinity of this curve is defined by $z=0$. The
curve, in the three dimensional $R \rightarrow 0$ limit, can be
described by (\ref{eq:282}), but with $\hbox{Vol}(E_u)=\infty$.
Next, we will see that this limit is equivalent to deleting the
points at infinity of (\ref{eq:287}), i. e., the points with
$z=0$. In fact, for $z \neq 0$ we cab define a new variable,
\begin{equation}
v=x-zu,
\label{eq:288}
\end{equation}
and write (\ref{eq:287}) as
\begin{equation}
zy^2 = x^2 v +z^2 x.
\label{eq:289}
\end{equation}
We can interpret (\ref{eq:289}) as defining a surface in the
projective space $\IP^3$, but (\ref{eq:289}) is in fact the
Atiyah-Hitchin space in homogeneous coordinates. Thus, we
conclude that the $R \rightarrow 0$ limit is equivalent to
deleting the points at infinity of the curves $E_u$ defined by
(\ref{eq:282}).
  
We can see this phenomena in a different way as follows. The
representation (\ref{eq:282}) of the Atiyah-Hitchin space is as
an elliptic fibration, so that we have selected one complex
structure. However, we can yet rotate in the space of complex
structures, preserving the one selected by the elliptic
fibration. This defines a $U(1)$ action. This $U(1)$ action must
act on $E_u$; however, this is impossible if $E_u$ is a compact
torus. But when we delete the point at infinity, and pass to the
projective curve (\ref{eq:289}), we have a well defined $U(1)$
action \cite{SW3d},
\begin{eqnarray}
x & \rightarrow & \lambda^2 x, \nonumber \\
y & \rightarrow & \lambda y, \nonumber \\
v & \rightarrow & \lambda^{-2} v.
\label{eq:290}
\end{eqnarray}
Only a ${\bf Z}_2$ subgroup of this action survives on $u$:
\begin{equation}
u \rightarrow \lambda^2 x - \lambda^{-2} v \equiv \hat{\lambda}
u,
\label{eq:291}
\end{equation}
which means
\begin{equation}
\lambda^2 = \lambda^{-2} = \hat{\lambda},
\end{equation}
i. e., $\lambda^4=1$ or $\hat{\lambda}^2=1$. This ${\bf Z}_2$
action moves $u \rightarrow -u$, and is the only part of the
$U(1)$ action surviving when we work in the four dimensional
limit. More simply, at $z=0$, i. e., at infinity, in the
projective sense, $v=x$, and we get $\lambda^2=\lambda^{-2}$, and
the ${\bf Z}_4$ symmetry of (\ref{eq:282}) becomes
\begin{eqnarray}
y & \rightarrow & iy, \nonumber \\
x & \rightarrow & -x, \nonumber \\
u & \rightarrow & -u.
\label{eq:292}
\end{eqnarray}
Notice also the relation between $\Lambda$ and the breaking of
$U(1)$. In fact, for $\Lambda=0$ we have
\begin{equation}
y^2 = x^3 -x^2 u,
\label{eq:293}
\end{equation}
which is invariant under 
\begin{eqnarray}
y & \rightarrow & \lambda^3 y, \nonumber \\
x & \rightarrow & \lambda^2 x, \nonumber \\
u & \rightarrow & \lambda^2 u.
\label{eq:294}
\end{eqnarray}

\subsection{Effective Superpotentials.}

Maybe the most spectacular result derived from the Seiberg-Witten
solution to $N\!=\!2$ supersymmetric theories is the first
dynamical proof of electric confinement. In order to properly
understand this proof, we need first to go through the recent
history of confinement. The simplest physical picture of
confinement is that of dual BCS superconductivity theory \cite{P3d,Man,tHconf}. In that
picture, a confining vacua is to be represented as the dual of
the standard superconducting vacua, which is characterized by the
condensation of Cooper pairs. In ordinary
superconductivity we find, under the name of Meisner effect, the
mechanism for magnetic confinement. In a superconducting vacua, a
monopole-antimonopole pair creates a magnetic flux
tube that confines them. The relativistic Landau-Ginzburg
description of superconductivity was first introduced by Nielsen
and Olesen \cite{NO}, where vortices in the Higgs phase are interpreted as
Meisner magnetic flux tubes. The order parameter of
the phase is the standard vacuum expectation value of the Higgs
field; in this model, simply a scalar coupled to the $U(1)$
electric-magnetic field. The confined monopoles would be $U(1)$
Dirac monopoles, and the magnetic string is characterized by the
Higgs mass of the photon. The dual version of this picture is in
fact easy to imagine. We simply consider a dual photon, or
dual $U(1)$ theory, now coupled to magnetic Higgs matter, a field
representing the magnetic monopoles with magnetic $U(1)$ charge,
and we look for a dual Higgs mechanism that, by a vacuum
expectation value of the monopole field, will induce a Higgs mass
for the dual magnetic photon. This mass gap will characterize the
confinement phase. As the reader may realize, this whole picture
of confinement is based on Higgs, or dual Higgs mechanisms for
abelian gauge theories; however, in standard QCD, we expect
confinement to be related to the very non abelian nature of the
gauge groups. Indeed, only non abelian gauge theories are
assymptotically free, and would possess the infrared slavery,
or confinement, phenomena. Moreover, in a pure non abelian gauge
theory, we do not have the right topology to define stable `t
Hooft-Polyakov abelian monopoles, so the extesion of the
superconductivity picture to the $N\!=\!0$ pure Yang-Mills
theory, or standard QCD, is far from being direct. Along the last
two decades, with `t Hooft and Polyakov as leaders, some pictures
for confinement have been sugested. Perhaps, the main steps in
the story are
\begin{itemize}
	\item[{i)}] $2+1$ Polyakov quantum electrodynamics \cite{P3d}.
	\item[{ii)}] `t Hooft ${\bf Z}(N)$ duality relations \cite{tHconf}.
	\item[{iii)}] `t Hooft twisted boundary conditions \cite{tHtw}.
	\item[{iv)}] `t Hooft abelian projection gauge \cite{tHap}.
\end{itemize}
Concerning $i)$, we have already described the relevant dynamics in
chapter I. Let us therefore now consider 
the other points. Concerning $ii)$, the general idea is
dealing with the topology underlying pure $SU(N)$ Yang-Mills
theory, namely
\begin{equation}
\Pi_1(SU(N)/U(1)) \simeq {\bf Z}(N).
\label{eq:II78}
\end{equation}
This is the condition for the existence of magnetic 
${\bf Z}(N)$ vortices. The `t Hooft loop $B(C)$ is the magnetic
analog of the Wilson loop $A(C)$, and was defined for creating a
${\bf Z}(N)$ magnetic flux tube along the path $C$. The Wilson
criteria for confinement, $A(C)$ going like the area, has
now its dual in $B(C)$ behaving like the perimeter, reproducing
again the picture that dual Higgs is equivalent to confinement.
The duality relations established by `t Hooft reduce to 
\begin{equation}
A(C)B(C')=e^{2\pi i \nu(C,C')/N}B(C')A(C),
\label{eq:II79}
\end{equation}
where $\nu(C,C')$ is the link number between the loops $C$ and
$C'$. From (\ref{eq:II78}), the different posible phases
compatible with duality were obtained. A way to make more
quantitative the previous picture was also introduced by `t
Hooft, by means of twisted boundary conditions in a box. Some of 
the main ingredients were already introduced in chapter I, but we
will come back to them later on. In what follows of this section
we will mainly be interested in the abelian projection gauge.
  
The idea of the abelian projection gauge was originally that of
defining a unitary gauge, i. e., a gauge absent of ghosts. The
simplest way to do it is first reducing the theory to an abelian
one, and then fixing the gauge, which is (in the abelian theory)
a certainly easier task. Using a formal notation, if $G$ is the
non abelian gauge group, and $L$ is its maximal abelian subgroup,
then the non abelian part is simply given by $G/L$, so that we
can take, as the degrees of freedom for the abelian gauge theory,
the space $R/(G/L)$, where $R$ generically represents the whole
space of gauge configurations. Now, the theory defined by
$R/(G/L)$, is an abelian theory, and we can fix the gauge, going
finally to the unitary gauge, characterized by $R/G=L\backslash
R/(G/L)$, Now, two questions arise, concerning 
the content of the intermediate abelian
theory, $R/(G/L)$, and the more important point of how such a theory
should be defined. In order to fix the non abelian part of the
gauge group, i. e., the piece $G/L$, `t Hooft used the following
trick \cite{tHap}: let $X$ be a field that we can think of as a functional of
$A$, $X(A)$, or an extra field that will be decoupled at the end.
For the time being, we simply think of $X$ as a functional,
$X(A)$. We will require $X(A)$ to transform under the adjoint
representation, i. e.,
\begin{equation}
X(A) \rightarrow g X(A)g^{-1}.
\label{eq:II80}
\end{equation}
Now, the gauge condition that fixes the non abelian part of the
gauge group is 
\begin{equation}
X(A) =  \left( \begin{array}{ccccc} \lambda_l &   &   &     &    \\
					      &   &   &     &    \\
					      &   & \ddots   &     &    \\
					      &   &   &     &    \\
					      &   &   &     & \lambda_N
	\end{array} \right). 
\label{eq:II81}
\end{equation}
  
Indeed, if $X(A)$ is diagonal, the residual group is just the
maximal abelian subgroup. Notice that $X(A)$ is playing a similar
role to a Higgs field in the adjoint representation, and
(\ref{eq:II81}) is what we will interpret as a vacuum expectation
value, breaking the $G$ symmetry to its maximal abelian subgroup.
As in the standard Higgs mechanism, now the degrees of freedom
are the diagonal parts of the gauge field, $A_{\mu}^{(ij)}$, that
transform as $U(1)$ charged particles. In addition, we have the $N$
scalars fields $\lambda$, appearing in (\ref{eq:II81}).
Summarizing, the particle content we get in the maximal abelian
gauge is
\begin{itemize}
	\item[{i)}] $N-1$ photons, $A^{(ii)}$.
	\item[{ii)}] $\frac {1}{2} N(N-1)$ charged particles,
$A^{(ij)}$.
	\item[{iii)}] $N$ scalar fields, $\lambda_i$.
\end{itemize}
Notice that (\ref{eq:II81}) does not require the $\lambda_i$ to
be constant; in fact, $\lambda_i$ are fields depending on the
spacetime position. Another important aspect of (\ref{eq:II81})
is that, by means of this maximal abelian gauge we are not
introducing, in principle, any form of potential for the
$\lambda_i$ fields, so that their expectation values are a priori
undetermined. Concerning the previous spectrum, charged particles
of type $ii)$ can be considered formally massive, with the mass
being proportional to $\lambda_i-\lambda_j$, as is the case in
the standard Higgs mechanism.
  
The spectrum $i)$, $ii)$ and $iii)$ is not complete. Extra
spectrum, corresponding to singularities of the maximal abelian
gauge, (\ref{eq:II81}), is also allowed. These singularities
correspond to points in spacetime, where
$\lambda_i(x)=\lambda_{i+1}(x)$, i. e., where two eigenvalues
coincide. We have impossed that $\lambda_i>\lambda_{i+1}$, i. e.,
the eigenvalues of (\ref{eq:II81}) are ordered. These
singularities are point-like in three dimensions, and $d-3$
dimensional for spaces of dimension $d$. It is easy to see that 
these singularities of the gauge (\ref{eq:II81}) are 
`t Hooft-Polyakov monopoles. Once we have this set of degrees of
freedom to describe the non abelian theory, we may proceed to
consider the phenomenum of confinement, following in essence the
same philosophy as in abelian superconductors. `t Hooft's rules
of construction are:
\begin{itemize}
	\item[{R1}] Eliminate the electric charges. This
means constructing an effective lagrangian, where the
``massive'' electric particles $A^{(ii)}$ have been integrated
out inside loops.
	\item[{R2}] Perform duality transformations on the
effective lagrangian obtained upon the above integration of the
electric charges, going to dual photons. These dual photons
should interact with the charged monopoles by ordinary vertices,
coupling the dual photon to two monopoles. The interaction
between monopoles is certainly not reduced to the the single
exchange of dual photons; there is in practice a missing link
connecting the dual photon-monopole vertices, and the effective
lagrangian, and which is played by the $\lambda$-fields: the
$\Gamma_{eff}$ action depends also on the $\lambda$-fields, that
have Yukawa coupling with the charged $A^{(ij)}$ particles,
running inside the loop. As we dualize, we should also take into
account duality on these fields $\lambda_i$. In fact, this should
be the most relevant part of our story, as it is the potential
interaction between monopoles and the dual $\lambda_i$ fields
what naturally leads to next rule.
	\item[{R3}] The expectation value $<M>$, for the theory
obtained in R2, must be computed. In fact, this vacuum
expectation value should be obtained after minimizing the theory
with respect to the $\lambda_i$ field values.
\end{itemize}
  
In spite of the beatiful physical structure underlying `t Hooft's
approach, this program is far from being of practical use in
standard QCD or pure Yang-Mills theory. However, progress in
lattice computations is being made at present.
  
After this introduction to `t Hooft's abelian projection gauge,
let us come back to the simpler example of $N\!=\!2$ pure
Yang-Mills theory to find out the validity of the above rules.
The careful reader wil have already found some similarities in
our discussion and the way the Seiberg-Witten solution for
$N\!=\!2$ supersymmetric Yang-Mills has been presented. In fact,
in the $N\!=\!2$ theory, the $X$ field can simply be interpreted
as the Higgs field in the adjoint, breaking $SU(2)$ to $U(1)$ on
generic points of the moduli (for a group of higher rank, $r$,
the breaking is down to $U(1)^r$). Moreover, we also have the
spectrum of `t Hooft-Polyakov monopoles and, according to degrees
of freedom, we are certainly quite close to the abelian
projection picture; however, we should be careful at this point.
In `t Hooft's abelian projection, it was not assumed at any
moment that we must be at a Higgs phase with well defined massive
monopoles. The type of monopoles we find in the abelian
projection gauge are not massive in the usual sense and,
moreover, they have not finite size but are simply point like
singularities.
  
Rule R1 is almost accomplished through the Seiberg-Witten
solution \cite{SW,SW2}. In fact, we can consider the effective lagrangian
obtained from $\Gamma_{eff}(A_{\mu}^0,a)$, where $A_{\mu}^0$
represents the photon, and $a$ is the scalar field in the
$N\!=\!2$ hypermultiplet (notice that this effective lagrangian
is constrained to be $N\!=\!2$ invariant). For each value of
$u=\frac {1}{2} <\hbox{tr} \phi^2>$, the vacuum expectation value
of the field $a$, in the perturbative regime, is simply
\begin{equation}
a(u) = \sqrt{2u}.
\label{eq:II82}
\end{equation}
  
The effective lagrangian contains only one loop logarithmic
contributions (see equation (\ref{eq:II35})), and instanton
effects. The instanton and multiinstanton contributions
contribute each with four fermionic zero modes, as we kill the
four zero modes associated with superconformal transformations.
The expansion of the effective lagrangian in perturbative and non
perturbative effects can be done in the weak coupling regime and,
if we know how to perform the duality trasnformation, we can
start obtaining non trivial information on the strong coupling
regime. Let us formally denote through $\Gamma_{eff}(A^D,a_D)$ the dual
effective lagrangian. In the dual perturbative regime, the
effective lagrangian is an expansion in one loop terms,
corresponding to light magnetic monopoles, and non perturbative
higher order terms. From the moduli space point of view, the dual
perturbative expansion should appear as a good description of the
infrared region, i. e., for values of $u$ such that the
electric constant is large, which are points at the neighbourhood
of $u \simeq \Lambda$, with $\Lambda$ the dynamically generated
scale. To complete the equivalent dual description, the
equivalent to expression (\ref{eq:II82}) for the dual variable $a_D$
should be constructed; impossing $N\!=\!2$
supersymmetry, we obtain that the dual theory has a coupling 
\begin{equation}
a_D M\tilde{M},
\label{eq:II83}
\end{equation}
of Yukawa type for monopoles. Then, $a_D$ is the mass of the
monopole, in the very same way as the mass of $W^{\pm}$
particles, in the standard Higgs mechanism, is given by $a$. We
can now write a general formula for electrically and magnetically
charged particles,
\begin{equation}
M(n_e,n_m)= |n_e a+n_m a_D|.
\label{eq:II84}
\end{equation}
Here, we have only motivated equation (\ref{eq:II84}) from
physical arguments but, as we will see, the mathematical and
supersymmetric meaning of (\ref{eq:II84}) goes far beyond the
scope of the simple argument we have used.
  
Coming back to our problem of discovering $a_D(u)$, a proper
description will require some results on K\"{a}hler geometry. In
fact, we know that the metric on the $u$ moduli, is certainly
K\"{a}hler with respect to the complex structure distinguished by
the elliptic fibration representation of the $N\!=\!4$ three
dimensional moduli space. If it has a K\"{a}hler structure, the
corresponding K\"{a}hler potential can be defined through
\begin{equation}
g_{u\bar{u}} = \hbox {Im} \left( \frac {\partial^2 K}{\partial u
\partial \bar{u}} \right).
\label{eq:II85}
\end{equation}
This K\"{a}hler potential can be read out from the effective
$N\!=\!2$ low energy action. In fact, as a general statement, the
metric on the moduli space is given by the quadratic terms of the
effective low energy lagrangian. Now, for $N\!=\!2$ the
lagrangian can be written in terms of the so called
prepotential as follows:
\begin{equation}
{\cal L} = \int d^4 \theta {\cal F}({\cal A}),
\label{eq:II86}
\end{equation}
where ${\cal A}$ is an $N\!=\!2$ superfield, which is holomorphic
or, in supersymmetric language, depends only on chiral
fields. The K\"{a}hler potential is derived from ${\cal
F}$ as
\begin{equation}
K = \hbox {Im} \left( \frac {\partial {\cal F}}{\partial {\cal
A}} \cdot \bar{\cal A} \right),
\label{eq:II87}
\end{equation}
from which (\ref{eq:II85}) becomes
\begin{equation}
g_{u\bar{u}} = \hbox {Im} \left( \frac {\partial a_D}{\partial u}
\frac {\partial \bar{a}}{\partial \bar{u}} \right),
\label{eq:II88}
\end{equation}
where we have defined
\begin{equation}
a_D \equiv \frac {\partial {\cal F}}{\partial a},
\label{eq:II89}
\end{equation}
in the sense of lower components. Using (\ref{eq:II86}) and
(\ref{eq:II87}) we get, for the metric,
\begin{equation}
d s^2 = \hbox {Im} \frac {\partial^2 {\cal F}}{\partial a^2} da
d\bar{a},
\label{eq:II90}
\end{equation}
and therefore we can identify $\frac {\partial^2 {\cal
F}}{\partial a^2}$ with $\tau(u)$ or, equivalently, 
\begin{equation}
\tau(u) = \frac {da_D}{da}.
\label{eq:II91}
\end{equation}
Notice that equation (\ref{eq:II91}) is perfectly consistent with
what we expect for the definition of $a_D$, as it provides the mass
of the monopole. In the perturbative regime, we know that it behaves
like $\hbox {Im} \tau \cdot a \simeq \frac {a}{g^2}$. Therefore,
(\ref{eq:II89}) is the right generalization. Fortunatelly, thanks
to (\ref{eq:II91}) and relation (\ref{eq:II62}), we get a
definite representation of $a_D(u)$ in terms of the
Seiberg-Witten elliptic fibration solution,
\begin{equation}
\frac {da_D}{du} = \oint_{\gamma} \varphi(z;u)dz,
\label{eq:II92}
\end{equation}
where $\varphi$ is the holomorphic differential of $E_u$, which
is given by
\begin{equation}
\varphi(z;u) = \frac {dx}{y}.
\label{eq:II93}
\end{equation}
Now that we have a candidate for $a_D(u)$, we can continue our
analysis following 't Hooft's rules (in fact, $a_D(u)$ or,
equivalently, how to define the dual scalar field $\lambda_i^D$,
was a missing part in `t Hooft's program). Next, we want to work
out the dynamics of the monopoles. Until now, we have used
$N\!=\!2$ dynamics, so that the fields $a$ and $a_D$ are part of
our original lagrangian, and not a gauge artifact, as in `t
Hooft's abelian projection gauge. However, if we softly break
$N\!=\!2$ to $N\!=\!1$ \cite{SW} adding a mass term for the scalar fields,
\begin{equation}
m \hbox{tr}\Phi^2,
\label{eq:II94}
\end{equation}
then for large enough $m$ the low energy theory is $N\!=\!1$,
where the interpretation of the fields $a$ and $a_D$ should
become closer and closer to the fields of the
abelian projection. The soft breaking term (\ref{eq:II94}) should
reproduce `t Hooft's hidden dynamics governing the
$\lambda$-fields. In fact, there is a simple procedure, discovered
by Seiberg and Witten, to do that. The effect of (\ref{eq:II94})
on the low energy description of the theory is to add a
superfield ${\cal U}$, with lower component $u$, such that
$<u>=<\hbox {tr} \phi^2>$, with superpotential
\begin{equation}
W=m \: {\cal U}.
\label{eq:II95}
\end{equation}
This extra term contains in fact the dynamics about $a_D$ fields
we are looking for, so we can write (\ref{eq:II95}) as
\begin{equation}
W=m \: {\cal U}(a_D),
\label{eq:II96}
\end{equation}
and interpret it as a lagrangian term for $a_D$. The monopole
dynamics is then controlled by a superpotential of type
\begin{equation}
W=a_DM \tilde{M} + m \: {\cal U}(a_D),
\label{eq:II97}
\end{equation}
where the first term is the $N\!=\!2$ Yukawa coupling. Now, in
order to fulfill rule R3, we only need to minimize the
superpotential (\ref{eq:II97}). Clearly, we get two minima with
monopole vacuum expectation value given by
\begin{equation}
<M> = \pm \left( \frac {\partial {\cal U}}{\partial a_D}
\right)^{1/2} m,
\label{eq:II98}
\end{equation}
which is the desired proof of confinement. `t Hooft's program is
then completed. In order to extend this approach to non
supersymmetric theories, we can still use the the trick of adding
a mass term for the $X$ field; however, because of the lack of
holomorphy, no translation of such procedure in the form of
(\ref{eq:II96}) is possible.
  
Instead of using the relation for ${\cal U}(a_D)$, we can try to
get a more direct geometrical interpretation of (\ref{eq:I95}):
let us work with the curve (\ref{eq:282}), and consider the
points $A$ and $B$ with $y=0$,
\begin{equation}
x^2-xu+\Lambda^2=0.
\label{y=0}
\end{equation}
Now, we can define the function
\begin{equation}
{\cal U}(x) = x + \frac {\Lambda^2}{x}.
\label{u}
\end{equation}
The purpose of this function is giving a value of ${\cal U}$, such that
$x$ is one of the crossing points. Obviously, $U(x)$ posseses two
minima, at
\begin{equation}
x= \pm \Lambda,
\end{equation}
and therefore the superpotential $m \: {\cal U}$ has two minina,
at $\pm \Lambda_1$, with $\Lambda_1$ the scale of the $N=1$
theory. Of course, the minima of ${\cal U}(x)$ take place when
the tow points $A$ and $B$ coincide, i. e., at the singular nodal
curves. Now, we can use the following heuristic argument to find
out what happens in the three dimensional $R \rightarrow 0$
limit. In projective coordinates, the region at infinity of
(\ref{eq:282}) is 
\begin{equation}
zy^2 = x^3 - zx^2 u + \Lambda^2 x z^2,
\label{p}
\end{equation}
at $z=0$. If we delete the infinity point, i. e., the
intersection of the projective curve $C$ defined by (\ref{p}) and
$H_{\infty}=\{(x,y,0)\}$, and we then put $x^3=0$ in (\ref{p}) we
get, instead of (\ref{u}) \cite{SW3d},
\begin{equation}
{\cal U}_{3D}(x) = \frac {\Lambda_{N=2}^2}{x},
\end{equation}
with $\Lambda_{N=2}^2$ the $N=2$ three dimensional scale.

\section{Chapter III}

Taking into account the enormous amount of good reviews and books
\cite{GSW,Vafa,Polchinski,Kiri} in string theory, we will reduce ourselves in this section to
simply stablishing some notation and motivating fundamental
relations as mass formulas.

\subsection{Bosonic String.}
\label{sec:III1}

\subsubsection{Classical Theory.}

Let us start considering classical bosonic string theory in flat
Minkowski spacetime. This physical system is characterized by the
lagrangian
\begin{equation}
{\cal L} = - \frac {T}{2} \int d^2 \sigma \sqrt{h} h^{\alpha
\beta} \partial_{\alpha} X \partial_{\beta} X,
\label{eq:III1}
\end{equation}
where $h^{\alpha \beta}$ is the worldsheet metric. The equations
of motion, with respect to $h^{\alpha \beta}$, imply that
\begin{equation}
T_{\alpha \beta} = - \frac {2}{T} \frac {1}{\sqrt{h}} \frac
{\delta S}{\delta h^{\alpha \beta}}=0.
\label{eq:III2}
\end{equation}
The parameter $T$ in (\ref{eq:III1}) has units of squared mass,
and can be identified with the string tension,
\begin{equation}
T= \frac {1}{2 \pi \alpha'}.
\label{eq:III3}
\end{equation}
Using the Weyl invariance of (\ref{eq:III1}), the gauge 
\begin{equation}
h_{\alpha \beta} = \eta_{\alpha \beta} = \left(
\begin{array}{cc} -1 & 0 \\ 0 & 1 \end{array} \right)
\label{eq:III4}
\end{equation}
can be chosen. In this gauge, the equations of motion for
(\ref{eq:III1}) become
\begin{equation}
\Box X = 0.
\label{eq:III5}
\end{equation}
Defining light cone coordinates,
\begin{eqnarray}
\sigma^- & = & \tau - \sigma, \nonumber \\
\sigma^+ & = & \tau + \sigma, 
\label{eq:III6}
\end{eqnarray}
the generic solution to (\ref{eq:III5}) can be written as
\begin{equation}
X^{\mu} = X_R^{\mu}(\sigma^-) + X_L^{\mu}(\sigma^+).
\label{eq:III7}
\end{equation}
  
Now, we will introduce open and closed strings. We will first
work out the case of the closed bosonic string; in this case, we
impose periodic boundary conditions,
\begin{equation}
X^{\mu}(\tau,\sigma)=X^{\mu}(\tau,\sigma+\pi).
\label{eq:III8}
\end{equation}
  
The solution to (\ref{eq:III5}), compatible with these boundary
conditions, becomes
\begin{eqnarray}
X_R^{\mu} & = & \frac {1}{2} x^{\mu} + \frac {1}{2}(2 \alpha')
p^{\mu}(\tau -\sigma) + i \sqrt{\frac {\alpha'}{2}} \sum_{n \neq0}
\frac {1}{n} \alpha_n^{\mu} e^{-2in(\tau-\sigma)}, \nonumber \\
X_L^{\mu} & = & \frac {1}{2} x^{\mu} + \frac {1}{2}(2 \alpha')
p^{\mu}(\tau +\sigma) + i \sqrt{\frac {\alpha'}{2}} \sum_{n \neq0}
\frac {1}{n} \tilde{\alpha}_n^{\mu} e^{-2in(\tau+\sigma)}. 
\label{eq:III9}
\end{eqnarray}
Using this Fourier decomposition we get, for the hamiltonian, 
\begin{equation}
H= \frac {1}{2} \left[ \sum_{- \infty}^{\infty} \alpha_{m-n}
\alpha_n + \sum_{-\infty}^{\infty}
\tilde{\alpha}_{m-n}\tilde{\alpha}_n \right],
\label{eq:III10}
\end{equation}
where we have used the notation
\begin{equation}
\alpha_0^{\mu} = \sqrt{\frac {\alpha'}{2}}p^{\mu}.
\label{eq:III11}
\end{equation}
Using now (\ref{eq:III2}), we get the classical mass formula
\begin{equation}
M^2 = \frac {2}{\alpha'} \sum_{n=1}^{\infty} (\alpha_{-n}
\alpha_n + \tilde{\alpha_{-n}}\tilde{\alpha_n} ).
\label{eq:III12}
\end{equation}
The constraint (\ref{eq:III2}) also implies that the left and
right contributions to (\ref{eq:III12}) are equal. Using the
standard quantization rules,
\begin{eqnarray}
[\alpha_m^{\mu},\tilde{\alpha}_n^{\nu}] & = & 0, \nonumber \\ 
\left[\alpha_m^{\mu},\alpha_n^{\nu}\right]         & = & m \delta_{m+n}\eta^{\mu \nu}, \nonumber \\
\left[\tilde{\alpha}_m^{\mu},\tilde{\alpha}_n^{\nu}\right] & = & m \delta_{m+n}\eta^{\mu \nu}, \nonumber \\
\left[x^{\mu},p^{\nu}\right] & = & i \eta^{\mu \nu},
\label{eq:III13}
\end{eqnarray}
and taking into account the normal ordering factors we get, for
$\alpha'= \frac {1}{2}$,
\begin{equation}
M^2= -8a + 8 \sum_{n=1}^{\infty}
\tilde{\alpha}_{-n}\tilde{\alpha}_n = -8a + 8 \sum_{n=1}^{\infty}
\alpha_{-n} \alpha_n.
\label{eq:III14}
\end{equation}
  
Two things are left free in deriving (\ref{eq:III14}), the
constant $a$, defining the zero point energy, and the number of
dimensions of the target space. The classical way to fix these
constants is impossing Lorenz invariance in the light cone gauge,
where physical degrees of freedom are reduced to transversal
oscillations. The result, for the closed bosonic string, is that
$a$ should equal one and the number of dimensions should be $26$. 
  
From (\ref{eq:III14}), we can easily deduce the spectrum of
massless states. First of all, we have a tachyon with no
oscillator modes, and squared mass negative $(-8)$. The massless
modes are of the type
\begin{equation}
\alpha_{-1}^{\mu} \alpha_{-1}^{\mu} |0>.
\label{eq:III15}
\end{equation}
  
To discover the meaning of these modes, we can see the way they
transform under $SO(24)$ in the light cone gauge; then, we get
three different types of particles: gravitons for the symmetric
and traceless part, a dilaton for the trace part and, finally,
the antisymmetric part.

\subsubsection{Background Fields.}

The simplest generalization of the worldsheet lagrangian
(\ref{eq:III1}) corresponds to including background fields. The
obvious is the $G^{\mu \nu}$ metric of the target spacetime,
\begin{equation}
S_1=- \frac {T}{2} \int d^2\sigma \sqrt{h} h_{\alpha \beta}
G^{\mu \nu} (X) \partial_{\alpha} X_{\mu} \partial_{\beta}
X_{\nu}.
\label{eq:III16}
\end{equation}
  
However, not any background $G^{\mu \nu}$ is allowed, since we
want to preserve Weyl invariance on the worldsheet. Scale
invariance, for the two dimensional system defined by
(\ref{eq:III16}) is equivalent, from the quantum field theory
point of view, to requiring a vanishing $\beta$-function. At one
loop, the $\beta$-function for (\ref{eq:III16}) is given by
\begin{equation}
\beta = - \frac {1}{2\pi} R,
\label{eq:III17}
\end{equation}
for $\alpha'=\frac {1}{2}$, and with $R$ the Ricci tensor of the
target spacetime. Therefore, the first condition we require on
allowed spacetime backgrounds is to be Ricci flat manifolds. We
will allow the addition of extra manifolds to (\ref{eq:III16}), namely the
spectrum of massless particles of the bosonic closed string,
\begin{equation}
S=S_1 - \frac {T}{2} \int d^2\sigma \epsilon^{\alpha \beta}
\partial_{\alpha} X^{\mu} \partial_{\beta} X^{\nu} B_{\mu \nu}(X)
+ \frac {1}{4} \int d^2 \sigma \sqrt{h} \Phi(X) R^{(2)},
\label{eq:III18}
\end{equation}
where $R^{(2)}$ in (\ref{eq:III18}) is the worldsheet curvature.
$\alpha'$ does not appear in the last term due to dimensional
reasons (the first two terms in (\ref{eq:III18}) contain the
$X^{\mu}$ field, which has length units). 
  
Notice that for a
constant dilaton field, the last term in (\ref{eq:III18}) is
simply
\begin{equation}
\chi \cdot \Phi,
\label{eq:III19}
\end{equation}
with $\chi$ the Euler number; in terms of the genus, $g$, for a
generic Riemann surface the Euler number is simply given by
\begin{equation}
\chi = 2-2g.
\label{eq:III20}
\end{equation}
Thus, the powers of $\Phi$ in the partition function behave like
$2-2g$. This topological number possesses a nice meaning in
string theory: it is equal to the number of vertices joining
three closed strings, needed to build up a Riemann surface of
genus $g$. This naturally leads to a precise physical meaning of
the dilaton background field: it is the string coupling constant,
\begin{equation}
g=e^{\Phi}.
\label{eq:III21}
\end{equation}
  
Once the background fields in (\ref{eq:III18}) have been added,
the condition of Weyl invariance generalizes to vanishing
$\beta$-funtions for $G$, $B$ and $\Phi$. At one loop, they are
\begin{eqnarray}
R_{\mu \nu} + \frac {1}{4} H_{\mu} ^{\lambda \rho}H_{\nu \lambda
\rho}- 2 D_{\mu} D_{\nu} \Phi & = & 0, \nonumber \\
D_{\lambda} H^{\lambda}_{\mu \nu} - 2 (D_{\lambda} \Phi) H_{\mu
\nu}^{\lambda} & = & 0, \nonumber \\
4 (D_{\mu} \Phi)^2 - 4 D_{\mu} D^{\mu} \Phi + R + \frac
{1}{12}H_{\mu \nu \rho} H^{\mu \nu \rho} + (D-26) & = & 0,
\label{eq:III22}
\end{eqnarray}
where $H_{\mu \nu \rho} = \partial_{\mu} B_{\nu \rho} +
\partial_{\rho} B_{\mu \nu} + \partial_{\nu} B_{\rho \mu}$.
  
\subsubsection{World Sheet Symmetries.}

Before ending this quick survey on the bosonic string, let us
mention an aspect of worldsheet symmetries. Worldsheet parity 
acts exchanging left and right oscillators,
\begin{equation}
\Omega: \alpha_n^{\mu} \leftrightarrow \tilde{\alpha}_n^{\mu}.
\label{eq:III23}
\end{equation}
  
Among massless states (\ref{eq:III15}), only the symmetric part (the graviton) 
is invariant under this transformation. We can now reduce the Hilbert 
space to states invariant under $\Omega$. The inmediate effect of this 
on the worldsheet geometry is that a one loop surface can be defined in 
two ways: the opposite $S^1$ boundaries of a cylinder can be glued 
preserving orientation, to generate a torus, or up to an $\Omega$ 
trasnformation, giving rise to a Klein bottle.

\subsubsection{Toroidal Compactifications.}

A torus is a Ricci flat manifold that can be used as target spacetime. Let us 
consider the simplest case, ${\bf R}^{25} \times S^1$, where the compact 
$S^1$ dimension is taken to be of radius $R$. Then, the coordinate 
$x^{25}$, living on this $S^1$, must satisfy
\begin{equation}
x^{25} \equiv x^{25} + 2 \pi n R.
\label{eq:III24}
\end{equation}
If we now include the identification (\ref{eq:III24}) in the mode expansion 
(\ref{eq:III9}) we get, for the right and left momenta,
\begin{eqnarray}
p_L & = & \frac {m}{2R} -nR, \nonumber \\
p_R & = & \frac {m}{2R} +nR, 
\label{eq:III25}
\end{eqnarray}
while the mass formula becomes
\begin{equation}
M^2 = 4 \left( \frac {m}{2R}- nR \right)^2 + 8(N-1) = 4 \left( 
\frac {m}{2R} +nR \right)^2 + 8 (\bar{N}-1),
\label{eq:III26}
\end{equation}
with $N$ and $\bar{N}$ the total level of left and right moving excitations, respectively. 
The first thing to be noticed, from (\ref{eq:III25}), is the invariance under the 
transformation
\begin{eqnarray}
T: R & \rightarrow & \frac {1}{2R}, \nonumber \\
   m & \rightarrow & n.
\label{eq:III27}
\end{eqnarray}
  
A nice way to represent (\ref{eq:III25}) is using a lattice of $(1,1)$ type, 
which will be referred to as $\Gamma^{1,1}$. 
This is an even lattice, as can be observed  
from (\ref{eq:III25}),
\begin{equation}
p_L^2-p_R^2 =2mn.
\label{eq:III28}
\end{equation}
If $\Pi$ is the spacelike $1$-plane where $p_L$ lives, then 
$p_R \in \Pi^{\perp}$. In fact, $p_L$ froms a $\theta$\footnote{$\theta$  
is the coordinate parametrizing the radius of the compact dimension.} angle with 
the positive axis of the $\Gamma^{1,1}$ lattice, while $p_R$ forms a 
negative angle, $- \theta$, and changes in $R$, which are simply 
changes in $\theta$ (or Lorentz rotations in the $\Gamma^{1,1}$ 
hyperbolic space), are changes in the target space preserving the $\beta=0$ 
condition, and therefore are what can be called the {\em moduli of the 
$\sigma$-model} (\ref{eq:III16}). Of course, no change arises in the 
spectrum upon rotations of the $\Pi$ and $\Pi^{\perp}$ planes. We have 
now obtained a good characterization of the moduli space for the 
string $\sigma$-model on a simple $S^1$ torus. However, in addition 
to rotations in $\Pi$ and $\Pi^{\perp}$, we should also take into account 
the symmetry (\ref{eq:III27}), representing rotations of the $\Gamma^{1,1}$ 
lattice. 
  
The previous discussion can be generalized to compactifications on higher 
dimensional tori, $T^{d}$ (i. e., working in a background spacetime 
${\bf R}^{26-d} \times T^d$). In this case, $(p_L,p_R)$ will 
belong to a lattice $\Gamma^{d,d}$, and the moduli space will be
given by \cite{Narain}
\begin{equation}
O(d,d;{\bf Z}) \backslash O(d,d) / O(d) \times O(d),
\label{eq:III29}
\end{equation}
where the $O(d,d;{\bf Z})$ piece generalizes the $T$-transformations (\ref{eq:III27}) 
to $T^d$. From now we will call these transformations $T$-duality \cite{T}. Notice also 
that the dimension of the moduli (\ref{eq:III29}) is $d \cdot d$, which is 
the number of massles degrees of freedom that have been used to define the 
background fields of the $\sigma$-model (\ref{eq:III18}). The manifold 
(\ref{eq:III29}) is the first example of moduli of a $\sigma$-model we find; these 
moduli spaces will be compared, in next section, to the $K3$ moduli 
described. 

\subsubsection{$\sigma$-Model $K3$ Geometry. A First Look at Quantum Cohomology.}
\label{k3}

The concept of moduli space introduced in previous paragraph, for the $\sigma$-model 
(\ref{eq:III18}), when the target space is a $T^d$ torus, leading to 
manifold (\ref{eq:III29}), can be generalized to more complicated 
spacetime geometries satisfying the constraints derived from conformal invariance, 
namely Ricci flat manifolds. This is a physical way to approach the theory 
of moduli spaces where, instead of working out the cohomology of the manifold, 
a string is forced to move on it, which allows to wonder about the moduli 
of the so defined conformal field theory. In order to properly
use this approach, let us first review some facts about $K3$
geometry.

Let us first recall the relation between supersymmetry 
and the number of complex structures. Let us think of a $\sigma$-model,
with target space ${\cal M}$. Now, we want this $\sigma$-model to
be invariant under some supersymmetry transformations. It turns
out that in order to make the $\sigma$-model, whose bosonic part
is given by
\begin{equation}
\eta^{\mu \nu} g_{ij}(\phi(x))\partial_{\mu}\phi^{i}
\partial_{\nu}\phi^{j},
\label{eq:b1}
\end{equation}
with $\eta$ the metric on spacetime, and $g$ the metric on the
target, invariant under $N\!=\!2$ supersymmetry we have to
require the manifold to be K\"{a}hler and, in order to be
$N\!=\!4$, to be hyperk\"{a}hler.
  
Let us now enter the description of the $K3$ manifold \cite{GrHa,Aspinwall,Persson}. To
characterize topologically $K3$, we will first obtain its Hodge
diamond. The first property of $K3$ is that the canonical class,
\begin{equation}
K \equiv - c_1(T),
\label{eq:b2}
\end{equation}
with $c_1(T)$ the first Chern class of the tangent bundle, $T$,
is zero,
\begin{equation}
K=0.
\label{eq:b3}
\end{equation}
Equation (\ref{eq:b3}) implies that there exists a holomorphic
$2$-form $\Omega$, everywhere non vanishing. Using the fact that only
constant holomorphic functions are globally defined, we easily
derive, from (\ref{eq:b3}), that
\begin{equation}
\hbox{dim } H^{2,0}=h^{2,0}=1.
\label{eq:b4}
\end{equation}
In fact, if there are two different $2$-forms $\Omega_1$ and
$\Omega_2$, then $\Omega_1/\Omega_2$ will be holomorphic and
globally defined, and therefore constant. 
  
The second important property characterizing $K3$ is 
\begin{equation}
\Pi_1 = 0,
\label{eq:b5}
\end{equation}
so that
\begin{equation}
h^{1,0}=h^{0,1}=0,
\label{eb:b6}
\end{equation}
as $b_1=h^{1,0}=h^{0,1}=0$, because of (\ref{eq:b5}). 
  
The Euler number can be now derived using Noether-Riemann
theorem, and property (\ref{eq:b3}), and it turns out to be $24$.
Using now the decomposition of the Euler number as an
alternating sum of Betti numbers, we can complete the Hodge
diamond,
\begin{equation}
24 = b_0 - b_1 + b_2 - b_3 +b_4 = 1 -0 +b_2 -0 +1,
\label{eq:b7}
\end{equation}
which implies that 
\begin{equation}
\hbox {dim } H^2 =22,
\label{eq:b8}
\end{equation}
and therefore, from (\ref{eq:b4}), we get 
\begin{equation}
\hbox {dim } H^{1,1} =h^{1,1}=20,
\label{eq:b9}
\end{equation}
leading to the Hodge diamond
\begin{equation}
\begin{array}{ccccc}  &   & 1  &   &    \\
					      & 0  &   & 0  &    \\
					    1  &   & 20 &   & 1  \\
					      & 0  &   & 0    &    \\
					      &   & 1 &     & 
	\end{array}
\label{eq:b10}
\end{equation}
  
Using Hirzebuch's pairing, we can give an inner product to the
$22$ dimensional space $H^2$. In homology terms, we have
\begin{equation}
\alpha_1 \cdot \alpha_2 = \# (\alpha_1 \cap \alpha_2),
\label{eq:b11}
\end{equation}
with $\alpha_1, \alpha_2 \in H^2(X,Z)$, and $\#(\alpha_1 \cap
\alpha_2)$ the number of oriented intersections. From the
signature complex,
\begin{equation}
\tau = \int_X \frac {1}{3} (c_1^2-2c_2) = - \frac {2}{3} \int_X
c_2 = - \frac {2\cdot24}{3} = -16, 
\label{eq:b12}
\end{equation}
we know that $H^2(X,Z)$ is a lattice of signature $(3,19)$. The
lattice turns out to be self dual, i. e., there exits a basis
$\alpha_i^*$ such that 
\begin{equation}
\alpha_i \cdot \alpha_j^* = \delta_{ij},
\label{eq:b13}
\end{equation}
and even,
\begin{equation}
\alpha \cdot \alpha \in 2 {\bf Z}, \: \: \: \: \: \: \forall
\alpha \in H^2(X,{\bf Z}).
\label{eq:b14}
\end{equation}
Fortunatelly, lattices with these characteristics are unique up
to isometries. In fact, the $(3,19)$ lattice can be represented
as
\begin{equation}
E_8 \perp E_8 \perp {\cal U} \perp {\cal U} \perp {\cal U},
\label{eq:b15}
\end{equation}
with ${\cal U}$ the hyperbolic plane, with lattice $(1,1)$, and
$E_8$ the lattice of $(0,8)$ signature, defined by the Cartan
algebra of $E_8$. The appearance of $E_8$ in $K3$ will be at the
very core of future relations between $K3$ and string theory,
mainly in connection with the heterotic string.
  
Next, we should separetely characterize the complex structure and
the metric of $K3$. Recall that this is exactly what we did in
our study of the moduli of $N\!=\!4$ supersymmetric three
dimensional Yang-Mills theories. Concerning the complex
structure, the proper tool to be used is Torelli's theorem, that
stablishes that the complex structure of a $K3$ marked surface\footnote{By 
a marked $K3$ surface we mean a specific map of $H^2(X,{\bf Z})$ into the 
lattice (\ref{eq:b15}), that we will denote, from now on, $\Gamma_{3,19}$.} is
completely determined by the periods of the holomorphic $2$-form,
$\Omega$. Thus, the complex structure is fixed by
\begin{itemize}
	\item[{i)}] The holomorphic form $\Omega$.
	\item[{ii)}] A marking.
\end{itemize}
To characterize $\Omega \in H^{2,0}(X,{\bf C})$, we can write
\begin{equation}
\Omega = x + iy,
\label{eq:b16}
\end{equation}
with $x$ and $y$ in $H^2(X,{\bf R})$, that we identify with the
space ${\bf R}^{3,19}$. Now, we know that
\begin{eqnarray}
\int_X \Omega \wedge \Omega & = & 0, \nonumber \\
\int_X \Omega \wedge \bar{\Omega} & > & 0, 
\label{eq:b17}
\end{eqnarray}
and we derive
\begin{eqnarray}
x \cdot y & = & 0, \nonumber \\
x \cdot x & = & y \cdot y.
\label{eq:b18}
\end{eqnarray}
Therefore, associated with $\Omega$, we define a plane of vectors
$v=nx+my$ which, due to (\ref{eq:b17}), is space-like, i. e.,
\begin{equation}
v \cdot v >0.
\label{eq:b19}
\end{equation}
The choice of (\ref{eq:b16}) fixes an orientation of the two
plane, that changes upon complex conjugation. Thus, the moduli
space of complex structures of $K3$, will reduce to simply the
space of oriented space-like $2$-planes in ${\bf R}^{3,19}$. To
describe this space, we can use a Grassmanian \cite{Aspinwall},
\begin{equation}
Gr = \frac {(O(3,19))^+}{(O(2)\times O(1,19))^+},
\label{eq:b20}
\end{equation}
where $(\: \:)^+$ stands for the part of the group preserving
orientation. If, instead of working with the particular marking
we have been using, we change it, the result turns out to be an
isometry of the $\Gamma^{3,19}$ lattice; let us refer to this
group by $O(\Gamma^{3,19})$. The moduli then becomes
\begin{equation}
{\cal M}^C = Gr/O^+(\Gamma^{3,19}).
\label{eq:b21}
\end{equation}
The group $O(\Gamma^{3,19})$ is the analog to the modular group, when we work out the moduli 
space of complex structures for a Riemann surface ($Sl(2,{\bf Z})$ for a torus).
 
Let us now make some comments on the distinguished complex
structure we have used in the study of the moduli of the three
dimensional $N\!=\!4$ theories. This complex structure is such
that the the elliptic curve is a $(1,1)$-form, and is
characterized by the $2$-form
\begin{equation}
\Omega = du \wedge \frac {dx}{y},
\label{eq:b22}
\end{equation}
with $\frac {dx}{y}$ the holomorphic differential on the elliptic
fiber. However, before entering a more detailed discussion on
this issue, let us consider the question of metrics. Once a
complex structure has been introduced, we have a Hodge
decomposition of $H^2$, as
\begin{equation}
H^2 = H^{2,0} \oplus H^{1,1} \oplus H^{0,2}.
\label{eq:b23}
\end{equation}
Thus, relative to a complex structure characterized by $\Omega$,
the K\"{a}hler form $J$ in $H^{1,1}$ is orthogonal to $\Omega$,
and such that
\begin{equation}
\hbox {Vol} = \int_X J \wedge J >0,
\label{eq:b24}
\end{equation}
which means that $J$ is represented by a space-like vector in
${\bf R}^{3,19}$ and, therefore, together with $\Omega$, spans
the whole three dimensional space-like subspace of ${\bf
R}^{3,19}$. Yau's theorem now shows how the metric is completely
determined by $J$ and $\Omega$, i. e., by a space-like $3$-plane
in ${\bf R}^{3,19}$. Thus, we are in a smilar position to the
characterization of the moduli space of complex structures, and
we end up with a Grassmannian manifold of three space-like planes
in ${\bf R}^{3,19}$,
\begin{equation}
Gr=O(3,19)/O(3) \times O(19).
\label{eq:b25}
\end{equation}
Now, we need to complete $Gr$ with two extra ingredients. One is
the volume of the manifold, that can change by dilatations, and
the other is again the modular part, corresponding to isometries
of $\Gamma^{3,19}$, so that finally we get
\begin{equation}
{\cal M}^{M} = O(\Gamma_{3,19})\backslash Gr \times {\bf R}^+.
\label{eq:b26}
\end{equation}

Hence, the moduli of the $\sigma$-model 
(\ref{eq:III18}), defined on a $K3$ surface, will contain the 
moduli of Einstein metrics on $K3$ (see equations (\ref{eq:b25}) and (\ref{eq:b26})). 
Now, the dimension of manifold (\ref{eq:III26}) is $58$. For the $\sigma$-model 
(\ref{eq:III18}) we must also take into account the moduli of $B$-backgrounds. 
In the string action, what we have is the integral, $\int B$, over the 
worldsheet, which now becomes a $2$-cycle of $K3$; thus, the moduli of $B$-backgrounds 
is given by the second Betti number of the $K3$ manifold, which is $22$. Finally, 
the dilaton field $\Phi$ has to be taken into account in (\ref{eq:III18}). 
As mentioned, if $\Phi$ is constant, as we will require, it counts the number of 
loops in the perturbation series, so we will not consider it as an extra moduli. 
More precisely, we will probe the $K3$ geometry working at tree level in 
string theory. Under these conditions, the $\sigma$ moduli space is 
of dimension \cite{Snew} 
\begin{equation}
58+22=80,
\label{eq:IIIa}
\end{equation}
and the natural guess is the manifold
\begin{equation}
{\cal M}^{\sigma} = O(4,20)/O(4) \times O(20).
\label{eq:IIIb}
\end{equation}
Naturally, this is not the final answer, as we have not divided yet by the 
equivalent to the $T$-duality trasnformations in the toroidal case, which are, 
for $K3$, isometries of the $H^{2}(X;{\bf Z})$ lattice, i. e.,
\begin{equation}
O(\Gamma^{3,19}).
\label{eq:IIIc}
\end{equation}
However, the final answer is not the quotient of (\ref{eq:IIIb}) by 
(\ref{eq:IIIc}), as an important symmetry from the point of view of 
conformal field theory is yet being missed: mirror symmetry. In
order to get a geometrical understanding of mirror symmetry \cite{mirror}, we
need first to define the Picard lattice.

Let us consider curves inside the $K3$ manifold. The Picard lattice is defined as
\begin{equation}
\hbox {Pic} (X) = H^{1,1}(X) \cap H^2(S,{\bf Z}),
\label{eq:b27}
\end{equation}
which means curves (i. e., $2$-cycles) holomorphically embedded
in $X$. By definition (\ref{eq:b27}), $\hbox {Pic}(X)$ defines a
sublattice of $H^2(S;{\bf Z})$. This Picard lattice has signature
$(18,t)$. Let us consider, as an example, an elliptic fibration
where the base is a $2$-cycle $B$, and $F$ is the fiber. The
Picard lattice defined by these two $2$-cycles is given by
\begin{eqnarray}
B \cdot B & = & -2, \nonumber \\
B \cdot F & = & 1, \nonumber \\
F \cdot F & = & 0,
\label{eq:b28}
\end{eqnarray}
which is a lattice of $(1,1)$ type. Self intersections are given
by the general expression
\begin{equation}
C \cdot C = 2(g-1),
\label{eq:b29}
\end{equation}
where $g$ is the genus, so that for $g=0$, the base space, we get
$-2$, and for the elliptic fiber, with $g=1$, we get $0$ for the
intersection. The intersection between the base and the fiber, $B
\cdot F$, reflects the nature of the fibration. Notice that
expression (\ref{eq:b29}) is consistent with the even nature of
the lattice $\Gamma_{3,19}$. Now, from (\ref{eq:b27}), it is
clear that the number of curves we have in $\hbox {Pic} (X)$
depends on the complex structure. Taking this fact into account, 
we can ask ourselves about the moduli space of complex structures
preserving a given Picard sublattice; for instance, we can be
interested in the moduli space of elliptic fibrations preserving
the structure of the fibration. As $\hbox {Pic} (X)$ are elements
in $H^{1,1}(X)$, they should be orthogonal to $\Omega$, so the
moduli we are looking for will be defined in terms of the
Grassmannian of space-like $2$-planes in ${\bf R}^{2,19-t}$, i.
e.,
\begin{equation}
Gr^P = O(2,19-t)/O(2) \times O(19-t),
\label{eq:b30}
\end{equation}
where we should again quotient by the corresponding modular
group. This modular group will be given by isometries of the
lattice $\Lambda$, called the transcendental lattice, and is simply
defined as the orthogonal complement to the Picard lattice. Thus,
$\Lambda$ is of $\Gamma^{2,19-t}$ type, and the moduli preserving the
Picard group is
\begin{equation}
{\cal M}^P = Gr^P/O(\Lambda).
\label{eq:b31}
\end{equation}
  
As is clear from (\ref{eq:b30}), the dimension of the moduli
space of complex structures preserving the Picard group, reduces
in an amount given by the value of $t$ for the Picard lattice. At
this point of the discussion, a question at the core of mirror
symmetry comes naturally to our mind, concerning the posibility
to define a manifold $X^*$ whose Picard group is the
transcendental lattice $\Lambda$ of $X$ \cite{Dolgachev}. In these terms, the
answer is clearly negative, as the Picard lattice is of signature
$(1,t)$, and $\Lambda$ is of signature $(2,19-t)$, so that we need
either passing from $\Lambda$ to a $(1,t')$ lattice, or generalize the concept of Picard
lattice, admiting lattices of signature $(2,t)$. It turns out
that both approaches are equivalent, but the second has a more
physical flavor; in order to get from $\Lambda$ a Picard lattice, 
what we can do is to introduce an isotropic vector $f$ in $\Lambda$,
and define the new lattice through
\begin{equation}
f^{\perp}/f,
\label{eq:b32}
\end{equation}
which is of $(1,18-t)$ type; now, the mirror manifold $X^*$ is
defined as the manifold possesing as Picard lattice the one
defined by (\ref{eq:b32}). The moduli space of the mirror
manifold is therefore given by the equivalent to expression
(\ref{eq:b30}),
\begin{equation}
Gr^{*P} = O(2,t+1)/O(2) \times O(t+1).
\label{eq:b33}
\end{equation}
Then, we observe that the dimension of the two moduli spaces sums
up to $20$, and that the dimension of the moduli space of the
mirror manifold is exactly given by the rank $t+1$ of the
Picard of the original moduli space.
  
A different approach will consist in definig the so called
quantum Picard lattice. Given a Picard lattice of signature
$(1,t)$, we define its quantum analog as the lattice of signature
$(2,t+1)$, obtained after multiplying by the hyperbolic lattice
$\Gamma^{1,1}$. So, the question of mirror will be that of given
a manifold $X$, with transcendental lattice $\Lambda$, finding a
manifold $X^*$ such that its quantum Picard lattice is precisely
$\Lambda$. Now, we observe that the quantum Picard lattices of
$X$ and $X^*$ produce a lattice of signature $(4,20)$. The
automorphisms $O(\Gamma^{4,20})$ will result of compossing the
$T$-duality transformations and mirror symmetry. Coming back to
(\ref{eq:IIIb}), and including mirror symmetry, we get, as
moduli space of the $\sigma$-model on $K3$,
\begin{equation}
O(4,20;{\bf Z}) \backslash O(4,20) / O(4) \times O(20).
\end{equation}
This concludes our analysis of $\sigma$-models on $K3$.

\subsubsection{Elliptically Fibered $K3$ and Mirror Symmetry.}

We are now going to consider singularities in the $K3$ manifold. 
Let $C$ be a rational curve in the $K3$ manifold; then, by equation
(\ref{eq:b29}), $C \cdot C=-2$. If the curve $C$ is holomorphically embedded it 
will be an element of the Picard lattice. Its volume is defined
as 
\begin{equation}
\hbox {Vol} (C) = J \cdot C,
\label{eq:b34}
\end{equation}
with $J$ the K\"ahler class. A singularity will appear whenever
the volume of $C$ goes zero, i. e., whenever the K\"ahler class
$J$ is orthogonal to $C$. Notice that this implies that $C$
should be orthogonal to the whole $3$-plane defined by $\Omega$
and $J$, as $C$ is in fact $(1,1)$, and therefore orthogonal to
$\Omega$.
  
Now, we can define the process of blowing up or down a curve $C$
in $X$. In fact, a way to blow up is simply changing the moduli
space of metrics $J$, until $J \cdot C$ becomes different from
zero. The opposite is the blow down of the curve. The other way
to get rid off the singularity is simply changing the complex
structure in such a way that the curve is not in 
$H^{1,1}$, i. e., the curve does not exist anymore.
  
We can have different types of singularities, according to how
many rational curves $C_i$ are orthogonal to $J$. The type of
singularity will be given by the lattice generated by these $C_i$
curves. Again, these lattices would be characterized by Dynkin
diagrams. 
  
Let us now consider an elliptically fibered $K3$ manifold,
\begin{equation}
E \rightarrow X \rightarrow B.
\end{equation}
Now, we can come back to Kodaira's analysis on elliptic
fibrations, as presented in chapter II. Elliptic singularities of
Kodaira type are characterized by the set of irreducible
components $X_i$ of the corresponding singularities. The Picard
lattice for these elliptic fibrations contains the $\Gamma^{1,1}$
lattice generated by the fiber and the base, and the
contribution of each singularity as given by the Shioda-Tate
formula \cite{Dolgachev}. Defining the Picard number $\rho(X)$ as $1+t$ for a
Picard lattice of type $(1,t)$ we get
\begin{equation}
\rho(X) = 2 + \sum_{\nu} \sigma(F_{\nu}),
\label{ST}
\end{equation}
where the sum is over the set of singularities, and where
$\sigma$ is given by $\sigma(A_{n-1})=n-1$, $\sigma(D_{n+4})=n+4$,
$\sigma(E_6)=6$, $\sigma(E_7)=7$, $\sigma(E_8)=8$, $\sigma(IV)=2$,
$\sigma(III)=1$, $\sigma(II)=0$. Equation (\ref{ST}) is true
provided the Mordell-Weyl group of sections is trivial.
 
As described in the previous section, the mirror map goes from a
manifold $X$, with Picard lattice of type $(1,t)$, to $X^*$, with
Picard lattice $(1,18-t)$ or, equivalently,
\begin{equation}
\rho(X) + \rho(X^*)= 20.
\end{equation}
Through mirror, we can then pass from an elliptically fibered $K3$
surface, with Picard number $\rho(X)=2$, which should for
instance have all its singularities of type $A_0$, to a $K3$
surface of Picard number $\rho(X^*)=18$, which should have $16$
singularities of $A_1$ type, or some other combination of
singularities.

\subsubsection{The Open Bosonic String.}

Repeating previous comments on closed strings for the open case is 
straightforward. The only crucial point is deciding the type of boundary 
conditions to be imposed. From (\ref{eq:III1}), we get boundary terms of 
the form
\begin{equation}
\frac {T}{2} \int \partial X^{\mu} \partial_n X_{\mu},
\label{eq:III31}
\end{equation}
with $\partial_n$ the normal boundary derivative. In order to avoid 
momentum flow away form the string, it is natural to imposse 
Neumann boundary conditions,
\begin{equation}
\partial_n X_{\mu} = 0.
\label{eq:III31b}
\end{equation}
Using these boundary conditions the mode expansion (\ref{eq:III9}) becomes, 
for the open string,
\begin{equation}
X^{\mu} (\sigma,\tau)=x^{\mu} + 2 \alpha' p^{\mu} \tau + i \sqrt{2 \alpha'} 
\sum_{n \neq 0} \frac {1}{n} \alpha_n^{\mu} e^{-i n \tau} \cos n \sigma,
\label{eq:III32}
\end{equation}
and the quantum mass formula (\ref{eq:III14}) is, for $\alpha'=\frac {1}{2}$,
\begin{equation}
M^2 = -2 +2 \sum_{n=1}^{\infty} \alpha_{-n} \alpha_n.
\label{eq:III33}
\end{equation}
  
Now, the first surprise arises when trying to generalize the $T$-duality 
symmetry, (\ref{eq:III27}), to the open string case.

\subsubsection{D-Branes.}

By introducing the complex coordinate 
\begin{equation}
z= \sigma^2 + i\sigma,
\label{eq:III34}
\end{equation}
with $\sigma^2 \equiv i \tau$, (\ref{eq:III32}) can be rewritten as
\begin{equation}
X^{\mu} (\sigma,\tau)=x^{\mu} -i \alpha' p^{\mu} \ln(z \bar{z}) + i 
\sqrt{\frac {\alpha'}{2}} \sum_{n \neq 0} \frac {1}{n} \alpha_n^{\mu} 
(z^{-n}+\bar{z}^{-n}). 
\label{eq:III35}
\end{equation}
  
Let us now consider the open string moving in ${\bf R^{25}} \times S^1$. 
Neumann boundary conditions in the compactified direction are
\begin{equation}
\partial_n X^{25} = 0.
\label{eq:III35b}
\end{equation}
Now, we will work out the way these boundary conditions modify under the 
$R \rightarrow \frac {1}{R}$ transformation \cite{GP}. To visualize the answer, 
we will consider the cylinder swept out by a time evolving closed 
string, both from the closed and open string pictures (in the open 
string picture the cylinder can be understood as an open string with both 
ends at the $S^1$ edges of the cylinder). In fact, from the 
open string point of view, the propagation of the string is at tree 
level, while the open string approach is a one loop effect. We will now 
assume that the $S^1$ boundary circles of the cylinder are in the $25$ 
direction. Recalling then what happens in the closed string case, under 
change (\ref{eq:III27}), the mode expansion (\ref{eq:III9}) turns 
(\ref{eq:III27}) equivalent to the change
\begin{equation}
\tilde{\alpha}^{25}_n \rightarrow - \tilde{\alpha}_n^{25}.
\label{eq:III36}
\end{equation}
In the $n=0$ case we get, from (\ref{eq:III11}) and (\ref{eq:III25}) 
(with $\alpha'=\frac {1}{2}$),
\begin{equation}
\alpha_0^{25} = \frac {m}{2R} -nR \rightarrow nR - \frac {m}{2R} = 
- \tilde{\alpha}_0^{25}.
\label{eq:III37}
\end{equation}
What this means is that the theory in the dual circle of radius $\frac {1}{2R}$ is 
equivalent to a theory on a circle of radius $R$, but written in terms  
of a new space coordinate $Y^{25}$, defined from $X^{25}$ by the change (\ref{eq:III36}). 
Now, it easy to see that 
\begin{equation}
\partial_{\alpha} Y^{25} = \epsilon_{\alpha \beta} \partial^{\beta} X^{25}.
\label{eq:III38}
\end{equation}
  
Returning now to the cylinder image described above, let us consider 
boundary conditions in the open string picture. From the closed string 
approach, they will be represented as
\begin{equation}
\partial_{\tau} X^{25} =0.
\label{eq:III39}
\end{equation}
Now, after performing the duality transformation
(\ref{eq:III27}), equation (\ref{eq:III38}) implies
\begin{equation}
\partial_{\sigma} Y^{25} =0,
\label{eq:III40}
\end{equation}
that, from the open string point of view, looks as Dirichlet
boundary conditions, so that the extreme points of the open
string do not move in time in the $25$ direction. Summarizing, we
observe that under $R \rightarrow \frac {\alpha'}{R}$, Neumann
and Dirichlet boundary conditions for the open string are
exchanged. Besides, the picture we get if the end points
of the open string do not move in the $25$ direction is that of
D-brane hypersurfaces, with fixed $25$ coordinate, where the open
string should end.
  
For a better understanding of the dynamical nature of these D-brane
hypersurfaces, and their physical meaning, the above approach
must be generalized to include several D-brane hypersurfaces; the
tool needed comes from the old fashioned primitive string theory,
interpreted as a meson model: the Chan-Paton factors \cite{Chan-Paton}.

\subsubsection{Chan-Paton Factors and Wilson Lines.}

Chan-paton factors are simply defined encoding the end points of
the open string with labels $i$, $j$, with $i,j=1,\ldots,N$. The
corresponding string states will be defined as $|k;i,j>$. Let us
now define a set of $N \times N$ matrices, $\lambda^{a}_{N \times
N}$, hermitian and unitary, which define the adjoint
representation of $U(N)$. We can now define the open string state
$|k;a>$ as
\begin{equation}
|k;a> = \sum_{i,j} \lambda^{a}_{i,j} |k;i,j>.
\label{eq:III41}
\end{equation}
The string states $|i,j>$ can now be easily interpreted in the
language of gauge theories. In order to do that, we will again
use the abelian projection introduced in previous chapter. In
the abelian projection gauge, states $|i,i>$ correspond to $U(1)$
photons, while $|i,j>$ states (non diagonal components of the
gauge field) correspond to charged massive particles. The way
they transform under the abelian $U(1)^N$ group is
\begin{equation}
|i,j> \rightarrow e^{i(\alpha_j-\alpha_i)}|i,j>,
\label{eq:III42}
\end{equation}
for the abelian transformation
\begin{equation}
	\left( \begin{array}{ccc} e^{i\alpha_1} &   &  \\
					  & \ddots &   \\
				       &  &   e^{i\alpha_N}
		\end{array} \right).
\label{eq:III43}
\end{equation}
  
As discussed in chapter II, to define an abelian projection
gauge, a field $X$ must be chosen to transform in the adjoint
representation; then, the gauge is fixed through imposing $X$ to
be diagonal. A simple example of field $X$ is a Wilson line. So,
let us assume we are working in ${\bf R}^{25} \times S^1$, and
define $X$ as the Wilson line in the $25$ compactified direction.
Choosing $X$ diagonal means taking $A^{25}$ in the abelian group
$U(1)^N$; a diagonal Wilson line is obtained from
\begin{equation}
A^{25} = \frac {1}{2 \pi R} \left( \begin{array}{ccc} \theta_1 &   &  \\
						  &  \ddots &   \\
						  &  &  \theta_N
		\end{array} \right),
\label{eq:III44}
\end{equation}
corresponding to a pure gauge
\begin{equation}
A^{25} = \partial_{25} \Lambda = \partial_{25} \frac {X^{25}}{2
\pi R} \left( \begin{array}{ccc} \theta_1 &  &   \\
						    & \ddots &    \\
						    &  &  \theta_N
		\end{array} \right).
\label{eq:III45}
\end{equation}
  
Now, $\{\theta_1,\ldots,\theta_N\}$ are the analogs to
$\{\lambda_1,\ldots,\lambda_N\}$, used in the standard abelian
projection. The effect of the Wilson line (\ref{eq:III44}) on a
charged state $|i,j>$ is transforming it in the way
(\ref{eq:III42}) defines, which in particular means that the
$p^{25}$ momentum of the $|i,j>$ state becomes
\begin{equation}
p^{25} = \frac {n}{R} + \frac {\theta_j-\theta_i}{2\pi R}.
\label{eq:III46}
\end{equation}
When moving from $R$ to $R'=\frac {1}{2R}$, the momentum
(\ref{eq:III46}) turns into a winding,
\begin{equation}
2 n R' + (\theta_j R' - \theta_i R') \frac {1}{\Gamma}.
\label{eq:III47}
\end{equation}
  
The geometrical meaning of (\ref{eq:III47}) is quite clear: the
open string can wind around the dual circle of radius $R'$ any
number of times, but its end points are fixed, as expected after
the $R \rightarrow R'$ duality transformation, to be in $\theta_j
R'$ and $\theta_i R'$ positions. Thus, the picture we get is that
of several D-brane hypersurfaces fixed in the dual circle to be
at positions $\theta_1 R', \ldots,\theta_N R'$, and the string
states of type $|i,j>$ are now living between the $i^{th}$ and $j^{th}$
D-brane hypersurface.
  
Using mass formula (\ref{eq:III26}), and equation
(\ref{eq:III46}) for the momentum, we observe that only
$\alpha_{-1}^{\mu} |i,i>$ states can be massless (the $U(1)$
photons), and the mass of the $\alpha_{-1}^{\mu} |i,j>$ states
goes like $\left( \frac {(\theta_i-\theta_j)R'}{\Gamma}
\right)^2$. Both of these states have the kinematical index $\mu$
in the uncompactified directions. We can also consider the
massless Kaluza-Klein states, $\alpha_{-1}^{25} |i,i>$, which can
be interpreted as scalars living on the $24$ dimensional space
defined by the D-brane hypersurface. However, this spectrum is
the abelian projected gauge spectrum for a $U(N)$ gauge theory,
now defined on the D-brane hypersurface. Therefore, two
complementary pictures arise,
\begin{itemize}
	\item The distribution of D-branes represents a new type
of background for string theory, where a $U(N)$ Wilson line has
been introduced in the internal or compactified $S^1$.
	\item The distribution of D-branes provides, for the
massless spectrum, a geometrical representation of a gauge theory
living on the worldvolume of the D-brane. Moreover, the spectrum
is presented as the abelian projection spectrum.
\end{itemize}
  
Of course, this second approach only takes into account, as is
usual in string theory, low energy degrees of freedom. Properly
speaking, what we are doing is embedding the gauge theory into
string theory in a new way. 
  
To end this first contact with
D-branes (for more details see, for instamce, \cite{Polchinski}, 
and references therein) we should, at least qualitatively, answer the question
possed above on the dynamical nature of D-branes. The simplest
answer will be obtained analizing the gravitational interactions
through the computation of the mass density, leading to the
tension of the D-brane hypersurface. A graviton, which is a
closed string state can couple a D-brane, defining an interaction
vertex. The disc coupling the graviton to the D-brane can be
interpreted in terms of open strings ending on its circle boundary.
Without performing any computation, we already know something on
the order of magnitude of the process: it is a process determined
by the topology of a disc, with half the Euler number of a
sphere, so the order in the string coupling constant, defined in
(\ref{eq:III21}), is $O(\frac {1}{g})$.
  
A more detailed discussion on D-branes needs the use of more
general string theories (superstring theories), which is what we
will discuss in next section.

\subsection{Superstring Theories.}

Superstrings correspond to the supersymmetric generalization of
the $\sigma$-model (\ref{eq:III1}). This is done adding the
fermionic term
\begin{equation}
S_F = \int d^2 \sigma i \bar{\psi}^{\mu} \rho^{\alpha}
\partial_{\alpha} \psi_{\mu},
\label{eq:III48}
\end{equation}
where $\psi^{\mu}$ are spinors, relative to the worldsheet, and
vectors with respect to the spacetime Lorentz group, $SO(1,D-1)$.
Spinors in (\ref{eq:III48}) are real Majorana spinors, and the
Dirac matrices $\rho^{\alpha}$, $\alpha=0,1$, are defined by
\begin{eqnarray}
\rho^0 & = & \left( \begin{array}{cc} 0 & -i \\
				      i & 0 \end{array} \right),
\nonumber \\
\rho^1 & = & \left( \begin{array}{cc} 0 & i \\
				      i & 0 \end{array} \right),
\label{eq:III49}
\end{eqnarray}
satisfying
\begin{equation}
\{\rho^{\alpha}, \rho^{\beta}\}=-2\eta^{\alpha \beta}.
\label{eq:III50}
\end{equation}
The supersymmetry transformations are defined by
\begin{eqnarray}
\delta x^{\mu} & = & \bar{\epsilon} \psi^{\mu}, \nonumber \\
\delta \psi^{\mu} & = & -i \rho^{\alpha} \partial_{\alpha}
x^{\mu} \epsilon,
\label{eq:III51}
\end{eqnarray}
with $\epsilon$ a constant anticonmuting spinor. Defining the
components
\begin{equation}
\psi^{\mu} = \left( \begin{array}{c} \psi_-^{\mu} \nonumber \\
\psi_+^{\mu} \end{array} \right),
\label{eq:III52}
\end{equation}
the fermionic lagrangian (\ref{eq:III48}) can be written as
\begin{equation}
S_F = \int d^2 \sigma ( \psi_-^{\mu} \partial_+ \psi_-^{\mu} +
\psi_+^{\mu} \partial_- \psi_+^{\mu}),
\label{eq:III53}
\end{equation}
with $\partial_{\pm} \equiv \frac {1}{2} (\partial_{\tau} \pm
\partial_{\sigma})$. As was the case for the bosonic string, we
need now to specify the boundary conditions for the fermion
fields, both in the open and closed string case. For open
strings, there are two posibilities:
\begin{eqnarray}
\hbox {Ramond} & : & \psi_+^{\mu} (\pi,\tau) = \psi_-^{\mu}(\pi,
\tau), \nonumber \\
\hbox {Neveu-Schwarz} & : & \psi_+^{\mu} (\pi,\tau) = - \psi_-^{\mu}(\pi,
\tau),
\label{eq:III54}
\end{eqnarray}
which produce the mode expansions 
\begin{eqnarray}
\hbox {Ramond} & : & \psi_{\mp}^{\mu} = \frac {1}{\sqrt{2}}
\sum_z d_n^{\mu} e^{-in(\tau \mp \sigma)},  \nonumber \\
\hbox {Neveu-Schwarz} & : & \psi_{\mp}^{\mu} = \frac
{1}{\sqrt{2}} \sum_{z+\frac {1}{2}} b_n^{\mu} e^{-in(\tau \mp
\sigma)}.
\label{eq:III55}
\end{eqnarray}
In the case of closed strings, we can impose either periodic or
antiperiodic boundary conditions for the fermions, obtaining
Ramond (R) or Neveu-Schwarz (NS) for both $\psi_{\pm}$ fields.
After quantization we get, following similar steps to those in
the bosonic case, that the critical dimension is $10$, and that
the mass formulas and normal ordering correlators are given by 
\begin{equation}
M^2 = 2 (N_L - \delta_L) = 2 (N_R - \delta_R),
\label{eq:III56}
\end{equation}
with $\delta= \frac {1}{2}$ in the NS sector, and $\delta=0$ in
the R sector. Using this formula, and the GSO projection, we
easily get the massless spectrum. For the closed string we get 
\begin{eqnarray}
\hbox {NS-NS sector} & : & b_{-1/2}^{\mu} b_{-1/2}^{\nu} |0>, \nonumber \\
\hbox {NS-R sector}  & : & b_{-1/2}^{\mu} |S>, \nonumber \\
\hbox {R-R sector}  & : &  |S> \otimes |S> .
\label{eq:spectrum}
\end{eqnarray}
The state $|S>$ corresponds to the Ramond vacua (recall
$\delta=0$ in the Ramond sector). 
  
The $d_0^{\mu}$ oscillators in (\ref{eq:III55}) define a Clifford
algebra,
\begin{equation}
\{d_0^{\mu},d_0^{\nu}\} = \eta^{\mu \nu},
\label{eq:III57}
\end{equation}
and therefore the $|S>$ vacua can be one of the two ${\bf 8}_S$,
${\bf 8}_{S'}$ spinorial representations of $SO(8)$. Depending on
what is the spinorial representation chosen we get, from
(\ref{eq:spectrum}), two different superstring theories. In the
chiral case, we choose the same chirality for the two fermionic
states in the NS-R and R-NS sectors. This will lead to two
gravitinos of equal chirality. Moreover, in the R-R sector we
get, for same chirality,
\begin{equation}
{\bf 8}_S \times {\bf 8}_S = {\bf 1} \oplus {\bf 28} \oplus {\bf 35}_S,
\label{eq:III58}
\end{equation}
corresponding to a scalar field being identified with the axion,
an antisymmetric field, and a $4$-form field. We will call this
superstring theory type IIB. In case we choose different
chiralities for the spinor representations associated with the
Ramond vacua, what we get is type IIA superstring theory, which
is also an $N\!=\!2$ theory, but this time with two gravitinos of 
different chirality; now, the R-R sector contains
\begin{equation}
{\bf 8}_S \otimes {\bf 8}_{S'} = {\bf 8}_V \oplus {\bf 56}_V, 
\label{eq:III59}
\end{equation}
i. e., a vector field and a $3$-form. These are the first two
types of superstring theories that we will consider.

\subsubsection{Toroidal Compactification of Type IIA and Type IIB
Theories. $U$-duality.}
\label{susy}

Before considering different compactifications of superstring
theories, we will first review some general results on the
maximum number of allowed supersymmetry, depending on the
spacetime dimension.

Spinors should be considered as representations of
$SO(1,d-1)$. Irreducible representations have dimension
\begin{equation}
2^{\left[ \frac {d+1}{2} \right]-1},
\label{eq:a1}
\end{equation}
where $[ \: ]$ stands for the integer part. Depending on the
dimension, the larger spinor can be real, complex or
quaternionic,
\begin{eqnarray}
{\bf R}, \: \: \hbox {if} \: \: d & = & 1,2,3 \: \: \hbox {mod} \: 8,
\nonumber \\
{\bf C}, \: \: \hbox {if} \: \: d & = & 0 \: \: \hbox {mod} \: 4, \nonumber
\\
{\bf H}, \: \: \hbox {if} \: \: d & = & 5,6,7 \: \: \hbox {mod} \: 8.
\label{eq:a2}
\end{eqnarray}
  
Using (\ref{eq:a1}) and (\ref{eq:a2}), we get the number of
supersymmetries listed in the table below\footnote{This table 
is constrained by the physical requirement that
particles with spin $>2$ do not appear.}.

\begin{center}

\begin{tabular}{|c|c|c|}     \hline\hline
	 {\bf Dimension  }      & ${\bf N}$  & {\bf Irreducible Representation}   \\ \hline
	 $11$                   & $1$        & ${\bf R}^{32}$   \\
	 $10$                   & $2$        & ${\bf R}^{16}$   \\
	 $9$                    & $2$        & ${\bf R}^{16}$   \\
	 $8$                    & $2$        & ${\bf C}^8$   \\
	 $7$                    & $2$        & ${\bf H}^8$   \\
	 $6$                    & $4$        & ${\bf H}^4$   \\ 
	 $5$                    & $4$        & ${\bf H}^4$   \\         
	 $4$                    & $8$        & ${\bf C}^2$   \\
	 $3$                    & $16$       & ${\bf R}^2$   \\ \hline\hline

\end{tabular}
\label{tab:a1}
\end{center}

The maximum number of supersymmetries in three dimensions is then
$16$. From the table it is also clear that through
standard Kaluza-Klein compactification, starting with six
dimensional $N\!=\!1$ supersymmetry leads to four dimensional
$N\!=\!2$, and three dimensional $N\!=\!4$ supersymmetry. We can
also notice that ten dimensional $N\!=\!1$ leads to $N\!=\!4$
supersymmetry in four dimensions.
  
It must be stressed that the counting of supersymmetries after
dimensional reduction is slightly more subtle if we compactify on
manifolds with non trivial topology. Here, the adequate concept
is the holonomy of the internal manifold; let us therefore recall
some facts on the concept of holonomy. Given a Riemannian
manifold ${\cal M}$, the holonomy group $H_{\cal M}$ is defined
as the set of transformations $M_{\gamma}$ associated with paths
$\gamma$ in ${\cal M}$, defined by parallel transport of vectors
in the tangent bundle. The connection used in this definition is
the Levi-Civita connection. In general, for a vector budle $E
\rightarrow {\cal M}$, the holonomy group $H_{\cal M}$ is defined
by the paralell transport of $v$ in the fiber, with respect to
the connection on $E$. The Ambrose-Singer theorem shows how the
holonomy is generated by the curvature.
  
Manifolds can be classified according to its holonomy group.
Therefore, we get \cite{susyholo}
\begin{itemize}
	\item $H_{\cal M}=O(d)$, for real manifolds of dimension
$d$.
	\item $H_{\cal M}=U(\frac {d}{2})$, for K\"{a}hler
manifolds.
	\item $H_{\cal M}=SU(\frac {d}{2})$, for Ricci flat
K\"{a}hler manifolds.
	\item  $H_{\cal M}=Sp(\frac {d}{4})$, for hyperk\"{a}hler
manifolds\footnote{Notice that any hyperk\"{a}hler manifold is
always Ricci flat.}.
\end{itemize}
  
The answer to the question of what the role of holonomy is in the
counting of the number of supersymetries surviving after
compactification is quite simple: let us suppose we are in
dimension $d$, so that the spinors are in $SO(1,d-1)$. Now, the
theory is compactified on a manifold of dimension $d_1$, down to
$d_2=d-d_1$. Supersymmetries in $d_2$ are associated with
representations of $SO(1,d_2-1)$, so we need to decompose an
irreducible representation of $SO(1,d-1)$, into
$SO(1,d_2-1) \times SO(d_1)$. Now, the holonomy group of the
internal manifold $H_{{\cal M}_{d_1}}$ will be part of $SO(d_1)$.
Good spinors in $d_2$ dimensions would be associated with
singlets of the holonomy group of the internal manifold. Let us
consider the simplest case, with $d_1=4$; then,
\begin{equation}
SO(4) = SU(2) \otimes SU(2)
\label{eq:a3}
\end{equation}
and, if our manifold is Ricci flat and K\"{a}hler, the holonomy
will be one of these $SU(2)$ factors. Therefore, we will need a
singlet with respect to this $SU(2)$. As an example, let us
consider the spinor in ten dimensions, with $N\!=\!1$; as we can
see from the above table, it is a ${\bf 16}$, that we can decompose
with respect to $SO(1,5) \times SU(2) \times SU(2)$ as
\begin{equation}
{\bf 16} = ({\bf 4},{\bf 2},{\bf 1})\otimes ({\bf 4},{\bf 1},{\bf
2}).
\label{eq:a4}
\end{equation}
Therefore, we only get one surviving supersymmetry in six
dimensions. This is a general result: if we compactify a ten
dimensional theory on a manifold of dimension four, with $SU(2)$
holonomy, we will get a six dimensional theory with only one
supersymmetry. However, if the compactification is on a torus
with trivial holonomy, two supersymmetries are obtained (the
maximum number of supersymmetries available).

\vspace{2 mm}

As the first contact with type IIA string theory we will then consider
its compactification on a $d$-dimensional torus, $T^d$. To start
with, let us work in the particular case $d=4$. From the above table, 
we learn that the number of
supersymmetries in six dimensions is $4$, as the holonomy of
$T^4$ is trivial. If we do not take into account the R-R fields,
the moduli of the string $\sigma$-model is exactly the one
described in section \ref{sec:III1},
\begin{equation}
O(4,4;{\bf Z}) \backslash O(4,4)/O(4) \times O(4),
\label{eq:III60}
\end{equation}
with the $T$-duality $O(4,4;{\bf Z})$ corresponding to changes of
the type $R_i \rightarrow \frac {\alpha'}{R_i}$, for the four
$S^1$ cycles compossing the torus. The situation becomes
different if we allow R-R background fields. In such a case, we
should take into account the possiblity of including Wilson lines
for the $A_{\mu}$ field (the ${\bf 8}_V$ in (\ref{eq:III59})),
and also a background for the $3$-form $A_{\mu \nu \rho}$ (the
${\bf 56}_V$ of (\ref{eq:III59})). The number of Wilson lines is
certainly $4$, one for each non contractible loop in $T^d$, so we
need to add $4$ dimensions to the $16$-dimensional space
(\ref{eq:III60}). Concerning an $A_{\mu \nu \rho}$ background,
the corresponding moduli is determined by $H_3(T^4)$, which
implies $4$ extra parameters. Finally, the dimension equals
\begin{equation}
16+4+4=24.
\label{eq:III61}
\end{equation}
Now, a new extra dimension coming form the dilaton field must be
added. It is important here to stress this fact: in the approach
in previous section to $\sigma$-model moduli space the dilaton
moduli has not been considered. This corresponds to interpreting
the dilaton as a string coupling constant, and allowing changes
only in the string. Anyway, this differentiation is rather
cumbersome. Adding the dilaton moduli to (\ref{eq:III61}), we get
a moduli space of dimension (\ref{eq:III25}), that can be written
as 
\begin{equation}
O(5,5;{\bf Z}) \backslash O(5,5)/O(5) \times O(5).
\label{eq:III62}
\end{equation}
The proposal of moduli (\ref{eq:III62}) for type IIA on $T^4$
already contains a lot of novelties. First of all, the modular
group $O(5,5;{\bf Z})$ now acts on the dilaton and the resting
Ramond fields. In fact, relative to the $O(4,4;{\bf Z})$
$T$-duality of toroidal compactifications, we have now an extra
symmetry which is $S$-duality \cite{MO,rabino,S,mr,HT,Duff,Town,Wsvd,S2,HS,GrMS,VW,HT2},  
\begin{equation}
g \rightarrow \frac {1}{g},
\label{eq:III63}
\end{equation}
with $g$ the string coupling constant. This new modular symmetry
is called in the physics literature $U$-duality \cite{HT}. The phenomena
found here resembles very much what arises from mirror symmetry
in the analysis of $K3$. There, the ``classical'' modular group was $O(\Gamma^{3,19}; 
{\bf Z})$, and quantum mirror symmetry creates the enhancement to
$O(\Gamma^{4,20};{\bf Z})$ where, in addition to $T$-duality, we
have mirror transformations. In the case of type IIA on $T^4$, it
is because we include the R-R backgrounds and the dilaton that
the modular symmetry $O(4,4;{\bf Z})$ is enhanced to the
$U$-duality symmetry. In spite of the analogies, the physical
meaning is different. To apreciate this, let us now consider type
IIA on $K3$. The dilaton moduli can be added, but the R-R fields
are not producing any new moduli. In fact, recall that
$\Pi_1(K3)=0$, and $H_3=0$, so the moduli of type IIA on $K3$ is
simply
\begin{equation}
O(4,20;{\bf Z}) \backslash O(4,20)/O(4) \times O(20) \: \: \times
{\bf R},
\label{eq:III64}
\end{equation}
with ${\bf R}$ parametrizing the dilaton, and the modular group
not acting on it. 
  
The way to interpret the moduli (\ref{eq:III62}) goes under the
name of M-theory. Before entering a more precise definition of
M-theory, the basic idea is thinking of (\ref{eq:III62}) simply
as the moduli of a toroidal compactification on $T^5$; however,
in order to obtain a six dimensional $N\!=\!4$ theory, we need to
start with some theory living in $11$ dimensions. The theory
satisfying this is M-theory, a theory whose low energy
supergravity description is well understood: it should be such
that through standard Kaluza-Klein compactification it gives the
field theory limit of type IIA strings; but this a theory known as eleven
dimensional type IIA supergravity.
  
Once we have followed the construction of the type IIA string
theory moduli on $T^4$, let us consider the general case of 
compactification on $T^d$. The dimension of the moduli is 
\begin{equation}
\hbox {dim } = d^2 +1+d+ \frac {d(d-1)(d-2)}{3\!},
\label{eq:III65}
\end{equation}
where $d^2$ is the NS-NS contribution, the $1$ sumand comes form
the dilaton, $d$ from the Wilson lines, and $\frac
{d(d-1)(d-2)}{3\!}$ from the $3$-form $A_{\mu \nu \rho}$. The
formula (\ref{eq:III65}) has to be completed, for $d \geq 5$, by
including dual scalars. For $d=5$, the dual to the $3$-form
$A_{\mu \nu \rho}$ is a scalar. The result is 
\begin{eqnarray}
\frac {d(d-1)(d-2)(d-3)(d-4)}{5\!} & \hbox {duals} & \hbox {to  }A_{\mu
\nu \rho}, \nonumber \\
\frac {d(d-1)\ldots(d-6)}{7\!} & \hbox {duals} & \hbox {to  } A_{\mu}.
\label{eq:III66}
\end{eqnarray}
The moduli spaces, according to the value of the dimension of the
compactification torus, are listed in the table below.

\begin{center}

\begin{tabular}{|c|c|}     \hline\hline
	 {\bf Dimension  }      & {\bf Moduli}    \\ \hline
	 $d=4$       & $O(5,5;{\bf Z}) \backslash O(5,5) / O(5) \times O(5)$    \\
	 $d=5$       & $E_{6,(6)}({\bf Z}) \backslash E_{6,(6)} / Sp(4)$       \\
	 $d=6$       & $E_{7,(7)}({\bf Z}) \backslash E_{7,(7)} / SU(8)$       \\
	 $d=3$       & $Sl(5,{\bf Z}) \backslash Sl(5) / SO(5)$       \\
	 $d=2$       & $Sl(3,{\bf Z}) \times Sl(2,{\bf Z})
\backslash Sl(3) / SO(3) \: \: Sl(2) / SO(2)$       \\ \hline
\hline
\end{tabular}
\label{tab:modulis}
\end{center}
  
For supergravity practitioners, the appearance of $E_6$ and
$E_7$ in this table should not be a surprise. 
  
Let us now see what happens in the type IIB case. The moduli on,
for instance, $T^4$, is again the $16$ dimensional piece coming
from the NS-NS sector; now, the R-R sector is determined by the
cohomology groups $H^0$, $H^2$ and $H^4$ (see equation
(\ref{eq:III58})). From the Hodge diamond for $T^4$,
\begin{equation}
\begin{array}{ccccc}  &   & 1  &   &    \\
					      & 2  &   & 2  &    \\
					    1  &   & 4 &   & 1  \\
					      & 2  &   & 2  &    \\
					      &   & 1 &     & 
	\end{array}
\label{eq:hodge}
\end{equation}
we get $8$ extra modulis, exactly the same number as in the type
IIA case. This is a general result for any $T^d$
compactification. The reason for this is that type IIA and type
IIB string theories are, after toroidal compactification, related
by $T$-duality. However, on a manifold as $K3$, with $\Pi_1=0$,
the moduli for IIA and IIB are drastically different, as can be
derived from direct inspection of the $K3$ Hodge diamond (see
equation (\ref{eq:b10})). Therefore, for type IIB we get, from
the R-R sector, $1$ coming from $H^0$, $22$ from $H^2$, and $1$
from $H^4$, which sums up a total of $24$ extra modulis to be
added to the $58+22$ of the NS-NS sector. Then, including the
dilaton,
\begin{equation}
\hbox {dim IIB}(K3) = 22+58+24+1=105.
\label{eq:III67}
\end{equation}
Therefore, the natural guess for the moduli is
\begin{equation}
O(5,21;{\bf Z}) \backslash O(5,21) / O(5) \times O(21).
\label{eq:III68}
\end{equation}
Here, something quite surprising is taking place. As we can see
from (\ref{eq:III64}), when type IIA is compactified on $K3$, we
do not find any appearance of $U$-duality or, in other words,
$S$-duality. By contrast, in the type IIB case we find a modular
group $O(5,2;{\bf Z})$, that contains the dilaton and, therefore,
the $S$-duality transformation. This is what can be called the
$S$-duality of type IIB string theory \cite{Schwarz}, which can already be
observed from equation (\ref{eq:III58}). In fact, the R-R and
NS-NS sectors both contain scalar fields and the antisymmetric
tensor.

\subsubsection{Heterotic String.}

The idea of ``heterosis'', one of the most beatiful and
productive ideas in the recent history of string theory \cite{Gross} was
motivated by two basic facts. First of all, the need to find a
natural way to define non abelian gauge theories in string
theory, without entering the use of Chan-Paton factors, and,
secondly, the sharpness of the gap in string theory between left
and right moving degrees of freedom. Here, we will concentrate on
some of the ideas leading to the construction of heterosis. In
the toroidal compactification of the bosonic string on $T^d$,
we have found that the momenta live in a $\Gamma^{d,d}$ lattice.
This is also true for the NS sector of the superstring. The
lattice $\Gamma^{d,d}$, where the momenta live, is even and self
dual. Taking into account the independence between left and right
sectors, we can think on the possibility to compactify the left
and right components on different tori, $T^{d_L}$ and  $T^{d_R}$,
and consider as the corresponding moduli the manifold
\begin{equation}
O(d_L,d_R;{\bf Z}) \backslash O(d_L,d_R) / O(d_L) \times O(d_R).
\label{eq:III69}
\end{equation}
  
Before trying to find out the consistency of this picture, let us
try to get a simple interpretation of moduli (\ref{eq:III69}).
The dimension of this moduli is $d_L \times d_R$, and we can
separate it into $d_L \times d_L + d_L \times (d_R -d_L)$. Let us
interpret the first part, $d_L \times d_L$, as the standard
moduli for compactifications on a torus $T^{d_L}$; then, the
second piece can be interpreted as the moduli of Wilson lines for
a gauge group
\begin{equation}
U(1)^{d_R-d_L}.
\label{eq:III70}
\end{equation}
With this simple interpretation, we already notice the interplay
in heterosis when working with a gauge group that can be
potentially non abelian, the gauge group (\ref{eq:III70}), and
differentiating left and right parts. When we were working with
type II string theory, and considered toroidal compactifications,
we were also adding, to the moduli space, the contribution of the
Wilson lines for the RR gauge field, $A_{\mu}$ (in case we are in
type IIA). However, in the case of type IIA on $T^4$, taking 
into account the Wilson lines did not introduce any 
heterosis asymmetry in the moduli of the kind (\ref{eq:III69}).
However, $T^4$ is not the only Ricci flat four dimensional
manifold; we can also consider $K3$ surfaces. It looks like if
$T^4$, $K3$, and its orbifold surface in between, $T^4/{\bf
Z}_2$, saturate all compactification manifolds that can be
thought in four dimensions. In the case of $K3$, the moduli of
type IIA string (see equation (\ref{eq:III64})) really looks like
the heterotic moduli, of the kind (\ref{eq:III69}), we are looking
for. Moreover, in this case, and based on the knowledge of the
lattice of the second cohomology group of $K3$ (see equation
(\ref{eq:b15})),
\begin{equation}
E_8 \perp E_8 \perp {\cal U} \perp {\cal U} \perp {\cal U},
\label{eq:III71}
\end{equation}
we can interpret the $16=d_R-d_L$ units as corresponding
precisely to Wilson lines of the $E_8 \times E_8$ gauge group
appearing in (\ref{eq:III71}). In other words, and following a
very distant path form the historical one, what we are suggesting
is interpreting moduli (\ref{eq:III64}), of type IIA on $K3$, as
some sort of heterosis, with $d_L=4$ and $d_R=20$. The magic of
numbers is in fact playing in our team, as the numbers we get for
$d_L$ and $d_R$ strongly suggest a left part, of critical
dimension $10$, and a right part, of precisely the critical
dimension of the bosonic string, $26$. This was, in fact, the
original idea hidden under heterosis: working out a string theory
looking, in its left components, as the standard superstring, and
in its right components as the $26$ dimensional bosonic string.
However, we are still missing something in the ``heterotic'' 
interpretation of (\ref{eq:III64}), which is the visualization,
from $K3$ geometry, of the gauge group. In order to see this,
some of the geometrical material introduced in subsection 
\ref{k3} will be needed; in terms of the concepts there
introduced, we would claim that the $(p_L,p_R)$ momentum is
living in the lattice $\Gamma^{4,20}$. We can then think that
$p_L$ is in the space-like $4$-plane where the holomorphic top
form $\Omega$, and the K\"ahler class $J$, are included. Recall
that they define a space-like $3$-plane. Now, momentum vectors,
orthogonal to this $4$-plane, can be considered; they are of the
type
\begin{equation}
(0,p_R).
\label{eq:III72a}
\end{equation}
Now, whenever $p_R^2=-2$, this vector will define a rational
curve inside $K3$, with vanishing volume (in fact, the volume is
given by $p_R \cdot J=0$). The points $p_R^2=-2$ will be at the
root lattice of $E_8 \times E_8$. Now, from the mass formulas
(\ref{eq:III26}) we easily observe that $p_R^2=-2$ is the
condition for massless vector particles. In fact, if we separate,
in the spirit of heterosis, the $p_R$ of a $26$ dimensional
bosonic string into $(p_R^{(16)},p_R^{(10)})$, we get, from
(\ref{eq:III26}),
\begin{equation}
M^2 = 4(p_R^{(16)})^2+8(N-1),
\label{eq:III73}
\end{equation}
so that $M^2=0$, for $N\!=\!0$, if $(p_R^{(16)})^2=2$. The sign
difference appears here because (recall subsection \ref{k3}) in
the $K3$ construction used for the second cohomology lattice, the
$E_8$ lattice was defined by minus the Cartan algebra of $E_8$.
Therefore, we observe that massless vector bosons in heterotic
string are related to rational curves in $K3$ of vanishing volume,
which allows to consider enhancement of symmetries when moving in
moduli space \cite{Wsvd,MK,Mayr}. Some of these rational curves can be blown up,
which would be the geometrical analog of the Higgs mechanism, or
either blown down, getting extra massless stuff. Moreover, for
elliptically fibered $K3$ surfaces, the different Kodaira
singularities reflect, in its Dynkin diagram, the kind of gauge
symmetry to be found. 
  
The above discussion summarizes what can be called the first
quasi-theorem on string equivalence \cite{HT,Wsvd},
\begin{itemize}
	\item[{}] {\bf Quasi-Theorem $1$} Type IIA string on $K3$ is
equivalent to $E_8 \times E_8$ heterotic string on $T^4$.
\end{itemize}
  
Previous arguments were so general that we can probably obtain
extra equivalences by direct inspection of the different $K3$
moduli spaces that have been discussed in subsection \ref{k3}. In
particular, let us consider the moduli space of complex
structures for an elliptically fibered $K3$ surface, a fact
represented, in terms of the Picard lattice, claming that it is
of $\Gamma^{1,1}$ type, generated by a section, and with the fiber
satisfying relations (\ref{eq:III28}). This moduli is
\begin{equation}
O(2,18;{\bf Z}) \backslash O(2,18) / O(2) \times O(18),
\label{eq:III72}
\end{equation}
where we have used equation (\ref{eq:b30}), and the fact that the
transcendental lattice is of type $(2,18)$. From the heterosis
point of view, it would be reasonable to interpret
(\ref{eq:III72}) as heterotic $E_8 \times E_8$ string,
compactified on a $2$-torus, $T^2$. In fact, we will have $4$
real moduli, corresponding to the K\"ahler class and complex
structure of $T^2$, and $16$ extra complex moduli associated to
the Wilson lines. However, now the type II interpretation of
(\ref{eq:III72}) is far from being clear, as (\ref{eq:III72}) is
just the part of the moduli space that is preserving the elliptic
fibration. Now, in order to answer how (\ref{eq:III72}) can be
understood as a type II compactification a similar problem
appears as we try to work out an heterotic interpretation of the
type IIB moduli on $K3$, given in (\ref{eq:III68}). A simple way
to try to interpret (\ref{eq:III72}), as some kind of type II
compactification, is of course thinking of an elliptically
fibered $K3$, where the volume of the fiber is fixed to be equal
zero; generically,
\begin{equation}
J \cdot F=0,
\label{eq:III73b}
\end{equation}
where $F$ indicates the class of the fiber. Now, we can think
that we are compactifying a type II string on the base space of
the bundle. However, this does not lead to (\ref{eq:III72}) for
the type IIA case, as the RR fields are in $H^1$ and $H^3$, which
will vanish. But what about type IIB? In this case, we have the
NS field $\phi$, and the R field $\chi$, and we should fix the
moduli of possible configurations of these fields on the base
space of the elliptic fibration. Here, type IIB $S$-duality,
already implicit in moduli (\ref{eq:III68}), can help enormously,
mainly because we are dealing with an ellipticaly fibered $K3$
manifold \cite{As,Louis,Klemm}. To proceed, let us organize the fields $\phi$ and
$\chi$ into the complex
\begin{equation}
\tau = \chi + i e^{-\phi},
\label{eq:III74}
\end{equation}
and identify this $\tau$ with the moduli of the elliptic fiber.
Then, the $18$ complex moduli dimension of (\ref{eq:III72})
parametrizes the moduli of complex structures of the elliptic
fibration, and therefore the moduli of $\tau$ field
configurations on the base space (provided $\tau$ and $\frac
{a\tau +b}{c\tau+d}$ are equivalent from the type IIB point of
view). These moduli parametrize then the type IIB
compactification on the base space $B$ (it is $\IP^1$; recall
that in deriving (\ref{eq:III72}) we have used a base space $B$ such
that $B \cdot B=-2$). There is still one moduli missing: the size
of the base space $B$, that we can identify with the heterotic
string coupling constant. Thus, we arrive to the following
quasi-theorem,
\begin{itemize}
	\item[{}] {\bf Quasi-Theorem $2$} Heterotic string on
$T^2$ is equivalent to type IIB string theory on the base space
of an elliptically fibered $K3$.
\end{itemize}
  
The previous discussion is known, in the physics literature,
under the generic name of F-theory \cite{F,F1,F2}.
  
\vspace{2 mm}
  
We have been considering, until now, type II strings on $K3$, and
compared them to heteotic string on a torus. To find out what is
the expected moduli for the heterotic string on $K3$, we can use
the following trick: if heterotic string on $T^2$ is type IIB on
the base space of an elliptically fibered $K3$, by quasi-theorem
$2$ heterotic string on an elliptically fibered $K3$ should
correspond to type IIB on the base space of an elliptically
fibered Calabi-Yau manifold. More precisely, type IIB string
should be compactified on the basis of an elliptic fibration,
which is now four dimensional, and that can be represented as a
fibration of a $\IP^1$ space over another $\IP^1$. This type of
fibrations are known in the literature as Hirzebruch spaces,
${\bf F}_n$. Hirzebruch spaces can simply be determined through
heterotic data, given by the $E_8 \times E_8$ bundle on the $K3$
manifold. The moduli of these bundles on $K3$ will put us in
contact with yet another interesting topic: small instantons.

\subsubsection{Heterotic Compactifications to Four Dimensions.}

Before considering some definite examples, let us simply
summarize the different supersymmetries we can get when
compactifying to three dimensions, depending on the holonomy of
the target manifold. In order to do that, we will need the results 
in subsection \ref{susy}, on the maximum number of
supersymmetries allowed for a given spacetime dimension.

\begin{center}

\begin{tabular}{|c|c|c|c|}     \hline\hline
	 {\bf Type of String}   & {\bf Target Manifold} & {\bf Holonomy} & {\bf Supersymmetry}  \\ \hline
	 II                     & $K3\times T^2$        & $SU(2)$        & $N\!=\!4$   \\
	 Heterotic              & $T^6$                 & Trivial        & $N\!=\!4$   \\ \hline
	 II                     & Calabi-Yau            & $SU(3)$        & $N\!=\!2$   \\
	 Heterotic              & $K3\times T^2$        & $SU(2)$        & $N\!=\!2$   \\ \hline
	 II                     & $B_{SU(4)}$           & $SU(4)$        & $N\!=\!1$   \\
	 Heterotic              & Calabi-Yau            & $SU(3)$        & $N\!=\!1$   \\ \hline\hline

\end{tabular}
\label{tab:sc}
\end{center}

In the table above we have not differentiated between type IIA
and type IIB\footnote{This will be relevant when discussing the third
line where, by $B_{SU(4)}$, we are thinking in the spirit of the
discussion in the last part of previous section, where a
Calabi-Yau fourfold of $SU(4)$ holonomy, elliptically fibered,
and with a zero volume fiber, is used for compactification.}. The
first two lines, corresponding to cases with $N\!=\!4$ and
$N\!=\!2$ supersymmetry in four dimensional spacetime, will be the
basic examples we will use to introduce the concept of dual pairs
of string compactifications down to four dimensions.
  
Before entering a discussion on the ingredients of this table, 
we yet need to consider the holonomy of the moduli
space. This holonomy will of course depend on the number of
supersymmetries and the type (real, complex or quaternionic) of
the representation. Hence, from subsection \ref{susy}, we can complete 
the table below.

\begin{center}

\begin{tabular}{|c|c|c|c|}     \hline\hline
	 {\bf Spacetime Dimension} & {\bf Supersymmetries} & {\bf Type}    & {\bf Holonomy}  \\ \hline
	 $d=6$                     & $N\!=\!2$             & ${\bf H}^{4}$ & $Sp(1)\oplus Sp(1)$  \\
	 $d=4$                     & $N\!=\!4$             & ${\bf C}^{2}$ & $U(4)$  \\
	 $d=4$                     & $N\!=\!2$             & ${\bf C}^{2}$ & $U(2)$  \\ \hline\hline

\end{tabular}
\label{tab:*}
\end{center}

Using this results, we can now decompose the tangent vectors
to the moduli according to its transformation rules with respect
to the holonomy group. Let us concentrate in the $d=4$ case. For
$U(4)$, we get
\begin{equation}
U(4) \simeq U(1) \oplus SO(6).
\label{eq:III75}
\end{equation}
The matter multiplets will contain $6$ (real) scalars each, i. e., the
number of dimensions we compactify. Then, if we have $m$ of these
matter multiplets, the part of the moduli on which the $SO(6)$
part of the holonomy group is acting should be
\begin{equation}
O(6,m) / O(6) \times O(m).
\label{eq:III76}
\end{equation}
The $U(1)$ part of (\ref{eq:III75}) will act on the supergravity
multiplet so we expect, just from holonomy arguments, a moduli of
type
\begin{equation}
O(6,m) / O(6) \times O(m) \: \: \times Sl(2)/U(1).
\label{eq:III77}
\end{equation}
Now, we need to compute $m$. For heterotic string, the answer is
clear: $m=22$, and the total dimension of (\ref{eq:III77}) will
be $134$. Let us now consider the case of type IIA. From the table, 
we see that we should consider $K3\times T^2$ as
compactification manifold. Let us then first compute the
dimension of the moduli space:
\begin{eqnarray}
\hbox {Moduli of metrics and B fields on }K3  & = & 80 \nonumber \\
\hbox {Moduli of metrics and B fields on }T^2 & = & 4 \nonumber \\
b_1(K3\times T^2) & = & 2 \nonumber \\
b_3(K3\times T^2) & = & 44 \nonumber \\
\hbox {Axion-Dilaton} & = & 2 \nonumber \\ 
\hbox {Duals in }{\bf R}^4\hbox { to }2-\hbox {forms} & = & 2
\label{eq:III78}
\end{eqnarray}
which sums up to $134$. Notice that the $44$ in $b_3(K3\times
T^2)$ is coming from the $3$-cycles obtained from one $S^1$ of
$T^2$, and the $22$ elements in $H^2(K3;{\bf Z})$. The $3$-form
of IIA can be compactified on the $S^1$ cycles of $T^2$ to give
$2$-forms in four dimensions. Now, the dual of a $2$-form in
${\bf R}^4$ is scalar, so we get the last two extra moduli.
  
Now, we need to compare the two moduli spaces. If we expect
$S$-duality in $N\!=\!4$ for the heterotic compactification, the
moduli, once we have taken into account the $O(6,22;{\bf Z})$
$T$-duality, will look like
\begin{equation}
O(6,22;{\bf Z}) \backslash O(6,22) / O(6) \times O(22) \: \:
Sl(2,{\bf Z}) \backslash Sl(2) /U(1).
\label{eq:III79}
\end{equation}
Now, we have a piece in IIA looking naturally as the second term
in (\ref{eq:III79}), namely the moduli of the $\sigma$-model on
$T^2$, where $Sl(2,{\bf Z})$ will simply be part of the
$T$-duality. Thus, it is natural to relate the moduli of IIA on
the torus with the part of the moduli in (\ref{eq:III77}) coming
form the supergravity multiplet.
  
Let us now consider dual pairs in the second line of our table. 
There is a simple way to visualize under what
general conditions on the Calabi-Yau manifold with $SU(3)$
holonomy such dual pairs can exist. In fact, imagine that $K3$ is
ellipticaly fibered in $K3 \times T^2$; then, what we get is a
fibration on $\IP^1$ of the $T^4$ tori. Now, heterotic on $T^4$
is equivalent to type IIA on $K3$, so we expect that the
Calabi-Yau manifold should be a $K3$ fibration on $\IP^1$,
and that duality works fiberwise. Therefore, from general
arguments, we expect to get heterotic-type II dual pairs with
$N=2$ if we use Calabi-Yau manifolds which are $K3$ fibrations \cite{KaV,Klemm}.
In order to get a more precise picture, we need again to work out
the holonomy, which is $U(2)$ in this case. In $N=2$ we have two types 
of multiplets, vector and hypermultiplets. The vector multiplet
contains two real scalars, and the hypermultiplet four real
scalars. Then, we decompose $U(2)$ into $U(1)\oplus Sp(1)$, and
the moduli into vector and hypermultiplet part.
  
Let us first consider type IIA string on the Calabi-Yau manifold.
The moduli will contain $h^{1,1}$ deformations of $B$ and $J$,
$h^{2,1}$ complex deformations and $b^3$ RR deformations ($b^1$
does not contribute, as we are working with a Calabi-Yau
manifold). The total number, in real dimension, is
\begin{equation}
2h^{1,1}+4(h^{2,1}+1),
\label{eq:III80}
\end{equation}
where we have used that $b^3=2(h^{2,1}+1)$, in real dimension.
From (\ref{eq:III80}) we conclude that we have $h^{1,1}$ vector
multiplets, and $h^{2,1}+1$ hypermultiplets. Notice that
$4(h^{2,1}+1)$ is counting the $2$ coming from the dilaton and
the axion so, for type II we have combined dilaton and axion into
an hypermutiplet. 
  
Now, let us consider heterotic string on $K3 \times T^2$. The
moduli we must now consider, of $E_8 \times E_8$ bundles on $K3$,
is much more elaborated than that of $T^4$, or $T^6$, that we
have worked out. Part of the difficulty comes from anomaly
conditions. However, we know, accordding to Mukai's theorem, that
the moduli of holomorphic bundles on $K3$ is quaternionic, i. e.,
hyperk\"ahler, and that the moduli of the $\sigma$-model on $K3$
is of dimension $80$. We have yet the moduli on $T^2$, that will
be a manifold of $O(2,m)/O(2) \times O(m)$ type, and therefore a good
candidate for representing the vector multiplet. Thus, we get
\begin{eqnarray}
\hbox{Type IIA hypermultiplets} & \leftrightarrow & K3 \hbox {
Heterotic}, \nonumber \\
\hbox {Vector multiplets}       & \leftrightarrow & T^2.
\label{eq:III81}
\end{eqnarray}
From our previous discussion we know that vector multiplets, in
type IIA are related to $h^{1,1}$. Working fiberwise on a
$K3$ fibered Calabi-Yau manifold we get, for $h^{1,1}$,
\begin{equation}
h^{1,1}=1+\rho,
\label{eq:III82}
\end{equation}
with $\rho$ the Picard number of the $K3$ manifold. 
Then, in order to get a dual pair in the sense of
(\ref{eq:III81}) we need $m$ in the heterotic to statisfy
\begin{equation}
m=\rho.
\label{eq:III83}
\end{equation}
  
In order to control the value of $m$, from the heterotic 
point of view, we need to watch out for possible Wilson lines
that can be defined on $T^2$ after the gauge group has been fixed
from the $K3$ piece. From (\ref{eq:III82}) (and this was the
logic for the identification (\ref{eq:III83})), the heterotic
dilaton-axion is related to the $1$ term contributing in
(\ref{eq:III82}), i. e., the $2$-cycle defined by the base space
of the $K3$-fibration. 
  
As can be observed from (\ref{eq:III83}), if we do not freeze
either the K\"ahler class or the complex structure of $T^2$, the
minimum value for $\rho$ is $2$. This is the contribution to the
Picard lattice of a Dynkin diagram of type $A_2$, i. e., $SU(3)$.
A possible line of work opens here, in order to identify the moduli spaces
of vector multiplets for type IIA theories with the quantum
moduli, defined according to Seiberg and Witten, for gauge
theories, with
\begin{equation}
\hbox {rank } G=\rho.
\label{eq:III84}
\end{equation}

%%%%%%%%%%%%%%%%%%%%%%%%%%%%%%%%%%%%%%%%%%%%%%%%%%%%%%%%%%%%%%%%%%
%%%%%%%%%%%%%%%%%%%%%%%%%%%%%%%%%%%%%%%%%%%%%%%%%%%%%%%%%%%%%%%%%%

\section{Chapter IV}

\subsection{M-Theory Compactifications.}

Wittgenstein used to say that ``meaning is use''. This is the
kind of philosophycal slogan able to make unhappy the platonic
mathematician, but it is in essence the type of game we are going
to play in order to begin the study of M-theory \cite{Mth,HT,Town,Wsvd,HT2}. More precisely,
we will start without saying what M-theory is from a
microscopical point of view, giving instead a precise meaning to
M-theory compactifications.
  
Recall that our first contact with the idea of M-theory was in
connection with the interpretation of the moduli of type IIA string theory on $T^4$. In
that case the moduli, after including RR fields, was of the type
\begin{equation}
O(5,5;{\bf Z}) \backslash O(5,5) / O(5) \times O(5).
\label{eq:IV1}
\end{equation}
The M-theory interpretation of moduli (\ref{eq:IV1}) can be
summarized according to the equivalence
\begin{equation}
\hbox {M-theory compactified on }T^5 \leftrightarrow \hbox{ IIA} \hbox {
on } T^4,
\label{eq:IV2}
\end{equation}
and therefore, more generically,
\begin{equation}
\hbox {M-theory compactified on }X \times S^1 \leftrightarrow \hbox { IIA} \hbox {
on } X.
\label{eq:IV3}
\end{equation}
Let us now put rule (\ref{eq:IV3}) into work. In fact, one
particular case of (\ref{eq:IV3}) will consist in considering
M-theory on a manifold of type $B\times S^1 \times S^1$. Then,
using $T$-duality, we can get
\[
\hbox {M-theory compactified on }B \times S^1 \times S^1(R) \leftrightarrow \]
\begin{equation}
\hbox{ IIA} \hbox {
on } B\times S^1(R) \leftrightarrow \hbox { IIB} \hbox { on } B\times
S^1(\frac {\alpha'}{R}).
\label{eq:IV4}
\end{equation}
From (\ref{eq:IV4}) we see that in the $R \rightarrow \infty$
limit, we get type IIB string theory on $B$ or, equivalently,
M-theory on $B\times S^1$, since the second $S^1$ becomes
uncompactified. This is in fact a very close example to the ones
described in previous sections, under the generic name of
F-theory compactifications. Namely, the $R \rightarrow \infty$
limit in (\ref{eq:IV4}) can be interpreted as defining a
compactification of type IIB string theory on the base space $B$ of an elliptic
fibration $B \times S^1 \times S^1$, in the limit where the
volume of the elliptic fiber becomes zero. Following that path,
we get an interesting equivalence between M-theory on $B\times
S^1 \times S^1$, as elliptic fibration, in the limit in which the
volume of the elliptic fiber goes zero, and type IIB on $B$. This
stands as a surprise, when compared to the result derived from
the compactification rule (\ref{eq:IV3}). In fact, if $B$ is, for
instance, of dimension $d$, then we should expect that the
compactification of an eleven dimensional theory, as M-theory, on
$B\times S^1 \times S^1$, should lead to $11-d-2$ dimensions.
However, type IIB, which is ten dimensional, would lead, when
compactified on $B$, to a $10-d$ dimensional theory, so that one
dimension is missing. Getting rid off this contradiction requires
knowledge of the microscopic nature of M-theory. The first thing
to be required on M-theory is of course to have, as low energy
limit, eleven dimensional supergravity. There is a connection
between type IIA string theory and eleven dimensional
supergravity, as the corresponding Kaluza-Klein dimensional
reduction on an internal $S^1$, which allows an identification of
the string theory spectrum with supergravity. In particular, the
RR field in ten dimensions comes from the $g_{11,\mu}$ component
of the metric, while the dilaton is obtained from $g_{11,11}$.
The precise relation, in what is known as the string frame,
is\footnote{We have identified $g_{11,11}=e^{2\gamma}$.}
\begin{equation}
e^{-2\phi} = e^{-3\gamma},
\label{eq:IV5}
\end{equation}
with $\phi$ the type IIA dilaton field. In terms of the radius $R$ of
the $S^1$,
\begin{equation}
R=e^{2\phi/3}.
\label{eq:IV6}
\end{equation}
Using now equation (\ref{eq:III21}), we get a relation between
the $R$ of the internal manifold, $S^1$, and the string coupling
constant of type IIA string theory,
\begin{equation}
R=g^{2/3}.
\label{eq:IV7}
\end{equation}
From (\ref{eq:IV7}) it is obvious that, as $R \rightarrow
\infty$, we properly enter the M-theory region when $g$ is large,
i. e., working in the strong coupling regime of string theory.
Historically, this beatiful simple argument was put forward in
$1995$ by Witten \cite{Wsvd}. It is astonishing that, with all the pieces
around, nobody was able before to make at least the comment
relating the $R$ of eleven dimensional supergravity with the
string coupling constant, and to derive from it such a striking 
conjecture as it is that strongly coupled IIA strings are described
by eleven dimensional supergravity. In fact, there are good
reasons for such a mental obstacle in the whole community: first of all, nobody
did worry about type IIA dynamics, as it was a theory with only
uninteresting pure abelian gauge physics. Secondly, the Kaluza-Klein modes coming from the
compactification on $S^1$, which have a mass of the order $\frac
{1}{R}$, are charged with respect to the $U(1)$
gauge field defined by the $g_{11,\mu}$ piece of the metric. But
this $A_{\mu}$ field in ten dimensional type IIA string is of RR
type, so before the discovery of D-branes, there was no
candidate in the string spectrum to be put in correspondence with
these Kaluza-Klein modes, which can now be identified with
D-$0$branes.
  
Witten's approach to M-theory can be the
conceptual key to solve the problem concerning the missing dimension: in fact, something in the spectrum
is becoming massless as the volume of the elliptic fiber, in the 
case of $B \times S^1 \times S^1$, is sent to zero. Moreover, the
object becoming massless can be, as suggested by Sethi and
Susskind, interpreted as a Kaluza-Klein mode of an opening
dimension as the volume of the elliptic fiber goes zero. To
understand the nature of this object we should look more
carefully at M-theory. This theory is expected to contain a
fundamental two dimensional membrane; if this membrane wraps the
$2$-torus $S^1 \times S^1$, its mass becomes zero as the volume
of the fiber goes zero. Then, all what is left is to relate the
area with the standard Kaluza-Klein formula for compactifications
on $S^1$, which leads to 
\begin{equation}
\frac {1}{R} \sim L_1 L_2
\label{eq:IV8}
\end{equation}
solving our problem on the adequate interpretation of
(\ref{eq:IV4}).
  
Let us now concentrate on a concrete example of
(\ref{eq:IV4}): we will choose $X=B\times S^1 \times S^1_R$ as
representing a Calabi-Yau fourfold of $SU(4)$ holonomy. After
compactification, $SU(4)$ holonomy implies a three dimensional
theory with $N=2$ supersymmetry should be expected. Moreover,
sending $R \rightarrow \infty$ leads to a four dimensional $N=1$
theory. In order to work out the spectrum of the three
dimensional theory, standard Kaluza-Klein techniques can be used.
Compactification on the $2$-cycles of $H^2(X;{\bf Z})$ of the
$3$-form $C_{\mu \nu \rho}$ of eleven dimensional supergravity 
leads to a vector in three dimensions. Moreover, the K\"ahler
class can also be used to generate real scalars, from each
$2$-cycle. Thus, let us assume $r= \hbox {dim} H^2(X;{\bf Z})$;
then, the previous procedure produces $r$ real scalars and $r$ 
vector fields. In order to define $r$ $N=2$ vector multiplets in
three dimensions, with these vector fields, another set of
$r$ scalars is yet needed, in order to build the complex fields.
These extra $r$ scalars can, as usual, be identified with the
duals, in three dimensions, of the $1$-form vector fields: the three dimensional 
dual photon. 
  
Our next task will be reproducing, using M-theory, the well known
instanton effects in $N=2$ supersymmetric gauge theories in three
dimensions.

\subsection{M-Theory Instantons.}
\label{sec:Witins}

In order to define instantons in three dimensions, we will use
$5$-branes wrapped on the $6$-cycles of a Calabi-Yau fourfold
$X$ \cite{Wsp}. The reason for using $6$-cycles is understood as follows:
the gauge bosons in three dimensions are obtained from the
integration of the $3$-form $C_{\mu \nu \rho}$ over $2$-cycles.
Thus, in order to define the dual photon, we should consider the
dual, in the Calabi-Yau fourfold $X$, of $2$-cycles, which are
$6$-cycles. However, not any $6$-cycle can be interpreted as an
instanton with topological charge equal one, and therefore no
$6$-cycle will contribute to the three dimensional
superpotential.
  
If we interpret a $5$-brane wrapped on a $6$-cycle $D$ of $X$ as
an instanton, we can expect a superpotential of the type
\begin{equation}
W=e^{-(V_D+i\phi_D)},
\label{eq:s1}
\end{equation}
with $V_D$ the volume of $D$ measured in units of the $5$-brane
tension, and $\phi_D$ the dual photon field, associated with the
cycle $D$. In order to get, associated to $D$, a superpotential
like (\ref{eq:s1}), we need
\begin{itemize}
	\item[{i)}] To define a $U(1)$ transformation with
respect to which three dimensional fermions are charged.
	\item[{ii)}] To associate with the $6$-cycle $D$ a
violation of $U(1)$ charge, in the adecuate amount.
	\item[{iii)}] To prove that this $U(1)$ symmetry is not
anomalous.
	\item[{iv}] To interpret $\phi_D$ as the corresponding
Goldstone boson.
\end{itemize}
  
Following these steps, we will extend to M-theory the instanton
dynamics of three dimensional $N=2$ gauge theories described in
chapter I. We will start defining the $U(1)$ transformation. Let
$D$ be $6$-cycle in the Calabi-Yau fourfold $X$, and let us
denote by $N$ the normal bundle of $D$ in $X$. Since $X$ is a
Calabi-Yau manifold, its canonical bundle is trivial, and
therefore we get
\begin{equation}
K_D \simeq N,
\label{eq:s2}
\end{equation}
with $K_D$ the canonical bundle of $D$. Locally, we can interpret
$X$ as the total space of the normal bundle. Denoting by $z$ the
coordinate in the normal direction, the $U(1)$ transformation can
be defined as
\begin{equation}
z \rightarrow e^{i \theta} z.
\label{eq:s3}
\end{equation}
The $U(1)$ transformation defined by (\ref{eq:s3}) is very likely
not anomalous, since it is part of the diffeomorphisms of the
elevean dimensional theory; thus, it is a good candidate for the
$U(1)$ symmetry we are looking for. Next, we need to get the
$U(1)$ charge of the three dimensional fermions. However, before
doing so, we will review some well known facts concerning
fermions and Dirac operators on K\"ahler manifolds.
  
We will consider a K\"ahler manifold of
complex dimension $N$. In holomorphic coordinates,
\begin{equation}
g_{ab}=g_{\bar{a}\bar{b}}=0.
\label{eq:IV10}
\end{equation}
In these coordinates, the algebra of Dirac matrices becomes 
\begin{eqnarray}
\{\gamma^{a},\gamma^{b}\} & = &
\{\gamma^{\bar{a}},\gamma^{\bar{b}}\}=0, \nonumber \\
\{\gamma^{a},\gamma^{\bar{b}}\} & = & 2 g^{a \bar{b}}.
\label{eq:IV11}
\end{eqnarray}
The $SO(2N)$ spinorial representations of (\ref{eq:IV11}) can be
obtained in the standard Fock approach: a vacuum state is defined
by condition 
\begin{equation}
\gamma^{a} |\Omega> = 0,
\label{eq:IV12}
\end{equation}
and $n$-particle states are defined by
\begin{equation}
\gamma^{\bar{a}}\gamma^{\bar{b}} \ldots \gamma^{\bar{n}}
|\Omega>.
\label{eq:IV13}
\end{equation}
A spinor field $\psi(z,\bar{z})$ on the K\"ahler manifold takes
values on the spinor bundle defined by this Fock representation:
\begin{equation}
\psi(z,\bar{z}) = \phi(z,\bar{z}) |\Omega> + \phi_{\bar{a}}
(z,\bar{z}) \gamma^{\bar{a}} |\Omega> +
\phi_{\bar{a}\bar{b}}(z,\bar{z}) \gamma^{\bar{a}}
\gamma^{\bar{b}} |\Omega> \ + \cdots
\label{eq:IV14}
\end{equation}
  
The spaces $\Omega^{0,q}$ of $(0,q)$-forms, 
generated by the Dirac operator, define the Dolbeaut cohomology of the K\"ahler manifold.
Using this notation, the two different chirality spinor bundles
are 
\begin{eqnarray}
S^+ & = & (K^{1/2} \otimes \Omega^{0,0}) \oplus (K^{1/2} \otimes
\Omega^{0,2}) \oplus (K^{1/2} \otimes \Omega^{0,4}) \oplus
\cdots, \nonumber \\
S^- & = & (K^{1/2} \otimes \Omega^{0,1}) \oplus (K^{1/2} \otimes
\Omega^{0,3}) \oplus (K^{1/2} \otimes \Omega^{0,5}) \oplus
\cdots, 
\label{eq:IV18}
\end{eqnarray}
and the change of chirality (the index for the Dirac operator on
the manifold $X$) will be given by the aritmetic genus,
\begin{equation}
\chi = \sum^{N} (-1)^n h_n,
\label{eq:IV19}
\end{equation}
where $h_n= \hbox {dim }\Omega^{0,n}$.
  
The previous comments can be readily applied to the case of a
six dimensional divisor $D$ in a Calabi-Yau fourfold $X$. Now, we
should take into account the normal budle $N$, to $D$, in $X$.
Using the fact that $X$ is Calabi-Yau, i. e., with trivial
canonical bundle, we conclude that $N$ is isomorphic to $K_D$,
the canonical bundle on $D$. The spinor bundle on $N$ will be
defined by
\begin{equation}
K_D^{1/2} \oplus K_D^{-1/2}.
\label{eq:IV20}
\end{equation}
In fact, in this case the complex dimension of $N$ is one, and
the vacum and filled states have, respectively, $U(1)$ charges
$\frac {1}{2}$ and $- \frac
{1}{2}$. On the other hand, the spinor budle on $D$ will be
defined by (\ref{eq:IV18}), with $K=K_D$. Thus, spinors on $D$
are, up to the $SO(3)$ spacetime part, taking values in the
positive and negative chirality bundles
\begin{eqnarray}
\hat{S}^+ & = & (K^{1/2} \oplus K^{-1/2}) \otimes [ (K^{1/2} \otimes
\Omega^{0,0}) \oplus (K^{1/2} \otimes \Omega^{0,2})], \nonumber \\
\hat{S}^- & = & (K^{1/2} \oplus K^{-1/2}) \otimes [ (K^{1/2} \otimes
\Omega^{0,1}) \oplus (K^{1/2} \otimes \Omega^{0,3})].
\label{eq:IV21}
\end{eqnarray}
  
Now, we are interested in a change of $U(1)$ charge, with the
$U(1)$ charge defined by the $\frac {1}{2}$ and $-\frac {1}{2}$
charges of the spinor bundle (\ref{eq:IV20}) on $N$. 
  
For spinors of a given chirality,
the change of $U(1)$ charge is given by 
\begin{equation}
\hbox {dim }(K \otimes \Omega^{0,0}) + \hbox {dim }(K \otimes
\Omega^{0,2}) - \hbox {dim } (\Omega^{0,0}) - \hbox {dim }
(\Omega^{0,2}).
\label{eq:IV23}
\end{equation}
Using now Serre's duality,
\begin{equation}
\hbox {dim }(K\otimes \Omega^{0,3-n}) = \hbox {dim
}(\Omega^{0,n}),
\label{eq:IV24}
\end{equation}
we get that the number of holomorphic $(0,k)$-forms is equal to
the number of holomorphic sections in $K \otimes \Omega^{0,3-k}$,
and therefore the number of fermionic zero modes with $U(1)$
charge equal $\frac {1}{2}$ is given by $h_3+h_1$, and the number
of fermionic zero modes with $- \frac {1}{2}$ $U(1)$ charge, is
given by $h_0+h_2$ (here we have used the Dirac operator
$\bar{\partial}+\bar{\partial}^*$, with $\bar{\partial}^*$ the
adjoint of $\bar{\partial}$. Thus, the index for the twisted spin
bundle (\ref{eq:IV21}) is given by the holomorphic Euler
characteristic,
\begin{equation}
\chi(D) = h_0 -h_1 +h_2 -h_3.
\end{equation}
  
Now, each of these fermionic zero modes is doubled once we tensor
with spinors in ${\bf R}^3$. In summary, for each $6$-cycle $D$
we get an effective vertex with a net change of $U(1)$ charge
equal to $\chi(D)$.
  
Therefore, in order to get the three dimensional in a three
dimensional $N=2$ theory, we need to look for $6$-cycles $D$,
with $\chi(D)=1$, as the net change of $U(1)$ charge in that case
is one, provided, as we did in (\ref{eq:IV20}), we normalize the
$U(1)$ charge of the fermions to be $\frac {1}{2}$. More
precisely, the number of fermionic zero modes for a three
dimensional instanton, defined by a $6$-cycle $D$, is $2
\chi(D)$.

\subsection{D-Brane Configurations in Flat Space.}

We will consider a D-brane of dimension $p$, in flat ten 
dimensional Minkowski space, and with a flat $p+1$ dimensional
worldvolume. The quantization of the open superstring ending on
the D-brane defines a low energy field theory, which is ten
dimensional $N=1$ supersymmetric Yang-Mills, with $U(1)$ gauge
group. The dimensional reduction of this theory to $p+1$
dimensions describes the massless excitations propagating on the
worldvolume of the $p$ dimensional D-brane. We will use as
worldvolume coordinates $x^0,x^1,\ldots,x^p$. The worldvolume
lagrangian will contain a $U(1)$ massless gauge field $A_i(x_s)$,
with $i,s=0,\ldots,p$, and a set of scalar fields $\phi_j(x_s)$,
$j=p+1,\ldots,9$, transforming in the adjoint representation. We
can geometrically interpret the set of fields $\phi_j(x_s)$ as
representing the ``location'' of the flat D-brane in transverse space. The simplest
generalization of the previous picture corresponds to
configurations of $k>1$ parallel D-$p$branes. In this case we have,
in addition to the massless excitations, a set of $k$ 
massive excitations corresponding to open strings
ending on different D-branes.
  
The field theory interpretation of this configuration of D-branes
would be that of a gauge thory with $U(k)$ gauge group, spontaneously broken to
$U(1)^k$, with the strings stretching between different D-branes
representing charged massive vector bosons. To get such an interpretation, we can
start with $N=1$ $U(k)$ supersymmetric Yang-Mills in ten
dimensions, and perform again dimensional reduction down to $p+1$
dimensions. In this case, we will get a set of scalar fields,
$X^{j}(x_s)$, with $j=p+1,\ldots,9$, which are now $k \times k$
matrices, transforming in the adjoint of $U(k)$. Moreover, the
kinetic term in ten dimensions produces a potential of the form
\begin{equation}
V= \frac {g^2}{2} \sum_{i,j=p+1}^9 \hbox {tr} [X^i,X^j]^2.
\label{eq:d1}
\end{equation}
  
As we have already observed in many examples before, this
potential possesses flat directions, correspoding to classical
vacumm states. These flat directions are defined by diagonal
$X^i$ matrices,
\begin{equation}
X^i = \left( \begin{array}{ccc} \lambda_1^i &   &   \\
				  & \ddots  &       \\
				  &  & \lambda_k^i  \end{array} \right).
\label{eq:d2}
\end{equation}
  
On each of these vacua, the $U(k)$ gauge symmetry is spontaneously broken to $U(1)^k$; thus, 
we can use these vacuum configurations to describe sets of $k$ parallel $p$ dimensional 
D-branes. In fact, as we observe for the simpler case of one D-brane, 
the set of scalars appearing by dimensional reduction has the 
geometrical interpretation of the position of the D-brane. In the case (\ref{eq:d2}), 
we can in fact consider $\lambda_l^i$ as defining the $i^{th}$-coordinate of the 
$l^{th}$-brane. This is consistent with the idea of interpreting the strings 
stretching between different D-branes as massive vector bosons. In fact, the mass 
of this string states would be
\begin{equation}
M \sim g | \lambda_l - \lambda_m |,
\label{eq:d3}
\end{equation}
for a $(l,m)$ string. This is, in fact, the Higgs mass of the corresponding 
massive charged boson. In summary, merging the previous comments into
a lemma: the classical moduli space of the worldvolume lagrangian of a
D-brane coincides with its transversal space. It is important
realizing that only the minima of the potential (\ref{eq:d1}), i.
e., the moduli space of the worldvolume lagrangian, is the one
possessing this simple geometrical interpretation. In
particular, the dimensional reduction, down to $p+1$ dimensions,
of $N=1$ $U(k)$ gauge theory, describes a set of $k$ parallel
branes, but its full fledged dynamics is described by the
complete matrix $X^i$, with non vanishing off diagonal terms. A
nice way to think about the meaning of (\ref{eq:d2}) is again in
terms of `t Hooft's abelian projection. In fact, we can think of
(\ref{eq:d2}) as a unitary gauge fixing, where we now allow
$\lambda_l^i$ to depend on the worldvolume coordinates. The case
of flat parallel D-branes corresponds to a Higgs phase, with
$\lambda_l^i$ constant functions on the worldvolume. Moreover, we
can even consider the existence of singularities, which will be
points where two eigenvalues coincide,
\begin{equation}
\lambda_l^i = \lambda_{l+1}^i, \: \: \forall i.
\label{eq:d4}
\end{equation}
It is quite obvious realizing that (\ref{eq:d4}) imposes three
constraints, so we expect, for $p$-dimensional D-branes, that on
a $p-3$-dimensional region of the worldvolume, two consecutive
D-branes can overlap. The $p-3$ region in the $p+1$ dimensional worldvolume
of the D-brane will represent, from the point of view of $p+1$
dynamics, a monopole, in the very same sense as is the case in
`t Hooft's abelian projection.

Next we will consider some brane configurations for type IIA and
type IIB string theory (some of the widely increasing refences are those 
from \cite{HW} to \cite{Wbqcd}). In order to define these configurations we will first
work out the allowed vertices for intersecting branes. Let us
start with a vertex of type $(p,1^F)$, corresponding to a
Dirichlet $p$-brane, and a fundamental string ending on the
D-brane worldvolume. In type IIA $p$ should be even, and odd for
type IIB. In fact, the RR fields for type IIA and type IIB string
theory are
\begin{eqnarray}
\hbox {IIA} & \rightarrow & A_{\mu} \: \: A_{\mu \nu \rho},
\nonumber \\
\hbox {IIB} & \rightarrow & \chi \: \: B_{\mu \nu} \: \: A_{\mu
\nu \rho \sigma}. 
\end{eqnarray}
The corresponding strength tensors are, respectively, two and
four-forms for type IIA, and one, three and five-forms for type
IIB. Thus, the sources are D-branes of dimensions zero and two,
for type IIA string theory, and one and three for type IIB. In
addition, we have the (Hodge) magnetic duals, which are six and
four D-branes for type IIA string theory, and five and three
D-branes for type IIB (notice that the threebrane in type IIB is
self dual). Besides, for the $\chi$ field in type IIB, the source
is a $-1$ extended object, and its dual is a D-7brane. 
  
Let us then start with a vertex of type $(p,1^F)$ in type IIB, i. e.,
with $p$ odd. We can use the $Sl(2,{\bf Z})$ duality symmetry of
type IIB strings to transform this vertex into a $(p,1)$ vertex,
between a D-$p$brane and a D-$1$brane, or D-string. By performing
$j$ T-duality transformations on the spacetime directions
orthogonal to the worldvolume of the D-brane and the D-string, we
pass form $(p,1^F)$ to a vertex $(p+j,1^F+j)$ of two D-branes,
sharing $j$ common worldvolume coordinates. If $j$ is even, we
end up with a vertex in type IIB, and if $j$ is odd with a vertex
in type IIA. Namely, through a T-duality transformation we pass
from type IIB string theory to type IIA. As an
example, we will consider the vertex $(3,1^F)$ in type IIB string
theory. After a S-duality transformation in the $Sl(2,{\bf Z})$
duality group of type IIB strings, and two T-duality
transformations, we get the vertex $(5,3)$ for branes. As we are
in type IIB, we can perform a duality transformation on it to
generate the vertex $(5^{NS},3)$, between the solitonic
Neveau-Schwarz fivebrane and a D-$3$brane.
  
Let us now consider some brane configurations build up using the
vertices $(5,3)$ and $(5^{NS},3)$ in type IIB theory \cite{HW}. In
particular, we will consider solitonic fivebranes, with
worldvolume coordinates $x^0,x^1,x^2,x^3,x^4$ and $x^5$, located
at some definite values of $x^6,x^7,x^8$ and $x^9$. 
It is convenient to organize the
coordinates of the fivebrane as $(x^6,\vec{\omega})$ where
$\vec{\omega}=(x^7,x^8,x^9)$. By construction of the vertex, the
D-$3$brane will share two worldvolume coordinates, in addition to
time, with the fivebrane. Thus, we can consider D-$3$branes with
worldvolume coordinates $x^0,x^1,x^2$ and $x^6$. If we put a
D-$3$brane in between two solitonic fivebranes, at $x^6_2$ and $x^6_1$ positions 
in the $x^6$ coordinate, then the
worldvolume of the D-$3$brane will be finite in the $x^6$
direction (see Figure $1$).

        %%%%%%%%%%%%%%%%%%%%%%%%%%%%%%%%%%%%%%%%%%%%%%%%%%%%%%%%%%%%%%%%%%
        %%%%%%%%%%%%%%%      Figure 1     %%%%%%%%%%%%%%%%%%%%%%%%%%%%%%%%
        %%%%%%%%%%%%%%%%%%%%%%%%%%%%%%%%%%%%%%%%%%%%%%%%%%%%%%%%%%%%%%%%%%

        \begin{figure}[htbp]
        \centering

        \begin{picture}(400,170)

                \put(150,60){\line(1,0){140}} \put(150,75){\line(1,0){140}}
                \put(150,90){\line(1,0){140}} \put(150,45){\line(1,0){140}} 

		\put(100,120){\vector(1,0){250}} \put(330,138){$x^6$-coordinate}                

		\put(80,92){NS-$5$brane}
		\put(304,92){NS-$5$brane}
                \put(200,15){D-$3$branes}
		
		\put(148,138){$x^6_1$} \put(288,138){$x^6_2$}

        \thicklines
                \put(150,20){\line(0,1){100}}
                \put(290,20){\line(0,1){100}}

        \end{picture}
                \caption{Solitonic fivebranes with $n$ Dirichlet threebranes stretching along them.}
        \end{figure}
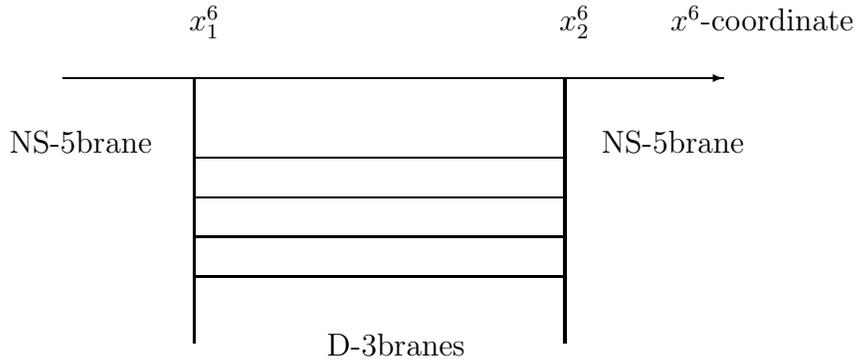

        %%%%%%%%%%%%%%%%%%%%%%%%%%%%%%%%%%%%%%%%%%%%%%%%%%%%%%%%%%%%%%%%%
        %%%%%%%%%%%%%%%%%%%%%%%%%%%%%%%%%%%%%%%%%%%%%%%%%%%%%%%%%%%%%%%%%

Therefore, the macroscopic physics, i. e., for scales 
larger than $|x^6_2-x^6_1|$, can be effectively described by a
$2+1$ dimensional theory. In order to unravel what kind of $2+1$
dimensional theory, we are obtaining through this brane
configuration, we must first work out the type of constraint
impossed by the fivebrane boundary conditions. In fact, the
worldvolume low energy lagrangian for a D-$3$brane is a $U(1)$
gauge theory. Once we put the D-$3$brane in between two solitonic
fivebranes we imposse Neumann boundary conditions, in the $x^6$
direction, for the fields living on the D-$3$brane worldvolume.
This means in particular that for scalar fields we imposse
\begin{equation}
\partial_6 \phi =0
\label{eq:f1}
\end{equation}
and, for gauge fields,
\begin{equation}
F_{\mu 6} =0, \: \: \: \: \: \: \mu=0,1,2.
\label{eq:f2}
\end{equation}
Thus, the three dimensional $U(1)$ gauge field, $A_{\mu}$, with
$\mu=0,1,2$, is unconstrained which already means that we can
interpret the effective three dimensional theory as a $U(1)$
gauge theory for one D-$3$brane, and therefore as a $U(n)$ gauge
theory for $n$ D-$3$branes. Next, we need to discover the amount
of supersymmetry left unbroken by the brane configuration. If we
consider Dirichlet threebranes, with worldvolume coordinates
$x^0,x^1,x^2$ and $x^6$, then we are forcing the solitonic
fivebranes to be at positions $(x^6,\vec{\omega}_1)$ and
$(x^6,\vec{\omega}_2)$, with $\vec{\omega}_1=\vec{\omega}_2$. In
this particular case, the allowed motion for the D-$3$brane is
reduced to the space ${\bf R}^3$, with coordinates $x^3,x^4$ and
$x^5$. These are the coordinates on the fivebrane worldvolume
where the D-$3$brane ends. Thus, we have defined on the
D-$3$brane three scalar fields. By condition (\ref{eq:f1}), the
values of these scalar fields can be constrained to be constant
on the $x^6$ direction. What this in practice means is that the
two ends of the of the D-$3$brane have the same $x^3,x^4$ and
$x^5$ coordinates. Now, if we combine these three scalar fields
with the $U(1)$ gauge field $A_{\mu}$, we get an $N=4$ vector
multiplet in three dimensions. Therefore, we can conclude that
our effective three dimensional theory for $n$ parallel
D-$3$branes suspended between two solitonic fivebranes (Figure $1$) is a gauge
theory with $U(n)$ gauge group, and $N=4$ supersymmetry. Denoting
by $\vec{v}$ the vector $(x^3,x^4,x^5)$, the Coulomb branch of
this theory is parametrized by the $v_i$ positions of the $n$
D-$3$branes (with $i$ labelling each brane). In addition, we
have, as discussed in chapter II, the dual photons for each
$U(1)$ factor. In this way, we get the hyperk\"ahler structure of
the Coulomb branch of the moduli. Hence, a direct way to get
supersymmetry preserved by the brane configuration is as follows.
The supersymmetry charges are defined as 
\begin{equation}
\epsilon_L Q_L + \epsilon_R Q_R,
\label{eq:f3}
\end{equation}
where $Q_L$ and $Q_R$ are the supercharges generated by the left
and right-moving worldsheet degrees of freedom, and $\epsilon_L$
and $\epsilon_R$ are ten dimensional spinors. Each solitonic $p$brane, 
with worldvolume extending along
$x^0,x^1,\ldots,x^p$, imposses the conditions
\begin{equation}
\epsilon_L = \Gamma_0 \ldots \Gamma_p \epsilon_L, \: \:
\epsilon_R = - \Gamma_0 \ldots \Gamma_p \epsilon_R,
\label{eq:f4}
\end{equation}
in terms of the ten dimensional Dirac gamma matrices, $\Gamma_i$;
on the other hand, the D-$p$branes, with worldvolumes extending
along $x^0,x^1,\ldots,x^p$, imply the constraint
\begin{equation}
\epsilon_L = \Gamma_0 \Gamma_1 \ldots \Gamma_p \epsilon_R.
\label{eq:f5}
\end{equation}
Thus, we see that NS solitonic fivebrane, with worldvolume
located at $x^0,x^1,x^2,x^3,x^4$ and $x^5$, and equal values of
$\vec{\omega}$, and Dirichlet threebranes with worldvolume along
$x^0,x^1,x^2$ and $x^6$, preserve eight supersymmetries on the
D-$3$brane worldvolume or, equivalently, $N=4$ supersymmetry on the
effective three dimensional theory. 
  
The brane array just described allows a simple computation of the
gauge coupling constant of the effective three dimensional
theory: by standard Kaluza-Klein reduction on the finite $x^6$
direction, after 
integrating over the (compactified) $x^6$ direction to reduce the lagrangian 
to an effective three dimensional lagrangian, the gauge coupling constant is given by
\begin{equation}
\frac {1}{g_3^2} = \frac {|x_6^2-x_6^1|}{g_4^2},
\label{eq:f6}
\end{equation}
in terms of the four dimensional gauge coupling constant.
Naturally, (\ref{eq:f6}) is a classical expression that is not
taking into account the effect on the fivebrane position at $x^6$
of the D-$3$brane ending on its worldvolume. In fact, we can
consider the dependence of $x^6$ on the coordinate $\vec{v}$,
normal to the position of the D-$3$brane. The dynamics of the
fivebranes should then be recovered when the Nambu-Goto action of
the solitonic fivebrane is minimized. Far from the influence of
the points where the fivebranes are located (at large values of
$x^3,x^4$ and $x^5$), the equation of motion is simply three
dimensional Laplace's equation,
\begin{equation}
\nabla^2 x^6(x^3,x^4,x^5)=0,
\label{eq:f7}
\end{equation}
with solution 
\begin{equation}
x^6(r)= \frac {k}{r} + \alpha,
\label{eq:f8}
\end{equation}
where $k$ and $\alpha$ are constants depending on the threebrane tensions, 
and $r$ is the spherical
radius at the point $(x^3,x^4,x^5)$. From (\ref{eq:f8}), it is
clear that there is a well defined limit as $r \rightarrow
\infty$; hence, the difference $x^6_2-x^6_1$ is a well defined
constant, $\alpha_2-\alpha_1$, in the $r \rightarrow \infty$
limit.
  
Part of the beauty of brane technology is that it allows to
obtain very strong results by simply performing geometrical brane
manipulations. We will now present one example, concerning our
previous model. If we consider the brane configuration from the
point of view of the fivebrane, the $n$ suspended threebranes
will look like $n$ magnetic monopoles. This is really suggesting
since, as described in chapter II, we know that the Coulomb
branch moduli space of $N=4$ supersymmetric $SU(n)$ gauge
theories is isomorphic to the moduli space of BPS monopole
configurations, with magnetic charge equal $n$. This analogy can
be put more precisely: the vertex $(5^{NS},3)$ can, as described
above, be transformed into a $(3,1)$ vertex. In this case, from
the point of view of the threebrane, we have a four dimensional
gauge theory with $SU(2)$ gauge group broken down to $U(1)$, and
$n$ magnetic monopoles. Notice that by passing from the
configuration build up ussing $(5^{NS},3)$ vertices, to that
build up with the $(3,1)$ vertex, the Coulomb moduli remains the
same.
  
Next, we will work out the same configuration, but now with the
vertex $(5,3)$ made out of two Dirichlet branes. The main difference with
the previous example comes from the boundary conditions
(\ref{eq:f1}) and (\ref{eq:f2}), which should now be replaced by
Dirichlet boundary conditions. We will choose as worldvolume
coordinates for the D-$5$branes $x^0,x^1,x^2,x^7,x^8$ and $x^9$,
so that they will be located at some definite values of
$x^3,x^4,x^5$ and $x^6$. As before, let us denote this positions
by $(\vec{m},x^6)$, where now $\vec{m}=(x^3,x^4,x^5)$. An
equivalent configuration to the one studied above will be now a
set of two D-$5$branes, at some points of the $x^6$ coordinate,
that we will again call $x^6_1$ and $x^6_2$, subject to 
$\vec{m}_1=\vec{m}_2$, with D-$3$branes stretching between them
along the $x^6$ coordinate, with worldvolume extending again
along the coordinates $x^0,x^1,x^2$ and $x^6$ (Figure $2$). Our task now will

        %%%%%%%%%%%%%%%%%%%%%%%%%%%%%%%%%%%%%%%%%%%%%%%%%%%%%%%%%%%%%%%%%%
        %%%%%%%%%%%%%%%      Figure 2     %%%%%%%%%%%%%%%%%%%%%%%%%%%%%%%%
        %%%%%%%%%%%%%%%%%%%%%%%%%%%%%%%%%%%%%%%%%%%%%%%%%%%%%%%%%%%%%%%%%%

        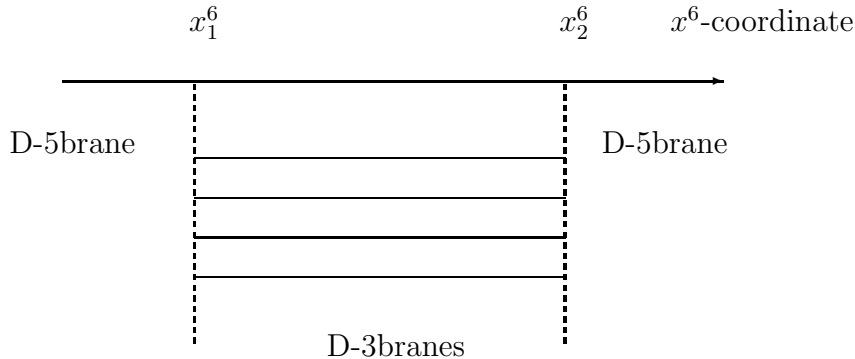
\begin{figure}[htbp]
        \centering

        \begin{picture}(400,170)

                \put(150,60){\line(1,0){140}} \put(150,75){\line(1,0){140}}
                \put(150,90){\line(1,0){140}} \put(150,45){\line(1,0){140}} 

		\put(100,119){\vector(1,0){250}} \put(330,138){$x^6$-coordinate}                

		\put(80,92){D-$5$brane}
		\put(304,92){D-$5$brane}
                \put(200,15){D-$3$branes}
		
		\put(148,138){$x^6_1$} \put(288,138){$x^6_2$}

        \thicklines
                \multiput(150,20)(0,4){25}{\line(0,1){2}}
                \multiput(290,20)(0,4){25}{\line(0,1){2}}

        \end{picture}
                \caption{Dirichlet threebranes extending between a pair of Dirichlet fivebranes (in dashed lines).}
        \end{figure}

        %%%%%%%%%%%%%%%%%%%%%%%%%%%%%%%%%%%%%%%%%%%%%%%%%%%%%%%%%%%%%%%%%
        %%%%%%%%%%%%%%%%%%%%%%%%%%%%%%%%%%%%%%%%%%%%%%%%%%%%%%%%%%%%%%%%%

be the description of the effective three dimensional theory on
these threebranes. The end points of the D-$3$branes on the
fivebrane worlvolumes will now be parametrized by values of
$x^7,x^8$ and $x^9$. This means that we have three scalar fields
in the effective three dimensional theory. The scalar fields
corresponding to the coordinates $x^3,x^4,x^5$ and $x^6$ of the
threebranes are forzen to the constant values where the
fivebranes are located. Next, we should consider what happens to
the $U(1)$ gauge field on the D-$3$brane worldvolume. Impossing
Dirichlet boundary conditions for this field is equivalent to 
\begin{equation}
F_{\mu \nu} =0, \: \: \: \: \mu, \nu=0,1,2,
\label{eq:f9}
\end{equation}
i. e., there is no electromagnetic tensor in the effective three dimensional field
theory. Before going on, it would be convenient summarizing the
rules we have used to impose the different boundary conditions.
Consider a D-$p$brane, and let $M$ be its worldvolume manifold,
and $B=\partial M$ the boundary of $M$. Neumann and Dirichlet
boundary conditions for the gauge field on the D-$p$brane
worldvolume are defined respectively by
\begin{eqnarray}
N & \longrightarrow & F_{\mu \rho} =0, \nonumber \\
D & \longrightarrow & F_{\mu \nu} =0,
\label{eq:f10}
\end{eqnarray}
where $\mu$ and $\nu$ are directions of tangency to $B$, and
$\rho$ are the normal coordinates to $B$. If $B$ is part of the
worldvolume of a solitonic brane, we will imposse Neumann
conditions, and if it is part of the worldvolume of a Dirichlet
brane, we will imposse Dirichlet conditions. Returning to
(\ref{eq:f9}), we see that on the three dimensional effective
theory, the only non vanishing component of the four dimensional
strenght tensor is $F_{\mu 6} \equiv \partial_{\mu}b$. Therefore,
all together we have four scalar fields in three dimensions or,
equivalently, a multiplet with $N=4$ supersymmetry. Thus, the
theory defined by the $n$ suspended D-$3$branes in between a pair
of D-$5$branes, is a theory of $n$ $N=4$ massless
hypermultiplets.
  
There exits a different way to interpret the theory, namely as a
magnetic dual gauge theory. In fact, if we perform a duality
transformation in the four dimensional $U(1)$ gauge theory, and
use magnetic variables $*F$, instead of the electric field $F$,
what we get in three dimensions, after impossing D-boundary
conditions, is a dual photon, or a magnetic $U(1)$ gauge theory.
  
The configuration chosen for the worldvolume of the Dirichlet and
solitonic fivebranes yet allows a different configuration with
D-$3$branes suspended between a D-$5$brane and a NS-$5$brane.
This is in fact consistent with the supersymmetry requirements
(\ref{eq:f4}) and (\ref{eq:f5}). Namely, for the Dirichlet
fivebrane we have
\begin{equation}
\epsilon_L = \Gamma_0 \Gamma_1 \Gamma_2 \Gamma_7 \Gamma_8
\Gamma_9 \epsilon_R. 
\label{eq:f11}
\end{equation}
The solitonic fivebrane imposses 
\begin{equation}
\epsilon_L = \Gamma_0 \ldots \Gamma_5 \epsilon_L, \: \:
\epsilon_R = - \Gamma_0 \ldots \Gamma_5 \epsilon_R,
\label{eq:f12}
\end{equation}
while the suspended threebranes imply
\begin{equation}
\epsilon_L = \Gamma_0 \Gamma_1 \Gamma_2 \Gamma_6 \epsilon_R,
\label{eq:f13}
\end{equation}
which are easily seen to be consistent. The problem now is that
the suspended D-$3$brane is frozen. In fact, the position
$(x^3,x^4,x^5)$ of the end point of the NS-$5$brane is equal to
the position $\vec{m}$ of the D-$5$brane, and the position 
$(x^7,x^8,x^9)$ of the end point on the D-$5$brane is forced to be
equal to the position $\vec{\omega}$ of the NS-$5$brane. The fact
that the D-$3$brane is frozen means that the theory defined on it
has no moduli, i. e., posseses a mass gap. 
  
Using the vertices between branes we have described so far we can
build quite complicated brane configurations. When Dirichlet
threebranes are placed to the right and left of a fivebrane, open
strings can connect the threebranes at different
sides of the fivebrane. They will represent hypermultiplets
transforming as $(k_1,\bar{k}_2)$, with $k_1$ and $k_2$ the number 
of threebranes to the left and right, respectively, of the fivebrane. In case the fivebrane is
solitonic, the hypermultiplets are charged with respect to an
electric group, while in case it is a D-$5$brane, they are
magnetically charged. Another possibility is that with a pair of
NS-$5$branes, with D-$3$branes extending between them, and also a
D-$5$brane located between the two solitonic fivebranes. A
massless hypermultiplet will now appear whenever the
$(x^3,x^4,x^5)$ position of the D-$3$brane coincides with the
$\vec{m}=(x^3,x^4,x^5)$ position of the D-$5$brane.
  
\vspace{2 mm}
  
So far we have used brane configurations for representing
different gauge theories. In these brane configurations we have
considered two different types of moduli. For the examples
described above, these two types of moduli are as follows: the
moduli of the effective three dimensional theory, corresponding
to the different positions where the suspended D-$3$branes can be
located, and the moduli corresponding to the different locations
of the fivebranes, which are being used as boundaries. This
second type of moduli specifies, from the point of view of the
three dimensional theory, different coupling constants; hence, we
can move the location of the fivebranes, and follow the changes
taking place in the effective three dimensional theory. Let us
then consider a case with two solitonic branes, and a Dirichlet
fivebrane placed between them. Let us now move the NS-$5$brane on
the left of the D-$5$brane to the right. In doing so, there is a
moment when both fivebranes meet, sharing a common value of
$x^6$. If the interpretation of the hypermultiplet we have
presented above is correct, we must discover what happens to the
hypermultiplet after this exchange of branes has been performed.
In order to maintain the hypermultiplet, a new D-$3$brane should
be created after the exchange, extending from the right solitonic
fivebrane to the Dirichlet fivebrane. To prove this we will need
D-brane dynamics at work. Let us start considering two
interpenetrating closed loops, $C$ and $C'$, and suppose
electrically charged particles are moving in $C$, while
magnetically charged particles move in $C'$. The linking number
$L(C,C')$ can be defined using the standard Wilson and `t Hooft
loops. Namely, we can measure the electric flux passing through
$C'$ or, equivalently, compute $B(C')$, or measure the magnetic
flux passing through $C$, i. e., the Wilson line $A(C)$. In both
cases, what we are doing is integrating over $C'$ and $C$ the
dual to the field created by the particle moving in $C$ and $C'$,
respectively. Let us now extend this simple result to the case of
fivebranes. A fivebrane is a source of $7$-form tensor field, and
its dual is therefore a $3$-form. We will call this $3$-form
$H_{NS}$ for NS-$5$branes, and $H_D$ for D-$5$branes. Now, let us
consider the worldvolume of the two fivebranes,
\begin{eqnarray}
{\bf R}^3 & \times & Y_{NS}, \nonumber \\
{\bf R}^3 & \times & Y_D.
\label{eq:f14}
\end{eqnarray}
We can now define the linking number as we did before, in the
simpler case of a particle:
\begin{equation}
L(Y_{NS},Y_D) = - \int_{Y_{NS}} H^D = \int_{Y_D} H^{NS}.
\label{eq:f15}
\end{equation}
The $3$-form $H^{NS}$ is locally $dB_{NS}$. Since we have no 
sources for $H^{NS}$, we can use $H^{NS}=dB_{NS}$ globally;
however, this requires $B$ to be globally defined, or gauge
invariant. In type IIB string theory, $B$ is not gauge invariant;
however, on a D-brane we can define the combination $B_{NS}-F_D$,
which is invariant, with $F_D$ the two form for the $U(1)$ gauge
field on the D-brane. Now, when the D-$5$brane and the
NS-$5$brane do not intersect, the linking number is obviously
zero. When they intersect, this linking number changes, which
means that (\ref{eq:f15}) should, in that case, be non vanishing.
Writing
\begin{equation}
\int_{Y_D} H^{NS} = \int_{Y_D} dB_{NS} - dF_D,
\label{eq:f16}
\end{equation}
we observe that the only way to get linking numbers would be
adding sources for $F_D$. These sources for $F_D$ are point like
on $Y_D$, and are therefore the D-$3$branes with worldvolume
${\bf R}^3 \times C$, with $C$ ending on $Y_D$, which is
precisely the required appearance of extra D-$3$branes.

\subsection{D-Brane Description of Seiberg-Witten Solution.}

In the previous example we have considered type IIB string theory and three and
fivebranes. Now, let us consider type IIA strings, where we have
fourbranes that can be used to define, by analogy with the
previous picture, $N=2$ four dimensional gauge theories \cite{Wm4}. The idea
will again be the use of solitonic fivebranes, with sets of
fourbranes in between. The only difference now is that the
fivebrane does not create a RR field in type IIA string theory
and, therefore, the physics of the two parallel solitonic
fivebranes does not have the interpretation of a gauge theory, as
was the case for the type IIB configuration above described \cite{Wm4}.

Let us consider configurations of infinite solitonic fivebranes,
with worldvolume coordinates $x^0,x^1,x^2,x^3,x^4$ and $x^5$,
located at $x^7=x^8=x^9=0$ and at some fixed value of the $x^6$
coordinate. In addition, let us introduce finite Dirichlet
fourbranes, with worldvolume coordinates $x^0,x^1,x^2,x^3$ and
$x^6$, which terminate on the solitonic fivebranes; thus, they
are finite in the $x^6$ direction. On the fourbrane worldvolume,
we can define a macroscopic four dimensional field theory, with
$N=2$ supersymmetry. This four dimensional theory will, as in the
type IIB case considered in previous section, be defined by
standard Kaluza-Klein dimensional reduction of the five
dimensional theory defined on the D-$4$brane worldvolume.
Then, the bare coupling constant of the four dimensional theory
will be
\begin{equation}
\frac {1}{g_4^2} = \frac {|x_6^2-x_6^1|}{g_5^2},
\label{eq:e1}
\end{equation}
in terms of the five dimensional coupling constant. Moreover, we
can interpret as classical moduli parameters of the effective
field theory on the dimensionally reduced worldvolume of the
fourbrane the coordinates $x^4$ and $x^5$, which locate the
points on the fivebrane worldvolume where the D-$4$branes
terminate.
  
In addition to the Dirichlet fourbranes and solitonic fivebranes, we can
yet include Dirichlet sixbranes, without any further break of
supersymmetry on the theory in the worldvolume of the fourbranes.
To prove this, we notice that each NS-$5$brane imposes the projections
\begin{equation}
\epsilon_L = \Gamma_0 \ldots \Gamma_5 \epsilon_L, \: \:
\epsilon_R = - \Gamma_0 \ldots \Gamma_5 \epsilon_R,
\label{eq:e2}
\end{equation}
while the D-$4$branes, with worldvolume localized at 
$x^0,x^1,x^2,x^3$ and $x^6$, imply
\begin{equation}
\epsilon_L = \Gamma_0 \Gamma_1 \Gamma_2 \Gamma_3 \Gamma_6
\epsilon_R.
\label{eq:e3}
\end{equation}
Conditions (\ref{eq:e2}) and (\ref{eq:e3}) can be recombined into
\begin{equation}
\epsilon_L = \Gamma_0 \Gamma_1 \Gamma_2 \Gamma_3 \Gamma_7
\Gamma_8 \Gamma_9 \epsilon_R,
\label{eq:e4}
\end{equation}
which shows that certainly sixbranes can be added with no
additional supersymmetry breaking.
  
The solitonic fivebranes break half of the supersymmetries, while
the D-$6$brane breaks again half of the remaining symmetry,
leaving eight real supercharges, which leads to four dimensional
$N=2$ supersymmetry.
  
As we will discuss later on, the sixbranes of type IIA string
theory can be used to add hypermultiplets to the effective
macroscopic four dimensional theory. In particular, the mass of
these hypermultiplets will become zero whenever the D-$4$brane
meets a D-$6$brane. 
  
One of the main achievements of the brane representations of
supersymmetric gauge theories is the ability to represent the
different moduli spaces, namely the Coulomb and Higgs branches,
in terms of the brane motions left free. For a configuration of
$k$ fourbranes connecting two solitonic fivebranes along the
$x^6$ direction, as the one we have described above, the Coulomb
branch of the moduli space of the four dimensional theory is
parametrized by the different positions of the transversal
fourbranes on the fivebranes. When $N_f$ Dirichlet sixbranes are added
to this configuration, what we are describing is the Coulomb
branch of a four dimensional field theory with $SU(N_c)$ gauge group (in case
$N_c$ is the number of D-$4$branes we are considering), with
$N_f$ flavor hypermultiplets. In this brane representation, the
Higgs branch of the theory is obtained when each fourbrane is
broken into several pieces ending on different sixbranes: the
locations of the D-$4$branes living between two D-$6$branes
determine the Higgs branch. However, we will mostly concentrate
on the study of the Coulomb branch for pure gauge theories.
  
As we know from the Seiberg-Witten solution of $N=2$
supersymmetric gauge theories, the classical moduli of the theory
is corrected by quantum effects. There are two types of effects
that enter the game: a non vanishing beta function (determined at 
one loop) implies the existence, in the assymptotically free regime, of a
singularity at the infinity point in moduli space, and strong
coupling effects, which imply the existence of extra
singularities, where some magnetically charged particles become
massless. The problem we are facing now is how to derive such a
complete characterization of the quantum moduli space of four
dimensional $N=2$ supersymmetric field theory directly from the
dynamics governing the brane configuration. The approach to be
used is completely different from a brane construction in type
IIA string theory to a type IIB brane configuration. In fact, in
the type IIB case, employed in the description of the preceding
section of three dimensional $N=4$ supersymmetric field theories,
we can pass from weak to strong coupling through the standard
$Sl(2,{\bf Z})$ duality of type IIB strings; hence, the essential
ingredient we need is to know how brane configurations transform
under this duality symmetry. In the case of type IIA string
theory, the situation is more complicated, as the theory is not
$Sl(2,{\bf Z})$ self dual. However, we know that the strong
coupling limit of type IIA dynamics is described by the eleven
dimensional M-theory; therefore, we should expect to recover the
strong coupling dynamics of four dimensional $N=2$ supersymmetric
gauge theories using the M-theory description of strongly coupled
type IIA strings.
  
Let us first start by considering weak coupling effects. The
first thing to be noticed, concerning the above described
configuration of $N_c$ Dirichlet fourbranes extending along the 
$x^6$ direction between two solitonic fivebranes, where only a
rigid motion of the  transversal fourbranes is allowed, is that
this simple image is missing the classical dynamics of the fivebranes. In
fact, in this picture we are assuming that the $x^6$ coordinate
on the fivebrane worldvolume is constant, which is in fact a very
bad approximation. Of course, one physical requirement we should
impose to a brane configuration, as we did in the case of the
type IIB configurations of the previous section, is that of
minimizing the total worldvolume action. More precisely, what we
have interpreted as Coulomb or Higgs branches in term of free
motions of some branes entering the configuration, should
correspond to zero modes of the brane configuration, i. e., to
changes in the configuration preserving the condition of
minimum worldvolume action (in other words, changes in the brane
configuration that do not constitute an energy expense). The coordinate $x^6$ 
can be assumed
to only depend on the ``normal'' coordinates $x^4$ and $x^5$,
which can be combined into the complex coordinate
\begin{equation}
v \equiv  x^4 + ix^5,
\label{eq:e5}
\end{equation}
representing the normal to the position of the transversal
fourbranes. Far away from the position of the fourbranes, the
equation for $x^6$ reduces now to a two dimensional
laplacian,
\begin{equation}
\nabla^2x^6(v)=0,
\label{eq:e6}
\end{equation}
with solution
\begin{equation}
x^6(v) = k \ln |v| + \alpha,
\label{eq:e7}
\end{equation}
for some constants $k$ and $\alpha$, that will depend on the solitonic and Dirichlet 
brane tensions. As we can see from (\ref{eq:e7}), the value of
$x^6$ will diverge at infinity. This constitutes, as a difference with 
the type IIB case, a first problem 
for the interpretation of equation (\ref{eq:e1}). In fact, in
deriving (\ref{eq:e1}) we have used a standard Kaluza-Klein
argument, where the four dimensional coupling constant is defined
by the volume of the internal space (in this ocasion, the $x^6$
interval between the two solitonic fivebranes). Since the Dirichlet four
branes will deform the solitonic fivebrane, the natural way to define
the internal space would be as the interval defined by the values
of the coordinate $x^6$ at $v$ equal to infinity, which is the
region where the disturbing effect of the four brane is very
likely vanishing, as was the case in the definition of the
effective three dimensional coupling in the type IIB case. 
However, equations (\ref{eq:e6}) and (\ref{eq:e7})
already indicate us that this can not be the right picture, since
these values of the $x^6$ coordinate are divergent. Let us
then consider a configuration with $N_c$ transversal
fourbranes. From equations
(\ref{eq:e1}) and (\ref{eq:e7}), we get, for large $v$,
\begin{equation}
\frac {1}{g_4^2} = - \frac {2 k N_c \ln (v)}{g_5^2},
\label{eq:e8}
\end{equation}
where we have differentiated the direction in which the
fourbranes pull the fivebrane. Equation (\ref{eq:e8}) can have a
very nice meaning if we interpret it as the one loop
renormalization group equation for the effective coupling
constant. In order to justify this interpretation, let us first
analyze the physical meaning of the parameter $k$. From equation
(\ref{eq:e7}), we notice that if we move in $v$ around a value
where a fourbrane is located (that we are assuming is $v=0$), we
get the monodromy transformation
\begin{equation}
x^6 \rightarrow x^6 + 2 \pi i k.
\label{eq:e9}
\end{equation}
This equation can be easily understood in M-theory, where we add
an extra eleventh dimension, $x^{10}$, that we use to define the
complex coordinate
\begin{equation}
x^6+ix^{10}.
\label{eq:e10}
\end{equation}
Now, using the fact that the extra coordinate is compactified on
a circle of radius $R$ we can, from (\ref{eq:e9}), identify $k$
with $R$. From a field theory point of view, we have a similar
interpretation of the monodromy of (\ref{eq:e8}), but now in
terms of a change in the theta parameter. Let us then consider
the one loop renormalization group equation 
for $SU(N_c)$ $N=2$ supersymmetric gauge theories without hypermultiplets,
\begin{equation}
\frac {4 \pi}{g_4^2(u)} = \frac {4\pi}{g_0^2} - \frac {2 N_c}{4 \pi} \ln 
\left( \frac {u}{\Lambda} \right),
\label{eq:e11}
\end{equation}
with $\Lambda$ the dynamically generated scale, and $g_0$ the
bare coupling constant. The bare coupling constant can be
absorved through a change in $\Lambda$; in fact, when going from
$\Lambda$ to a new scale $\Lambda'$, we get
\begin{equation}
\frac {4 \pi}{g_4^2(u)} = \frac {4\pi}{g_0^2} - \frac {2 N_c}{4 \pi} \ln 
\left( \frac {u}{\Lambda'} \right) - \frac {2 N_c}{4 \pi} \ln 
\left( \frac {\Lambda'}{\Lambda} \right).
\end{equation}
Thus, once we fix a reference scale $\Lambda_0$, the dependence
on the scale $\Lambda$ of the bare coupling constant is given by
\begin{equation}
- \frac {2 N_c}{4 \pi} \ln 
\left( \frac {\Lambda}{\Lambda_0} \right).
\label{beta}
\end{equation}
  
It is important to distinguish the dependence on $\Lambda$ of the
bare coupling constant, and the dependence on $u$ of the
effective coupling. In the brane configuration approach, the
coupling constant defined by (\ref{eq:e8}) is the bare coupling
constant of the theory, as determined by the definite brane
configuration. Hence, it is (\ref{beta}) that we should compare
with (\ref{eq:e8}); naturally, some care is needed concerning
units and scales. Once we interpret $k$ as the radius of the
internal $S^1$ of M-theory we can, in order to make contact with
(\ref{beta}), identify $g_5^2$ with the radius of $S^1$, which in
M-theory units is given by
\begin{equation}
R = g l_s,
\end{equation}
with $g$ the string coupling constant, and $l_s$ the string
length, $\frac {1}{\alpha'}$. Therefore, (\ref{eq:e1}) should be modified
to
\begin{equation}
\frac {1}{g_4^2} = \frac {x^6_2-x^6_1}{gl_s} = - 2 N_c \ln (v),
\label{eq:e12}
\end{equation}
which should be dimensionless. Then, we should interpret $v$ in
(\ref{eq:e12}) as a dimensionless variable or, equivalently, as
$\frac {v}{R}$, with $R$ playing the role of natural unit of the
theory. Then, comparing (\ref{beta}) and (\ref{eq:e12}), $\frac
{v}{R}$ becomes the scale $\frac {\Lambda}{\Lambda_0}$ in the
formula for the bare coupling constant. In summary, $v$ fixes the
scale of the theory. From the previous discussion, an equivalent
interpretation follows, where $R$ fixes $\Lambda_0$, and
therefore changes in the scale are equivalent to changes in the
radius of the internal $S^1$.
  
Defining
now an adimensional complex variable,
\begin{equation}
s \equiv (x^6+ix^{10})/R,
\label{eq:e13}
\end{equation}
and a complexified coupling constant,
\begin{equation}
\tau = \frac {4 \pi i}{g^2} + \frac {\theta}{2 \pi},
\label{eq:e14}
\end{equation}
we can generalize (\ref{eq:e12}) to
\begin{equation}
-i \tau_{\alpha}(v) = s_{2}(v) -s_{1}(v),
\label{eq:e15}
\end{equation}
for the simple configuration of branes defining a pure gauge
theory. Now, we can clearly notice how the monodromy, as we move
around $v=0$, means a change $\theta \rightarrow + 2 \pi N_c$.

Let us now come back, for a moment, to the bad behaviour of
$x^6(v)$ at large values of $v$. A possible way to solve this
problem is modifying the configuration of a single pair of
fivebranes, with $N_c$ fourbranes extending between them, to
consider a larger set of solitonic fivebranes. Labelling this
fivebranes by $\alpha$, with $\alpha=0,\ldots,n$, the corresponding $x^6_{\alpha}$
coordinate will depend on $v$ as follows:
\begin{equation}
x^6(v)_{\alpha} = R \sum_{i=1}^{q_L} \ln |v-a_i| - R \sum_{j=1}^{q_R} \ln |v-b_j|,
\label{eq:e16}
\end{equation}
where $q_L$ and $q_R$ represent, respectively, the number of
D-$4$branes to the left and right of the $\alpha^{th}$ fivebrane.
As is clear from (\ref{eq:e16}), a good behaviour at large $v$
will only be possible if the numbers of fourbranes to the right
and left of a fivebrane are equal, $q_L=q_R$, which somehow
amounts to compensating the perturbation created by the
fourbranes at the sides of a fivebrane. The four dimensional
field theory represented now by this brane array will have a
gauge group $\prod_{\alpha} U(k_{\alpha})$, where $k_{\alpha}$ is the number
of transversal fourbranes between the $\alpha-1$ and
$\alpha^{th}$ solitonic fivebranes. Now, minimization of the
worldvolume action will require not only taking into account the
dependence of $x^6$ on $v$, but also the fourbrane positions on
the NS-$5$brane, represented by $a_i$ and $b_j$ in
(\ref{eq:e16}), on the four dimensional worldvolume coordinates
$x^0,x^1,x^2$ and $x^3$. Using (\ref{eq:e16}), and the Nambu-Goto
action for the solitonic fivebrane, we get, for the kinetic
energy,
\begin{equation}
\int d^4 x d^2 v \sum_{\mu = 0} ^{3} \partial_{\mu} x^6
(v,a_i(x^{\mu}),b_j(x^{\mu})) \partial^{\mu} x^6
(v,a_i(x^{\mu}),b_j(x^{\mu})).
\label{eq:e17}
\end{equation}
Convergence of the $v$ integration implies
\begin{equation}
\partial_{\mu}(\sum_i a_i -\sum_j b_j)=0
\label{eq:e18}
\end{equation}
or, equivalently,
\begin{equation}
\sum_i a_i - \sum_j b_j = \hbox{constant}.
\label{eq:e19}
\end{equation}

This ``constant of motion'' is showing how the average of the
relative position between left and right fourbranes must be hold
constant. Since the Coulomb branch of the $\prod_{\alpha}
U(k_{\alpha}) $ gauge theory will be associated with different
configurations of the transversal fourbranes, constraint
(\ref{eq:e19}) will reduce the dimension of this space. As we
know from our general discussion on D-branes, the $U(1)$ part of
the $U(k_{\alpha})$ gauge group can be associated to the motion
of the center of mass. Constraint (\ref{eq:e19}) implies that the
center of mass is frozen in each sector. With no semi-infinite
fourbranes to the right, we have that $\sum_i a_i=0$; now, this
constraint will force the center of mass of all sectors to
vanish, which means that the field theory we are describing is
$\prod_{\alpha} SU(k_{\alpha})$, instead of $\prod_{\alpha}
U(k_{\alpha})$. The same result can be derived if we include
semi-infinite fourbranes to the left and right of the first and 
last solitonic fivebranes: as they are infinitely massive, we can
assume that they do not move in the $x^4$ and $x^5$ directions.
An important difference will appear if we consider periodic
configurations of fivebranes, upon compactification of the $x^6$
direction to a circle: in this case, constraint (\ref{eq:e19}) is
now only able to reduce the group to $\prod_{\alpha}
SU(k_{\alpha}) \times U(1)$, leaving alive a $U(1)$ factor.
  
Hypermultiplets in this gauge theory are understood as strings
connecting fourbranes on different sides of a fivebrane;
therefore, whenever the positions of the fourbranes to the left and
right of a solitonic brane become coincident, a massless
hypermultiplet arises. As the hypermultiplets are charged under
the gauge groups at both sides of a certain $\alpha+1$ fivebrane,
they will transform as $(k_{\alpha},\bar{k}_{\alpha+1})$.
  
However, as the position of the fourbranes on both sides of a
fivebrane varies as a function of $x^0, x^1,x^2$ and $x^3$, the
existence of a well defined hypermultiplet can only be
accomplished thanks to the fact that its variation rates on
both sides are the same, as follows again from (\ref{eq:e18}):
$\partial_{\mu}(\sum_i a_{i,\alpha}) = \partial_{\mu}(\sum_j
a_{j,\alpha+1})$. The definition of the bare massses comes then
naturally from constraint (\ref{eq:e19}):
\begin{equation}
m_{\alpha} = \frac {1}{k_{\alpha}} \sum_i a_{i,\alpha} - \frac
{1}{k_{\alpha+1}} \sum_j a_{j,\alpha+1}.
\label{eq:e20}
\end{equation}
  
With this interpretation, the constraint (\ref{eq:e19}) becomes
very natural from a physical point of view: it states that the
masses of the hypermultiplets do not depend on the spacetime
position.
  
The consistency of the previous definition of hypermultiplets can
be checked using the previous construction of the one-loop beta
function. In fact, from equation (\ref{eq:e15}), we get, for
large values of $v$,
\begin{equation}
-i \tau_{\alpha}(v) =  (2
k_{\alpha}-k_{\alpha-1}-k_{\alpha+1})\ln v.
\label{eq:e21}
\end{equation}
  
The number $k_{\alpha}$ of branes in the $\alpha^{th}$ is, as we
know, the number of colours, $N_c$. Comparing with the beta
function for $N=2$ supersymmetric $SU(N_c)$ gauge theory with
$N_f$ flavors, we conclude that
\begin{equation}
N_f = k_{\alpha-1}+ k_{\alpha+1},
\label{eq:e22}
\end{equation}
so that the number of fourbranes (hypemultiplets) at both sides
of a certain pair of fivebranes, $k_{\alpha+1}+k_{\alpha-1}
\equiv N_f$, becomes the number of flavors. 
  
Notice, from (\ref{eq:e20}), that the mass of all the
hypermultiplets associated with fourbranes at both sides of a
solitonic fivebrane are the same. This implies a global flavor
symmetry. This global flavor symmetry is the gauge symmetry of
the adjacent sector. This explains the physical meaning of
(\ref{eq:e20}).

Let us now come back to equation (\ref{eq:e12}). What we need in
order to unravel the strong coupling dynamics of our effective
four dimensional gauge theory is the $u$ dependence of the
effective coupling constant, dependence that will contain non
perturbative effects due to instantons. It is from this
dependence that we read the Seiberg-Witten geometry of the
quantum moduli space. Strong coupling effects correspond to $u$
in the infrared region, i. e., small $u$ or, equivalently, large
$\Lambda$. From our previous discussion of (\ref{eq:e12}), we
conclude that the weak coupling regime corresponds to the type
IIA string limit, $R \rightarrow 0$, and the strong coupling
regime to the M-theory reime, at large values of $R$ (recall that
changes of scale in the four dimensional theory correspond to
changes of the radius of the internal $S^1$). This explains our
hopes that M-theory could describe the strong coupling regime of
the four dimensional theory). We will then see now how M-theory
is effectively working.

\subsubsection{M-Theory and Strong Coupling.}

From the M-theory point of view, the brane configuration we are
considering can be interpreted in a different way. In particular,
the D-$4$branes we are using to define the four dimensional
macroscopic gauge theory can be considered as fivebranes wrapping
the eleven dimensional $S^1$. Moreover, the trick we have used to
make finite these fourbranes in the $x^6$ direction can be
directly obtained if we consider fivebranes with worldvolume
${\bf R}^4 \times \Sigma$, where ${\bf R}^4$ is parametrized by
the coordinates $x^0,x^1,x^2$ and $x^4$, and $\Sigma$ is two
dimensional, and embedded in the four dimesional space of
coordinates $x^4,x^5,x^6$ and $x^{10}$. If we think in purely
classical terms, the natural guess for $\Sigma$ would be a
cylinder with the topology $S^1 \times [x^6_2,x^6_1]$, for a
configuration of $k$ D-$4$branes extending along the $x^6$
direction between two solitonic fivebranes. This is however a
very naive compactification, because there is no reason to
believe that a fivebrane wrapped around this surface will
produce, on the four dimensional worldvolume ${\bf R}^4$, any
form of non abelian gauge group. In fact, any gauge field on
${\bf R}^4$ should come from integrating the chiral antisymmetric
tensor field $\beta$ of the M-theory fivebrane worldvolume, on
some one-cycle of $\Sigma$. If we wnat to reproduce, in four
dimensions, some kind of $U(k)$ or $SU(k)$ gauge theory, we
should better consider a surface $\Sigma$ with a richer first
homology group. However, we can try to do something better when
including the explicit dependence of the $x^6$ coordinate on $v$.
In this case, we will get a picture that is closer to the right
answer, but still far away from the true solution. Including the
$v$ dependence of the $x^6$ coordinate leads to a family of
surfaces, parametrized by $v$, $\Sigma_v$, defined by $S^1 \times
[x^6_2,x^6_1](v)$. The nice feature about this picture is that
$v$, which is the transverse coordinate of $\Sigma$ in the space
${\cal Q}$, defined by the coordinates $x^4,x^5,x^6$ and
$x^{10}$, becomes now similar to the moduli of $\Sigma_v$;
however, we have yet the problem of the of the genus or, in more
general terms, the first homology group of $\Sigma$. The reason
for following the previous line of thought, is that we are trying
to keep alive the interpretation of the $v$ coordinate as moduli,
or coordinate of the Coulomb branch. This is, in fact, the reason
giving rise to the difficulties with the genus, as we are using
just one complex coordinate, independently of the rank of the
gauge group, something we are forced to do because of the
divergences in equation (\ref{eq:e7}).
  
The right M-theory approach is quite different. In fact, we must
try to get $\Sigma$ directly from the particular brane
configuration we are working with, and define the Colomb branch
of the theory by the moduli space of brane configurations. Let us
then define the single valued coordinate $t$,
\begin{equation}
t \equiv \exp - s,
\label{eq:e24}
\end{equation}
and define the surface $\Sigma$ we are looking for through
\begin{equation}
F(t,v)=0.
\label{eq:e25}
\end{equation}
From the classical equations of motion of the fivebrane we know
the assymptotic behaviour for very large $t$,
\begin{equation}
t \sim v^k,
\label{eq:e26}
\end{equation}
and for very small $t$,
\begin{equation}
t \sim v^{-k}.
\label{eq:e27}
\end{equation}
Conditions (\ref{eq:e26}) and (\ref{eq:e27}) imply that $F(t,v)$
will have, for fixed values of $t$, $k$ roots, while two
different roots for fixed $v$. It must be stressed that the
assymptotic behaviour (\ref{eq:e26}) and (\ref{eq:e27})
corresponds to the one loop beta function for a field theory with
gauge group $SU(k)$, and without hypermultiplets. A function
satisfying the previous conditions will be of the generic type 
\begin{equation}
F(t,v)=A(v)t^2+B(v)t+C(v),
\label{eq:e28}
\end{equation}
with $A$, $B$ and $C$ polynomials in $v$ of degree $k$. From
(\ref{eq:e26}) and (\ref{eq:e27}), the function (\ref{eq:e28}) 
becomes 
\begin{equation}
F(t,v)=t^2+B(v)t+\hbox { constant},
\label{cons}
\end{equation}
with one undetermined constant. In order to kill this constant,
we can rescale $t$ to $t/\hbox{constant}$. The meaning of this
rescaling can be easily understood in terms of of the one loop
beta function, written as (\ref{eq:e26}) and (\ref{eq:e27}). In
fact, these equations can be read as
\begin{equation}
s = - k \ln \left( \frac {v}{R} \right),
\end{equation}
and therefore the rescaling of $R$ goes like
\begin{equation}
s \rightarrow - k \ln \left( \frac {v}{R'} \frac {R'}{R} \right)
\end{equation}
or, equivalently,
\begin{equation}
t \rightarrow t \left( \frac {R'}{R} \right) ^k.
\end{equation}
Thus, and based on the above discussion on the definition of the
scale, we observe that the constant in (\ref{cons}) defines the
scale of the theory. With this interpretation of the constant, we
can get the Seiberg-Witten solution for $N=2$ pure gauge
theories, with gauge group $SU(k)$. If $B(v)$ is chosen to be
\begin{equation}
B(v) =v^k +u_2 v^{k-2} + u_3 v^{k-3} + \cdots + u_k,
\label{eq:e29}
\end{equation}
we finally
get the Riemann surface
\begin{equation}
t^2 + B(v) t +1=0,
\label{eq:e30}
\end{equation}
a Riemann surface of genus $k-1$, which
is in fact the rank of the gauge group. 
Moreover, we can now try to visualize this Riemann surface as the
worldvolume of the fivebrane describing our original brane
configuration: each $v$-plane can be compactified to ${\bf P}^1$,
and the transversal fourbranes cna be interpreted as gluing
tubes, which clearly represents a surface with $k-1$ handles.
This image corresponds to gluing two copies of ${\bf P}^1$, with
$k$ disjoint cuts on each copy or, equivalently, $2k$ branch
points. Thus, as can be observed from (\ref{eq:e30}), to each
transversal D-$4$brane there correspond two branch points and one
cut on ${\bf P}^1$.
  
If we are interested in $SU(k)$ gauge theories with
hypermultiplets, then we should first replace (\ref{eq:e26}) and
(\ref{eq:e27}) by the corresponding relations,
\begin{equation}
t \sim v^{k-k_{\alpha-1}},
\label{eq:e31}
\end{equation}
and
\begin{equation}
t \sim v^{-k-k_{\alpha+1}},
\label{eq:e32}
\end{equation}
for $t$ large and small, respectively. These are, in fact, the
relations we get from the beta functions for these theories. If
we take $k_{\alpha_1}=0$, and $N_f=k_{\alpha+1}$, the curve
becomes
\begin{equation}
t^2 + B(v) t + C(v)=0,
\label{eq:e33}
\end{equation}
with $C(v)$ a polynomial in $v$, of degree $N_f$, parametrized by
the masses of the hypermultiplets,
\begin{equation}
C(v) = f \prod _{j=1}^{N_f} (v-m_j),
\label{eq:e34}
\end{equation}
with $f$ a complex constant.
  
Summarizing, we have been able to find a moduli of brane
configurations reproducing four dimensional $N=2$ supersymmetric
$SU(k)$ gauge theories. The exact Seiberg-Witten solution is
obtained by reduction of the worldvolume fivebrane dynamics on
the surface $\Sigma_{\vec{u}}$ defined at (\ref{eq:e30}) and
(\ref{eq:e32}). Obviously, reducing the fivebrane dynamics to
${\bf R}^4$ on $\Sigma_{\vec{u}}$ leads to an effective coupling constant
in ${\bf R}^4$, the $k-1 \times k-1$ Riemann matrix $\tau(\vec{u})$ of
$\Sigma_{\vec{u}}$.
  
\vspace{2 mm}

Before finishing this section, it is important to stress some
peculiarities of the brane construction. First of all, it should
be noticed that the definition of the curve $\Sigma$, in terms of
the brane configuration, requires working with uncompactified
$x^4$ and $x^5$ directions. This is part of the brane philosophy,
where we must start with a particular configuration in flat
spacetime. A different approach will consist in directly working
with a spacetime ${\cal Q} \times R^7$, with ${\cal Q}$ some
Calabi-Yau manifold, and consider a fivebrane worldvolume $\Sigma
\times {\bf R}^4$, with ${\bf R}^4 \subset {\bf R}^7$, and
$\Sigma$ a lagrangian submanifold of ${\cal Q}$. Again, by Mc
Lean's theorem, the $N=2$ theory defined on ${\bf R}^4$ will
have a Coulomb branch with dimension equal to the first Betti
number of $\Sigma$, and these deformations of $\Sigma$ in ${\cal
Q}$ will represent scalar fields in the four dimensional theory.
Moreover, the holomorphic top form $\Omega$ of ${\cal Q}$ will
define the meromorphic $\lambda$ of the Seiberg-Witten solution.
If we start with some Calabi-Yau manifold ${\cal Q}$, we should
provide some data to determine $\Sigma$ (this is what we did in
the brane case, with ${\cal Q}$ non compact and flat. If, on the
contrary, we want to select $\Sigma$ directly from ${\cal Q}$, we
can only do it in some definite cases, which are those related to
the {\em geometric mirror construction} \cite{SYZ,Mor}. Let us then recall some
facts about the geometric mirror. The data are
\begin{itemize}
	\item The Calabi-Yau manifold ${\cal Q}$.
	\item A lagrangian submanifold $\Sigma \rightarrow {\cal
Q}$.
	\item A $U(1)$ flat bundle on $\Sigma$.
\end{itemize}
  
The third requirement is equivalent to interpreting $\Sigma$ as a
D-brane in ${\cal Q}$. This is a crucial data, in order to get
from the above points the structure of abelian manifold of the
Seiberg-Witten solution. Namely, we frist use Mc Lean's theorem
to get the moduli of deformations of $\Sigma \rightarrow {\cal
Q}$, preserving the condition of lagrangian submanifold. This
space is of dimension $b_1(\Sigma)$. Secondly, on each of these
points we fiber the jacobian of $\Sigma$, which is of dimension
$g$. This family of abelian varieties defines the quantum moduli
of a gauge theory, with $N=2$ supersymmetry, with a gauge group
of rank equal $b_1(\Sigma)$. Moreover, this family of abelian
varieties is the moduli of the set of data of the second and third points
above, i. e., the moduli of $\Sigma$ as a D-$2$brane. In some
particular cases, this moduli is ${\cal Q}$ itself or, more
properly, the geometric mirror of ${\cal Q}$. This will be the
case for $\Sigma$ of genus equal one, i. e., for the simple
$SU(2)$ case. In this cases, the characterization of $\Sigma$ in
${\cal Q}$ is equivalent to describing ${\cal Q}$ as an elliptic
fibration. The relation between geometric mirror and T-duality
produces a completely different physical picture. In fact, we
can, when $\Sigma$ is a torus, consider in type IIB a threebrane
with classical moduli given by ${\cal Q}$. After T-duality or
mirror, we get the type IIA description in terms of a fivebrane.
In summary, it is an important problem to understand the relation
of quantum mirror between type IIA and type IIB string theory,
and the M-theory strong coupling description of type IIA strings.
  
\subsection{Brane Description of $N=1$ Four Dimensional Field
Theories.}

In order to consider field theories with $N=1$ supersymmetry, the
first thing we will study will be $R$-symmetry. Let us then
recall the way $R$-symmetries were defined in the case of four
dimensional $N=2$ supersymmetry, and three dimensional $N=4$
supersymmetry, through compactification of six dimensional $N=1$
supersymmetric gauge field theories. The $U(1)_R$ in four
dimensions, or $SO(3)_R$ in three dimensions, are simply the
euclidean group of rotations in two and three dimensions,
respectively. Now, we have a four dimensional space ${\cal Q}$,
parametrized by coordinates $t$ and $v$, and a Riemann surface
$\Sigma$, embedded in ${\cal Q}$ by equations of the type
(\ref{eq:e28}). To characterize $R$-symmetries, we can consider
transformations on ${\cal Q}$ which transform non trivially its
holomorphic top form $\Omega$. The unbroken $R$-symmetries will
then be rotations in ${\cal Q}$ preserving the Riemann surface
defined by the brane configuration. If we consider only the
assymptotic behaviour of type (\ref{eq:e26}), or (\ref{eq:e31}),
we get $U(1)_R$ symmetries of type 
\begin{eqnarray}
t & \rightarrow \lambda^k t, \nonumber \\
v & \rightarrow \lambda v. 
\label{eq:e35}
\end{eqnarray}
This $U(1)$ symmetry is clearly broken by the curve
(\ref{eq:e30}). This spontaneous breakdown of the $U(1)_R$
symmetry is well understood in field theory as an instanton
induced effect. If instead of considering ${\cal Q}$, we take the
larger space $\hat{\cal Q}$, containing the $x^7,x^8$ and $x^9$
coordinates, we see that the $N=2$ curve is invariant under
rotations in the $(x^7,x^8,x^9)$ space.
  
Let us now consider a brane configuration which reproduces $N=1$
four dimensional theories \cite{Wbqcd}. We will again start in type IIA string
theory, and locate a solitonic fivebrane at $x^6=x^7=x^8=x^9=0$
with, as usual, worldvolume coordinates $x^0,x^1,x^2,x^3,x^4$ and
$x^5$. At some definite value of $x^6$, say $x^6_0$, we locate
another solitonic fivebrane, but this time with worldvolume
coordinates $x^0,x^1,x^2,x^3,x^7$ and $x^8$, and $x^4=x^5=x^9=0$.
As before, we now suspend a set of $k$ D-$4$branes in between. They
will be parametrized by the positions $v=x^4+ix^5$, and
$w=x^7+ix^8$, on the two solitonic fivebranes. The worldvolume
coordinates on this D-$4$branes are, as in previous cases,
$x^0,x^1,x^2$ and $x^3$. The effective field theory defined by
the set of fourbranes is macroscopically a four dimensional gauge
theory, with coupling constant
\begin{equation}
\frac {1}{g^2} = \frac {x^6_0}{g l_s}.
\label{eq:e36}
\end{equation}
Moreover, now we have only $N=1$ supersymmetry, as no massless bosons
can be defined on the four dimensional worldvolume
$(x^0,x^1,x^2,x^3)$. In fact, at the line $x^6=0$ the only
possible massless scalar would be $v$, since $w=0$ and $x^9=0$,
so that we project out $x^9$ and $w$. On the other hand, at
$x^6_0$ we have $v=0$ and $x^9=0$ and, therefore, we have
projected out all massless scalars. Notice that by the same
argument, in the case of two solitonic fivebranes located at
different values of $x^6$ but at $x^7=x^8=x^9=0$, we have one
complex massless scalar that is not projected out, which leads to
$N=2$ supersymmetry in four dimensions. The previous discussion
means that $v$, $w$ and $x^9$ are projected out as four
dimensional scalar fields; however, $w$ and $v$ are still
classical moduli parameters of the brane configuration. 
  
Now, we return to a comment already done in previous section: each of the
fourbranes we are suspending in between the solitonic fivebranes
can be interpreted as a fivebrane wrapped around a surface
defined by the eleven dimensional $S^1$ of M-theory, multiplied
by the segment $[0,x^6_0]$. Classically, the four dimensional
theory can be defined through dimensional reduction of the
fivebrane worldvolume on the surface $\Sigma$. The coupling
constant will be given by the moduli $\tau$ of this surface,
\begin{equation}
\frac {1}{g^2} = \frac {2 \pi R}{S},
\label{eq:e37}
\end{equation}
with $S$ the length of the interval $[0,x^6_0]$, in M-theory
units. In $N=1$ supersymmetric field theories, on the contrary of
what takes place in the $N=2$ case, we have not a classical
moduli and, therefore, we can not define a wilsonian coupling
constant depending on some mass scale fixed by a vacuum
expectation value. This fact can produce some problems, once we
take into account the classical dependence of $x^6$ on $v$ and
$w$. In principle, this dependence should be the same as that in
the case studied in previous section,
\begin{eqnarray}
x^6 & \sim & \tilde{k} \ln v, \nonumber \\
x^6 & \sim & \tilde{k} \ln w.
\label{eq:e38}
\end{eqnarray}
Using the $t$ coordinate defined in (\ref{eq:e24}), equations
(\ref{eq:e38}) become
\begin{eqnarray}
t & \sim & v^k, \nonumber \\
t & \sim & w^k,
\label{eq:e39}
\end{eqnarray}
for large and small $t$, respectively, or, equivalently, $t \sim
v^k$, $t^{-1} \sim w^k$.
Now, we can use these relations in (\ref{eq:e36}). Taking into
account the units, we can write
\begin{equation}
\frac {1}{g^2} \sim N_c [\ln \frac {v}{R}+\ln \frac {w}{R}],
\label{eq:e41}
\end{equation}
with $k \equiv N_c$. As we did in the $N=2$ case, we can try to
compare (\ref{eq:e41}) with the one loop beta function for $N=1$
supersymmetric $SU(N_c)$ pure Yang-Mills theory,
\begin{equation}
\Lambda = \mu \exp - \frac {8 \pi^2}{3 N_c g(\mu)^2}.
\label{n=1}
\end{equation}
In order to get the scale from (\ref{eq:e41}) we impose
\begin{equation}
v = \zeta w^{-1},
\label{eq:e42}
\end{equation}
with $\zeta$ some constant with units of $(\hbox {length})^2$.
Using (\ref{eq:e42}) and (\ref{eq:e41}) we get
\begin{equation}
\frac {1}{g^2} \sim N_c \ln \left( \frac {\zeta}{R^2} \right). 
\end{equation}
In order to make contact with (\ref{n=1}) we must impose
\begin{equation}
\frac {\zeta}{R^2} = (\Lambda R)^3,
\end{equation}
where we have used $\frac {1}{R}$ in order to measure $\Lambda$.
Using (\ref{eq:e42}), we get the curve associated to four
dimensional $N=1$ field theory,
\begin{eqnarray}
t & = & v^k, \nonumber \\
\zeta^k t^{-1} & = & w^k \nonumber \\
v & = & \zeta w^{-1}.
\label{eq:e44}
\end{eqnarray}
The curve defined by (\ref{eq:e44}) will only depend on
$\zeta^k$. The different
set of brane configurations compatible with (\ref{eq:e44}) are
given by values of $\zeta$, with fixed $\zeta^k$. These $N_c$
roots parametrize the $N_c$ different vacua
predicted by $\hbox {tr} (-1)^F$ arguments. It is important to
observe that the coupling constant $\frac {1}{g^2}$ we are
defining is the so called wilsonian coupling. We can interpret it
as a complex number with $\hbox {Im } \frac {1}{g^2} \equiv
\frac {\theta}{8 \pi^2}$. Hence, the value of $\hbox {Im }
\zeta^k$ fixes the $\theta$ parameter of the four dimensional
theory.

For a given value of $\zeta$, (\ref{eq:e44}) defines a Riemann
surface of genus zero, i. e., a rational curve. This curve is now
embedded in the space of $(t,v,w)$ coordinates. We will next
observe that these curves, (\ref{eq:e44}), are the result of
``rotating'' \cite{Barbon} the rational curves in the Seiberg-Witten solution,
corresponding to the singular points. However, before doing that
let us comment on $U(1)_R$ symmetries. As mentioned above, in
order to define an $R$-symmetry we need a transformation on
variables $(t,v,w)$ not preserving the holomorphic top form,
\begin{equation}
\Omega = dv \wedge dw \wedge \frac {dt}{t} R.
\label{eq:e45}
\end{equation}
A rotation in the $w$-plane, compatible with the assymptotic
conditions (\ref{eq:e39}), and defining an
$R$-symmetry, is
\begin{eqnarray}
v & \rightarrow & v, \nonumber \\
t & \rightarrow & t, \nonumber \\
w & \rightarrow & e^{2 \pi i/k} w.
\label{eq:e46}
\end{eqnarray}
Now, it is clear that this symmetry is broken spontaneously by
the curve (\ref{eq:e44}). More interesting is an exact $U(1)$
symmetry, that can be defined for the curve (\ref{eq:e44}):
\begin{eqnarray}
v & \rightarrow e^{i \delta} v, \nonumber \\
t & \rightarrow e^{i \delta k} t, \nonumber \\
w & \rightarrow e^{-i \delta} w.
\label{eq:e47}
\end{eqnarray}
As can be seen from (\ref{eq:e45}), this is not an $R$-symmetry,
since $\Omega$ is invariant. Fields charged with respect to this
$U(1)$ symmetry should carry angular momentum in the $v$ or $w$
plane, or linear momentum in the eleventh dimension interval (i.
e., zero branes) The fields of $N=1$ SQCD do not carry any of
these charges, so all fields with $U(1)$ charge should be
decoupled from the $N=1$ SQCD degrees of freedom. This is
equivalent to the way we have projected out fields in the
previous discussion on the definition of the effective $N=1$ four
dimensional field theory.
  
\subsubsection{Rotation of Branes.}

A different way to present the above construction is by
performing a rotation of branes. We will now concentrate on this
procedure. The classical configuration of NS-$5$branes with
worldvolumes extending along $x^0,x^1,x^2,x^3,x^4$ and $x^5$, can
be modified to a configuration where one of the solitonic
fivebranes has been rotated, from the $v=x^4+ix^5$ direction, to
be also contained in the $(x^7,x^8)$-plane, so that, by moving it
a finite angle $\mu$, it is localized in the $(x^4,x^5,x^7,x^8)$
space. Using the same notation as in previous section, the brane
configuration, where a fivebrane has been moved to give rise to
an angle $\mu$ in the $(v,w)$-plane, the rotation is equivalent
to impossing
\begin{equation}
w= \mu v.
\label{eq:e48}
\end{equation}
In the brane configuration we obtain, points on the rotated
fivebrane are parametrized by the $(v,w)$ coordinates in the
$(x^4,x^5,x^7,x^8)$ space. We can therefore imposse the following
assymptotic conditions \cite{Oz}:
\begin{eqnarray}
t & = & v^k, \: \: \: \: \: \: \: w=\mu v, \nonumber \\
t & = & v^{-k}, \: \: \: \: \: \: w=0, 
\label{eq:e48b}
\end{eqnarray}
respectively for large and small $t$. Let us now assume that this
brane configuration describes a Riemann surface, $\hat{\Sigma}$,
embedded in the space $(x^6,x^{10},x^4,x^5,x^7,x^8)$, and let us
denote by $\Sigma$ the surface in the $N=2$ case, i. e., for
$\mu=0$. In these conditions, $\hat{\Sigma}$ is simply the graph
of the function $w$ on $\Sigma$. We can interpret (\ref{eq:e48})
as telling us that $w$ on $\Sigma$ posseses a simple pole at
infinity, extending holomorphically over the rest of the Riemann
surface. If we imposse this condition, we get that the projected
surface $\Sigma$, i. e., the one describing the $N=2$ theory, is
of genus zero. In fact, it is a well known result in the theory
of Riemann surfaces that the order of the pole at infinity
depends on the genus of the surface in such a way that for genus
larger than zero, we will be forced to replace (\ref{eq:e48}) by
$w=\mu v^{a}$ for some power $a$ depending on the genus. A
priori, there is no problem in trying to rotate using,
instead of $w=\mu v$, some higher pole modification of the type
$w=\mu v^{a}$, for $a>1$. This would provide $\Sigma$ surfaces
with genus different from zero; however, we would immediately
find problems with equation (\ref{eq:e41}), and we will be unable
to kill all dependence of the coupling constant on $v$ and $w$.
Therefore, we conclude that the only curves that can be rotated
to produce a four dimensional $N=1$ theory are those with zero
genus. This is in perfect agreement with the physical picture we
get from the Seiberg-Witten solution. Namely, once we add a soft
breaking term of the type $\mu \hbox {tr} \phi^2$, the only
points remaining in the moduli space as real vacua of the theory
are the singular points, where the Seiberg-Witten curve
degenerates.

\subsubsection{QCD Strings and Scales.}

In all our previous discussion we have not been careful enough in
separating arguments related to complex or {\em holomorphic
structure}, and those related to {\em K\"ahler structure}. The
M-theory description contains however relevant information on
both aspects. For instance, in our previous derivation of curves,
we were mostly interested in reproducing the complex structure of the Seiberg-Witten 
solution, as is, for instance, the moduli dependence on vacuum
expectation values, i. e., the effective wilsonian coupling
constant. However, we can also ask ourselves on BPS masses and,
in that case, we will need the definite embedding of $\Sigma$ in
the ambient space ${\cal Q}$, and the holomorphic top form
defined on ${\cal Q}$. As is clear from the fact that we
are working in M-theory, the holomorphic top form on ${\cal Q}$
will depend explicitely on $R$, i. e., on the string coupling
constant, and we will therefore find BPS mass formulas that will
depend explicitely on $R$. We will discuss this type of
dependence on $R$ first in the case of $N=1$ supersymmetry. The
$N=1$ four dimensional field theory we have described contains,
in principle, two parameters. One is the constant $\zeta$
introduced in equation (\ref{eq:e42}) which, as we have already
mentioned, is, because of (\ref{eq:e41}), intimately connected
with $\Lambda$, and the radius $R$ of the eleven dimensional
$S^1$. Our first task would be to see what kind of four
dimensional dynamics is dependent on the particular value of $R$,
and in what way. The best example we can of course use is the
computation of gaugino-gaugino condensates. In order to do that,
we should try to minimize a four dimensional suerpotential for the $N=1$
theory. Following Witten, we will define this superpotential $W$
as an holomorphic function of $\Sigma$, and with critical points
precisely when the surface $\Sigma$ is a holomorphic curve in
${\cal Q}$. The space ${\cal Q}$ now is the one with coordinates
$x^4,x^5,x^6,x^7,x^8$ and $x^{10}$ (notice that this second
condition was the one used to prove that rotated curves are
necesarily of genus equal zero). Moreover, we need to work 
with a holomorphic curve because of $N=1$ supersymmetry. A
priori, there are two different ways we can think about this
superpotential: maybe the simplest one, from a physical point of
view, is as a functional defined on the volume of $\Sigma$, where
this volume is given by
\begin{equation}
\hbox {Vol} (\Sigma)= J . \Sigma,
\label{eq:e49}
\end{equation}
with $J$ the K\"ahler class of ${\cal Q}$. The other posibility is defining 
\begin{equation}
W(\Sigma) = \int_B \Omega,
\label{eq:e50}
\end{equation}
with $B$ a $3$-surface such that $\Sigma=\partial B$, and
$\Omega$ the holomorphic top form in ${\cal Q}$. Definition
(\ref{eq:e50}) automatically satisfies the condition of being
stationary, when $\Sigma$ is a holomorphic curve in ${\cal Q}$.
Notice that the holomorphy condition on $\Sigma$ means, in
mathematical terms, that $\Sigma$ is an element of the Picard
lattice of ${\cal Q}$, i. e., an element in $H_{1,1}({\cal Q})
\cap H_2({\cal Q})$. This is what allows us to use
(\ref{eq:e49}), however, and this is the reason for temporarily
abandoning the approach based on (\ref{eq:e49}). What we require
to $W$ is being stationary for holomorphic curves, but it should,
in principle, be defined for arbitrary surfaces $\Sigma$, even
those which are not part of the Picard group. Equation
(\ref{eq:e50}) is only well defined if $\Sigma$ is contractible,
i. e., if the homology class of $\Sigma$ in $H_2({\cal Q};{\bf
Z})$ is trivial. If that is not the case, a reference surface
$\Sigma_0$ needs to be defined, and (\ref{eq:e50}) is modified to
\begin{equation}
W(\Sigma) - W(\Sigma_0) = \int _B \Omega,
\label{q:e51}
\end{equation}
where now $\partial B= \Sigma \cup \Sigma_0$. For simplicity, we
will assume $H_3({\cal Q};{\bf Z})=0$. From physical arguments we
know that the set of zeroes of the superpotential should be
related by ${\bf Z}_k$ symmetry, with $k$ the number of
transversal fourbranes. Therefore, if we choose $\Sigma_0$ to be
${\bf Z}_k$ invariant, we can write $W(\Sigma_0)=0$, and
$W(\Sigma)=\int_B \Omega$. Let us then take $B$ as the complex
plane multiplied by an interval $I=[0,1]$, and let us first map
the complex plane into $\Sigma$. Denoting $r$ the coordinate on
this complex plane, $\Sigma$, as given by (\ref{eq:e44}), is
defined by
\begin{eqnarray}
t & = & r^k, \nonumber \\
v & = & r,   \nonumber \\
w & = & \zeta r^{-1}.
\label{eq:e52}
\end{eqnarray}
Writing $r=e^{\rho}e^{i \theta}$, we can define $\Sigma_0$ as
\begin{eqnarray}
t & = & r^k, \nonumber \\
v & = & f(\rho) r,   \nonumber \\
w & = & \zeta f(-\rho) r^{-1},
\label{eq:e53}
\end{eqnarray}
with $f(\rho)=1$ for $\rho>2$, and $f(\rho)=0$ for $\rho<1$. The
${\bf Z}_k$ transformation $t\rightarrow t, \: w \rightarrow e^{2
\pi i /k}w$ and $v \rightarrow v$, is a symmetry of
(\ref{eq:e53}) if, at the same time, we perform the
reparametrization of the $r$-plane
\begin{eqnarray}
\rho \rightarrow \rho, \nonumber \\
\theta \rightarrow \theta + b(\rho),
\label{eq:e54}
\end{eqnarray}
with $b(\rho)=0$ for $\rho\geq1$, and $b(\rho)=- \frac {2\pi}{k}$
for $\rho\leq -1$. Thus, the $3$-manifold entering the
definition of $B$, is given by
\begin{eqnarray}
t & = & r^k, \nonumber \\
v & = & g(\rho,\sigma) r,   \nonumber \\
w & = & \zeta g(-\rho,\sigma) r^{-1},
\label{eq:e55}
\end{eqnarray}
such that for $\sigma=0$ we have $g=1$, and for $\sigma=1$, we
get $g(\rho)=f(\rho)$. Now, with
\begin{equation}
\Omega = R dv \wedge dw \wedge \frac {dt}{t},
\label{eq:e56}
\end{equation}
we get
\begin{equation}
W(\Sigma) = kR \int_B dv \wedge dw \wedge \frac {dr}{r}.
\label{eq:e57}
\end{equation}
The dependence on $R$ is already clear from (\ref{eq:e57}). In
order to get the dependence on $\zeta$ we need to use
(\ref{eq:e55}),
\begin{equation}
W(\Sigma) = k R \zeta \int d \sigma d \theta d \rho \left( \frac
{\partial g_+}{\partial \sigma} \frac {\partial g_-}{\partial
\rho}- \frac {\partial g_+}{\partial \rho} \frac {\partial g_-}{\partial
\sigma} \right),
\label{eq:e58}
\end{equation}
for $g_{\pm}=g(\pm \rho,\sigma)$. Thus we get
\begin{equation}
W(\Sigma) \sim k R \zeta,
\label{eq:e59}
\end{equation}
Notice that the superpotential (\ref{eq:e59}) is given in units $(\hbox {length})^3$, 
as corresponds to the volume of a $3$-manifold. In order to make contact 
with the gaugino-gaugino condensate, we need to obtain $(\hbox {length})^{-3}$ units. 
We can do this multiplying by $\frac {1}{R^6}$; thus, we get
\begin{equation}
<\lambda \lambda> \sim k R \zeta \frac {1}{R^6} \sim \Lambda^3,
\end{equation}
where we have used equation (\ref{eq:e56}). A different 
way to connect $\zeta$ with $\Lambda$ is defining, in the M-theory context, 
the QCD string and computing its tension. Following Witten, we will then try an interpretation of $\zeta$ 
independent of (\ref{eq:e41}), by computing in terms of $\zeta$ the tension 
of the QCD string. We will then, to define the tension, consider the QCD string as a membrane, 
product of a string in ${\bf R}^4$, and a string living in ${\cal Q}$. Let us then 
denote by $C$ a curve in ${\cal Q}$, and assume that $C$ ends on $\Sigma$ in 
such a way that a membrane wrapped on $C$ defines a string in ${\bf R}^4$  \footnote{Notice 
that if we were working in type IIB string theory, we would have the option to 
wrap a threebrane around $\Sigma$, in order to define a string on ${\bf R}^4$.}. Moreover, 
we can simply think of $C$ as a closed curve in ${\cal Q}$, going around the 
eleven dimensional $S^1$,
\begin{eqnarray}
t & = & t_0 \exp (-2 \pi i \sigma), \nonumber \\
v & = & t_0^{1/k}, \nonumber \\
w & = & \zeta v^{-1}.
\label{eq:e60}
\end{eqnarray}
This curve is a non trivial element in $H_1({\cal Q}; {\bf Z})$, and a 
membrane wrapped on it will produce an ordinary type IIA string; however, we 
can not think that the QCD string is a type IIA string. If ${\cal Q} = {\bf R}^3 
\times S^1$, then $H_1({\cal Q};{\bf Z}) = {\bf Z}$, and curves of type 
(\ref{eq:e60}) will be the only candidates for non trivial $1$-cycles in 
${\cal Q}$. However, we can define QCD strings using cycles in the relative 
homology, $H_1({\cal Q}/\Sigma; {\bf Z})$, i. e., considering non trivial 
cycles ending on the surface $\Sigma$. To compute $H_1({\cal Q}/\Sigma; 
{\bf Z})$, we can use the exact sequence
\begin{equation}
H_1(\Sigma;{\bf Z}) \rightarrow H_1({\cal Q};{\bf Z}) \stackrel{\imath}{\rightarrow} 
H_1({\cal Q}/\Sigma; {\bf Z}),
\label{eq:e61}
\end{equation}
which implies
\begin{equation}
H_1({\cal Q}/\Sigma;{\bf Z}) = H_1({\cal Q};{\bf Z})/ \imath H_1(\Sigma;{\bf Z}).
\label{eq:e62}
\end{equation}
The map $\imath$ is determined by the map defining $\Sigma$ ($t=v^k$), and thus 
we can conclude that, very likely,
\begin{equation}
H_1({\cal Q}/\Sigma;{\bf Z})={\bf Z}_k.
\label{eq:e63}
\end{equation}
A curve in $H_1({\cal Q}/\Sigma;{\bf Z})$ can be defined as follows:
\begin{eqnarray}
t & = & t_0, \nonumber \\
v & = & t_0^{1/k} e^{2 \pi i \sigma/k}, \nonumber \\
w & = & \zeta v^{-1},
\label{eq:e64}
\end{eqnarray}
with $t_0^{1/k}$ one of the $k$ roots. The tension of (\ref{eq:e64}), by construction, 
is independent of $R$, because $t$ is fixed. Using the metric on ${\cal Q}$, 
the length of (\ref{eq:e64}) is given by
\begin{equation}
\left( \frac {\zeta^2 t^{-2/n}}{n^2} + \frac {t^{2/n}}{n^2} \right)^{1/2},
\label{eq:e65}
\end{equation}
and its minimum is obtained when $t^{2/n}=\zeta$. Thus, the length of the QCD string 
should be
\begin{equation}
\frac {|\zeta|^{1/2}}{n},
\label{eq:e66}
\end{equation}
which has the right length units, as $\zeta$ behaves as $(\hbox
{length})^2$. In order to define the tension we need to go to
$(\hbox {length})^{-1}$ units, again using $\frac {1}{R^2}$.
Then, if we identify this tension with $\Lambda$, we get
\begin{equation}
\Lambda \sim \frac {|\zeta|^{1/2}}{n} \frac {1}{R^2}
\end{equation}
or, equivalently,
\begin{equation}
\Lambda^2 \sim \Lambda^3 R.
\end{equation}
Thus, consistency with QCD results requires $\Lambda \sim \frac
{1}{R}$. These are not good news, as they imply that the theory
we are working with, in order to match QCD, posseses $0$-brane
modes, with masses of the order of $\Lambda$, and therefore we
have not decoupled the M-theory modes.

Next, we would like to compare the superpotential described above with the 
ones obtained using standard instanton techniques in M-theory. However, 
before doing that we will conclude this brief review on brane configurations 
with the description of models with $N=4$ supersymmetry.

\subsubsection{$N=2$ Models with Vanishing Beta Function.}

Let us come back to brane configurations with $n+1$ solitonic fivebranes, with 
$k_{\alpha}$ Dirichlet fourbranes extending between the $\alpha^{th}$ pair 
of NS-5branes. The beta function, derived in (\ref{eq:e21}), is
\begin{equation}
-2k_{\alpha}+k_{\alpha+1}+k_{\alpha-1},
\label{eq:e68}
\end{equation}
for each $SU(k_{\alpha})$ factor in the gauge group. In this section, we 
will compactify the $x^6$ direction to a circle of radius $L$. Impossing the 
beta function to vanish in all sectors immediately implies that all $k_{\alpha}$ 
are the same. Now, the compactification of the $x^6$ direction does 
not allow to eliminate all $U(1)$ factors in the gauge group: one of them 
can not be removed, so that the gauge group is reduced from $\prod_{\alpha=1}^n 
U(k_{\alpha})$ to $U(1) \times SU(k)^n$. Moreover, using the definition 
(\ref{eq:e20}) of the mass of the hypermultiplets we get, for periodic 
configurations, 
\begin{equation}
\sum_{\alpha} m_{\alpha} = 0.
\label{eq:e69}
\end{equation}
The hypermultiplets are now in representations of type $k \otimes \bar{k}$, and 
therefore consists of a copy of the adjoint representation, and a neutral singlet.
  
Let us consider the simplest case, of $N=2$ $SU(2) \times U(1)$ four dimensional theory, 
with one hypermultiplet in the adjoint representation \cite{Wm4}. The corresponding brane 
configuration contains a single solitonic fivebrane, and two Dirichlet 
fourbranes. The mass of the hypermultiplet is clearly zero, and the corresponding 
four dimensional theory has vanishing beta function. A geometric procedure 
to define masses for the hypermultiplets is a fibering of the $v$-plane on the 
$x^6$ $S^1$ direction, in a non trivial way, so that the fourbrane positions are 
identified modulo a shift in $v$,
\begin{eqnarray}
x^6 & \rightarrow & x^6 + 2 \pi L, \nonumber \\
v   & \rightarrow & v + m,
\label{eq:e70}
\end{eqnarray}
so that now, the mass of the hypermultiplet, is the constant $m$ appearing in   
(\ref{eq:e70}), as $\sum_{\alpha} m_{\alpha} = m$.
  
From the point of view of M-theory, the $x^{10}$ coordinate has also been 
compactified on a circle, now of radius $R$. The $(x^6,x^{10})$ space has the 
topology of $S^1 \times S^1$. This space can be made non trivial if, when going 
around $x^6$, the value of $x^{10}$ is changed as follows:
\begin{eqnarray}
x^6 & \rightarrow & x^6 + 2 \pi L, \nonumber \\
x^{10}  & \rightarrow & x^{10} + \theta R,
\label{eq:e71}
\end{eqnarray}
and, in addition, $x^{10} \rightarrow x^{10} + 2 \pi R$. Relations (\ref{eq:e71}) 
define a Riemann surface of genus one, and moduli depending on $L$ and $\theta$ 
for fixed values of $R$. $\theta$ in (\ref{eq:e71}) can be understood as the 
$\theta$-angle of the four dimensional field theory: the $\theta$-angle can 
be defined as
\begin{equation}
\frac {x^{10}_1-x^{10}_2}{R},
\label{eq:e72}
\end{equation}
with $x^{10}_2=x^{10}(2 \pi L)$, and $x^{10}_1=x^{10}(0)$. Using (\ref{eq:e71}), we 
get $\theta$ as the value of (\ref{eq:e72}). This is the bare $\theta$-angle of 
the four dimensional theory.
  
A question inmediately appears concerning the value of the bare coupling 
constant: the right answer should be
\begin{equation}
\frac {1}{g^2} = \frac {2 \pi L}{R}.
\label{eq:e73}
\end{equation}
It is therefore clear that we can move the bare coupling constant of the 
theory keeping fixed the value of $R$, and changing $L$ and $\theta$. Let 
us now try to solve this model for the massless case. The solution will 
be given by a Riemann surface $\Sigma$, living in the space $E \times C$, where 
$E$ is the Riemann surface defined by (\ref{eq:e71}), and $C$ is the $v$-plane. Thus, 
all what we need is defining $\Sigma$ through an equation of the type
\begin{equation}
F(x,y,z)=0,
\label{eq:e74}
\end{equation}
with $x$ and $y$ restricted by the equation of $E$,
\begin{equation}
y^2 = (x-e_1(\tau))(x-e_2(\tau))(x-e_3(\tau)),
\label{eq:e75}
\end{equation}
with $\tau$ the bare coupling constant defined by (\ref{eq:e72}) and (\ref{eq:e73}) \cite{DW}. In case 
we have a collection of $k$ fourbranes, we will require $F$ to be a polynomial 
of degree $k$ in $v$,
\begin{equation}
F(x,y,z)=v^k-f_1(x,y)v^{k-1}+ \cdots
\label{eq:e76}
\end{equation}
  
The moduli parameters of $\Sigma$ are, at this point, hidden in the functions 
$f_i(x,y)$ in (\ref{eq:e76}). Let us denote $v_i(x,y)$ the roots of (\ref{eq:e76}) 
at the point $(x,y)$ in $E$. Notice that (\ref{eq:e76}) is a spectral curve defining 
a branched covering of $E$, i. e., (\ref{eq:e76}) can be interpreted as a 
spectral curve in the sense of Hitchin's integrable system \cite{H}. If $f_i$ has a pole 
at some point $(x,y)$, then the same root $v_i(x,y)$ should go to infinity. These 
poles have the interpretation of locating the position of the solitonic 
fivebranes. In the simple case we are considering, with a single fivebrane, the 
Coulomb branch of the theory will be parametrized by meromorphic functions on $E$ 
with a simple pole at one point, which is the position of the fivebrane. As 
we have $k$ functions entering (\ref{eq:e76}), the dimension of the Coulomb 
branch will be $k$, which is the right one for a theory with $U(1)\times SU(k)$ 
gauge group. 
  
Now, after this discussion of the model with massless hypermultiplets, we 
will introduce the mass. The space where now we need to define $\Sigma$ is 
not $E \times C$, but the non trivial fibration defined through
\begin{eqnarray}
x^6 & \rightarrow & x^6 + 2 \pi L, \nonumber \\
x^{10} & \rightarrow & x^{10} + \theta R, \nonumber \\
v & \rightarrow & v + m
\label{eq:e77}
\end{eqnarray}
or, equivalently, the space obtained by fibering $C$ non trivially on $E$. We 
can flat this bundle over all $E$, with the exception of one point $p_0$. 
Away from this point, the solution is given by (\ref{eq:e76}). If we write 
(\ref{eq:e76}) in a factorized form,
\begin{equation}
F(x,y,z)=\prod_{i=1}^k(v-v_i(x,y)),
\label{eq:e78}
\end{equation}
we can write $f_1$ in (\ref{eq:e76}) as the sum
\begin{equation}
f_1 =\sum_{i=1}^k v_i(x,y);
\label{eq:e79}
\end{equation}
therefore, $f_1$ will have poles at the positions of the fivebrane. The mass 
of the hypermultiplet will be identified with the residue of the differential 
$f_1 \omega$, with $\omega$ the abelian differential, $\omega = \frac {dx}{y}$. As 
the sum of the residues is zero, this means that at the point at infinity, that 
we identify with $p_0$, we have a pole with residue $m$.

\subsection{M-Theory and String Theory.}

In this section we will compare the M-theory description of $N=2$ and $N=1$ 
four dimensional gauge theories, with that obtained in string theory upon 
performing the point particle limit \cite{KKLMV,GHL1,KLMVW,GHL2}. Let us then return for a moment to the 
brane representation of $N=2$ four dimensional gauge theories. In the 
M-theory approach, we will consider M-theory on flat spacetime, ${\bf R}^7 \times {\cal Q}$, 
with ${\cal Q}= {\bf R}^5 \times S^1$. The $S^1$ stands for the (compactified)  
eleventh dimension, with the radius $R$ proportional to the string coupling constant. 
The brane configuration in ${\bf R}^7 \times {\cal Q}$ turns out to be 
equivalent to a solitonic fivebrane, with worldvolume $\Sigma \times {\bf R}^4$, where 
$\Sigma$ is a complex curve in ${\cal Q}$, defined by
\begin{equation}
F(t,v)=0.
\label{eq:e80}
\end{equation}
This is equivalent to defining an embedding
\begin{equation}
\Phi : \Sigma \hookrightarrow {\cal Q}.
\label{eq:e81}
\end{equation}
If $\Sigma$ is a lagrangian manifold of ${\cal Q}$, then we can interpret the 
moduli space of the effective four dimensional $N=2$ theory as the space of 
deformations of $\Phi$ in (\ref{eq:e81}) preserving the condition of 
lagrangian submanifold\footnote{Recall that a lagrangian manifold is defined 
by the condition that 
\[ \int_{\Sigma} \Phi^* (\Omega) = \hbox {Vol } (\Sigma), \]
with $\Phi^*$ such that $\Phi^*(\omega)=0$ (where $\omega$ is the K\"ahler class 
of ${\cal Q}$), and $\Omega$ the holomorphic top form of ${\cal Q}$.}. By 
Mc Lean's theorem, we know that the dimension of this space of deformations 
is $b_1(\Sigma)$, in agreement with the existing relation between the genus of 
$\Sigma$ and the rank of the gauge group in the effective four dimensional 
theory. It is important keeping in mind that in the M-theory approach two 
ingredients are being used: the curve defined by (\ref{eq:e80}), and the holomorphic 
top form $\Omega$ of ${\cal Q}$, which explicitely depends on the radius $R$ 
of the eleventh dimension. This will be very important, as already noticed 
in the discussion of the $N=1$ superpotentials, because an explicit dependence 
on the string coupling constant will be induced in the BPS mass formulas.
  
A different approach to (\ref{eq:e80}) and (\ref{eq:e81}) is that
based on geometric engineering \cite{ge}. In this case, the procedure is
based on the following set of steps:
\begin{itemize}
	\item[{1.}] String theory is compactified on a Calabi-Yau
threefold $X$, with the apropiate number of vector multiplets in
four dimensions.
	\item[{2.}] A point corresponding to classical enhancement of gauge 
symmetry in the moduli space of the Calabi-Yau
threefold must be localized.
	\item[{3.}] A rigid Calabi-Yau threefold is defined by
performing a point particle limit.
	\item[{4.}] The rigid Calabi-Yau manifold is used to
define the Seiberg-Witten surface $\Sigma$.
	\item[{5.}] Going form type IIB to type IIA string theory
represents a brane configuration corresponding to an ALE space
with singularity of some Dynkin type into a set of fivebranes
that can be interpreted as a fivebrane with worldvolume $\Sigma
\times {\bf R}^4$.
	\item[{6.}] The BPS states are defined through the
meromorphic one-form $\lambda$, derived from the the Calabi-Yau
holomorphic top form, in the rigid point particle limit.
\end{itemize}
  
As we can see from the previous set of steps, that we will
explicitely show at work in one definite example, the main
difference between both approaches is at the level of the
meromorphic form in Seiberg-Witten theory. There is also an
important difference in the underlying philosophy, related to the
implicit use in the string approach, described in the above
steps, of the heterotic-type II dual pairs, driving us to the
choice of a particular Calabi-Yau manifold. The most elaborated
geometric engineering approach uses, instead of a certain
heterotic-type II dual pair, a set of local geometrical data,
determined by the type of gauge symmetry we are interested on,
and generalizes mirror maps to this set of local data. In all
these cases, the four dimensional field theory we are going to
obtain will not depend on extra parameters, as the string
coupling constant. On the other hand, the M-theory approach,
where field theories are obtained depending explicitely on the
string coupling constant, might be dynamically rich enough as to
provide a direct explanation of phenomena that can not be easily
understood in the more restricted context of the point particle
limit of string theory.
  
Next, we will follow steps $1$ to $6$ through an explicit
example \cite{KLMVW}. In order to obtain a field theory with gauge group
$SU(n)$ we should start with a Calabi-Yau manifold with
$h_{2,1}=n$, and admiting the structure of a $K3$-fibered
threefold (see chapter II for definitions, and additional
details). We will consider the $SU(3)$ case, corresponding to a
Calabi-Yau manifold whose mirror is the weighted projective space
$\IP_{1,1,2,8,12}^{24}$,
\begin{equation}
\frac {1}{24}(x_1^{24} +x_2^{24}) + \frac {1}{12} x_3^{12} +
\frac {1}{2} x_5^{2} - \psi_0 x_1x_2x_3x_4x_5 - \frac {1}{6}
(x_1x_2x_3)^6- \frac {1}{12} (x_1x_2)^{12} =0.
\label{eq:e82}
\end{equation}
In order to clearly visualize (\ref{eq:e82}) as a $K3$-fibration
we will perform the change of variables
\begin{equation}
x_1/x_2 \equiv \hat{z} ^{1/12} b^{-1/24}, \: \: \: \: \: \:
x_1^{2} \equiv x_0 \hat{z}^{1/12}, 
\label{eq:e83}
\end{equation}
so that (\ref{eq:e82}) can be rewriten in the form
\begin{equation}
\frac {1}{24} (\hat{z} + \frac {b}{\hat{z}} +2)x_0^{12} + \frac
{1}{12}x_3^{12} + \frac {1}{3}x_4^3+ \frac {1}{2} x_5^2 + \frac
{1}{6 \sqrt{c}} (x_0 x_3)^6 + \left( \frac {a}{\sqrt{c}}
\right)^{1/6} x_0 x_3 x_4 x_5 =0,
\label{eq:e84}
\end{equation}
which represents a $K3$ surface, fibered over a $\IP^1$ space 
parametrized by the coordinate $z$. Parameters in (\ref{eq:e84})
are related to those in (\ref{eq:e82}) through
\begin{equation}
a = - \psi_0^6/\psi_1, \: \: \: \: b=\psi_2^{-2}, \: \: \: \:
c=\psi_2/\psi_1^2.
\label{eq:e85}
\end{equation}
The parameter $b$ can be interpreted as the volume of $\IP^1$:
\begin{equation}
- \log b = \hbox {Vol }(\IP^1).
\label{eq:e86}
\end{equation}
Next, we should look for the points $\hat{z}$ in $\IP^1$
over which the $K3$ surface is singular. The discriminant can be
written as
\begin{equation}
\Delta_{K3} = \prod_{i=0}^2
(\hat{z}-e_i^+(a,b,c))(\hat{z}-e_i^-(a,b,c)), 
\label{eq:e87}
\end{equation}
where 
\begin{eqnarray}
e_0^{\pm} & = & -1 \pm \sqrt{1-b}, \nonumber\\
e_1^{\pm} & = & \frac {1-c \pm \sqrt{(1-c)^2-bc^2}}{c}, \nonumber\\
e_2^{\pm} & = & \frac {(1-a)^2-c \pm \sqrt{((1-a)^2-c)^2-bc^2}}{c}.
\label{eq:e88}
\end{eqnarray}
The Calabi-Yau manifold will be singular whenever two roots $e_i$
coalesce, as
\begin{equation}
\Delta_{\hbox {Calabi-Yau}}=\prod_{i<j}(e_i-e_j)^2.
\label{eq:e89}
\end{equation}
We will consider the  singular point in the moduli space
corresponding to $SU(3)$ symmetry. Around this point we will
introduce new coordinates, through
\begin{eqnarray}
a & = & -2 (\alpha' u)^{3/2}, \nonumber \\
b & = & \alpha' \Lambda^6, \nonumber \\
c & = & 1 - \alpha'^{3/2}(-2 u^{3/2} +3\sqrt{3} v).
\label{eq:e90}
\end{eqnarray}
Going now to the $\alpha' \rightarrow 0$ limit in (\ref{eq:e89}),
we get a set of roots $e_i(u,v;\Lambda^6)$ on a $z$-plane, with
$z$ defined in $\alpha'^{3/2} z \equiv \hat{z}$:
\begin{eqnarray}
e_0 & = & 0, \: \: e_{\infty} = \infty, \nonumber \\
e_1^{\pm} & = & 2u^{3/2} + 3\sqrt{3}v \pm
\sqrt{(2u^{3/2}+3\sqrt{3}v)^2- \Lambda^6}, \nonumber \\
e_2^{\pm} & = & - 2u^{3/2} + 3\sqrt{3}v \pm
\sqrt{(2u^{3/2}-3\sqrt{3}v)^2- \Lambda^6}.
\label{eq:e91}
\end{eqnarray}
Now, we can use (\ref{eq:e91}) as the definition of a Riemann
surface $\Sigma$, defined by the Calabi-Yau data at the singular
$SU(3)$ point, and in the point particle limit. There exits a
natural geometrical picture for understanding the parameters $u$
and $v$ in (\ref{eq:e90}), which is the definition of the blow
up, in the moduli space of complex structures of
$\IP_{1,1,2,8,12}^{24}$, of the $SU(3)$ singular point. From this
point of view, the parameters $u$ and $v$ in (\ref{eq:e90}) will
be related to the volume of the set of vanishing two-cycles
associated with a {\em rational} singularity, i. e., an orbifold
singularity of type $A_{n-1}$ (in the case we are considering,
$n=3$). These vanishing cycles, as is the case with rational
singularities, are associated with Dynkin diagrams of non affine
type. The branch points (\ref{eq:e91}) on the $z$-plane define
the curve
\begin{equation}
y^2 = \prod_i (x-e_i(u,v;\Lambda^6)), 
\label{eq:e93}
\end{equation}
which can also be represented as the vanishing locus of a
polynomial $F(x,z)=0$, with $F$ given by \cite{int1,int2}
\begin{equation}
F(x,z)=z + \frac {\Lambda^6}{z} + B(x),
\label{eq:e94}
\end{equation}
where $B(x)$ is a polynomial in $x$ of degree three; in the general
case of $SU(n)$ theories, the polynomial will be of degree $n$. 
  
This has exactly the same look as what we have obtained using
brane configurations, with the space ${\cal Q}$ replaced by the
$(x,z)$ space. The difference is that now we are not considering
the $(x,z)$ space as a part of spacetime, and $\Sigma$ as
embedded in it, but we use $\Sigma$ as defined in (\ref{eq:e94})
to define a Calabi-Yau space in a rigid limit by the equation
\begin{equation}
F(x,z)+y^2+w^2=0,
\label{eq:e95}
\end{equation}
which defines a threefold in the $(x,y,z,w)$ space. And, in
addition, we think of (\ref{eq:e95}) as a Calabi-Yau
representation of the point particle limit. In order to get the
meromorphic one-form $\lambda$, and the BPS states, we need to
define a map from the third homology group, $H_3(CY)$, of the
Calabi-Yau manifold, into $H_1(\Sigma)$. This can be done as
follows. The three-cycles in $H_3(CY)$ of the general type $S^2
\times S^1$, with $S^2$ a vanishing cycle of $K3$, correspond to
$S^1$ circles in the $z$-plane. The three-cycles with the
topology of $S^3$ can be interpreted as a path from the north to
the south pole of $S^3$, starting with a vanishing two-cycle, and
ending at another vanishing two-cycle of $K3$. This corresponds,
in the $z$-plane, to paths going from $e_i^+$ to $e_i^-$. Once we
have defined this map,
\begin{equation}
f: H_3(CY) \longrightarrow H_1(\Sigma),
\label{eq:e96}
\end{equation}
we define 
\begin{equation}
\lambda (f(C)) = \Omega (C),
\label{eq:e97}
\end{equation}
with $\Omega$ the holomorphic top form. 
  
A similar analysis can be done for computing the mass of BPS
states, and the meromorphic one-form $\lambda$ in the brane
framework. In fact, we can consider a two-cycle $C$ in ${\cal Q}$
such that
\begin{equation}
\partial C \subset \Sigma,
\label{eq:e98}
\end{equation}
or, in other words, $C \in H_2({\cal Q}/\Sigma;{\bf Z})$. The
holomorphic top form on ${\cal Q}$ is given by
\begin{equation}
\Omega = R \frac {dt}{t} \wedge dv,
\label{eq:e99}
\end{equation}
and thus the BPS mass will be given by
\begin{equation}
M \sim R \int_C \frac {dt}{t} \wedge dv = R \int_{\partial C}
\frac {dt}{t} v(t),
\label{eq:e100}
\end{equation}
with $v(t)$ given by
\begin{equation}
F(t,v)=0,
\label{eq:e101}
\end{equation}
for the corresponding Seiberg-Witten curve, $\Sigma$. Notice that
the same analysis, using (\ref{eq:e97}) and the holomorphic top
form for (\ref{eq:e95}) will give, by contrast to the brane case,
a BPS mass formula independent of $R$.
  
Next, we will compare the brane construction and geometric
engineering in the more complicated case of $N=1$ \cite{KV,BJPSV}.

\subsection{Local Models for Elliptic Fibrations.}
\label{sec:elliptic}

Let $V$ be an elliptic fibration,
\begin{equation}
\Phi: V \rightarrow \Delta,
\label{eq:r1}
\end{equation}
with $\Delta$ an algebraic curve, and $\Phi^{-1}(a)$, with $a$
any point in $\Delta$, an elliptic curve. Let us denote
$\{a_{\rho}\}$ the finite set of points in $\Delta$ such that
$\Phi^{-1}(a_{\rho})={\cal C}_{\rho}$ is a singular fiber. Each singular
fiber ${\cal C}_{\rho}$ can be written as
\begin{equation}
{\cal C}_{\rho} = \sum_i n_{i \rho} \Theta_{i \rho},
\label{eq:r2}
\end{equation}
where $\Theta_{i \rho}$ are non singular rational curves, with
$\Theta_{i \rho}^2 =-2$, and $n_{i \rho}$ are integer numbers.
Different types of singularities are characterized by
(\ref{eq:r2}) and the intersection matrix $(\Theta_{i \rho} .
\Theta_{j \rho})$. All different types of Kodaira singularities
satisfy the relation
\begin{equation}
{\cal C}_{\rho}^2=0.
\label{eq:r3}
\end{equation}
  
Let $\tau(u)$ be the elliptic modulus of the elliptic fiber at
the point $u \in \Delta$. For each path $\alpha$ in
$\Pi_1(\Delta')$, with $\Delta'= \Delta- \{a_{\rho}\}$, we can
define a monodromy transformation $S_{\alpha}$, in $Sl(2,{\bf
Z})$, acting on $\tau(u)$ as follows:
\begin{equation}
S_{\alpha} \tau(u) = \frac {a_{\alpha} \tau(u) +
b_{\alpha}}{c_{\alpha} \tau(u) + d_{\alpha}}.
\label{eq:r4}
\end{equation}
Each type of Kodaira singularity is characterized by a particular
monodromy matrix. 
  
In order to define an elliptic fibration \cite{Kodaira}, the starting point will
be an algebraic curve $\Delta$, that we will take, for
simplicity, to be of genus zero, and a meromorphic function
${\cal J}(u)$ on $\Delta$. Let us assume ${\cal J}(u) \neq
0,1,\infty$ on $\Delta' =\Delta-\{a_{\rho}\}$. Then, there
exists multivalued holomorphic function $\tau(u)$, with $\hbox
{Im } \tau(u)>0$, satisfying ${\cal J}(u) = j(\tau(u))$, with $j$
the elliptic modular $j$-function on the upper half plane. As
above, for each $\alpha \in \Pi_1(\Delta')$ we define a monodromy
matrix $S_{\alpha}$, acting on $\tau(u)$ in the form defined by
(\ref{eq:r4}). Associated to these data we will define an elliptic 
fibration, (\ref{eq:r1}). In order to do that, let us first define 
the universal covering $\tilde{\Delta}'$, of $\Delta'$, and let
us identify the covering transformations of $\tilde{\Delta}'$
over $\Delta'$, with the elements in $\Pi_1(\Delta')$. Denoting
by $\tilde{u}$ a point in $\tilde{\Delta}'$, we define, for each
$\alpha \in \Pi_1(\Delta')$, the covering transformation
$\tilde{u} \rightarrow \alpha \tilde{u}$, by
\begin{equation}
\tau(\alpha \tilde{u}) = S_{\alpha} \tau(\tilde{u});
\label{eq:r5}
\end{equation}
in other words, we consider $\tau$ as a single valued holomorphic
function on $\tilde{\Delta}'$. Using (\ref{eq:r4}), we define
\begin{equation}
f_{\alpha}(\tilde{u}) = (c_{\alpha} \tau(\tilde{u}) +
d_{\alpha})^{-1}.
\label{eq:r6}
\end{equation}
Next, we define the product $\tilde{\Delta}' \times {\bf C}$ and,
for each $(\alpha,n_1,n_2)$, with $\alpha \in \Pi_1(\Delta')$,
and $n_1,n_2$ integers, the automorphism
\begin{equation}
g(\alpha,n_1,n_2) : (\tilde{u},\lambda) \rightarrow (\alpha
\tilde{u}, f_{\alpha} (\tilde{u}) (\lambda+n_1 \tau
(\tilde{u})+n_2)).
\label{eq:r7}
\end{equation}
Denoting by ${\cal G}$ the group of automorphisms (\ref{eq:r7}),
we define the quotient space
\begin{equation}
B' \equiv (\tilde{\Delta}' \times {\cal C}) / {\cal G}.
\label{eq:r8}
\end{equation}
This is a non singular surface, since $g$, as defined by
(\ref{eq:r7}), has no fixed points in $\tilde{\Delta}'$. From
(\ref{eq:r7}) and (\ref{eq:r8}), it is clear that $B'$ is an
elliptic fibration on $\Delta'$, with fiber elliptic curves of
elliptic modulus $\tau(u)$. Thus, by the previous construction, we
have defined the elliptic fibration away from the singular points
$a_{\rho}$. 
  
Let us denote $E_{\rho}$ a local neighbourhood of the point
$a_{\rho}$, with local coordinate $t$, and such that
$t(a_{\rho})=0$. Let $S_{\rho}$ be the monodromy associated with
a small circle around $a_{\rho}$. By ${\cal U}_{\rho}$ we will denote
the universal covering of $E_{\rho}'=E_{\rho}-a_{\rho}$, with
coordinate $\rho$ defined by
\begin{equation}
\rho= \frac {1}{2 \pi i} \log t.
\label{eq:r9}
\end{equation}
The analog of (\ref{eq:r5}) will be
\begin{equation}
\tau(\rho+1) = S_{\rho} \tau(\rho).
\label{eq:r10}
\end{equation}
If we go around the points $a_{\rho}$, $k$ times, we should act
with $S_{\rho}^k$; hence, we parametrize each path by the winding
number $k$. The group of automorphisms (\ref{eq:r7}), reduced to
small closed paths around $a_{\rho}$, becomes
\begin{equation}
g(k,n_1,n_2)(\rho,\lambda) =
(\rho+k,f_k(\rho)[\lambda+n_1\tau(\rho)+n_2]).
\label{eq:r11}
\end{equation}
Denoting by ${\cal G}_{\rho}$ the group (\ref{eq:r11}), we define
the elliptic fibration around $a_{\rho}$ as
\begin{equation}
({\cal U}_{\rho} \times {\bf C})/{\cal G}_{\rho}.
\label{eq:r12}
\end{equation}
  
Next, we will extend the elliptic fibration to the singular point
$a_{\rho}$. We can consider two different cases, depending on the
finite or infinite order of $S_{\rho}$.

\subsection{Singularities of Type $\hat{D}_4$: ${\bf Z}_2$
Orbifolds.}

Let us assume $S_{\rho}$ is of finite order,
\begin{equation}
(S_{\rho})^m = {\bf 1}_d.
\label{eq:r13}
\end{equation}
In this case, we can extend (\ref{eq:r12}) to the singular
points, simply defining a new variable $\sigma$ as
\begin{equation}
\sigma^m = t.
\label{eq:r14}
\end{equation}
Let us denote $D$ a local neighbourhood in the $\sigma$-plane of
the point $\sigma=0$, and define the group $G_D$ of automorphisms
\begin{equation}
g(n_1,n_2):(\sigma,\lambda) = (\sigma,\lambda+n_1
\tau(\sigma)+n_2),
\label{eq:r15}
\end{equation}
and the space
\begin{equation}
F=(D \times {\bf C})/G_D.
\label{eq:r16}
\end{equation}
Obviously, $F$ defines an elliptic fibration over $D$, with fiber
$F_{\sigma}$ at each point $\sigma \in D$, an elliptic curve of
modulus $\tau(\sigma)$. From (\ref{eq:r13}) and (\ref{eq:r6}), it
follows that
\begin{equation}
f_k(\sigma) = 1,
\label{eq:r17}
\end{equation}
with $k=O(m)$. Thus, we can define a normal subgroup ${\cal N}$
of ${\cal G}_{\rho}$ as the set of transformations
(\ref{eq:r11}):
\begin{equation}
g(k,n_1,n_2):(\rho,\lambda) \rightarrow (\rho+k,\lambda+n_1
\tau(\rho)+n_2).
\label{eq:r18}
\end{equation}
Comparing now (\ref{eq:r15}) and (\ref{eq:r18}), we get
\begin{equation}
({\cal U}_{\rho} \times {\bf C})/{\cal N} = (D' \times {\bf
C})/G_D \equiv F-F_0.
\label{eq:r19}
\end{equation}
Using (\ref{eq:r18}) and (\ref{eq:r11}) we get
\begin{equation}
{\cal C} = {\cal G}/{\cal N},
\label{eq:r20}
\end{equation}
with ${\cal C}$ the cyclic group of order $m$, defined by
\begin{equation}
g_k:(\sigma,\lambda) \rightarrow (e^{2 \pi i k/m} \sigma,
f_k(\sigma) \lambda).
\label{eq:r21}
\end{equation}
From (\ref{eq:r20}) and (\ref{eq:r19}), we get the desired
extension to $a_{\rho}$, namely
\begin{equation}
F/{\cal C} = ({\cal U}_{\rho} \times {\bf C})/{\cal G}_{\rho}
\cup F_0/{\cal C}.
\label{eq:r22}
\end{equation}
Thus, the elliptic fibration extended to $a_{\rho}$, in case
$S_{\rho}$ is of finite order, is defined by $F/{\cal C}$. Now,
$F/{\cal C}$ can have singular points that we can regularize. The
simplest example corresponds to
\begin{equation}
S_{\rho} = \left( \begin{array}{cc} -1 & 0 \\ 0 & -1 \end{array}
\right),
\label{eq:r23}
\end{equation}
i. e., a parity transformation. In this case, the order is $m=2$,
and we define $\sigma$ by $\sigma^2 \equiv t$. The cyclic 
group (\ref{eq:r21}) in this case simply becomes
\begin{equation}
(\sigma, \lambda) \rightarrow (- \sigma, -\lambda),
\label{eq:r24}
\end{equation}
since from (\ref{eq:r23}) and (\ref{eq:r6}) we get $f_1=-1$. At the point 
$\sigma=0$ we have four fixed points, the standard ${\bf Z}_2$ 
orbifold points,
\begin{equation}
(0,\frac {a}{2}\tau(0)+ \frac {b}{2}),
\label{eq:r25}
\end{equation}
with $a,b=0,1$. The resolution of these four singular points will produce 
four irreducible components, $\Theta^1, \ldots, \Theta^4$. In addition, 
we have the irreducible component $\Theta_0$, defined by the curve itself 
at $\sigma=0$. Using the relation $\sigma^2=t$, we get the $\hat{D}_4$ cycle, 
\begin{equation}
{\cal C} = 2 \Theta_0 + \Theta^1 + \Theta^2 + \Theta^3 + \Theta^4,
\label{eq:r26}
\end{equation}
with $(\Theta_0,\Theta^1)=(\Theta_0,\Theta^2)=(\Theta_0,\Theta^3)=
(\Theta_0,\Theta^4)=1$. In general, the four external points of $D$-diagrams 
can be associated with the four ${\bf Z}_2$ orbifold points of the torus. 

\subsection{Singularities of Type $\hat{A}_{n-1}$.}

We will now consider the case
\begin{equation}
S_{\rho} = \left( \begin{array}{cc} 1 & n \\ 0 & 1 \end{array} \right), 
\label{eq:r27}
\end{equation}
which is of infinite order. A local model for this monodromy can be defined by 
\begin{equation}
\tau(t) = \frac {1}{2 \pi i} n \log t.
\label{eq:r28}
\end{equation}
Using the variable $\rho$ defined in (\ref{eq:r9}), we get, for the group 
${\cal G}_{\rho}$ of automorphisms,
\begin{equation}
g(k,n_1,n_2):(\rho,\lambda) \rightarrow (\rho+k,\lambda+n_1n \rho+n_2),
\label{eq:r29}
\end{equation}
and the local model for the elliptic fibration, out of the singular point,
\begin{equation}
({\cal U}_{\rho} \times {\bf C})/{\cal G}_{\rho},
\label{eq:r30}
\end{equation}
i. e., fibers of the type of elliptic curves, with elliptic modulus $n\rho$. 
A simple way to think about these elliptic curves is in terms of cyclic 
unramified coverings \cite{Fay}. Let us recall that a cyclic unramified covering, 
$\Pi : \hat{C} \rightarrow C$, of order $n$, of a curve $C$ of genus $g$, is 
a curve $\hat{C}$ of genus 
\begin{equation}
\hat{g}=ng+1-n.
\label{eq:r31}
\end{equation}
Thus, for $g=1$, we get $\hat{g}=1$, for arbitrary $n$. Denoting by $\tau$ 
the elliptic modulus of $C$, in case $g=1$, the elliptic modulus of $\hat{C}$ 
is given by
\begin{equation}
\hat{\tau}=n \tau.
\label{eq:r32}
\end{equation}
Moreover, the generators $\hat{\alpha}$ and $\hat{\beta}$ of $H_1(\hat{C};{\bf Z})$ 
are given in terms of the homology basis $\alpha$, $\beta$ of $C$ as
\begin{eqnarray}
\Pi \hat{\alpha} & = & \alpha, \nonumber \\
\Pi \hat{\beta}  & = & n \beta, 
\label{eq:r33}
\end{eqnarray}
with $\Pi$ the projection $\Pi: \hat{C} \rightarrow C$. From (\ref{eq:r32}) 
and (\ref{eq:r29}), we can interpret the elliptic fibration (\ref{eq:r30}) 
as one with elliptic fibers given by $n$-cyclic unramified coverings of a 
curve $C$ with elliptic modulus $\rho$ or, equivalently, $\frac {1}{2 \pi i} 
\log t$. There exits a simple way to define a family of elliptic curves, 
with elliptic modulus given by $\frac {1}{2 \pi i} \log t$, which is the plumbing 
fixture construction. Let $D_0$ be the unit disc around $t=0$, and let 
$C_0$ be the Riemann sphere. Define two local coordinates, $z_a: {\cal U}_a 
\rightarrow D_0$, $z_b : {\cal U}_b \rightarrow D_0$, in disjoint 
neigbourhoods ${\cal U}_a$, ${\cal U}_b$, of two points $P_a$ and $P_b$ of 
${\cal C}_0$. Let us then define
\[
W=\{(p,t) | t \in D_0, p \in C_0-{\cal U}_a - {\cal U}_b, \hbox { or } 
p \in {\cal U}_a, \hbox { with } |z_a(p)|>|t|, \hbox { or } \]
\begin{equation} 
p \in {\cal U}_b, \hbox { with } |z_b(p)|>|t| \},
\end{equation}
and let $S$ be the surface
\begin{equation}
S=\{ xy=t;(x,y,t)\in D_0 \times O_0 \times D_0 \}.
\end{equation}
We define the family of curves through the following identifications
\begin{eqnarray}
(p_a,t) \in W \cap {\cal U}_a \times D_0 & \simeq & (z_a(p_a), \frac {t}{z_a(p_a)}, t) 
\in S, \nonumber \\
(p_b,t) \in W \cap {\cal U}_b \times D_0 & \simeq & (\frac {t}{z_b(p_b)}, 
{z_b(p_b)}, t) \in S.
\end{eqnarray}
For each $t$ we get a genus one curve, and at $t=0$ we get a nodal curve by 
pinching the non zero homology cycles. The pinching region is characterized 
by
\begin{equation}
xy=t,
\label{eq:r34}
\end{equation}
which defines a singularity of type $A_0$. The elliptic modulus of the curves 
is given by
\begin{equation}
\tau(t) = \frac {1}{2\pi i} \log t + C_1 t + C_2,
\label{eq:r35}
\end{equation}
for some constants $C_1$ and $C_2$. We can use an appropiate choice of 
coordinate $t$, such that $C_1=C_2=0$. The singularity at $t=0$ is a singularity 
of type $\hat{A}_0$, in Kodaira's classification, corresponding to
\begin{equation}
S_{\rho} = \left( \begin{array}{cc} 1 & 1 \\ 0 & 1 \end{array} \right).
\label{eq:r36}
\end{equation}
Using now (\ref{eq:r32}) and (\ref{eq:r35}) we get, for the cyclic covering 
of order $n$, the result (\ref{eq:r28}), and the group (\ref{eq:r29}). The 
pinching region of the cyclic unramified covering is given by
\begin{equation}
xy=t^n,
\label{eq:r37}
\end{equation}
instead of (\ref{eq:r34}), i. e., for the surface defining the $A_{n-1}$ 
singularity, ${\bf C}^2/{\bf Z}_n$. Now, we can proceed to the resolution 
of the singularity at $t=0$. The resolution of the singularity (\ref{eq:r37}) 
requires $n-1$ exceptional divisors, $\Theta_1, \ldots, \Theta_{n-1}$. In 
addition, we have the rational curve $\Theta_0$, defined by the complement 
of the node. Thus, we get, at $t=0$,
\begin{equation}
{\cal C} =\Theta_0 + \cdots + \Theta_{n-1},
\label{eq:r38}
\end{equation}
with $(\Theta_0,\Theta_1)=(\Theta_0,\Theta_{n-1})=1$, and $(\Theta_i,
\Theta_{i+1})=1$, which is the $\hat{A}_{n-1}$ Dynkin diagram. The group 
of covering transformations of the $n^{th}$ order cyclic unramified 
covering is ${\bf Z}_n$, and the action over the components (\ref{eq:r38}) 
is given by
\begin{eqnarray}
\Theta_i & \rightarrow & \Theta_{i+1}, \nonumber \\
\Theta_{n-1} & \rightarrow & \Theta_0.
\label{eq:r39}
\end{eqnarray}

\subsection{Singularities of Type $\hat{D}_{n+4}$.}

This case is a combination of the two previous examples. Through the same 
reasoning as above, the group ${\cal G}_{\rho}$ is given, for
\begin{equation}
S_{\rho} = \left( \begin{array}{cc} -1 & -n \\ 0 & -1 \end{array} \right).
\label{eq:r40}
\end{equation}
by 
\begin{equation}
g(k,n_1,n_2) : (\rho,\lambda) \rightarrow (k+\rho, (-1)^k (\lambda +n_1 n \rho +n_2)).
\label{eq:r41}
\end{equation}
Using a new variable $\sigma^2=t$, what we get is a set of irreducible components 
$\Theta_0, \ldots \Theta_{2n}$, with the identifications $\Theta_i 
\rightarrow \Theta_{2n-i}$. In addition, we get the four fixed ${\bf Z}_2$ 
orbifold points described above. The singular fiber is then given by
\begin{equation}
{\cal C} = 2 \Theta_0 + \cdots + 2 \Theta_n + \Theta^1 + \Theta^2 + 
\Theta^3 + \Theta^4,
\label{eq:r42}
\end{equation}
with the intersections of the $\hat{D}_{n+4}$ affine diagram. It is easy to see 
that in this case we also get
\begin{equation}
({\cal C})^2 =0.
\label{eq:r43}
\end{equation}
Defing the genus of the singular fiber by $C^2=2g-2$, we conclude that $g=1$, 
for all singularities of Kodaira type. Notice that for rational singularities, 
characterized by non affine Dynkin diagrams of ADE type \cite{Artin}, we get self intersection 
${\cal C}^2=-2$, which corresponds to genus equal zero.

\subsection{Decompactification and Affinization.}

The general framework in which we are working in order to get four dimensional 
$N=1$ gauge theories is that of M-theory compactifications on elliptically 
fibered Calabi-Yau fourfolds, in the limit $\hbox {Vol }(E)=0$, with $E$ the elliptic fiber. 
As described above, we can interpolate between $N=2$ supersymmetry in three 
dimensions, and $N=1$ in four dimensions, by changing the radius $R$ 
through 
\begin{equation}
\hbox {Vol }(E) = \frac {1}{R}.
\label{eq:r44}
\end{equation}
The three dimensional limit then corresponds to $\hbox {Vol }(E) \rightarrow 
\infty$, and the four dimensional to $\hbox {Vol } (E) \rightarrow 0$. Now, 
we will work locally around a singular fiber of Kodaira $\hat{A}\hat{D}
\hat{E}$ type. As we know, for the Calabi-Yau fourfold $X$,
\begin{equation}
E \rightarrow X \stackrel{\Pi}{\rightarrow} B,
\label{eq:r45}
\end{equation}
the locus $C$ in $B$, where the fiber is singular, is of codimension one 
in $B$, i. e., of real dimension four. Let us now see what happens to the 
singular fiber in the three dimensional limit. In this case, we have 
$\hbox {Vol }(E)=\infty$. A possible way to represent this phenomenon is 
by simply extracting the point at infinity. In the case of $\hat{A}_{n-1}$ 
singularities, as described in previous subsection, taking out the point at infinity 
corresponds to decompactifying the irreducible component $\Theta_0$, that 
was associated with the curve itself. As was clear in this case, we then pass 
from the affine diagram, $\hat{A}_{n-1}$, to the non affine, $A_{n-1}$. More 
generally, as the elliptic fibration we are considering possesses a global section, 
we can select the irreducible component we are going to decompactify as the one 
intersecting with the basis of the elliptic fibration. When we decompactify, 
in the $\hbox {Vol }(E)=0$ limit, what we are doing, at the level of the 
fiber, is precisely compactifying the extra irreducible component, 
which leads to the affine Dynkin diagram.

\subsection{M-Theory Instantons and Holomorphic Euler Characteristic.}

Using the results of reference \cite{Wsp} a vertical instanton in a Calabi-Yau 
fourfold, of the type (\ref{eq:r45}), will be defined by a divisor $D$ 
of $X$, such that $\Pi(D)$ is of codimension one in $B$, and with holomorphic 
Euler characteristic
\begin{equation}
\chi(D,{\cal O}_D)=1.
\label{eq:r46}
\end{equation}
It is in case (\ref{eq:r46}) that we have two fermionic zero modes \cite{Wsp}, and we 
can define a superpotential contribution associated to $D$. For $N$, the 
normal bundle to $D$ in $X$, which is locally a complex line bundle on $D$, we 
define the $U(1)$ transformation
\begin{equation}
t \rightarrow e^{i \alpha}t,
\label{eq:r47}
\end{equation}
with $t$ a coordinate of the fiber of $N$. The two fermionic zero modes have 
$U(1)$ charge equal one half. Associated to the divisor $D$, we can define 
a scalar field $\phi_D$ that, together with $\hbox {Vol }(D)$ defines the 
imaginary and real parts of a chiral superfield. Under $U(1)$ rotations 
(\ref{eq:r47}), $\phi_D$ transforms as
\begin{equation}
\phi_D \rightarrow \phi_D + \chi(D) \alpha.
\label{eq:r48}
\end{equation}
In three dimensions, this is precisely the transformation of the dual photon field 
as Goldstone boson \cite{AHW}. However, transformation (\ref{eq:r48}) has perfect sense, 
for vertical instantons, in the four dimensional decompactification limit. 
  
Let us now consider an elliptically fibered Calabi-Yau fourfold, with 
singular fiber of $\hat{A}_{n-1}$ type, over a locus $C$ of codimension 
one in $B$. We will assume that the singular fiber is constant over $C$. Moreover, 
in the geometrical engineering spirit, we will impose
\begin{equation}
h_{1,0}(C)=h_{2,0}(C)=0
\label{eq:r49}
\end{equation}
and, thus, $\Pi_1(C)=0$. This prevents us from having non trivial transformations 
on the fiber by going, on $C$, around closed loops, since all closed loops 
are contractible. In addition, we will assume, based on (\ref{eq:r49}), 
that $C$ is an Enriques surface. After impossing these assumptions, we will 
consider divisors $D_i$, with $i=0, \ldots, n-1$, defined by the fibering over 
$C$, in a trivial way, of the irreducible components $\Theta_i$ of the 
$\hat{A}_{n-1}$ singularity \cite{KV}. Ussing the Todd representation of the holomorphic 
Euler class \cite{Fulton},
\begin{equation}
\chi(D) = \frac {1}{24} \int_{D_i} c_1 (\Theta_i) c_2(C)
\label{eq:r50}
\end{equation}
we get, for $C$ an Enriques surface,
\begin{equation}
\chi(D_i) =1.
\label{eq:r51}
\end{equation}
Interpreting now the $t$ variable (\ref{eq:r47}) on the fiber of the normal 
bundle $N$ of $D$ in $X$ as the $t$ variable used in our previous 
description of Kodaira singularities of type $\hat{A}_{n-1}$, we can derive the 
transformation law, under the ${\bf Z}_n$ subgroup of $U(1)$, of the scalar 
fields $\phi_{D_i}$ associated to these divisors. Namely, from (\ref{eq:r39}) 
we get
\begin{equation}
{\bf Z}_n: \phi_{D_i} \rightarrow \phi_{D_{i+1}},
\label{eq:r52}
\end{equation}
with the ${\bf Z}_n$ transformation being defined by
\begin{equation}
t \rightarrow e^{2 \pi i/n}t.
\label{eq:r53}
\end{equation}
Using now (\ref{eq:r48}), we get
\begin{equation}
{\bf Z}_n : \phi_{D_i} \rightarrow \phi_{D_i} + \frac {2 \pi}{n}.
\label{eq:r54}
\end{equation}
Combining (\ref{eq:r52}) and (\ref{eq:r54}) we get, modulo $2 \pi$,
\begin{equation}
\phi_{D_j} = \frac {2 \pi j}{n} + c,
\label{eq:r55}
\end{equation}
with $j=0,\ldots,n-1$, and $c$ a constant independent of $j$.
  
Let us now consider the divisor ${\cal D}$ obtained by fibering over $C$ the 
singular fiber ${\cal C} = \sum_{j=0}^{n-1} \Theta_j$, defined in (\ref{eq:r38}). In 
this case we need to be careful in order to compute (\ref{eq:r50}). If we 
naively consider the topological sum of components $\Theta_j$ in (\ref{eq:r50}), 
we will get the wrong result $\chi({\cal D})=n$. This result would be correct 
topologically, but not for the holomorphic Euler characteristic we are 
interested in. In fact, what we should write in (\ref{eq:r50}) for 
$\int c_1(\sum_{j=0}^{n-1} \Theta_j)$ is $2(1-g(\sum_{j=0}^{n-1} \Theta_j))$, 
with $g$ the genus of the cycle (\ref{eq:r38}), as defined by 
${\cal C}^2=2g-2$, with ${\cal C}^2$ the self intrsection of the cycle (\ref{eq:r38}) 
which, as for any other Kodaira singularity, is zero. Thus, we get $g=1$, 
and \cite{Gomez} 
\begin{equation}
\chi({\cal D})=0.
\label{eq:r56}
\end{equation}
We can try to intepret the result (\ref{eq:r56}) in terms of the fermionic 
zero modes of each component $\Theta_i$, and the topology of the cycle. In fact, 
associated to each divisor $D_i$ we have, as a consequence of (\ref{eq:r51}), 
two fermionic zero modes. In the case of the $\hat{A}_{n-1}$ singularity, 
we can soak up all zero modes inside the graph, as shown in Figure $1$,

%%%%%%%%%%%%%%%%%%%%%%%%%%%%%%%%%%%%
%%%%%   Figure 3   %%%%%%%%%%%%%%%%%
%%%%%%%%%%%%%%%%%%%%%%%%%%%%%%%%%%%%

\begin{figure}[ht]
\def\epsfsize#1#2{.6#1}
\centerline{\epsfbox{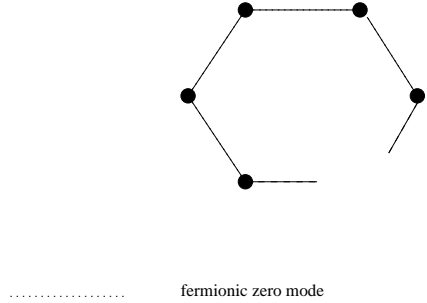}}
\caption{Soaking up of zero modes for $\hat{A}_{n-1}$.}
\end{figure}

%%%%%%%%%%%%%%%%%%%%%%%%%%%%%%%%%%%%
 
where from each node, representing one $\Theta_i$, we have two 
fermionic zero mode lines. The soaking up of fermionic zero modes represented 
in the figure is an heuristic interpretation of the result (\ref{eq:r56}).

\subsection{$\theta$-Parameter and Gaugino Condensates.}

We will, in this section, only consider singularities of
$\hat{A}_{n-1}$ and $\hat{D}_{n+4}$ type. In both cases, and for
each irreducible component $\Theta_i$, we get a divisor $D_i$,
with $\chi(D_i)=1$. Associated to this divisor, we can get a
superpotential term of the order \cite{Wsp}
\begin{equation}
\int d^2 \theta e^{-(V(D_i))+ i \phi_{D_i}}),
\label{eq:r60}
\end{equation}
where $V(D_i)$ means the volume of the divisor $D_i$. As
explained above, we are using vertical instanton divisors $D_i$,
defined by a trivial fibering of $\Theta_i$ over the singular
locus $C \subset B$, satisfying conditions (\ref{eq:r49}). In
order to get the four dimensional $N=1$ limit, we will take the
limit $\hbox {Vol }(E)=\frac {1}{R} \rightarrow 0$. Since the
singular fibers are, topologically, the union of irreducible
components (see (\ref{eq:r38}) and (\ref{eq:r42})), we can write
\begin{equation}
\hbox {Vol }(\Theta_i) = \frac {1}{R \hbox {Cox}},
\label{eq:r61}
\end{equation}
with $\hbox {Cox}$ the Coxeter number of the corresponding
singularity, which equals the total number of irreducible
components. Therefore, we will define $\hbox {Vol }(D_i)$ as
\begin{equation}
\hbox {Vol }(D_i) = \lim_{R \rightarrow \infty} \hbox {Vol }(C)
\frac {1}{R \hbox {Cox}}.
\label{eq:r62}
\end{equation}
If we first consider the $N=2$ supersymmetric three dimensional
theory obtained by compactifying M-theory on the Calabi-Yau
fourfold $X$, i. e., in the limit $R \rightarrow 0$, we know that
only the divisor $\Theta_0$, for the $\hat{A}_{n-1}$ case, is decompactified, passing from the
affine diagram describing an elliptic singularity to the non
affine diagram describing a rational, Artin like, singularity \cite{Artin}. In
that case, the volumes of the $\Theta_i$ components, for $i \neq
0$, are free parameters, corresponding to the Coulomb branch of
the $N=2$ three dimensional theory. In the three dimensional
theory, the factor $\hbox{Vol }(C)$ corresponds to the bare
coupling constant in three dimensions,
\begin{equation}
\hbox {Vol }(C)= \frac {1}{g_3^2},
\label{eq:r63}
\end{equation}
and $\hbox {Vol }(D_i)= \frac {1}{g_3^2} \chi_i$, for $i \neq 0$,
with $\chi_i$ the three dimensional Coulomb branch coordinates.
In the four dimensional case, we must use (\ref{eq:r62}), that
becomes
\begin{equation}
\hbox {Vol }(D_i) = \lim_{R \rightarrow \infty} \frac {1}{g_3^2}
\frac {1}{R \hbox {Cox}} = \frac {1}{g_4^2 \hbox {Cox}}.
\label{eq:r64}
\end{equation}
  
Let us now concentrate on the $\hat{A}_{n-1}$ case, where $\hbox
{Cox}=n$. Using  (\ref{eq:r60}) we get the following
superpotential for each divisor $D_j$,
\begin{equation}
\exp - \left( \frac {1}{g_4^2 {n}} + i \left( \frac {2
\pi j}{ {n}}+ { c } \right) \right).
\label{eq:r65}
\end{equation}
Let us now fix the constant $c$  in (\ref{eq:r65}). In order to do
that, we will use the transformation rules (\ref{eq:r48}). From
the four dimensional point of view, these are the transformation
rules with respect to the $U(1)_R$ symmetry. From (\ref{eq:r48})
we get, that under $t \rightarrow e^{i \alpha}t$,
\begin{equation}
\sum_{i=0} ^{\hbox {n-1}} \phi_{D_i} \rightarrow
\sum_{i=0}^{\hbox {n-1}} \phi_{D_i} + n \alpha.
\label{eq:r66}
\end{equation}
This is precisely the transformation rule under $U(1)_R$ of the
$N=1$ $\theta$-parameter,
\begin{equation}
\theta \rightarrow \theta +n \alpha.
\label{eq:r67}
\end{equation}
In fact, (\ref{eq:r66}) is a direct consequence of the $U(1)$
axial anomaly equation: if we define $\theta$ as
\begin{equation}
\frac {\theta}{32 \pi^2} F \tilde{F},
\label{eq:r69}
\end{equation}
the anomaly for $SU(n)$ is given by
\begin{equation}
\partial_{\mu} j^{\mu}_5 = \frac {n}{16 \pi^2} F
\tilde{F}.
\label{eq:r70}
\end{equation}
The factor $2$, differing (\ref{eq:r69}) from (\ref{eq:r70}),
reflects the fact that we are assigning $U(1)_R$ charge $\frac
{1}{2}$ to the fermionic zero modes. Identifying the $\theta$-parameter 
with the topological sum $\sum_{i=0}^{n-1} \phi_{D_i}$ we get 
that the constant $c$ in (\ref{eq:r65}) is simply
\begin{equation}
\hbox {c} = \frac {\theta}{n},
\label{eq:r71}
\end{equation}
so that we then finally obtain the superpotential
\begin{equation}
\exp - \left( \frac {1}{g_4^2 n} + i \left( \frac {2
\pi j}{n} + \frac {\theta}{n} \right) \right)
\simeq \Lambda^3 e^{2 \pi ij/n} e^{i \theta/n}, 
\label{eq:r72}
\end{equation}
with $j=0, \ldots , n-1$, which is the correct value
for the gaugino condensate.
  
Let us now try to extend the previous argument to the
$\hat{D}_{n+4}$ type of singularities. Defining again the four
dimensional $\theta$-parameter as the topological sum of
$\phi_{D_i}$ for the whole set of irreducible components we get,
for the cycle (\ref{eq:r42}), the transformation rule
\begin{equation}
\theta \rightarrow \theta + \hbox {Cox }. \alpha
\label{eq:h1}
\end{equation}
where now the Coxeter for $\hat{D}_{n+4}$ is $2n+6$. Interpreting
$\hat{D}_{n+4}$ as $O(N)$ gauge groups, with $N=2n+8$, we get
$\hbox {Cox}(\hat{D}_{n+4})=N-2$. Since $\theta$ is defined
modulo $2 \pi$ we get that for $\hat{D}_{n+4}$ singularities the
value of $\phi_{D_i}$, for any irreducible component, is
\begin{equation}
\frac {2 \pi k}{N-2} + \frac {\theta}{N-2},
\label{eq:h2}
\end{equation}
with $k=1, \ldots, N-2$. However, now we do not know how to
associate a value of $k$ to each irreducible component $\Theta_i$
of the $\hat{D}_{n+4}$ diagram. Using (\ref{eq:h2}), we get a set
of $N-2$ different values for the gaugino condensate for $O(N)$
groups:
\begin{equation}
\exp \left( - \frac {1}{g_4^2(N-2)} + i \left( \frac {2 \pi
k}{N-2} + \frac {\theta}{N-2} \right) \right),
\label{eq:h3}
\end{equation}
with $k=1,\ldots,N-2$. However, we still do not know how to
associate to each $\Theta_i$ a particular value of $k$. A
possibility will be associating consecutive values of $k$ to
components with non vanishing intersection; however, the topology
of diagrams of type $D$ prevents us from doing that globally.
Notice that the problem we have is the same sort of puzzle we
find for $O(N)$ gauge groups, concerning the number of values for
$<\lambda \lambda>$, and the value of the Witten index, which in
diagramatic terms is simply the number of nodes of the diagram.
In order to unravel this puzzle, let us consider more closely the
way fermionic zero modes are soaked up on a $\hat{D}_{n+4}$
diagram. We will use the cycle (\ref{eq:r42}); for the components
$\Theta^1$ to $\Theta^4$, associated to the ${\bf Z}_2$ orbifold
points, we get divisors with $\chi=1$. Now, for the components $2
\Theta_0, \ldots, 2 \Theta_n$ we get, from the Todd
representation of the holomorphic Euler characteristic,
\begin{equation}
\chi=4.
\label{eq:h4}
\end{equation}
The reason for this is that the cycle $2 \Theta$, with
$\Theta^2=-2$, has self intersection $-8$. Of course,
(\ref{eq:h4}) refers to the holomorphic Euler characteristic of
the divisor obtained when fibering over $C$ any of the cycles $2
\Theta_i$, with $i=0, \ldots, n$. Equation (\ref{eq:h4}) implies
$8$ fermionic zero modes, with the topology of the soaking up of zero modes of
the $\hat{D}_{n+4}$ diagram, as represented in Figure
$2$.

%%%%%%%%%%%%%%%%%%%%%%%%%%%%%%%%%%%%
%%%%%   Figure 4   %%%%%%%%%%%%%%%%%
%%%%%%%%%%%%%%%%%%%%%%%%%%%%%%%%%%%%

\begin{figure}[ht]
\def\epsfsize#1#2{.6#1}
\centerline{\epsfbox{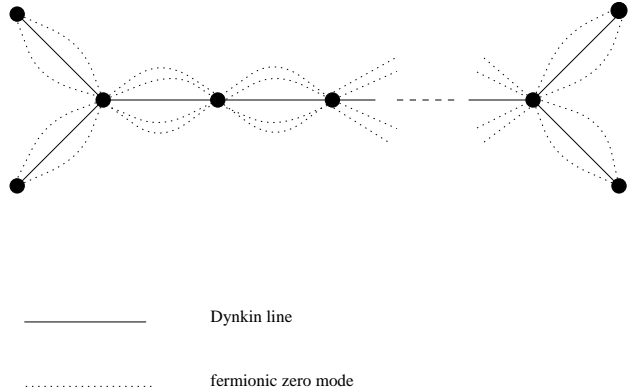}}
\caption{Soaking up of zero modes for $\hat{D}_{n+4}$.}
\end{figure}

%%%%%%%%%%%%%%%%%%%%%%%%%%%%%%%%%%%%

 Notice that the contribution to $\chi$ of $2 \Theta$ is different form that 
of $(\Theta_1 + \Theta_2)$, with $(\Theta_1 . \Theta_2)=0$;
namely, for the first case $\chi=4$, and $\chi=2$ for the second.
For the $\hat{D}_{n+4}$ diagram, we can define: $i)$ The Witten
index $\hbox {tr }(-1)^F$, as the number of nodes, i. e., $5+n$;
$ii)$ The Coxeter number, which is the number of irreducible
components, i. e., $2n+6$ and $iii)$ The number of intersections
as represented by the dashed lines in Figure $4$, i. e.,
$8+4n$. From the point of view of the Cartan algebra, used to
define the vacuum configurations in \cite{Wind}, we can only feel
the number of nodes. The $\theta$-parameter is able to feel the
Coxeter number; however, we now find a new structure related to
the intersections of the graph. In the Witten index case, the
nodes corresponding to cycles $2 \Theta_i$, with $i=0, \ldots ,n$
contribute with one, in the number of $<\lambda \lambda>$ values
with two, and in the number of intersections with four. This
value four calls for an orientifold interpretation of these
nodes. The topological definition of the $\theta$-parameter
implicitely implies the split of this orientifold into two
cycles, a phenomena recalling the F-theory description \cite{Sen}
of the Seiberg-Witten splitting \cite{SW}. Assuming this
splitting of the orientifold, the only possible topology for the
soaking up of zero modes is the one represented in Figure $3$,
where the ``splitted orientifold'' inside the box is associated
to four zero modes, corresponding to $\chi=2$ for a cycle
$\Theta_1 + \Theta_2$, with $\Theta_1 . \Theta_2=0$. 

%%%%%%%%%%%%%%%%%%%%%%%%%%%%%%%%%%%%
%%%%%   Figure 5   %%%%%%%%%%%%%%%%%
%%%%%%%%%%%%%%%%%%%%%%%%%%%%%%%%%%%%

\begin{figure}[ht]
\def\epsfsize#1#2{.6#1}
\centerline{\epsfbox{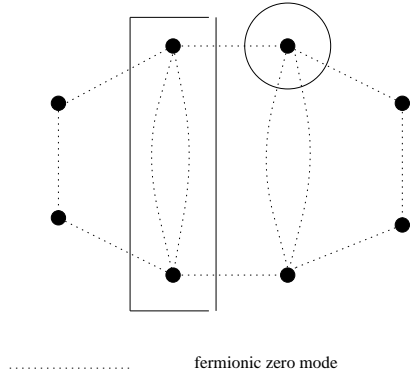}}
\caption{Orientifold splitting.}
\end{figure}

%%%%%%%%%%%%%%%%%%%%%%%%%%%%%%%%%%%%

On the
other hand, each node surrounded by a circle in Figure $5$ represents
itself the disconnected sum of two non singular rational curves;
thus, we represent each ``orientifold'' mode by four rational
curves, with the intersections depicted inside the box of Figure
$5$. When we forget about internal lines in Figure $5$, we
get the cyclic ${\bf Z}_{2n+6} \equiv {\bf Z}_{N-2}$ structure of
equation (\ref{eq:h3}). It is clear that much more is necessary
in order to reach a complete description of the $O(N)$ vacuum
structure.

\subsection{Domain Walls and Intersections.}

The discussion in the previous section already raises the problem
known as $\theta$-puzzle. In fact, and discussing again only the
$SU(n)$ case, the transformation law (\ref{eq:r67}) together with the
very definition fo the $\theta$-angle as the topological sum
$\sum_{i=0}^{n-1} \phi_{D_i}$ would imply that $\theta$ is the
scalar field $\phi_{\cal D}$ of the $6$-cycle associated to the $\hat{A}_{n-1}$ cycle, ${\cal
C}= \sum_{i=0}^{n-1} \Theta_i$. On the basis of (\ref{eq:r48}),
this will be equivalent to saying that $\chi({\cal D})=n$,
instead of zero. This is, in mathematical terms, the
$\theta$-puzzle. The mathematical solution comes from the fact
that $\chi({\cal D})=0$. In this section we will relate this
result, on the value of the holomorphic Euler chareacteristic, to
the appearance of domain walls \cite{ds,kss,Smilga}. To start
with, let us consider a cycle ${\cal C} = \Theta_1 + \Theta_2$,
with $(\Theta_1 . \Theta_2)=1$. The self intersection can be
expressed as
\begin{equation}
({\cal C}. {\cal C})=-2-2+2,
\label{eq:r76}
\end{equation}
where the $-2$ contributions come from $\Theta_1^2$ and
$\Theta_2^2$, and the $+2$ comes from the intersection between
$\Theta_1$ and $\Theta_2$. As usual, we can consider ${\cal C}$ 
trivially fibered on an Enriques surface. The holomorphic Euler
chracteristic of the corresponding six cycle can be written as 
\begin{equation}
\chi = \frac {1}{2} (- {\cal C}^2).
\label{eq:h5}
\end{equation}
Using now the decomposition (\ref{eq:r76}) we get two
contributions of one, coming from the components $\Theta_1$ and
$\Theta_2$, considered independently, and a contribution of $-1$
from the intersection term $+2$ in (\ref{eq:r76}). In this sense,
the intersection term can be associated to two fermionic zero
modes, and net change of chiral charge oposite to that of the
$\Theta_i$ components. When we do this for the cycle ${\cal C}$
of $\hat{A}_{n-1}$ singularities, we get that each intersection
is soaking up two zero modes, leading to the result that
$\chi({\cal C})=0$. A graphical way to represent equation
(\ref{eq:r76}) is presented in Figure $4$.

%%%%%%%%%%%%%%%%%%%%%%%%%%%%%%%%%%%%
%%%%%   Figure 6   %%%%%%%%%%%%%%%%%
%%%%%%%%%%%%%%%%%%%%%%%%%%%%%%%%%%%%

\begin{figure}[ht]
\def\epsfsize#1#2{.6#1}
\centerline{\epsfbox{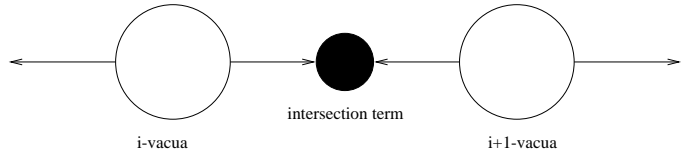}}
\caption{Intersection term.}
\end{figure}

%%%%%%%%%%%%%%%%%%%%%%%%%%%%%%%%%%%%

Now, we will wonder about the physical interpretation of the
intersection terms leading to $\chi({\cal C})=0$ for all Kodaira
singularities. The simplest, and most natural answer, is
certainly domain walls extending between different vacua, or
values of $<\lambda \lambda>$.

From the point of view of zero mode counting, the ``intersection
term'' behaves effectively as an anti-instanton with two
fermionic zero modes. One of these fermionic zero modes, let us
say $\psi_{j,j+1}$, is associated to the intersection of
$\Theta_{j}$ with $\Theta_{j+1}$ and the other $\psi_{j+1,j}$
with the intersection of $\Theta_{j}$ and $\Theta_{j+1}$. Thus, extending naively the 
computation done for irreducible components, the contribution of
the black box in Figure $4$ should be of the order
\begin{equation}
\Lambda^3 e^{2 \pi ij/n} (1-e^{2 \pi i/n}).
\label{eq:h10}
\end{equation}
In result (\ref{eq:h10}), interpreted as the contribution of the
intersection term, the most surprising fact is the appearance of
$\Lambda^3$, since now we are geometrically considering simply a
point; the factor $\Lambda^3$ in the computation of the gaugino
condensate comes from the volume of the divisor. In the same way
as we interpret M-theory instantons as fivebranes wrapped on the
six-cycles used to define the instanton, we can think of the
intersection terms as fivebranes wrapped on the cycle $C \times
\{(\Theta_i . \Theta_{i+1})\}$, i. e., the product of the singular
locus $C$ and the intersection point. The fivebrane wrapped on
this cycle defines, in four dimensions, a domain wall, let us say
interwining between the vacua $i$, at $x_3=+ \infty$, and the
vacua $i+1$, at $x_3=- \infty$, where the coordinate $x_3$ is
identified with the unwrapped direction. It is in this sense that
we should use (\ref{eq:h10}) to define the energy density, or
tension, of the domain wall. In the four dimensional limit,
$\hbox {Vol }(E)$ goes zero as $\frac {1}{R}$; moreover, the
local engineering approach works in the limit where the volume of
the singular locus $C$ is very large, so that we can very likely
assume that intersection terms behave like (\ref{eq:h10}), with
$\Lambda^3$, but only in the four dimensional limit. Cyclicity of
the $\hat{A}_{n-1}$ diagram allows us to pass from the $j$ to the
$j+1$ vacua in two different ways: $n-1$ steps, or a single one.
The sum of both contributions should define the physical domain
wall; thus the energy density will behave as
\begin{equation}
n \Lambda^3 |e^{2 \pi ij/n} (1-e^{2 \pi i/n})|.
\label{eq:h11}
\end{equation}
  
The extension of the previous argument to the case of $O(N)$
groups is certainly more involved, due to the topology of the
$\hat{D}$ diagram, and the presence of orientifolds. It would
certainly be interesting studying the interplay between
orientifolds and domain walls in this case.
  
Finally, we will say some words on the QCD string. In reference
\cite{Wbqcd}, the geometry of QCD strings is intimately related
to the topological fact that
\begin{equation}
H_1(Y/Z;{\bf Z}) = {\bf Z}_n,
\label{eq:r77}
\end{equation}
where $\Sigma$ is a rational curve associated to the
configuration of fourbranes, and $Y=S^1 \times {\bf R}^5$ is the
ambient space where $\Sigma$ is embedded (see \cite{Wbqcd} for
details). The QCD string is then associated to a partially
wrapped membrane on a non trivial element of $H_1(Y/\Sigma;{\bf
Z})$. Recall that $H_1(Y/\Sigma;{\bf Z})$ is defined by
one-cycles in $Y$, with boundary on $\Sigma$. The previous
discussion was done for $SU(N)$ gauge groups. Using our model of
$\hat{A}_{n-1}$ singularities, described in section 2, the
analog in our framework of (\ref{eq:r77}) is equation
(\ref{eq:r33}). Then we can, in the same spirit as in reference
\cite{Wbqcd}, associate the QCD string to paths going from $p_k$
to $p_{k+1}$, where $p_k$ are the intersection points,
\begin{equation}
\Theta_k . \Theta_{k-1} = p_k.
\label{eq:r78}
\end{equation}
Geometrically, it is clear that the tension of this QCD string is
the square root of the domain wall tension. By construction, the
QCD string we are suggesting here ends on domain walls, i. e., on
intersection points.

To end up, let us include some comments on the existence of
extra vacua, as suggested in \cite{KS}. It is known
that the strong coupling computation of $<\lambda \lambda>$ does
not coincide with the weak coupling computation; more precisely\cite{NSVZ2},
\begin{equation}
<\lambda \lambda>_{\hbox {sc}} < \: <\lambda \lambda>_{\hbox
{wc}}.
\label{eq:r75}
\end{equation}
In the framework of M-theory instanton computations, the
numerical factors will depend in particular on the moduli of
complex structures
of the Calabi-Yau fourfold. In the strong coupling regime we must consider
structures preserving the elliptic fibration structure and the
Picard lattice. In the weak coupling regime, where the
compuatation is performed in the Higgs phase, the amount of
allowed complex structures contributing to the value of $<\lambda
\lambda>$ is presumably larger. Obviously, the previous argument is only
suggesting a possible way out of the puzzle (\ref{eq:r75}).

Equally, at a very speculative level, the extra vacua, with no
chiral symmetry breaking, could be associated to the cycle ${\cal
D}$ defining the singular fiber, a cycle that we know leads to
$\chi=0$, and therefore does not produce any gaugino condensate.
Notice that any other cycle with $\chi \neq 0$ will lead, if
clustering is used, to some non vanishing gaugino condensates, so
that ${\cal D}$ with $\chi=0$ looks like a possible
candidate to the extra vacua suggested in \cite{KS}. If this
argument is correct this extra vacua will appears for any Kodaira
singularity i. e. in any ADE $N=1$ four dimensional gauge theory.

It is important to stress that the $\theta$-puzzle is not
exclusive of $N=1$ gluodynamics. In the $N=0$ case the
Witten-Veneziano formula \cite{Wi,Ve} for the $\eta'$ mass 
also indicates a dependence of the vacuum 
energy on $\theta$ in terms of $\frac {\theta}{N}$, 
which means a set of entangled "vacuum" states. 
In our approach to $N=1$ the origin of this entanglement is due to the fact 
that $\chi=0$ for the singular cycle. In fact, $\chi({\cal D})=0$ means that the set 
of divisors $D_i$, plus the intersections, i. e., the domain walls, are 
invariant under $U(1)$, as implied by equation (\ref{eq:r48}). If 
we naively think of something similar in $N=0$
and we look for the origin of vacuum entanglement in intersections we maybe should think in
translating the topology of intersections into topological properties 
of abelian proyection gauges \cite{tHap}.

\newpage

%%%%%%%%%%%%%%%%%%%%%%%%%%%%%%%%%%%%%%%%%%%%%%%%%%%%%%%%%
%%%%%%%%%%%%%%%%%%%%%%%%%%%%%%%%%%%%%%%%%%%%%%%%%%%%%%%%%

\appendix

\section{M(atrix) Theory.}

\subsection{The Holographic Principle.}

The holomorphic principle was originally suggested by `t Hooft in
\cite{holo}. Let us work first in four dimensional spacetime. Let
$S^{(2)}$ be a surface with the topology of two sphere in ${\bf
R}^3$, and let us wonder about how many orthogonal quantum states
can $S^{(2)}$ contain: we will find an upper bound for this number
of states. In order to do this, we will use the
Bekenstein-Hawking relation between the entropy and the horizon
area of the black hole. Let us then call ${\cal N}$ the number of states
inside $S^{(2)}$; the entropy can be defined as
\begin{equation}
\exp S = {\cal N}.
\label{eq:t1}
\end{equation}
If $S^2$ is the horizon of a black hole, we have
\cite{Bekenstein}
\begin{equation}
S \sim \frac {1}{4} \frac {A}{l_p^2},
\label{eq:t2}
\end{equation}
with $A$ the area of the horizon, in Planck length units. Now,
let us translate all the physical information contained inside
the $S^{(2)}$ surface, in terms of states of $q$-bits, defined as
quantum systems of two states. The number $n$ of $q$-bits we need
is given by
\begin{equation}
N=2^n.
\label{eq:t3}
\end{equation}
Using (\ref{eq:t1}) to (\ref{eq:t3}) we then get
\begin{equation}
n= \frac {1}{4 \ln 2} \frac {A}{l_p^2},
\label{eq:t4}
\end{equation}
which is essentially the number of cells, of area $l_p^2$,
covering the surface $S^{(2)}$. What we learn from this is that all
three dimensional physics inside $S^{(2)}$ can be described using
states of $q$-bits, living on the two dimensional surface
$S^{(2)}$. We will call these $q$-bits the holographic degrees of
freedom. What we need now is the two dimensional dynamics
governing these two dimensional degrees of freedom, able to
reproduce, in holographic projection, the three dimensional
physics taking place $S^{(2)}$. We can even consider, instead of
$S^{(2)}$, an hyperplane of dimension two, dividing space into two
regions. The extension, to this extreme situation, of the
holomorphic principle, will tell us that the $3+1$ dynamics can
be described in terms of some $2+1$ dynamics for the holomorphic
degrees of freedom living on the hypersurface.
  
This picture of the holographic principle allows to introduce
M(atrix) theory \cite{BFSS} as the holographic projection of M-theory. In
this case, we will pass from eleven dimensional to ten
dimensional physics. What we will need, in order to formulate
M(atrix) theory, will be
\begin{itemize}
	\item[{i)}] An explicit definition of the holographic
projection.
	\item[{ii)}] Identifying the holomorphic degrees of
freedom.
	\item[{iii)}] Providing a ten dimensional dynamics for
these degrees of freedom.
\end{itemize}
  
The conjectured answer in \cite{BFSS} to $i)$, $ii)$ and $iii)$
are
\begin{itemize}
	\item[{i)}] The infinite momentum frame.
	\item[{ii)}] D-$0$branes as degrees of freedom.
	\item[{iii)}] The dynamics is implemented through the
worldvolume of the lagrangian of D-$0$branes.
\end{itemize}
  
These set of conjectures define M(atrix) theory at present.
The idea of the infinite momentum frame is boosting in the
eleventh direction in eleven dimensional spacetime, in such a way
that $p_{11}$, the eleventh component of the momentum, becomes
larger than any scale in the problem. In this frame, we
associate, with an eleven dimensional massles system of momentum
$\vec{p}=(p_{11},p_{\perp})$, a ten dimensional galilean system
with mass $p_{11}$, and energy
\begin{equation}
E=\frac {p_{\perp}^2}{2p_{11}}.
\label{eq:m3}
\end{equation}
If we introduce an infrared cut off, by compactifying the
eleventh dimension on a circle $S^1$, of radius $R$, the $p_{11}$
is measured in units of $\frac {1}{R}$. Then,
\begin{equation}
p_{11} = \frac {n}{R}.
\label{eq:m2}
\end{equation}
  
We will interpret $n$ as the number of partons which are
necessary to describe a system with value of $p_{11}$ given by
(\ref{eq:m2}). These partons are the ten dimensional degrees of
freedom we are going to consider as holographic variables.
  
Using (\ref{eq:t4}), we can define the ten dimensional size of an
eleven dimensional massless particle, in eleven dimensions, with
some given $p_{11}$. In fact, $n=p_{11}R$ is the number of
holographic degrees of freedom which, by (\ref{eq:t4}), means
that
\begin{equation}
\frac {r^9}{l_p^9} \sim p_{11}R,
\label{eq:t7}
\end{equation}
and the radius $r$, characterizing the size, will be
$(p_{11}R)^{1/9}l_p$. Now, we can look for objects in ten
dimensions with mass equal to the mass of a parton, i. e., $\frac
{1}{R}$. Natural candidates are D-$0$branes. From this, it seems
natural to conjecture that the worldvolume dynamics of
D-$0$branes will be a good candidate for the holographic
description of M-theory.

M(atrix) theory would hence simply be defined as the worldvolume theory of D-$0$branes. 
As for any other type of D-branes, this worldvolume theory is defined 
as the dimensional reduction down to $0+1$ dimensions of ten 
dimensional Yang-Mills with $N=1$ supersymmetry. If we consider a set 
of $N$ D-$0$branes, we have to introduce matrices $X^{i}$, with 
$i=1,\ldots,9$. As usual, the diagonal part of this matrices can be interpreted 
in terms of the classical positions of the $N$ D-$0$branes, and the off 
diagonal terms as representing the exchange of open strings. Thus, the 
worldvolume lagrangian we get for $N$ D-$0$branes is $U(N)$ Yang-Mills 
quantum mechanics. Using units in which $\alpha'=1$, the bosonic part of 
this lagrangian is simply
\begin{equation}
{\cal L} = \frac {1}{2 g} [ \hbox { tr} \dot{X}^{i} \dot{X}^{i} - 
\frac {1}{2} \hbox { tr} [X^{i},X^{j}]^2 ],
\label{eq:m1}
\end{equation}
in units where $l_p=1$, and with $g$ the string coupling
constant. In this definition we have simply consider D-$0$branes in a ten dimensional 
type IIA string theory. However, we know that D-$0$branes are in fact Kaluza-Klein 
modes of an eleven dimensional theory named M-theory. In the eleven dimensional 
spacetime, the D-$0$branes have a momentum $p_{11}$ given by
\begin{equation}
p_{11} = \frac {1}{R},
\label{eq:m}
\end{equation}
with $R$ the radius of the eleventh dimension. The way to relate (\ref{eq:m1}) 
to the physics of the partons defined in the infinite momentum
frame is observing that the kinetic term in (\ref{eq:m1})
coincides with the equation (\ref{eq:m3}) for $p_{11}$ given in
(\ref{eq:m}). In fact, using  the relation 
\begin{equation}
R= g l_s,
\label{eq:m4}
\end{equation}
and choosing units where $l_s=1$, we notice that (\ref{eq:m1}) is precisely 
the galilean lagrangian for particles of mass $\frac {1}{g}$. Thus, we will 
interpret the worldvolume dynamics of D-$0$branes, (\ref{eq:m1}), as the 
infinite momentum frame of the M-theory D-$0$branes.
 
Our main task now will consist in deriving the brane spectrum 
directly from the M(atrix) lagrangian (\ref{eq:m1}), interpreting the different branes as collective excitations of D-$0$branes. In 
order to achive this, it will be necessary to work in the $N \rightarrow 
\infty$ of (\ref{eq:m1}). Using the relation 
\begin{equation}
l_s=g^{-1/3}l_p,
\end{equation}
between the string length, and the Plank scale, we can pass to
Plank units by defining $Y=\frac {X}{g^{1/3}}$. In $Y$ variables, 
and with $l_p=1$, we get
\begin{equation}
{\cal L} = \hbox { tr} \left[ \frac {1}{2R} D_t Y^{i} D_t Y^{i} - \frac 
{1}{4} R [Y^{i},Y^{j}]^2 \right],
\label{eq:m5}
\end{equation}
where $D_t = \partial_t +i A$, with $A$ equal the $A_0$ piece of the ten 
dimensional Yang-Mills theory. In order to get (\ref{eq:m1}), going to 
the temporal gauge $A_0=0$ is all what is needed.
  
Now, some of the ingredients introduced in chapter I will be needed; namely, the matrices 
$P$ and $Q$ defined in (\ref{eq:I79}). In terms of this basis of matrices, 
any matrix $Z$ can be written as
\begin{equation}
Z= \sum_{n,m=1}^N z_{n,m} P^n Q^m.
\label{eq:m6}
\end{equation}
Taking into account that 
\begin{equation}
PQ = QP e^{2 \pi i/N},
\label{eq:m7}
\end{equation}
we can define 
\begin{equation}
P= e^{i \hat{p}}, \: \: \: \: Q=e^{i \hat{q}},
\label{eq:m8}
\end{equation}
with 
\begin{equation}
[\hat{p},\hat{q}] = \frac {2 \pi i}{N}.
\label{eq:m9}
\end{equation}
Replacing (\ref{eq:m8}) in (\ref{eq:m6}) we get
\begin{equation}
Z = \sum_{n,m} z_{n,m} e^{in \hat{p}} e^{im \hat{q}},
\label{eq:m10}
\end{equation}
which looks like the Fourier transform of a function $Z(p,q)$. The only 
difference is that this function is defined on a quantum phase space defined 
by $\hat{p}$ and $\hat{q}$ variables satisfying (\ref{eq:m9}). In the 
$N \rightarrow \infty$ limit, we can interpret this quantum space as 
classical, since in this limit $[\hat{p},\hat{q}]=0$. Thus, in the 
$N \rightarrow \infty$ limit, the matrices can be replaced by functions 
$Z(p,q)$, as defined by (\ref{eq:m10}). The matrix operations become, 
in this limit,
\begin{eqnarray}
\hbox {tr} Z & \rightarrow & N \int Z(p,q) dp dq, \nonumber \\
{[}X,Y{]} & \rightarrow & \frac {1}{N} [ \partial_q X \partial_p Y- \partial_q Y 
\partial_p X ],
\label{eq:m11}
\end{eqnarray} 
i. e., the conmutator becomes the Poisson bracket. Now, we can use 
(\ref{eq:m11}) in (\ref{eq:m5}); what we then get is
\begin{equation}
{\cal L} = \frac {p_{11}}{2} \left[ \int dp dq \dot{Y}^{i}(p,q) \dot{Y}^{i}(p,q) 
- \frac {1}{P_{11}} \int dp dq [ \partial_q Y^{i} \partial_p Y^{j} - 
\partial_{q} Y^{j} \partial_p Y^{i}] \right],
\label{eq:m12}
\end{equation}
where $p_{11}= \frac {N}{R}$. The interest of (\ref{eq:m12}) is that this result 
coincides with the eleven diemensional lagrangian for the eleven dimensional 
supermembrane in the light cone frame. Notice that $i$ in (\ref{eq:m12}) goes from 
$1$ to $9$, which can be interpreted as the transversal directions to the 
supermembrane worldvolume. The previous result is alraedy a good indication 
of the consistency of M(atrix) theory as a microscopic description of M-theory. 
Next, we will try to define toroidal compactifications of (\ref{eq:m1}).

\subsection{Toroidal Compactifications.}

The definition of toroidal compactifications \cite{toroidal} of M(atrix) theory is quite 
simple. We will consider the worldvolume of lagrangian of D-$0$branes, starting 
with ten dimensional $N=1$ supersymmetric Yang-Mills in ${\bf R}^9 \times S^1$. 
In order to clear up the procedure, we will keep all indices for a while, 
so that we will write 
\begin{equation}
X^{i}_{k,l} 
\label{eq:m13}
\end{equation}
for the matrix $X^{i}$. The indices $k$ and $l$ will hence label different 
D-$0$branes. Now, if we force D-$0$branes to live in ${\bf R}^9 \times S^1$, 
and interpret $S^1$ as
\begin{equation}
{\bf R}/\Gamma,
\label{eq:m14}
\end{equation}
with $\Gamma$ a one dimensional lattice defined by a vector $\vec{e}= 
2 \pi R$, we can think of copies of each D-$0$brane, parametrized by 
integers $n$, depending on the cell of ${\bf R}/\Gamma$ where they are. Then, 
(\ref{eq:m13}) should be changed to
\begin{equation}
X^{i}_{k,m;l,m}.
\label{eq:m15}
\end{equation}
We can now forget about the indices $k$ and $l$, to write $X^{i}_{n m}$, where 
$n$ and $m$ are integers. The lagrangian (\ref{eq:m1}) then becomes
\begin{equation}
{\cal L} = \frac {1}{2g} [ \hbox {tr} \dot{X}^{i}_{mn} \dot{X}^{i}_{mn} + 
\frac {1}{2} \hbox {tr}(X^{i}_{mq} X^j_{qn} - X^{j}_{mq} X^{i}_{qn} )
(X^{i}_{nr} X^j_{rm} - X^{j}_{nr} X^{i}_{rm} )].
\label{eq:m16}
\end{equation}
Now, we should imposse symmetry with with respect to the action
of $\Gamma$, which implies
\begin{eqnarray}
X^{i}_{mn} & = & X^{i}_{m-1 \; n-1} \; \; i>1, \nonumber \\
X^{1}_{mn} & = & X^{1}_{m-1 \; n-1} \; \; m \neq n, \nonumber \\
X^{1}_{mn} & = & 2 \pi R {\bf I} + X^{i}_{m-1 \; n-1}.
\label{eq:m17}
\end{eqnarray}
The meaning of (\ref{eq:m17}) is that the coordinate $X^1$ is
periodic, so that the difference in $X^1$ for $n$ D-$0$branes,
and $n+1$ D-$0$branes, is simply the length of the compactified
direction. Using (\ref{eq:m17}), matrices can be simply labelled
by one index, $X^{i}_{0,n} \equiv X^{i}_n$, and the lagrangian
(\ref{eq:m16}) becomes
\begin{equation}
{\cal L} = \frac {1}{2g} [ \sum_{i=1}^{9} \hbox {tr}
\dot{X}^{i}_n \dot{X}^{i}_{-n}- \sum_{j=2}^9 \hbox {tr} S^{j}_n
(S^j_n)^+- \frac {1}{2} \sum_{j,k=2} ^9 \hbox {tr} T^{jk}_n
(T^{jk}_n)^+],
\label{eq:m18}
\end{equation}
where 
\begin{eqnarray}
S^j_n & = & \sum_q ([X^1_q,X^j_{n-q}])- 2 \pi R n X^j_n,
\nonumber \\
T^{jk}_n & = & \sum_q [X^j_q,X^k_{n-q}].
\label{eq:m19}
\end{eqnarray}
  
Once we get lagrangian (\ref{eq:m14}), we can compare it with the
worldvolume lagrangian for D-$1$branes. In fact, for D-$1$branes
the worldvolume lagrangian is $1+1$ dimensional super Yang-Mills
theory, with gauge fields $A^1$ and $A^0$, and matter fields
$Y^j$ (with $j=2, \ldots, 9$) in the adjoint representation. We
can then work in the temporal gauge, fixing $A^0=0$. On the other
hand, performing T-duality on $S^1$ takes form D-$0$branes to
D-$1$branes. Hence, on the dual $S^1$ the worldvolume lagrangian
for D-$1$branes should coincide with that of D-$0$branes in ${\bf
R}^9 \times S^1$, i. e, with lagrangian (\ref{eq:m18}). The
D-$1$brane worldvolume lagrangian in the dual $S^1$, with radius
$R'= \frac {1}{2 \pi R}$,
\begin{equation}
{\cal L} = \int dx \frac {dt}{2 \pi R'} \frac {1}{2g} [ \hbox{ tr
} \dot{Y}^{i} \dot{Y}^{i} + \hbox { tr } \dot{A}^1 \dot{A}^1 +
\frac {1}{2} \hbox { tr } [Y^{i},Y^{j}]^2- \hbox { tr }
[\partial_1 Y^{i} - i[A^1,Y^{i}]]^2,
\label{eq:m20}
\end{equation}
can be compared with (\ref{eq:m19}) if we just interpret
$X^{i}_n$ as the Fourier modes of $Y^{i}(x)$, and $X^{1}_n$ as
the Fourier modes of $A^1(x)$:
\begin{eqnarray}
A^1(x) & = & \sum_n e^{inx/R'} X^1_n, \nonumber \\
Y^{i}(x) & = & \sum_n e^{inx/R'} X^{i}_n.
\label{eq:m21}
\end{eqnarray}
Hence, we can readily induce the following result: M(atrix)
theory compactified on $T^d$ is equivalent to $d+1$
supersymmetric Yang-Mills on the dual $\hat{T}^d \times {\bf R}$,
with ${\bf R}$ standing for the time direction, and the
supersymmetric Yang-Mills theory defined through dimensional
reduction from $N=1$ ten dimensional Yang-Mills theory. This is a
surprising result, connecting M(atrix) compactifications with
Yang-Mills theories, a relation with far reaching consequences,
some of which we will consider in what follows.

\subsection{M(atrix) Theory and Quantum Directions.}

We can then represent M(atrix) theory, compactified on $T^d$, as
supersymmetric Yang-Mills theory defined on $\hat{T}^d \times {\bf
R}$, i. e., as the worldvolume lagrangian of $d$ D-branes wrapped
on the dual torus, $\hat{T}^d$. Let us then work out some simple
cases. We will first compactify M(atrix) on $T^4$, which will be
an interesting case concerning the U-duality symmetry.
  
Let $L_i$ be the lengths of $T^4$. The dual torus $\hat{T}$
will then be defined with sides of length
\begin{equation}
\Sigma_i = \frac {l_s^2}{L_i},
\label{eq:m22}
\end{equation}
with $l_s$ the string length. In terms of the eleven dimensional 
Planck scale, $l_P$, we have
\begin{equation}
\frac {l_P^3}{R} = l_s^2,
\label{eq:m23}
\end{equation}
and therefore
\begin{equation}
\Sigma_i = \frac {l_p^3}{L_i R}.
\label{eq:m24}
\end{equation}
  
Let us now consider the infinite momentum frame energy of a state
with one unit of $p_{11}$, and one unit of momentum in some
internal direction, $L_i$,
\begin{equation}
E = \frac {p_{\perp}}{2 P_{11}} = \frac {R}{2 L_i^2}.
\label{eq:A1}
\end{equation}
This state corresponds, in supersymmetric Yang-Mills, to a gauge
configuration with a non trivial Wilson line $A(C_i)$ (recall
that in the toroidal compactification the compactified components
$X^{i}$ behave as Yang-Mills fields. This non trivial Wilson line
means a flux through $C_i$. This energy is given by
\begin{equation}
\frac {g_{SYM}^2 \Sigma_i^2}{\Sigma_1 \Sigma_2 \Sigma_3
\Sigma_4}.
\label{eq:A2}
\end{equation}
Identiying (\ref{eq:A1}) and (\ref{eq:A2}) we get
\begin{equation}
g_{SYM}^2 = \frac {R}{2 L_i^2} \frac {\Sigma_ 1 \Sigma_2 \Sigma_3
\Sigma_4}{\Sigma_i^2}.
\label{eq:A3}
\end{equation}
Using (\ref{eq:m24}) we get
\begin{equation}
g_{SYM}^2 = \frac {R^3 \Sigma_1 \Sigma_2 \Sigma_3 \Sigma_4}{2
l_p^6} = \frac {l_p^6}{2 L_1 L_2 L_3 L_4 R},
\label{eq:A4}
\end{equation}
which means that $g^2$, as expected in $4+1$ dimensions, has
units of length.

From the definition of M(atrix)
compactifications, we expect that M(atrix) on $T^4$ will
reproduce type IIA string theory on $T^4$ that, has been derived
in chapter III, is invariant under the U-duality group,
$Sl(5,{\bf Z})$. Thus, our task is to unravel this U-duality
invariance, considering supersymmetric Yang-Mills on $\hat{T}^4
\times {\bf R}$. From (\ref{eq:A4}), we observe a clear
$Sl(4,{\bf Z})$ invariance of the gauge theory. These
transformations exchange all $\Sigma_i$, leaving their product
invariant. In order to extend this symmetry to $Sl(5,{\bf Z})$,
an extra dimension $\Sigma_5$ needs to be defined. A way to do
this is using as such direction the coupling constant itself in
$4+1$ directions that, as can be clearly seen from
(\ref{eq:A4}), has dimensions of length. In this way, we can
think that M(atrix) on $T^4$ is described by a $5+1$ dimensional
theory, with space dimensions a torus $T^5$, of dimensions
$\Sigma_i$, with $i=1,2,3,4$, and $\Sigma_5=\frac
{l_p^6}{L_1L_2L_3L_4 R}$. This is exactly the same picture we
have in M-theory, understood as the strong coupled limit of type
IIA string theory. There, we associated the RR D-$0$branes with
Kaluza-Klein modes of the extra dimension. In the gauge theory
context we should look for objects in $4+1$ dimensions, that can
be interpreted as Kaluza-Klein modes of the extra dimension
required by $U$-duality. As candidates to these states, we can
use instantons. Instantons are associated with the $\Pi_3$
homotopy group of the gauge group so that, in $4+1$ dimensions,
they look like particles. Moreover, their mass is given by $\frac
{1}{g^2}$, with the gauge coupling constant (recall that $\frac
{1}{g^2}$ is the action for the instanton in $3+1$ dimensions).
Therefore, using (\ref{eq:A4}), we get the desired result,
namely that instantons ar the Kaluza-Klein modes of the extra
dimension.
  
We can, in fact, try to understand what kind of dynamics is
playing the role here, using string theory language. The
supersymmetric Yang-Mills theory on $T^4$, with gauge group
$U(N)$, can be interpreted as the worldvolume lagrangian for $N$
fourbranes of type IIA, wrapped around $T^4$. In M-theory, we can
interpret this fourbranes as fivebranes partially wrapped in
around the internal eleventh dimension. When we move to strong
coupling, we open the extra direction, and we effectively get a
$5+1$ dimensional gauge theory. If this is the correct picture,
we can check it by comparison of the mass of the instanton and
the expected mass of the wrapped around $T^4$ and the internal
eleventh dimension. The energy of the fivebrane would then be
\begin{equation}
E= \frac {L_1L_2L_3L_4R}{l_p^6},
\label{eq:m26}
\end{equation}
which is exactly the mass of the instanton,
\begin{equation}
\frac {1}{g^2} \equiv \frac {1}{\hat{g}} \frac {L_1L_2L_3L_4
R}{l_p^6}.
\label{eq:m27}
\end{equation}
In order to understand the effect described above, it would be
convenient to discuss briefly the scales entering the theory.
Using relations (\ref{eq:A1}) to (\ref{eq:A4}), and $g l_s =R$ we
get, for generic dimension $d$,
\begin{equation}
g_{SYM}^2 = \frac {R^{3-d} l_s^{3d-6}}{L_i^d g^{d-3}}.
\label{eq:*}
\end{equation}
It is clear form (\ref{eq:*}) that for $d \leq 3$ the limit of
string coupling constant equal zero gives a weak coupled
supersymmetric Yang-Mills theory. However, a barrier appears in
$d=4$. In fact, for $d \geq 4$ the limit $ g \rightarrow 0$ leads
to strong coupling in the field theory. One of these strong
copling effects is the generation of the quantum dimension needed
for $U$-duality.

\vspace{20 mm}
  
\begin{center}
{\bf Acknowledgments}
\end{center}

This work is partially supported by European Community grant 
ERBFMRXCT960012, and by grant AEN-97-1711.

\newpage

%%%%%%%%%%%%%%%%%%%%%%%%%%%%%%%%%%%%%%%%%%%%%%%%%%%%%%%%%%%%%
%%%%%%%%%%%%%%%%%%%%%%%%%%%%%%%%%%%%%%%%%%%%%%%%%%%%%%%%%%%%%


\begin{thebibliography}{99}

\bibitem{D1931} P. A. M. Dirac, ``Quantized Singularities in the 
Electromagnetic Field'', Proc. Roy. Soc. Lond. {\bf A133} (1931), 60.

\bibitem{WY} T. T. Wu and C. N. Yang, ``Concept of Nonintegrable
Phase Factors and Global Formulation of Gauge Fields'', Phys.
Rev. {\bf D12} (1975), 3845.

\bibitem{GO} P. Goddard and D. Olive, Prog. Rep. Phys. {\bf 41}
(1978), 1357.

\bibitem{GNO} P. Goddard, J. Nuyts and D. Olive, ``Gauge Theories and 
Magnetic Charge'', Nucl. Phys. {\bf B125} (1977), 1.

\bibitem{MO} C. Montonen and D. Olive, ``Magnetic Monopoles as
Gauge Particles ?'', Phys. Lett. {\bf B72} (1977), 177.

\bibitem{GG} H. Georgi and S. Glashow, ``Unified Weak and Electromagnetic 
Interactions Without Neutral Currents'', Phys. Rev. Lett. {\bf 28} (1972),
1494.

\bibitem{tp} G. `t Hooft, ``Magnetic Monopoles in Unified 
Gauge Theories'', Nucl. Phys. {\bf B79} (1974), 276.

A. M. Polyakov, ``Particle Spectrum in the Quantum Field 
Theory'', JETP Lett. {\bf 20} (1974), 194.

\bibitem{PS} M. K. Prasad and C. M. Sommerfeld, ``An Exact Classical Solution for the 
`t Hooft Monopole and the Julia-Zee Dyon'', Phys. Rev. Lett. {\bf 35} (1975), 760.

\bibitem{B} E. B. Bogomolny, ``Stability of Classical Solutions'', 
Sov. J. Nucl. Phys. {\bf 24} (1976), 449.

\bibitem{BPST} A. A. Belavin, A. M. Polyakov, A. S. Swartz and
Y. S. Tyupkin, ``Pseudoparticle Solutions of the Yang-Mills
Equations'', Phys. Lett. {\bf B59} (1975), 85.

\bibitem{tHpseudo} G. `t Hooft, ``Computation of the Quantum
Effects due to a Four Dimensional Pseudoparticle'', Phys. Rev.
{\bf D14} (1977), 3432.

\bibitem{JR} R. Jackiw and C. Rebbi, ``Degree of Freedom in Pseudoparticle Physics'',
Phys. Lett. {\bf B67} (1977), 189.

\bibitem{At}  M. F. Atiyah N. Hitchin and I. M. Singer,
``Self-Duality in Four Dimensional Riemannian Geometry'', Proc. Roy. Soc. Lond. 
{\bf A362} (1978), 475.

\bibitem{AHS} M. F. Atiyah and I. M. Singer, ``Index of Elliptic
Operators I'', Ann. Math. {\bf 87} (1968), 485.

M. F. Atiyah and G. B. Segal, ``Index of Elliptic
Operators II'', Ann. Math. {\bf 87} (1968), 531.

M. F. Atiyah and I. M. Singer, ``Index of Elliptic
Operators III'', Ann. Math. {\bf 87} (1968), 546.

M. F. Atiyah and I. M. Singer, ``Index of Elliptic
Operators IV'', Ann. Math. {\bf 93} (1971), 119.

M. F. Atiyah and I. M. Singer, ``Index of Elliptic
Operators V'', Ann. Math. {\bf 87} (1971), 139.

\bibitem{cl} G. `t Hooft, ``Symmetry Breaking Through 
Bell-Jackiw Anomalies'', Phys. Rev. Lett. {\bf 37} (1976), 8.

\bibitem{jackiw} R. Jackiw and C. Rebbi, ``Vacuum Periodicity in a Yang-Mills 
Quantum Theory'', Phys. Rev. Lett. {\bf 37} (1976), 172.

\bibitem{CDG} C. G. Callan, R. Dashen and D. J. Gross, ``Towards
a Theory of the Strong Interactions'', Phys. Rev. {\bf D17} (1978), 2717.

\bibitem{Wdyon} E. Witten, ``Dyons af Charge $e\theta/2 \pi$'',
Phys. Lett. {\bf B86} (1979), 283.

\bibitem{Rubakov} V. A. Rubakov, ``Adler-Bell-Jackiw Anomaly and
Fermionic-Number Breaking in the Presence of a Magnetic
Monopole'', Nucl. Phys. {\bf B203} (1982), 311.

\bibitem{tHtw} G. `t Hooft, ``A Property of Electric and Magnetic Charges
in Non Abelian Gauge Theories'', Nucl. Phys. {\bf B153} (1979), 141.

\bibitem{tHcmp} G. `t Hooft, ``Some Twisted Self-Dual Solutions for the 
Yang-Mills Equations on a Hypertorus'', Commun. Math. Phys. {\bf 81} (1981), 267.

\bibitem{Wind} E. Witten, ``Constraints on Supersymmetry Breaking'', 
Nucl. Phys. {\bf B202} (1982), 253.

\bibitem{P3d} A. M. Polyakov, ``Quark Confinement and Topology of
Gauge Groups'', Nucl. Phys. {\bf B120} (1977), 429.

\bibitem{AHW} I. Affleck, J. A. Harvey and E. Witten, ``Instantons and 
Supersymmetry Breaking in $2+1$ Dimensions'', Nucl. Phys. {\bf B206} (1982), 413.

\bibitem{Callias} C. Callias, ``Axial Anomalies and Index
Theorems on Open Spaces'', Comm. Math. Phys. {\bf 62} (1978),
213.

\bibitem{WB} J. Wess and J. Bagger, ``{\em Supersymmetry and
Supergravity}'', Princeton University Press, Princeton, $1984$.

\bibitem{NSVZ} V. A. Novikov, M. A. Shifman, A. I. Vainshtein and
V. I. Zakharov, ``Exact Gell-Mann-Low Function of Supersymmetric
Yang-Mills Theories From Instanton Calculus'', 
Nucl. Phys. {\bf B229} (1983), 381.

\bibitem{SVZ1} V. A. Novikov, M. A. Shifman, A. I. Vainshtein and
V. I. Zakharov, ``Instanton Effects in Supersymmetric Theories'', 
Nucl. Phys. {\bf B229} (1983), 407.

\bibitem{SVZ2} M. A. Shifman, A. I. Vainshtein and V. I.
Zakharov, ``On Gluino Condensation in Supersymmetric Gauge
Theories. $SU(N)$ and $O(N)$ Gauge Groups'', 
Nucl. Phys. {\bf B296} (1988), 445.

\bibitem{A} D. Amati, K. Konishi, Y. Meurice, G. C. Rossi and G. Veneziano, 
``Non Perturbative Aspects in Supersymmetric Gauge Theories'', Phys. Rep. 
{\bf 162} (1988), 169.

\bibitem{ADS1} I. Affleck. M. Dine and N. Seiberg, ``Dynamical 
Supersymmetry Breaking in Supersymmetric QCD'', Nucl. Phys.
{\bf B241} (1984), 493.

\bibitem{ADS2} I. Affleck. M. Dine and N. Seiberg, ``Dynamical 
Supersymmetry Breaking in Four-Dimensions and its 
Phenomenological Implications'', Nucl. Phys.
{\bf B256} (1985), 557.

\bibitem{SV} M. A. Shifman and A. I. Vainshtein, ``On Gluino
Condensation in Supersymmetric Gauge Theories, $SU(N)$ and $O(N)$
Gauge Groups'', Nucl. Phys. {\bf B296} (1988), 445.

\bibitem{CG} E. Cohen and C. G\'{o}mez, ``Chiral Symmetry Breaking in Supersymmetric 
Yang-Milss'', Phys. Rev. Lett. {\bf 52} (1984), 237.

\bibitem{SW} N. Seiberg and E. Witten, ``Electric-Magnetic Duality, Monopole
Condensation and Confinement in $N\!=\!2$ Supersymmetric Yang-Mills Theory'',
Nucl. Phys. {\bf B426} (1994), 19. 

\bibitem{SW2} N. Seiberg and E. Witten, ``Monopoles, Duality and
Chiral Symmetry Breaking in $N=2$ Supersymmetric QCD'', Nucl.
Phys. {\bf B431} (1994), 484. 

\bibitem{SW3d} N. Seiberg and E. Witten, ``Gauge Dynamics and Compactification
to Three Dimensions'', {\bf hep-th/9607163}.

\bibitem{Shifwils} M. Shifman and A. I. Vainshtein, ``On Holomorphic Dependence 
and Infrared Effects in Supersymmetric Gauge
Theories'', Nucl. Phys.
{\bf B359} (1991), 571.

\bibitem{Sbeta} N. Seiberg, ``Supersymmetry and Nonperturbative
Beta Functions'', Phys. Lett. {\bf B206} (1988), 75.

\bibitem{Wilson} K. G. Wilson and J. Kogut, ``The Renormalization
Group and the $\epsilon$-Expansion'', Phys. Rep. {\bf 12} (1974),
75.

\bibitem{AB} L. Adler and W. A. Bardeen, ``Absence of Higher
Order Corrections in the Anomalous Axial Vector divergence
Equation'', Phys. Rev. {\bf 182} (1969), 1517.

\bibitem{AH} M. Atiyah and N. Hitchin, ``{\em The Geometry and
Dynamics of Magnetic Monopoles}'', Princeton University Press,
Princeton, $1988$.

\bibitem{Arnold} V. I. Arnold, ``{\em Singularity Theory}'',
London Math. Soc. Lecture Note Series {\bf 53}, Cambridge
University Press, Cambridge, 1981.

\bibitem{Artin} M. Artin, ``On Isolated Rational Singularities of
Surfaces'', Amer. J. Math. {\bf 88} (1966), 129.

\bibitem{Kodaira} K. Kodaira, Ann. Math. {\bf 77}, 3 (1963), 563.

\bibitem{rep1} A. Klemm, W. Lerche, S. Theisen and S.
Yankielowicz, ``Simple Singularities and $N=2$ Supersymmetric
Yang-Mills'', Phys. Lett. {\bf B344} (1995), 169.

\bibitem{rep2} P. Argyres and A. Faraggi, ``The Vacuum Structure
and Spectrum of $N=2$ Supersymmetric $SU(N)$ Gauge Theory'',
Phys. Rev. Lett. {\bf 73} (1995), 3931.

\bibitem{rep3} A. Hanany and Y. Oz, ``On the Quantum Moduli Space
of Vacua of $N=2$ Supersymmetric $SU(N_c)$ Gauge Theories'', Nucl.
Phys. {\bf B452} (1995), 283.

\bibitem{rep4} P. Argyres, M. Plesser and A. Shapere, ``The
Coulomb Phase of $N=2$ Supersymmetric QCD'', Phys. Rev. Lett.
{\bf 75} (1995), 1699.

\bibitem{rep5} U. Danielsson and B. Sundborg, ``The Moduli Space
and Monodromies of $N=2$ Supersymmetric $SO(2r+1)$ Yang-Mills
Theory'', Phys. Lett. {\bf B358} (1995), 273.

\bibitem{rep6} A. Brandhuber and K. Landsteiner, ``On the
Monodromies of $N=2$ Supersymmetric Yang-Mills Theory with Gauge
Group $SO(2n)$, Phys. Lett. {\bf B358} (1995), 73.

\bibitem{IS} K. Intriligator and N. Seiberg, ``Lectures 
on Supersymmetric Gauge Theories and Electric-Magnetic Duality'', 
Nucl. Phys. Proc. Supl. {\bf 45} (1996).

\bibitem{Reviews} C. G\'omez and R. Hern'andez, ``Electric-Magnetic Duality 
and Effective Field Theories'', {\bf hep-th/9510023}; published
in ``{\bf Advanced School on Effective Theories}'', F. Cornet and M.
J. Herrero eds. World Scientific (1997).
  
A. Bilal, ``Duality in $N=2$ SUSY $SU(2)$ 
Yang-Mills Theory: A Pedagogical Introduction to the 
Work of Seiberg and Witten'', {\bf 9601007}.
  
W. Lerche, ``Introduction to Seiberg-Witten Theory and its Stringy
Origin'', Nucl. Phys. Proc. Supl. {\bf B55} (1997), 83.
  

L. \'Alvarez-Gaum\'e and F. Zamora, ``Duality in Quantum Field
Theory (and String Theory)'', {\bf hep-th/9709180}.

\bibitem{GMS} O. J. Ganor, D. R. Morrison and N. Seiberg,
``Branes, Calabi-Yau Spaces, and Toroidal Compactification of the
$N=1$ Six-Dimensional $E_8$ Theory'', Nucl. Phys. {\bf B487}
(1997), 93.

\bibitem{Man} Mandelstam, ``Vortices and Quark Confinement 
in Non Abelian Gauge Theories'', Phys. Rep. {\bf 23} (1976), 245.

\bibitem{tHconf} G. `t Hooft, ``On the Phase Transition 
Towards Permanent Quark Confinement'', Nucl. Phys. {\bf B138} (1978), 1.

\bibitem{NO} H. B. Nielsen and P. Olesen, ``Vortex Line Models for 
Dual Strings'', Nucl. Phys. {\bf B61} (1973),
45.

\bibitem{tHap} G. `t Hooft, ``Topology of the Gauge Condition and
New Confinement Phases in Non Abelian Gauge Theories'', Nucl. Phys. {\bf B190}
(1981), 455.

\bibitem{GSW} M. B. Green, J. H. Schwarz and E. Witten,
``{\em Superstring Theory}'', Cambridge University Press,
Cambridge, $1987$.

\bibitem{Polchinski} J. Polchinski, ``What is String Theory?'',
{\bf hep-th/9411028}.

\bibitem{Vafa} C. Vafa, ``Lectures on Strings and Dualities'',
{\bf hep-th/9702201}.

\bibitem{Kiri} E. Kiritsis, ``Introduction to Superstring Theory'', {\bf hep-th/9709062}.

\bibitem{Fay} J. D. Fay, ``Theta Functions on Riemann Surfaces'', Lecture Notes 
in Mathematics, {\bf 352}, Springer-Verlag, 1973.

\bibitem{Narain} K. S. Narain, ``New Heterotic String Theories in
Uncompactified Dimensions $<10$'', Phys. Lett. {\bf B169} (1986),
41; K. S. Narain, M. H. Samadi and E. Witten, 
``A Note on the Toroidal Compactification of Heterotic String
Theory'', Nucl. Phys. {\bf B279} (1987), 369.

\bibitem{T} For reviews see A. Giveon, M. Porrati and E. Rabinovici, ``Target Space 
Duality in String Theory'', Phys. Rep.
{\bf 244} (1994), 77.

E. \'{A}lvarez, L. \'{A}lvarez-Gaum\'{e} and Y. Lozano, ``An Introduction 
to T-Duality in String Theory'', {\bf hep-th/9410237}.

\bibitem{GrHa} P. Griffiths and J. Harris, ``{\em Principles of 
Algebraic Geometry}'', Wiley-Interscience, 1978.

\bibitem{Aspinwall} P. A. Aspinwall, ``$K3$ Surfaces and String
Duality'', {\bf hep-th/9611137}.

\bibitem{Persson} U. Persson, Lecture Notes in Mathematics, 1124,
Springer-Verlag.

\bibitem{Snew} N. Seiberg, ``Observations on the Moduli Space of Superconformal Field 
Theories'', Nucl. Phys. {\bf B303} (1988), 286.

\bibitem{mirror} B. Greene and M. R Plesser, ``Duality in Calabi-Yau 
Moduli Space'', Nucl Phys. {\bf B338} (1990), 15.
  
P. Candelas, M. Lynker and R. S. Schimmrigk, ``Calabi-Yau 
Manifolds in Weighted P(4)'', Nucl. Phys. {\bf B341} (1990), 383.  
  
P. Candelas, X. de la Ossa, P. Green and L. Parkes, ``A Pair of 
Calabi-Yau Manifolds as an Exactly Soluble Superconformal 
Theory'', Nucl. Phys. {\bf B359} (1991), 21.
  
V. Batyrev, ``Dual Polyhedra and Mirror Symmetry for Calabi-Yau
Hypersurfaces in Toric Varieties'', {\bf alg-geom/9410003}.
  
For a series of references, see S. -T. Yau ed., ``Essays on
Mirror Manifolds'', International Press, Hong-Kong, 1992.

\bibitem{Dolgachev} I. V. Dolgachev, ``Mirror Symmetry for Lattice Polarized 
$K3$-Surfaces'', {\bf alg-geom/9502005}.

\bibitem{GP} J. Polchinski and Y. Cai, ``Consistency of Open
Superstring Theories'', Nucl. Phys. {\bf B296} (1988), 91.
  

M. Green, ``Space-Time Duality and Dirichlet String 
Theory'',  Phys. Lett. {\bf B266} (1991), 325.
  
  
J. Polchinski, ``Dirichlet Branes and Ramond-Ramond
Charges'', Phys. Rev. Lett. {\bf 75} (1995), 4724.
  
  
J. Polchinski, S. Chaudhuri and C. V.
Johnson, ``Notes on D-Branes'', {\bf hep-th/9602052}, and ``TASI Lectures 
on D-Branes'', {\bf hep-th/9611050}.

\bibitem{Chan-Paton} J. C. Paton and H. M. Chan, ``Generalized
Veneziano Model with Isospin'', Nucl. Phys. {\bf B10} (1969),
516.

\bibitem{susyholo} N. J. Hitchin, A. Karlhede, U. Lindstr\"om, M.
Rocek, ``Hyperk\"ahler Metrics and Supersymmetry'', Comm. Math.
Phys. {\bf 108} (1987), 535.

\bibitem{rabino} J. Cardy and E. Rabonovici, ``Phase Structure of $Z(P)$ Models 
in the Presence of a Theta Parameter'', Nucl. Phys. {\bf B205} (1982), 1.

\bibitem{S} A. Font, L. Iba\~{n}ez, D. L\"{u}st and F. Quevedo,
``Strong-Weak Coupling Duality and Non-Perturbative Effects in String theory'', 
Phys. Lett. {\bf B249} (1990), 35.

\bibitem{mr} A. Sen, ``Strong-Weak Coupling Duality in Four 
Dimensional Field theory'', Int. J. Mod. Phys. {\bf A9} (1994), 3707.

\bibitem{HT} C. Hull and P. Townsend, ``Unity of Superstring 
Dualities'', Nucl. Phys. {\bf B438} (1995), 109.

\bibitem{Duff} M. Duff, ``Strong/Weak Coupling Duality from the Dual String'', 
Nucl. Phys. {\bf B442} (1995), 47.

\bibitem{Town} P. Townsend, ``The Eleven-Dimensional Supermembrane Revisited'', 
Phys. Lett. {\bf B350} (1995), 184.

\bibitem{Wsvd} E. Witten, ``String Theory Dynamics in Various
Dimensions'', Nucl. Phys. {\bf B443} (1995), 85. 

\bibitem{S2} A. Sen, ``String-String Duality Conjecture in Six Dimensions and
Charged Solitonic Strings'', Nucl. Phys. {\bf B450} (1995), 103.

\bibitem{HS} J. Harvey and A. Strominger, ``The Heterotic String is a Soliton'', 
Nucl. Phys. {\bf B449} (1995), 535.

\bibitem{GrMS} B. Greene, D. Morrison and A. Strominger, ``Black Hole Condensation and the
Unification of String Vacua'', Nucl. Phys. {\bf B451} (1995),
109.

\bibitem{VW} C. Vafa and E. Witten, ``A One-Loop Test Of String Duality'', 
Nucl. Phys. {\bf B447} (1995), 261.

\bibitem{HT2} C. Hull and P. Townsend, ``Enhanced Gauge Symmetries in Superstring Theories'',
Nucl. Phys. {\bf B451} (1995), 525.

\bibitem{Schwarz} J. Schwarz, ``An $Sl(2,{\bf Z})$ Multiplet of 
Type IIB Superstrings'', Phys. Lett. {\bf B360} (1995), 13.

\bibitem{Gross} D. J. Gross, J. A. Harvey E. Martinec and R.
Rohm, ``Heterotc String'', Phys. Rev. Lett. {\bf 54} (1985), 502; 
``Heterotic String Theory (I). The Free Heterotic
String.'', Nucl. Phys. {\bf B256} (1985), 253; ``Heterotic String Theory (II). The 
Interacting Heterotic String.'', Nucl. Phys. {\bf B267} (1986), 75. 

\bibitem{MK} S. Katz, D. R. Morrison and M. R. Plesser, ``Enhanced Gauge 
Symmetry in Type II String Theory'', Nucl. Phys. {\bf 477} (1996), 105.

\bibitem{Mayr} A. Klemm and P. Mayr, ``Strong Coupling Singularities and Non-abelian Gauge 
Symmetries in $N=2$ String Theory'', Nucl. Phys. {\bf B469}, 37.
(1996), 

\bibitem{As} P. S. Aspinwall, ``Enhanced Gauge Symmetries and $K3$ Surfaces'', 
Phys. Lett. {\bf B357} (1995), 329..

\bibitem{Louis} P. S. Aspinwall and J. Louis, ``On the Ubiquity of $K3$ Fibrations 
in String Duality'', Phys. Lett. {\bf B369} (196), 233..

\bibitem{Klemm} A. Klemm, W. Lerche and P. Mayr,``$K3$-Fibrations and 
Heterotic-Type II String Duality'', Phys. Lett. {\bf B357} (1995), 313.

\bibitem{F} C. Vafa, ``Evidence for F-Theory'', Nucl. Phys. {\bf
B469} (1996), 403.

\bibitem{F1} D. R. Morrison and C. Vafa, ``Compactifications of
F-Theory on Calabi-Yau Threefolds. I.'', Nucl. Phys. {\bf B473}
(1996), 74. 

\bibitem{F2} D. R. Morrison and C. Vafa, ``Compactifications of
F-Theory on Calabi-Yau Threefolds. II.'', Nucl. Phys. {\bf B476}
(1996), 437. 

\bibitem{KaV} S. Kachru and C. Vafa, ``Exact Results for $N=2$
Compactifications of Heterotic Strings'', Nucl. Phys. {\bf 450},
(1995), 69.

\bibitem{Mth} M. J. Duff, P. Howe, T. Inami and K. S. Stelle,
``Superstrings in $D=10$ from Supermembranes in $D=11$'', Phys.
Lett. {\bf B191} (1987), 70.

M. J. Duff, R. Minasian and J. T. Liu, ``Duality Rotations in
Membrane Theory'', Nucl. Phys. {\bf B347} (1990), 394.
  
M. J. Duff, R. Minasian and J. T. Liu, ``Eleven Dimensional
Origin of String/String Duality: A One Loop Test'', Nucl. Phys.
{\bf B452} (1995), 261.
  

\bibitem{Wsp} E. Witten, ``Non-Perturbative Superpotentials in String Theory'',  
Nucl. Phys. {\bf B474} (1996), 343. 

\bibitem{HW} A. Hanany and E. Witten, ``Type IIB Superstrings,
BPS Monopoles and Three Dimensional Gauge Dynamics'', Nucl. Phys.
{\bf B492} (1997), 152. 

\bibitem{Eli} S. Elitzur, A. Giveon and D. Kutasov, ``Branes and
$N=1$ Duality in String Theory'', Phys. Lett. {\bf B400} (1997),
269.

\bibitem{Boer} J. de Boer, K. Hori, Y. Oz and Z. Yin, ``Branes
and Mirror Symmetry in $N=2$ Supersymmetric Gauge Theories in
Three Dimensions'', {\bf hep-th/9702154}.

\bibitem{Wm4} E. Witten, ``Solutions of Four-Dimensional Fields
Theories Via M-Theory'', Nucl. Phys. {\bf B500} (1997), 3.

\bibitem{Evans} N. Evans, C. V. Johnson and A. D. Shapere,
``Orientifolds, Branes and Duality of $4D$ Field Theories'', 
{\bf hep-th/9703210}.

\bibitem{Eli2} S. Elitzur, A. Giveon, D. Kutasov, E. Ravinovici
and A. Schwimmer, ``Brane Dynamics and $N=1$ Supersymmetric Gauge 
Theory'', {\bf hep-th/9704104}.

\bibitem{Barbon} J. L. F. Barbon, "Rotated Branes and $N=1$ 
Duality", Phys. Lett. {\bf B402} (1997), 59.

\bibitem{Brod} J. Brodie and A. Hanany, ``Type IIA Superstrings, 
Chiral Symmetry, and $N=1$ $4D$ Gauge Theory Dualities'', {\bf hep-th/9704043}.

\bibitem{Cobi} A. Brandhuber, J. Sommenschein, S. Theisen and S.
Yanckielowicz, ``Brane Configurations and $4D$ Field Theories'', {\bf hep-th/9704044}.

\bibitem{HH} O. Aharony and A. Hanany, ``Branes, Superpotentials and 
Superconformal Fixed Points'', {\bf hep-th/9704170}.

\bibitem{Tatar} R. Tartar, ``Dualities in $4D$ Theories with Product Gauge Groups 
from Brane Configurations'', {\bf hep-th/9704198}.

\bibitem{BK} I. Brunner and A. Karch, ``Branes and Six Dimensional Fixed 
Points'', {\bf hep-th/9705022}.

\bibitem{Kol} B. Kol, ``5d Field Theories and M Theory'', 
{\bf hep-th/9705031}.

\bibitem{MMM} A. Marshakov, M. Martellini and A. Morozov, ``Insights and Puzzles 
from Branes: $4d$
SUSY Yang-Mills from $6d$ Models'', {\bf hep-th/9706050}.

\bibitem{espe} K. Landsteiner, E. L\'opez and D. A. Lowe, ``$N=2$ 
Supersymmetric Gauge Theories, Branes and Orientifolds'', {\bf hep-th/9705199}.

\bibitem{Br} A. Brandhuber, J. Sommenschein, S. Theisen and S.
Yanckielowicz, {\bf hep-th/9705232}.
 
\bibitem{Oz} K. Hori, H. Ooguri and Y. Oz, "Strong Coupling 
Dynamics of Four Dimensional $N=1$ Gauge Theory fron M theory 
Fivebrane" {\bf hep-th/9706082}.  

\bibitem{Wbqcd} E. Witten, ``Branes and the Dynamics of QCD'',
{\bf hep-th/9706109}.

\bibitem{SYZ} A. Strominger, S. -T. Yau and E. Zaslow, ``Mirror
Symmetry is $T$-Duality'', Nucl. Phys. {\bf B479} (1996), 243.

\bibitem{Mor} D. R. Morrison, ``The Geometry Underlying Mirror
Symmetry'', {\bf alg-geom/9608006}.

\bibitem{DW} R. Donagi and E. Witten, ``Supersymmetric Yang-Mills
and Integrable Systems'', Nucl. Phys. {\bf B460} (1996), 299.

\bibitem{H} N. Hitchin, ``The Self-Duality Equations on a Riemann
Surface'', Proc. London Math. Soc. {\bf 55} (1987), 59; ``Stable
Bundles and Integrable Systems'', Duke Math. J. {\bf 54} (1987),
91.

\bibitem{KKLMV} S. Kachru, A. Klemm, W. Lerche P. Mayr and C:
Vafa, ``Non Perturbative Results on the Point Particle Limit of
$N=2$ Heterotic String Compactifications'', Nucl. Phys. {\bf
B459} (1996), 537. 

\bibitem{GHL1} C. G\'omez, R. Hern\'andez and E. L\'opez,
``$S$-Duality and the Calabi-Yau Interpretation of the $N=4$ to
$N=2$ Flow'', Phys. Lett. {\bf B386} (1996), 115.

\bibitem{KLMVW} A. Klemm, W. Lerche, P. Mayr, C. Vafa and N.
Warner, ``Self-Dual Strings and $N=2$ Supersymmetric Field
Theory'', Nucl. Phys. {\bf B477}, (1996), 746. 

\bibitem{GHL2} C. G\'omez, R. Hern\'andez and E. L\'opez,
``$K3$-Fibrations and Softly Broken $N=4$ Supersymmetric Gauge
Theories'', {\bf hep-th/9608104}.

\bibitem{KV} S. Katz and C. Vafa, ``Geometric Engineering of 
Quantum Field Theories'', Nucl. Phys. {\bf B497} (1997), 173, and ``Geometric Engineering of
$N=1$ Quantum Field Theories'', Nucl. Phys. {\bf B497} (1997), 196.

\bibitem{ge} S. Katz, A. Klemm, C. Vafa, ``Geometric Engineering of Quantum Field Theories'', Nucl. Phys. {\bf B497} (1997), 173

\bibitem{BJPSV} M. Bershadsky, A. Johansen, T. Pantev, V. Sadov and C. Vafa, 
``F-Theory, Geometric Engineering and $N\!=\!1$ Dualities'', {\bf hep-th/9612052}.

\bibitem{int1} A. Gorskii, I. Krichever, A. Marshakov, A. Mironov
and A. Morozov, ``Integrability and Exect Seiberg-Witten Solution'', 
Phys. Lett. {\bf B355} (1995), 466.

\bibitem{int2} E. Martinec and N. P. Warner, ``Integrable Systems and Supersymmetric 
Gauge Theories'', Nucl. Phys. {\bf B459} (1996), 97.

\bibitem{Gomez} C. G\'omez, ``Elliptic Singularities,
$\theta$-Puzzle and Domain Wall'', {\bf hep-th/9711074}.

\bibitem{ds} G. Dvali and M. Shifman, ``Domain Walls in 
Strongly Coupled Theories'', Phys. Lett. {\bf B396} (1997), 64.

\bibitem{kss} A. Kovner, M. Shifman and A. Smilga,``Domain Walls in Supersymmetric 
Yang-Mills Theories'', {\bf hep-th/9706089}.

\bibitem{Smilga} A. Smilga and A. Vaselov, ``Complex BPS 
Domain Walls and Phase Transition in Mass in Supersymmetric QCD'', {\bf hep-th/9706217} 
and ``Domain Walls Zoo in Supersymmetric QCD'', {\bf hep-th/9710123}.

\bibitem{KS} A. Kovner and M. Shifman, ``Chirally Symmetric Phase of
Supersymmetric Gluodynamics'', {\bf hep-th/9702174}.

\bibitem{NSVZ2} V. A. Novikov, M. A. Shifman, A. I. Vainshtein and V. I. Zakharov, 
``Supersymmetric Instanton Calculus (Gauge
Theories with Matter)'', Nucl. Phys. {\bf B260} (1985), 157.

\bibitem{Fulton} W. Fulton, ``{\em Intersection Theory}'',
Springer-Verlag, $1980$.

\bibitem{Sen} A. Sen, ``F-Theory and Orientifolds'', 
Nucl. Phys. B475 (1996), 562.

\bibitem{Wi} E. Witten, ``Current Algebra Theorems for 
the $U(1)$ 'Goldstone Boson' '', Nucl. Phys. B156 (1979), 269.

\bibitem{Ve} G. Veneziano, ``$U(1)$ without Instantons'', Nucl. Phys. B159 (1979), 213.

\bibitem{holo} G. `t Hooft, ``Dimensional Reduction in Quantum
Gravity'', {\bf gr-qc/9310026}.

\bibitem{Bekenstein} J. D. Bekenstein, ``Black Holes and
Entropy'', Phys. Rev. {\bf D7} (1973), 2333.

\bibitem{BFSS} T. Banks, W. Fischler, S. H. Shenker and L.
Susskind, ``M-Theory as a Matrix Model: A Conjecture'', Phys.
Rev. {\bf D55} (1997), 5112. 

\bibitem{RevMat} T. Banks, ``Matrix Theory'', {\bf
hep-th/9710231}.

\bibitem{toroidal}  W. Taylor, ``D-Brane Field 
Theory on Compact Spaces'', Phys.Lett. {\bf B394} (1997), 283.
  
O. J. Ganor, S. Ramgoolam, W. Taylor IV, ``Branes, Fluxes and Duality in
M(atrix)-Theory'', Nucl. Phys. {\bf B492} (1997), 191.

\bibitem{Rozali} M. Rozali, ``Matrix Theory and U-Duality in Seven
Dimensions'', Phys. Lett. {\bf B400} (1997), 260.

\bibitem{Srotinv} S. Sethi and L. Susskind, ``Rotational
Invariance in the M(atrix) Formulation of Type IIB Theory'',
Phys. Lett. {\bf B400} (1997), 265.

\bibitem{SenM} A. Sen, ``D$0$ Branes on $T^n$ and Matrix
Theory'', {\bf hep-th/9709220}.

\bibitem{SeiMat} N. Seiberg, ``Why is the Matrix Model
Correct?'', {\bf hep-th/9710009}.

\vspace{20 mm}

 








\end{thebibliography}
\end{document}